\documentclass[
reprint,
superscriptaddress,
amsmath,amssymb,
aps,
prb,
twocolumn
]{revtex4-2}
\usepackage{tikz}
\usepackage{amsfonts,amssymb}
\usetikzlibrary{arrows.meta, positioning, shapes.geometric}
\usepackage{graphicx}
\usepackage{dcolumn}
\usepackage{bm}
\usepackage[breaklinks]{hyperref}
\hypersetup{colorlinks=true, linkcolor=blue, citecolor=blue, filecolor=blue, urlcolor=blue}
\graphicspath{{figures}}
\usepackage{bm}

\usepackage{graphicx,subfigure}
\usepackage{epsfig}
\usepackage{bm}
\usepackage{dcolumn}
\usepackage{color}
\usepackage{physics}
\usepackage{array}
\usepackage{booktabs}
\usepackage{tabularx}
\usepackage{ragged2e}
\usepackage{verbatim}

\newcolumntype{Y}{>{\RaggedRight\arraybackslash}X}
\newcolumntype{P}[1]{>{\RaggedRight\arraybackslash}p{#1}}

\makeatletter
\newcommand*{\rom}[1]{\expandafter\@slowromancap\romannumeral #1@}
\makeatother
\usepackage{lipsum}
\usepackage{subfigure}
\usepackage{dsfont}
\usepackage{enumitem}
\usepackage{tikz}

\usepackage{amsmath} 
\usepackage{braket}      
\usepackage{tikz}
\usepackage{quantikz}    
\usepackage[normalem]{ulem}

\begin{document}
\title{Simulating Condensed Matter Physics on Quantum Hardware} 
\author{Ruizhe Shen}
\email{e0554228@u.nus.edu}
\affiliation{Department of Physics, National University of Singapore, Singapore 117551}
\author{Tianqi Chen}
\affiliation{Institute of High Performance Computing (IHPC), Agency for Science, Technology and Research (A*STAR),
1 Fusionopolis Way, No.~16-16 Connexis, Singapore 138632\looseness=-1}
\affiliation{Bioinformatics Institute, Agency for Science, Technology and Research (A*STAR), 30 Biopolis Street, No.~07-01 Matrix, Singapore 138671\looseness=-1}
\author{Tommy Tai}
\affiliation{Department of Physics, Massachusetts Institute of Technology, Cambridge, Massachusetts 02142, USA}
\author{Jin Ming Koh}
\affiliation{Department of Physics, Harvard University, Cambridge, Massachusetts 02138, USA}
\author{Pouyan Ghaemi}
\affiliation{Physics Department, City College of the City University of New York, New York 10031, USA}
\affiliation{Physics Program, Graduate Center of City University of New York, New York 10031, USA}
\author{Ching Hua Lee}
\email{phylch@nus.edu.sg}
\affiliation{Department of Physics, National University of Singapore, Singapore 117551}

\date{\today}
\begin{abstract}  
Quantum hardware platforms are getting increasingly sophisticated in their ability to simulate condensed matter, including but not limited to strongly-correlated, topological, and non-equilibrium phenomena. This review surveys recent progress in quantum-hardware-based simulations of condensed matter, primarily emphasizing gate-based digital quantum computer simulation, with analog experiments discussed as complementary benchmarks. We first review major hardware platforms, including superconducting qubits, trapped-ions, ultracold atoms, Rydberg arrays, photonic systems, and moire quantum materials. We then introduce the basic ingredients of digital quantum simulation. Building on this foundation, we discuss representative applications to condensed-matter physics, spanning ground-state problems, strongly correlated matter, topological phases, non-equilibrium dynamics, open-system physics, and high-energy-physics-inspired simulations. Finally, we summarize key methodological tools used in state-of-the-art quantum-simulation workflows. We emphasize that present noisy quantum simulations serve not only as near-term demonstrations, but also as prototypes for the encodings, diagnostic protocols and error-control strategies required for future fault-tolerant quantum simulation.  
\end{abstract}
\pacs{}    
\maketitle

\tableofcontents

\section{Introduction }\label{sec0}

Condensed matter physics seeks to understand how simple microscopic constituents,
such as electrons, spins, and atoms, give rise to the rich collective behavior observed
in quantum materials and synthetic many-body platforms
\cite{anderson1972more,coleman2015introduction,bloch2008many,bloch2012quantum,
georgescu2014quantum}. Across different geometries and dimensionalities, these
constituents can organize into phases with sharply distinct macroscopic properties,
including magnetic order and quantum criticality
\cite{anderson1950antiferromagnetism,auerbach2012interacting,sachdev1999quantum}, and topological magnetic textures such as magnetic skyrmions~\cite{yu2010real, chen2024allelectrical, chen2024tailoring},
unconventional superconductivity \cite{vonhoegen2026imaging, kirtley1995symmetry} and strongly correlated electron behavior
\cite{lee2006doping,keimer2015from,scalapino2012common},
interaction-driven metal--insulator transitions
\cite{imada1998metal}, symmetry-protected topological phases with protected
boundary excitations, as well as intrinsically topologically ordered phases
\cite{hasan2010colloquium,qi2011topological,wen1990topological,
chen2010local,levin2006detecting,levin2005colloquium,senthil2015symmetry,
nayak2008non}, and a broad class of non-equilibrium phenomena generated by Floquet engineering \cite{Wang2013,Mahmood2016,McIver2020,Merboldt2025,Choi2025,Zhou2023} 
quenches, periodic driving, thermalization dynamics, or dissipation
\cite{polkovnikov2011colloquium,d2016quantum,eisert2015quantum,
aoki2014nonequilibrium,eckardt2017colloquium,daley2014quantum}.

What makes the exploration of condensed matter physics difficult is that many of
the most interesting regimes are inherently many-body in character
\cite{anderson1972more,dagotto2005complexity,coleman2015introduction,bloch2008many}.
When interactions are weak, phases can often be understood in terms of nearly
independent quasiparticles or simple mean-field order parameters. In contrast,
in strongly interacting systems, the collective state cannot be reduced to a
collection of weakly coupled constituents: correlations extend over many sites,
competing tendencies can frustrate simple ordering, and qualitatively new
physics can appear. In this setting, quantum entanglement is not a small
correction but an organizing principle, providing a compact language for how
information and correlations are distributed across the system
\cite{amico2008entanglement,eisert2010colloquium}. In topologically ordered
systems, this viewpoint becomes especially concrete, because long-range
entanglement can itself diagnose phases beyond the Landau symmetry-breaking
paradigm
\cite{levin2006detecting,kitaev2006topological}. As a result, many central questions in condensed matter physics, ranging from
the microscopic nature of correlated ground states, phase transitions, and
entanglement structure
\cite{sachdev1999quantum,hastings2007area,eisert2010colloquium,
white1992density,schollwock2011density,vidal2004efficient,
orus2014practical,troyer2005computational}
to non-equilibrium real-time dynamics, thermalization, and open-system evolution
\cite{lieb1972finite,polkovnikov2011colloquium,eisert2015quantum,
d2016quantum,deutsch1991quantum,srednicki1994chaos,
rigol2008thermalization,nandkishore2015many,Abanin2019,
lindblad1976generators,gorini1976completely,daley2014quantum,
diehl2008quantum,verstraete2009quantum,breuer2016colloquium,
rivas2014quantum}, are intrinsically difficult at both the analytical and
computational levels.

This intrinsic complexity translates into a fundamental bottleneck for classical computation. At the most basic level, the Hilbert space of an interacting quantum system grows exponentially with system size, so that a state of $N$ spin-$1/2$ degrees of freedom might require $\mathcal{O}(2^N)$ complex amplitudes to specify exactly~\cite{feynman2018simulating,nielsen2010quantum,coleman2015introduction}. For fermionic matter, this limitation is encountered as the sign problem in stochastic methods. For real-time dynamics, the challenge is frequently entanglement growth: even when a ground state can be efficiently represented (for example, by a low-entanglement ansatz), unitary evolution typically generates entanglement rapidly~\cite{calabrese2005evolution,vidal2004efficient}. Despite these fundamental obstacles, classical numerical techniques nevertheless remain necessary and have achieved remarkable success:  tensor-network methods exploit limited entanglement to obtain controlled results in one dimension, and for certain two-dimensional geometries~\cite{white1992density,schollwock2011density,verstraete2004renormalization}; quantum Monte Carlo can provide essentially exact benchmarks when the sign problem is absent~\cite{loh1990sign,troyer2005computational,gull2011continuous}; and dynamical mean-field theory (and its cluster extensions) offers a powerful nonperturbative framework~\cite{georges1996dynamical,maier2005quantum}. The key point, however, is that each method relies on assumptions that do not hold universally, which is why a broadly scalable classical description of generic strongly correlated quantum matter remains elusive.

Quantum hardware provides a fundamentally different route to studying quantum matter: it represents the many-body wavefunction directly in a controllable physical system, so that entanglement and interference are native resources rather than computational obstacles~\cite{georgescu2014quantum,preskill2018quantum,houck2012chip}. In a quantum simulator, the microscopic degrees of freedom of interest (spins, fermionic modes, gauge fields, or effective qubits) are encoded into well-controlled quantum elements, and the target model is implemented either by engineering the system’s continuous-time dynamics (analog simulation)~\cite{bloch2008many,blatt2012quantum,georgescu2014quantum} or by compiling the desired evolution into sequences of programmable gates (digital simulation)~\cite{lloyd1996universal,childs2018toward,abrams1997simulation}. Consequently, quantum hardware opens a practical pathway to exploring regimes where classical methods struggle most, such as strongly correlated fermions, frustrated magnetism, and far-from-equilibrium dynamics~\cite{hart2015observation,bernien2017probing,gross2017quantum,polkovnikov2011colloquium}.

A further important direction is that quantum hardware provides not only a
new tool for traditional condensed-matter Hamiltonians; it also
makes it possible to study forms of many-body physics whose natural language
comes from quantum information itself. Concepts such as
entanglement spectra,  operator spreading, and scrambling
have become central diagnostics of quantum phases and non-equilibrium dynamics
\cite{amico2008entanglement,eisert2010colloquium,perezgarcia2007matrix,
cirac2021matrix,kitaev2006topological,levin2006detecting,chen2010local,
pollmann2010entanglement,XuSwingle2024Scrambling}. These ideas are deeply
connected to condensed-matter physics: for example, topological order and symmetry-protected topological (SPT) phases are
distinguished by nonlocal entanglement structures rather than conventional
symmetry-breaking order parameters.  In this sense, quantum simulation
increasingly bridges condensed-matter physics and quantum information science.

This review is organized around three complementary layers of quantum
simulation for condensed-matter physics. First, we survey the main experimental
platforms, including superconducting qubits, trapped-ions, ultracold atoms in
optical lattices, Rydberg atom arrays, photonic quantum simulators, moiré
quantum materials, and commercially available quantum hardware. We also
introduce the basic building blocks of digital quantum simulation.  We then survey representative
applications to condensed-matter physics. Finally, we discuss
the key methodological tools underlying these studies.  Together, these topics connect
existing noisy intermediate-scale quantum (NISQ) experiments with the longer-term goal of fault-tolerant quantum
simulation.

\section{Overview }

\subsection{Scope and organization}

In a prescient observation, Richard Feynman noted that “nature isn’t classical, dammit, and if you want to make a simulation of nature, you’d better make it quantum mechanical...”~\cite{feynman2018simulating,trabesinger2012quantum}. This insight laid the conceptual foundation for quantum simulation. By encoding many-body wavefunctions into qubit registers and implementing Hamiltonian dynamics through quantum circuits, one can use a controllable quantum system to emulate the behavior of target quantum systems. By leveraging superposition and entanglement as computational resources, quantum simulation is not merely a faster computational strategy but a paradigm shift in how complex quantum matter is modeled and explored~\cite{lloyd1996universal,preskill2018quantum}.

In this review, we present a pedagogical overview of quantum computing
approaches to condensed-matter simulation, spanning hardware platforms,
circuit primitives, and algorithmic techniques for studying equilibrium,
non-equilibrium, and topological phenomena
\cite{blatt2012quantum,bloch2008many,montanaro2016quantum}. Our emphasis is on physical motivation and practical implementation, with the goal of equipping condensed-matter physicists to critically assess both the promise and the current limitations of cutting-edge quantum-simulation experiments, especially those performed on programmable universal quantum processors~\cite{preskill2018quantum,arute2019quantum}. While most existing studies have
been performed on NISQ devices, which remain
limited in scale, coherence, connectivity, and accuracy, the techniques
developed in these experiments are not restricted to the NISQ era. Many of the
same ingredients, such as Hamiltonian encodings, Trotterized time evolution,
variational state preparation, and  mid-circuit
operations, will form the practical
building blocks of future fault-tolerant quantum simulations.

From this perspective, present-day  programmable devices provide not only
near-term experimental testbeds, but also prototypes for the logical-level
workflows expected in error-corrected quantum processors. In the
fault-tolerant regime, longer coherent circuit depths and logical error
suppression will enable more systematic use of quantum algorithms for quantitatively accurate simulation of condensed-matter systems
\cite{berry2015simulating,childs2012hamiltonian,low2019hamiltonian,
gilyen2019quantum}. Thus, the NISQ demonstrations reviewed here could identify useful encodings, circuit
structures, measurement strategies, and error-control principles that can be
carried forward to fault-tolerant quantum simulation.

\subsection{Physical Platforms for Quantum Simulation Experiments}
We first review the state of the art across a broad range of quantum simulation
platforms, even though most of this work subsequently focuses on advances in
universal quantum computing hardware
\cite{feynman2018simulating,lloyd1996universal,georgescu2014quantum,
Altman2021quantumSimulators}. Experimental platforms for quantum simulation
differ fundamentally in their native degrees of freedom, interaction mechanisms,
connectivity graphs, control primitives, and measurement capabilities
\cite{bloch2012quantum,gross2017quantum,blatt2012quantum,browaeys2020many,
houck2012chip,aspuru2012photonic,flamini2018photonic}. These physical
distinctions determine the classes of many-body Hamiltonians that can be
realized efficiently and the observables that can be accessed in practice.
Recent experiments further illustrate the rapid convergence between analog
quantum simulators and programmable digital processors, including Rydberg-array
studies of non-equilibrium collective dynamics [see Figs.~\ref{fig:scar}(a)
and~\ref{fig:stringdynamics}(b)], trapped-ion simulations of
digital quantum magnetism and fermionic dynamics [see Figs.~\ref{fig:time}(a,b)
and~\ref{fig:lgt}(d)], superconducting-processor
realizations of non-equilibrium topological phases [see Figs.~\ref{fig:floquet1}
and~\ref{fig:digital_topological_phenomena}], and scalable photonic
hardware demonstrations
\cite{manovitz2025quantum,Haghshenas2026DigitalMagnetism,
Alam_2025_fermionic_dynamics,QianScience2025,psiquantum2025manufacturable}.

\begin{figure*}
    \centering
    \includegraphics[width=0.8\linewidth]{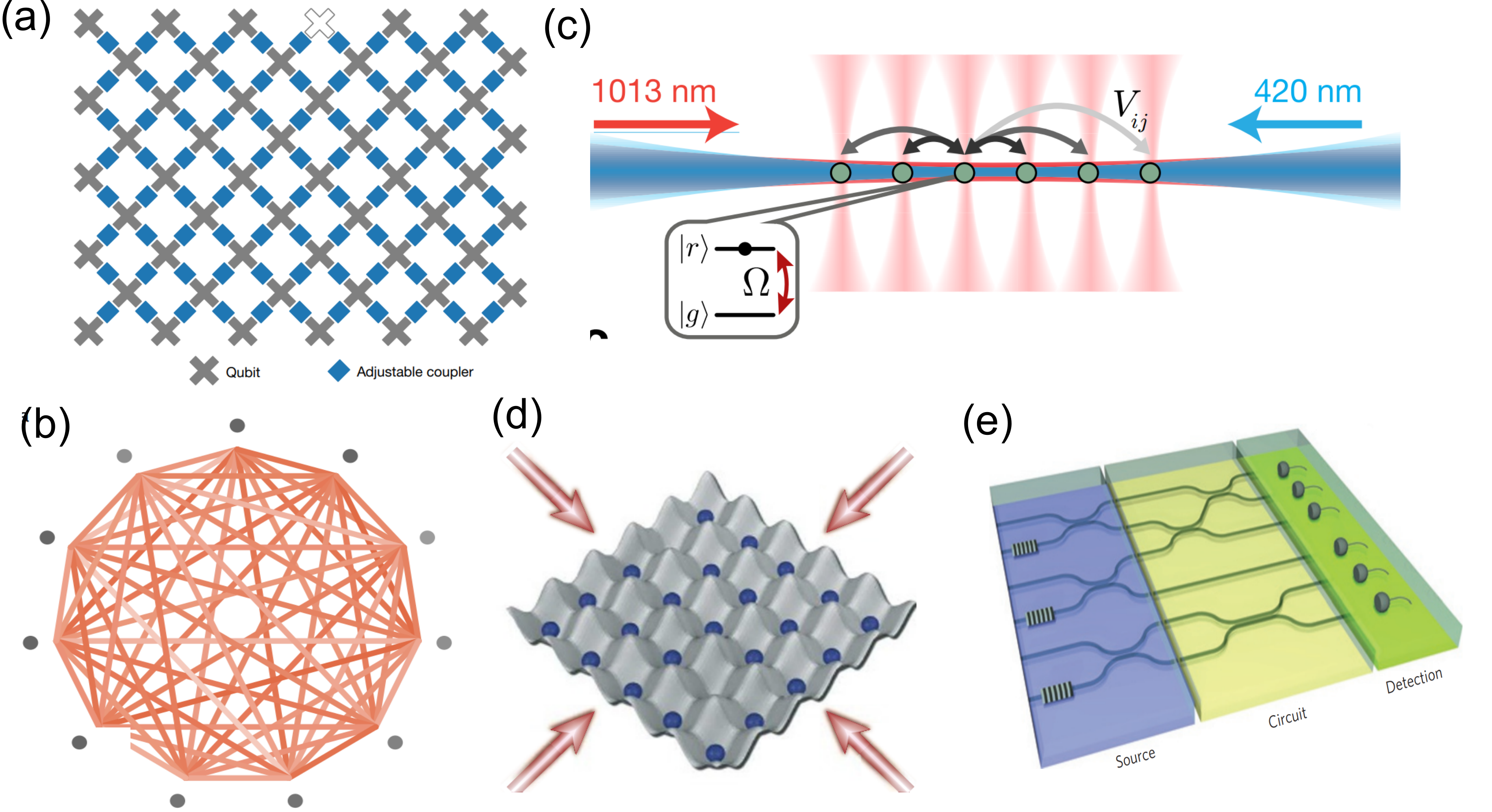}
    \caption{Representative platforms for quantum simulation. (a) Superconducting qubits arranged in a two-dimensional lattice with tunable couplers~\cite{arute2019quantum}, (b) Trapped-ion processor with all-to-all qubit connectivity~\cite{wright2019benchmarking}, (c) Rydberg atoms in optical tweezers with interactions among excited levels~\cite{bernien2017probing}, (d)  Ultracold atoms in an optical lattice, where laser fields create periodic trapping potentials~\cite{bloch2008quantum}, (e)  Photonic quantum simulator comprising single-photon sources, integrated optical circuits, and detection modules~\cite{aspuru2012photonic}. 
    }
    \label{fig:placeholder}
\end{figure*}

A useful organizing principle is the distinction between native, or analog,
quantum simulation and gate-based, or digital, quantum simulation
\cite{cirac2012goals,bloch2008many,georgescu2014quantum,
Altman2021quantumSimulators}. The former uses the natural Hamiltonian of a
well-controlled quantum platform to emulate the target many-body problem directly.
Representative examples include bosons in optical lattices [see
Fig.~\ref{fig:placeholder}(d)]
\cite{jaksch1998cold,greiner2002quantum,bloch2008many,bloch2012quantum,
gross2017quantum}, effective spin models in trapped-ion and neutral-atom
platforms [see Figs.~\ref{fig:placeholder}(b,c), \ref{fig:scar}(a),
and~\ref{fig:time}(a,b)]~\cite{porras2004effective,blatt2012quantum,kim2010quantum,
saffman2010quantum,browaeys2020many,bernien2017probing,scholl2021quantum,
ebadi2021quantum}, and long-range couplings mediated by collective modes such
as phonons or cavity photons~\cite{porras2004effective,ritsch2013cold}. Recent
solid-state analog simulators based on atomically engineered quantum-dot arrays
further broaden this landscape by targeting low-temperature correlated-electron
physics in large two-dimensional lattices~\cite{donnelly2026large}. While the
native dynamics of such engineered Hamiltonians can be explored directly,
questions of control, calibration, benchmarking, and verification remain central
challenges~\cite{hauke2012can,eisert2015quantum,daley2014quantum,
Altman2021quantumSimulators}. Gate-based digital simulation, by contrast, trades
physical naturalness for algorithmic universality: the target evolution is
compiled into sequences of elementary quantum gates, enabling controlled
Trotterized or algorithmic approximations, flexible access to nonlocal
observables, and simulations of models that are difficult to realize through
direct Hamiltonian engineering
\cite{lloyd1996universal,nielsen2010quantum,preskill2018quantum}. Recent
digital demonstrations on programmable quantum processors, including
superconducting and trapped-ion hardware [see Figs.~\ref{fig:floquet1},
\ref{fig:digital_topological_phenomena}, and~\ref{fig:lgt}(d)], illustrate how this gate-based route
is increasingly being used to probe non-equilibrium many-body dynamics,
topological phenomena, and fermionic or spin models in regimes approaching or
exceeding the reach of exact classical simulation
\cite{kim2023evidence,QianScience2025,Alam_2025_fermionic_dynamics,
Haghshenas2026DigitalMagnetism}. These choices entail tradeoffs among
controllability, scalability, coherence time, connectivity, measurement access,
and physical realism, and they influence whether a given platform is best suited
to ground-state preparation, real-time dynamics, or non-equilibrium phenomena.

Although much of this review focuses on gate-based digital quantum
hardware, analog quantum simulators have also played a central role in advancing
many-body quantum simulation
\cite{cirac2012goals,hauke2012can,georgescu2014quantum,
Altman2021quantumSimulators,daley2022practical}. Platforms such as ultracold
atoms, trapped-ions, superconducting analog circuits, and Rydberg tweezer arrays
can often realize target Hamiltonians, constraints, or dissipative processes
more directly than fully digital Trotterized circuits
\cite{bloch2008many,bloch2012quantum,gross2017quantum,
monroe2021programmable,houck2012chip,browaeys2020many,diehl2008quantum,
diehl2010dynamical,barreiro2011open}. This capability is especially valuable
for problems such as lattice gauge theories, constrained many-body dynamics,
non-equilibrium phases, and driven-dissipative systems
\cite{zohar2015quantum,banuls2020simulating,martinez2016real,kokail2019self,
bernien2017probing,scholl2021quantum,ebadi2021quantum,manovitz2025quantum}.
Analog and near-analog platforms have also provided important conceptual input
for digital quantum simulation. For example, the observation of long-lived
coherent dynamics in Rydberg-atom arrays motivated the identification of
quantum many-body scars as a mechanism for weak ergodicity breaking in
constrained Hilbert spaces
\cite{bernien2017probing,turner2018weak,turner2018quantum,ho2019periodic,
choi2019emergent,lin2019exact,Serbyn2021}. This development has subsequently
inspired digital simulations of scar dynamics and constrained models on
programmable quantum processors
\cite{zhang2023manybodyscarring,chen2022errormitigatedscars,
gustafson2023preparing,desaules2024robust,shen2024enhanced,
logaric2026dynamical}.

Importantly, in several platforms, the boundary between digital and analog
simulation is not sharply defined. Rydberg arrays provide a particularly clear
example
\cite{saffman2010quantum,saffman2016quantum,browaeys2020many,
Henriet2020quantumcomputing}. Their native blockade interactions offer an
analog route to constrained spin models, quantum dimer models, gauge constraints,
and string dynamics
\cite{bernien2017probing,labuhn2016tunable,scholl2021quantum,
ebadi2021quantum,semeghini2021probing,samajdar2021quantum,li2022quantum,
zeng2025quantum,gonzalez2025observation,bombieri2026u1}, while local
addressing, programmable pulse sequences, projective readout, and Rydberg
blockade gates also enable digital or digital--analog protocols
\cite{levine2018high,graham2022multi,evered2023high,bluvstein2024logical,
bluvstein2025fault,silverio2022pulser,wurtz2023aquila}. Such hybrid digital--analog
strategies   
represent a promising route for combining native many-body evolution with
gate-level programmability
\cite{arrazola2016digital,lamata2017digital,monroe2021programmable,
Tacchino2020UniversalQuantumSimulators,Fauseweh2024DigitalManyBody}.

In this section, we survey major established and emerging experimental platforms
for the quantum simulation of condensed-matter systems, highlighting their
distinct capabilities and emphasizing the physical intuition behind each
approach~\cite{cirac2012goals,georgescu2014quantum,Altman2021quantumSimulators}.
Trapped-ion platforms realize high-fidelity control and tunable long-range spin
interactions mediated by collective phonon modes, enabling early quantum-magnet
demonstrations and subsequent studies of frustration, long-range correlation
spreading, and entanglement propagation [see Figs.~\ref{fig:placeholder}(b),
\ref{fig:time}(a,b), and~\ref{fig:lgt}(d)]
\cite{porras2004effective,friedenauer2008simulating,kim2010quantum,
islam2011onset,britton2012engineered,blatt2012quantum,
richerme2014nonlocal,jurcevic2014quasiparticle}.
Ultracold neutral atoms in optical lattices provide direct analog realizations of
Bose-- and Fermi--Hubbard physics, quantum magnetism, and site-resolved many-body
correlations through quantum-gas microscopy
\cite{jaksch1998cold,greiner2002quantum,bloch2008many,bloch2012quantum,
gross2017quantum,jordens2008mott,schneider2008metallic,esslinger2010fermi,
bakr2009quantum,sherson2010single,cheuk2015quantum,hart2015observation,
mazurenko2017cold,brown2017spin}. Rydberg-atom arrays extend this neutral-atom
toolkit by combining strong interactions, programmable geometry, and single-atom
control, enabling constrained spin models, many-body scars, antiferromagnetic
ordering, spin-liquid probes, and non-equilibrium coarsening dynamics
\cite{saffman2010quantum,labuhn2016tunable,bernien2017probing,
browaeys2020many,ebadi2021quantum,scholl2021quantum,semeghini2021probing,
manovitz2025quantum}. Disordered and quasiperiodic cold-atom platforms further
provide controlled access to Anderson and many-body localization phenomena
\cite{schreiber2015observation,choi2016exploring,smith2016many,
luschen2017signatures}, while cavity-mediated systems and emerging solid-state
analog simulators broaden the accessible interaction mechanisms and lattice
architectures~\cite{ritsch2013cold,tang2020simulation,donnelly2026large}.
Complementing these analog platforms, superconducting quantum processors,
trapped-ion quantum computers, and photonic quantum circuits have emerged as
leading hardware for digital quantum simulation
\cite{houck2012chip,barends2015digital,smith2019simulating,kim2023evidence,
QianScience2025,Alam_2025_fermionic_dynamics,aspuru2012photonic,kok2007linear,
flamini2018photonic,broome2013photonic,spring2013boson}, where time evolution
is synthesized through sequences of elementary gates. Gate-based architectures
enable Trotterized time evolution~\cite{lloyd1996universal,suzuki1992general,
berry2007efficient}, variational quantum simulation
\cite{li2017efficient,yuan2019theory,cerezo2021variational}, and hybrid
analog--digital protocols~\cite{arrazola2016digital,lamata2017digital,
Tacchino2020UniversalQuantumSimulators,Fauseweh2024DigitalManyBody}. Together,
these platforms form a complementary toolkit of quantum simulators for exploring
the rich landscape of quantum matter.

\subsubsection{Superconducting Qubits}\label{superconducting}

Superconducting qubits are among the leading platforms for quantum computation
and quantum simulation, owing to their scalability, nanosecond-scale gate speeds,
and compatibility with mature microfabrication and integration technologies
\cite{clarke2008superconducting,devoret2013superconducting,gu2017microwave,
krantz2019quantum,kjaergaard2020superconducting,wendin2017quantum}. They are
artificial atoms realized in macroscopic electrical circuits that exploit
superconducting phase coherence and the Josephson effect
\cite{josephson1962possible,makhlin2001quantum,blais2004cavity,
koch2007charge}. In standard two-level qubit operation, computation is encoded in the two
lowest eigenstates of a nonlinear superconducting circuit mode, most commonly a
transmon or related Josephson circuit. More generally, however, the same
anharmonic spectrum contains higher excited levels that can be coherently
addressed and read out. These levels are usually treated as leakage states in
qubit-based processors, but they have also been deliberately used to implement
superconducting qutrits and higher-dimensional qudits
\cite{bianchetti2010control,peterer2015coherence,blok2021quantum,
goss2022high,liu2023performing,wang2025high}. Despite involving a macroscopic
number of superconducting Cooper pairs, the relevant circuit degree of freedom is
a collective quantum coordinate described by conjugate variables: the
superconducting phase difference across the Josephson element, $\phi$, and the
Cooper-pair number, $n$, satisfying
\begin{equation}
[\phi,n]=i .
\end{equation}
Different superconducting-qubit designs correspond to different operating
regimes of the same underlying nonlinear circuit, optimized for coherence,
control, addressability, and noise resilience. The charge qubit, or
Cooper-pair-box qubit, is one of the earliest superconducting-qubit designs. It
consists of a small superconducting island connected to a reservoir through a
Josephson junction, with the logical states encoded in different island-charge
states, i.e., different numbers of excess Cooper pairs
\cite{nakamura1999coherent,nakamura2001rabi}. Because the transition frequency
depends strongly on the offset charge, the original charge qubit is highly
sensitive to charge noise and is therefore limited in coherence. The transmon
qubit suppresses this sensitivity by shunting the Josephson junction with a
large capacitance, thereby increasing the ratio $E_J/E_C$ and making the
transition frequency exponentially insensitive to charge fluctuations while
retaining sufficient anharmonicity for qubit control
\cite{koch2007charge,schreier2008suppressing}. This noise resilience, together
with compatibility with microwave control and circuit-QED readout, has made the
transmon and its variants the standard choice in many current superconducting
quantum processors. The flux qubit instead encodes information in the direction
of a persistent current circulating in a superconducting loop, with the qubit
frequency tunable by an external magnetic flux; this architecture is naturally
sensitive to flux noise, although operation near flux sweet spots can mitigate
first-order dephasing~\cite{mooij1999josephson,bylander2011noise}. The phase
qubit uses a current-biased Josephson junction whose energy levels are quantized
in a tilted washboard potential~\cite{martinis2002rabi,cooper2004observation}.
While charge, flux, and phase qubits all played important roles in the early
development of superconducting quantum information processing, transmon-based
architectures and related capacitively shunted designs now dominate large-scale
gate-based superconducting processors because of their improved coherence,
control, and scalability.

Nevertheless, all designs realize the same underlying principle: a quantized nonlinear mode of the superconducting condensate whose lowest two eigenstates, $|0\rangle$ and $|1\rangle$, form the computational basis. The Josephson junction introduces the nonlinearity in $\phi$, which is required to make the spectrum of a LC circuit anharmonic, allowing $|0\rangle$ and $|1\rangle$ to be isolated from unwanted higher excited states and hence addressed selectively~\cite{josephson1962possible,koch2007charge}. Coherent control is achieved using microwave-frequency electromagnetic fields resonant with the qubit transition, enabling precise rotations of the qubit state within the computational basis. High-fidelity state readout is typically performed by coupling the qubit dispersively to a microwave resonator, such that the qubit state shifts the resonator frequency, allowing the qubit to be inferred from the reflected or transmitted microwave signal without direct excitation~\cite{blais2004cavity,wallraff2004strong}.

Interactions between superconducting qubits arise through engineered
electromagnetic couplings, including direct capacitive or inductive elements,
shared microwave resonators, and tunable coupler circuits
\cite{wallraff2004strong,majer2007coupling,blais2021circuit,chen2014qubit,
mckay2016universal,yan2018tunable}. Depending on the architecture, these
couplings can generate effective exchange, longitudinal, dispersive, or
parametrically activated two-qubit interactions, which may be always on,
refocused, resonantly activated, or dynamically switched by external microwave
and flux controls. This flexibility makes superconducting circuits useful for
both analog and gate-based quantum simulation
\cite{houck2012chip,roushan2014observation,xu2018emulating,wang2024realization}.
In the gate-based mode used by most contemporary superconducting quantum
processors, target Hamiltonians are implemented by discretizing the time-evolution
operator into sequences of calibrated single- and two-qubit gates, as in
Trotterized, variational, and more general circuit-based simulation protocols
\cite{lloyd1996universal,suzuki1992general,berry2007efficient,
barends2015digital,kandala2017hardware,arute2019quantum,kim2023evidence,
QianScience2025}.

Hence, superconducting quantum processors provide a flexible and programmable
platform for digitally emulating complex quantum Hamiltonians. By embedding
model graphs into fixed superconducting-device connectivity, such as heavy-hex
or related planar coupling graphs, and by engineering interactions through
microwave pulses, calibrated two-qubit gates, and tunable couplers
[FIG.~\ref{fig:placeholder}(a)], these systems enable digital simulations of a
wide range of spin, fermionic, gauge-theory, and topological models with
nearest-neighbor interactions and, after compilation or circuit routing,
effective longer-range couplings [see Figs.~\ref{fig:spt}, \ref{fig:qite},
\ref{fig:fqhe2}, and~\ref{fig:digital_topological_phenomena}]
\cite{arute2019quantum,barends2015digital,kim2023evidence,
hayata2024floquet,QianScience2025,shen2026observation,ilcic2026observation}.
A key strength of superconducting architectures is the combination of fast
single- and two-qubit gates, scalable microwave control, high-throughput
calibration, and hardware-aware compilation. Recent technical advances include
processor-level benchmarking at the 100-qubit scale, layer-fidelity metrics for
large connected subgraphs, improved noise characterization, and error-mitigation
protocols that can be applied to large circuit volumes
\cite{cross2019validating,kim2023scalable,berg2023probabilistic,
mckay2023benchmarking,abughanem2024ibm}. In parallel, superconducting devices
now support controlled dissipation, mid-circuit measurement, reset, and
real-time feedforward, enabling simulations of open quantum systems,
measurement-driven dynamics, and dynamic-circuit protocols [see
Fig.~\ref{fig:adaptive}]
\cite{vijay2012stabilizing,riste2013deterministic,campagne2013persistent,
google2023measurement,gupta2024probabilistic,carrera2024combining}. These
capabilities have enabled experiments on processors with more than 100 physical
qubits, including utility-scale kicked-Ising dynamics on a 127-qubit Eagle
processor, real-time classically linked dynamic circuits spanning up to 142
qubits across two 127-qubit QPUs, Floquet lattice-gauge simulations using more
than 100 qubits on a 156-qubit Heron processor, and recent 100--156-qubit
quantum-simulation experiments on IBM hardware
\cite{kim2023evidence,carrera2024combining,hayata2024floquet,
shen2026observation,ilcic2026observation}. As of 2026, state-of-the-art
superconducting devices include IBM Heron-class processors with 156 qubits and
rapidly improving calibration, connectivity-aware compilation, dynamic-circuit
support, and error-mitigation workflows
\cite{ibm_quantum_2026,abughanem2024ibm,mckay2023benchmarking,
google2025quantum}. Representative superconducting-processor simulations discussed later in this
review are highlighted in Figs.~\ref{fig:spt}, \ref{fig:qite},
\ref{fig:fqhe2}, and~\ref{fig:digital_topological_phenomena}.

Despite these rapid advances, several aspects of superconducting-qubit platforms
remain active areas of development. Finite coherence times continue to place
practical limits on achievable circuit depth and accessible dynamical timescales
\cite{clarke2008superconducting,devoret2013superconducting,
kjaergaard2020superconducting}, particularly for simulations of low-energy,
long-time, or weakly driven phenomena. Scaling to larger and more connected
qubit arrays while maintaining high gate fidelity, suppressing residual
couplings, controlling crosstalk, and preserving calibration stability remains
an ongoing challenge, especially when simulating lattice models whose physical
connectivity differs from the hardware graph
\cite{gambetta2017building,chen2014qubit,arute2019quantum,
cross2019validating,mckay2023benchmarking}.

At the processor level, recent benchmarks and demonstrations on
100-qubit-scale superconducting devices have made clear that system-wide
performance depends not only on isolated gate fidelities, but also on
simultaneous-gate operation, connectivity, readout performance, crosstalk-aware
calibration, and the fidelity of entire circuit layers
\cite{cross2019validating,kim2023evidence,mckay2023benchmarking,
abughanem2024ibm,ibm_quantum_2026}. In addition, the indirect, gate-based
construction of many-body Hamiltonians introduces compilation and routing
overheads that can obscure the physical structure and increase the effective circuit
depth at large system sizes. Error suppression, error mitigation, and
dynamic-circuit protocols can partly extend the useful regime of noisy
superconducting processors, but their sampling overheads, noise-model
assumptions, latency constraints, and calibration requirements also become
important scaling considerations
\cite{kim2023scalable,berg2023probabilistic,gupta2024probabilistic,
carrera2024combining}. As superconducting platforms evolve toward improved
coherence, connectivity, control electronics, calibration automation, and
error-corrected operation
\cite{google2025quantum}, they may enable selected large-scale digital
simulations that challenge the reach of exact classical methods.

\subsubsection{Trapped Ion Qubits}\label{ion}

Trapped-ion systems constitute one of the most precise and well-controlled
quantum-information platforms, distinguished by their long coherence times,
high-fidelity state preparation and readout, and accurate coherent control
\cite{leibfried2003quantum,blatt2012quantum,monroe2013scaling,
home2009complete,roos2004control}. In these systems, individual atomic ions,
such as $^{171}$Yb$^+$, $^{43}$Ca$^+$, and $^{40}$Ca$^+$, are confined using
static and radio-frequency electric fields and manipulated using laser,
microwave, or Raman fields
\cite{paul1990electromagnetic,olmschenk2007manipulation,harty2014high}.
The qubit is encoded in internal atomic states of a single ion, typically in
long-lived hyperfine, Zeeman, or optical-transition states. Because the ions are
spatially localized charged particles held in ultra-high-vacuum traps, their
internal states form well-isolated and reproducible quantum degrees of freedom
\cite{haffner2008quantum,wineland2011quantum,benhelm2008towards}. These states
are protected by atomic structure and can exhibit coherence times that greatly
exceed those of most solid-state qubits. Coherent single-qubit control is then
implemented by driving the relevant internal transition directly or indirectly,
depending on the ion species and qubit encoding.

Different trapped-ion implementations primarily involve different choices of
atomic species and qubit encoding, optimized for coherence, optical accessibility,
state preparation and readout, and gate control. Hyperfine clock-state qubits can
offer exceptional stability against magnetic-field fluctuations, whereas optical
qubits provide direct optical access to narrow internal transitions, with
performance set by laser phase stability, magnetic-field noise, and motional
control~\cite{leibfried2003quantum,home2009complete,benhelm2008towards}.
In standard trapped-ion quantum-information implementations, the qubit is encoded
in internal atomic states of individual ions, while collective motional modes
serve primarily as a quantum bus for entangling operations rather than as the
qubit degree of freedom itself. Coherent single-qubit control is implemented by
driving the relevant internal transition using microwave fields, Raman laser
beams, or resonant optical pulses, depending on the chosen ion species and qubit
encoding.

Most trapped-ion architectures rely on
a common physical principle: internal atomic two-level systems are coupled
through shared quantized motional modes
\cite{molmer1999multiparticle,leibfried2003experimental,roos2008ion}. The ions
confined in a trap form a Coulomb-coupled crystal whose collective vibrational
modes act as quantum buses mediating interactions between spatially separated
qubits. Although these motional modes are collective, qubit selectivity is
achieved at the control level. In practice, tightly focused laser beams, Raman
tones, microwave fields, or microwave near-field gradients are applied to
selected ions, while the drive frequencies, phases, amplitudes, and detunings
are chosen relative to the relevant carrier and motional-sideband transitions.
For example, in a Mølmer--Sørensen-type gate, bichromatic fields generate a
spin-dependent force on the addressed ions; when the motional trajectory closes
in phase space, the phonons disentangle from the qubits and the addressed ions
acquire an effective entangling phase. Ions that are not addressed, or are
spectrally or spatially weakly coupled to the applied tones, ideally remain
spectators, up to residual off-resonant excitation, mode participation, and
optical or microwave crosstalk. In this way, shared phonon modes can mediate
programmable pairwise or collective interactions between distant ions within an ion
chain. State readout is typically performed through state-dependent fluorescence,
where the qubit state is inferred from bright or dark photon-scattering
statistics~\cite{nagourney1986shelved,bergquist1986observation}.

A key strength of trapped-ion qubits lies in their all-to-all connectivity [FIG.~\ref{fig:placeholder} (b)]~\cite{wright2019benchmarking}, which enables highly-tunable long-range effective spin–spin couplings that can be engineered by adjusting laser detunings, intensities, and trap parameters. This naturally enables the simulation of interacting spin models with variable
interaction range and power-law decays [see Fig.~\ref{fig:strong}]
~\cite{friedenauer2008simulating,britton2012engineered}. These interactions, arising from continuous physical dynamics, place trapped-ions closer to the programmable-analog end of the quantum simulation spectrum. At the same time, trapped-ion platforms also support fully digital operation by disentangling the collective motional modes from the qubit state at the end
of each gate operation.

Although gate operations in trapped-ion systems are generally slower than those
in superconducting circuits, the relevant timescales remain well matched to the
long coherence of atomic qubits. As an order-of-magnitude comparison, typical
trapped-ion single-qubit gates operate on microsecond timescales and two-qubit
entangling gates on tens to hundreds of microseconds, whereas superconducting
processors commonly implement single-qubit gates in tens of nanoseconds and
two-qubit gates in roughly $10^2$ ns
\cite{wineland2011quantum,blatt2012quantum,monroe2013scaling,
harty2014high,ballance2016high,gaebler2016high,
krantz2019quantum,kjaergaard2020superconducting,arute2019quantum}.
This slower gate speed is partly offset by long coherence times, high-fidelity
state preparation and readout, and flexible long-range connectivity.
Consequently, trapped-ions remain one of the most powerful platforms for both
analog and digital quantum simulation.
Moreover, the ability to engineer spin--spin couplings that are large compared
with the relevant decoherence rates makes trapped-ions well suited for probing
non-equilibrium and critical spin dynamics, including the onset of quantum phase
transitions, frustrated Ising dynamics, Floquet phases, and the preparation of
highly entangled states [see Fig.~\ref{fig:strong}]
\cite{islam2011onset,kim2010quantum,zhang2017observation}. Recent advances
further show that trapped-ion devices support high-fidelity projective
measurements, individual-qubit readout, and mid-circuit measurement or
feed-forward capability, enabling measurement-based hybrid circuits,
measurement-induced phases, and qubit-reuse protocols [see Fig.~\ref{fig:anyon}]
\cite{noel2022measurement,iqbal2024topological,decross2023qubit}.

Despite these strengths, several aspects of trapped-ion platforms remain active areas of development. Scaling to large ion numbers introduces challenges associated with motional mode crowding and increased sensitivity to heating and noise~\cite{porras2004effective,blatt2012quantum,monroe2013scaling,kielpinski2002architecture}. Engineering strictly local interactions in large ion chains remains nontrivial, particularly when simulating lattice models with geometric locality. Continued advances in trap design, laser control, and modular architectures are therefore central to extending trapped-ion quantum simulators towards larger and more complex many-body systems. Representative trapped-ion examples discussed below are highlighted in
Figs.~\ref{fig:strong} and~\ref{fig:anyon}.

\subsubsection{Ultracold Atoms in Optical Lattices }\label{optical}

\newcommand{\tcell}[2]{\parbox[t]{#1}{\raggedright #2\strut}}
\begin{table*}[!t]
	\centering
	\small
	\renewcommand{\arraystretch}{1.18}
	\setlength{\tabcolsep}{4pt}
	\caption{
		Representative quantum-simulation platforms for condensed-matter and many-body physics up to early 2026.
		The table summarizes their natural capabilities, best-matched problems, advantages, and main limitations.
	}
	\label{tab:platforms_cm}
	\begin{tabular}{|l|l|l|l|l|}
		\hline
		\tcell{2.55cm}{\textbf{Platform}} &
		\tcell{3.35cm}{\textbf{Key features}} &
		\tcell{3.35cm}{\textbf{Best-matched problems}} &
		\tcell{3.25cm}{\textbf{Advantages}} &
		\tcell{3.15cm}{\textbf{Main limitations}} \\
		\hline
		
		\tcell{2.55cm}{\textbf{Superconducting qubit processors}\\
			(Sect.~\ref{superconducting})} &
		\tcell{3.35cm}{Fast gates and readout; two-dimensional chip layouts; mature compilation, calibration, and error-mitigation workflows; dynamic circuits with mid-circuit measurement, reset, and feed-forward.} &
		\tcell{3.35cm}{Digital spin and fermion models; Floquet dynamics; lattice gauge theories; monitored and adaptive circuits; topological-code experiments.} &
		\tcell{3.25cm}{High throughput; flexible gate-level programmability; strong software ecosystem and calibration infrastructure.} &
		\tcell{3.15cm}{Finite coherence and circuit depth; routing overhead; crosstalk, leakage, readout errors, and calibration drift.} \\
		\hline
		
		\tcell{2.55cm}{\textbf{Trapped ions}\\
			(Sect.~\ref{ion})} &
		\tcell{3.35cm}{Long coherence times; high-fidelity gates; high logical connectivity; native long-range spin interactions; high-quality measurement and reset; qubit reuse and qudit encodings.} &
		\tcell{3.35cm}{Long-range spin models; gauge-theory simulations; topological order; monitored dynamics; qudit-based boson and gauge-field encodings.} &
		\tcell{3.25cm}{Excellent coherence and control; flexible connectivity; compact qudit representations; high-fidelity mid-circuit measurement and feed-forward.} &
		\tcell{3.15cm}{Slower gates and measurements; scaling requires complex laser, motional-mode, shuttling, and calibration control.} \\
		\hline
		
		\tcell{2.55cm}{\textbf{Ultracold atoms in optical lattices}\\
			(Sect.~\ref{optical})} &
		\tcell{3.35cm}{Native Hubbard, Aubry--Andr\'e, and spin-exchange Hamiltonians; large atom numbers; tunable interactions, disorder, and geometry; quantum-gas microscopy.} &
		\tcell{3.35cm}{Hubbard physics; transport and thermalization; localization and MBL-like dynamics; synthetic gauge fields; topological bands.} &
		\tcell{3.25cm}{Direct analog realization of canonical many-body models; large system sizes; site-resolved probes of density, spin, and correlations.} &
		\tcell{3.15cm}{Limited gate-level programmability; finite temperature and entropy; slower experimental cycle times; restricted measurement basis compared with digital processors.} \\
		\hline
		
		\tcell{2.55cm}{\textbf{Rydberg tweezer arrays}\\
			(Sect.~\ref{rydberg})} &
		\tcell{3.35cm}{Large reconfigurable one- and two-dimensional arrays; strong Rydberg blockade; local addressing; analog and digital--analog operation.} &
		\tcell{3.35cm}{Constrained spin models; quantum scars; frustrated Ising models; optimization; gauge constraints and string dynamics.} &
		\tcell{3.25cm}{Flexible geometry; natural blockade constraints; strong interactions; single-site preparation and readout; large programmable analog system sizes.} &
		\tcell{3.15cm}{Atom loss; finite Rydberg lifetime; blockade leakage and laser errors; challenging deep universal gate operation across large arrays.} \\
		\hline
		
		\tcell{2.55cm}{\textbf{Photonic platforms}\\
			(Sect.~\ref{photonic})} &
		\tcell{3.35cm}{Discrete- and continuous-variable modes; linear optics; integrated photonic circuits; cluster-state and measurement-based resources.} &
		\tcell{3.35cm}{Boson sampling; quantum walks; photonic topological lattices; measurement-based simulation; networked and fault-tolerant photonic architectures.} &
		\tcell{3.25cm}{Room-temperature operation; low dephasing during propagation; natural compatibility with quantum networking and modular architectures.} &
		\tcell{3.15cm}{Photon loss; source and detector inefficiency; probabilistic entangling gates without strong nonlinearities; large overheads for scalable fault tolerance.} \\
		\hline
		
		\tcell{2.55cm}{\textbf{Moir\'e quantum-material emulators}\\
			(Sect.~\ref{moir})} &
		\tcell{3.35cm}{Twisted or lattice-mismatched van der Waals materials; narrow bands; tunable filling, bandwidth, topology, and interactions.} &
		\tcell{3.35cm}{Strongly correlated electrons; Hubbard-like physics; correlated insulators; superconductivity; Chern and fractional Chern phases.} &
		\tcell{3.25cm}{Direct solid-state realization of interacting-electron phases; continuous electrostatic tunability; access to real electronic materials phenomena.} &
		\tcell{3.15cm}{Not a universal programmable quantum processor; disorder, strain, and sample dependence; limited coherent real-time control and measurement programmability.} \\
		\hline
		
	\end{tabular}
\end{table*}

Ultracold atoms in optical lattices provide a paradigmatic platform for analog
quantum simulation, especially for lattice Hamiltonians in which itinerant
bosons or fermions occupy periodic potentials generated by optical standing
waves
\cite{jaksch1998cold,greiner2002quantum,bloch2008many,lewenstein2007ultracold,
esslinger2010fermi,lewenstein2012ultracold}. In this setting, the relevant
degrees of freedom are the occupations of lattice orbitals, together with the
internal spin or hyperfine states of the atoms, and the simulator directly
realizes microscopic Hamiltonian dynamics rather than implementing the evolution
through a sequence of discrete gates.
This distinction, however, should not be interpreted as a fundamental separation
between cold-atom platforms and gate-based quantum processing. In particular,
neutral-atom arrays based on optical tweezers and Rydberg-mediated interactions
can be operated in both analog and digital modes: analog protocols exploit
native Rydberg blockade or van der Waals interactions to realize spin models,
whereas digital protocols encode qubits in long-lived atomic states and use
transient Rydberg excitation to implement entangling gates
\cite{saffman2010quantum,saffman2016quantum,browaeys2020many,
Henriet2020quantumcomputing,levine2018high,evered2023high}. Recent
neutral-atom experiments have demonstrated programmable gate-model operation,
parallel high-fidelity entangling gates, logical qubits, mid-circuit readout,
nondestructive measurement, and architectures aimed at fault-tolerant universal
quantum computation
\cite{graham2022multi,bluvstein2024logical,radnaev2025universal,
bluvstein2025fault}. Thus, cold-atom systems should be viewed not as purely
analog simulators, but as a family of platforms ranging from direct Hamiltonian
emulators to increasingly programmable universal quantum processors.

The periodic optical potential gives rise to quantized Bloch bands, and at
sufficiently low temperatures the dynamics can often be restricted to the lowest
band, or to a controlled set of bands, leading to lattice Hamiltonians written
in second-quantized form. Each lattice site supports a local Hilbert space with
variable particle number and, when relevant, internal spin or hyperfine degrees
of freedom. This structure closely mirrors the many-body lattice degrees of
freedom encountered in quantum materials, including bosonic and fermionic
Hubbard-type models, rather than being restricted to a fixed array of qubits
\cite{jaksch1998cold,hofstetter2002high,lewenstein2007ultracold,
jordens2008mott,schneider2008metallic,esslinger2010fermi}. 
As a result, optical-lattice simulators naturally realize many-body physics
beyond effective spin models, including Bose-Hubbard dynamics [see Fig.~\ref{fig:pre} below] and
non-equilibrium correlation spreading
\cite{greiner2002quantum,cheneau2012light}, charge transport in Fermi-Hubbard
systems~\cite{brown2019bad}, superexchange-driven spin dynamics, spin transport,
and quantum magnetism
\cite{trotzky2008time,greif2013short,hart2015observation,
brown2017spin,mazurenko2017cold}, and orbital or higher-band physics in which
the local orbital structure of the lattice sites becomes an active many-body
degree of freedom
\cite{wirth2011orbital,olschlager2013interaction,kock2015observing,
li2016physics}. 

The high degree of tunability allows for the exploration of a wide range of
physical regimes, from weakly interacting superfluids to strongly correlated
Mott insulators~\cite{jaksch1998cold,greiner2002quantum}. Optical lattices can
be engineered in one, two, or three dimensions. Three-dimensional lattices
provided landmark demonstrations of the bosonic superfluid--Mott transition and
fermionic Mott physics
\cite{greiner2002quantum,jordens2008mott,schneider2008metallic}, while
subsequent optical-lattice experiments resolved short-range antiferromagnetic
correlations in Hubbard systems
\cite{greif2013short,hart2015observation}. Although cubic lattices are the most
direct extension to three dimensions, lower-dimensional optical lattices offer
particularly flexible geometric design: square, triangular, honeycomb, kagome,
and other superlattice or synthetic-gauge-field geometries can be engineered
using appropriately arranged laser configurations
[FIG.~\ref{fig:placeholder}(d)]
\cite{bloch2008many,bloch2012quantum,gross2017quantum,
becker2010ultracold,struck2011quantum,tarruell2012creating,jo2012ultracold,
aidelsburger2013realization,miyake2013realizing}. These capabilities make
optical lattices powerful tools for simulating strongly correlated systems,
including quantum magnetism, non-equilibrium relaxation, correlation spreading,
and transport processes
\cite{simon2011quantum,greif2013short,trotzky2659probing,
cheneau2012light}.

Interactions in optical-lattice systems arise from direct atomic collisions,
typically short-range contact interactions whose strength can be tuned using
Feshbach resonances. Combined with quantum tunneling between neighboring lattice
sites, these interactions naturally lead to Bose--Hubbard and Fermi--Hubbard
models~\cite{chin2010feshbach,jaksch1998cold} (also see Figs.~\ref{fig:strong},\ref{fig:mbl} below). This enables the exploration of
a wide variety of correlated regimes, including superfluid--Mott-insulator
transitions and fermionic Mott physics
\cite{greiner2002quantum,stoferle2004transition,jordens2008mott}. Cold-atom
simulators also allow control over the filling fraction and provide
density-resolved probes of Hubbard physics, enabling studies of spin correlations
and magnetism as a function of local density
\cite{sherson2010single,hart2015observation,mazurenko2017cold}. Using
laser-assisted tunneling and artificial gauge fields, optical lattices can
emulate topological band structures and synthetic magnetic-field physics
\cite{jaksch2003creation,dalibard2011colloquium,goldman2014light,
aidelsburger2011experimental}. More broadly, Floquet engineering and controlled
dissipation provide routes to non-equilibrium and
non-Hermitian phenomena, including Floquet fractionalization mechanisms, band non-linearity transitions, 
non-Hermitian pumping, and non-Hermitian skin-effect physics
\cite{lee2018floquet,qin2023light,qin2024light,liang2022dynamic,shen2025non,qin2024kinked,zhao2025two,hu2026boundary}.

The ability to engineer synthetic magnetic fields and perform site-resolved
preparation and detection in optical lattices has opened a controlled route to
quantum-Hall-related physics with ultracold atoms. For example, Raman-assisted
tunneling can realize Harper--Hofstadter-type lattice Hamiltonians with complex
hopping phases
\cite{jaksch2003creation,dalibard2011colloquium,cooper2019topological,
aidelsburger2013realization,miyake2013realizing}. In a recent small-system
realization, L\'eonard \emph{et al.} prepared a lattice version of a bosonic
$\nu=1/2$ Laughlin state with two ultracold atoms on a $4\times4$ square optical
lattice under a controlled synthetic flux~\cite{leonard2023realization}. The
experiment observed several hallmark signatures of Laughlin-type fractional
quantum Hall physics, including suppressed two-body interactions, a vortex
structure in density correlations, and a fractional Hall response. Beyond
lattice implementations, rapid rotation, and engineered confinement have enabled
few-body Laughlin physics in continuum-like ultracold-atom settings, including
the preparation of a Laughlin state of two rapidly rotating fermions in an
optical tweezer~\cite{lunt2024realization}. Although current realizations remain
at modest particle numbers, these experiments already capture key few-body
signatures of fractional quantum Hall matter and establish concrete pathways
toward larger systems through improved loading, cooling, coherent control, and
state-resolved detection.

Moving forward, key challenges for optical lattice platforms include achieving sufficiently low temperatures to access magnetic ordering and low-energy fermionic phases, as well as limited measurement and feedback capabilities during real-time evolution~\cite{bloch2008many,esslinger2010fermi,gross2017quantum}. While quantum gas microscopy has enabled site-resolved measurements, most observables are still accessed at the end of an experimental run~\cite{bakr2009quantum,gross2017quantum,cheuk2015quantum}. Ongoing advances in cooling techniques, synthetic gauge fields, and measurement protocols continue to expand the scope of optical lattice quantum simulators, reinforcing their central role as faithful analog emulators of condensed-matter systems.

\subsubsection{Rydberg atoms in optical tweezer arrays}\label{rydberg}

Rydberg systems based on neutral-atom platforms can be viewed as a specialized branch of the broader cold-atom optical-lattice paradigm. As in
conventional optical-lattice simulators, neutral atoms are confined by optical
potentials generated by laser fields, but the role of atomic motion is
different. In standard optical lattices, atoms are itinerant and can tunnel
between neighboring sites, leading naturally to Bose-- or Fermi--Hubbard
Hamiltonians with hopping and interaction terms. In Rydberg neutral-atom arrays,
by contrast, the atoms are typically pinned at fixed, programmable positions,
often by optical tweezers, so intersite tunneling is negligible in
experiments. The relevant dynamics is instead generated by coherent driving
between internal atomic states and by strong, long-range interactions between
Rydberg excitations. Programmable tweezer arrays therefore provide a spatial
register for realizing effective spin models with highly controllable geometry
and interaction range, allowing individual atoms to be arranged into one- or
two-dimensional geometries
\cite{labuhn2016tunable,bernien2017probing,barredo2018synthetic,barredo2016atom,bloch2008many}.
This makes Rydberg arrays a particularly flexible and scalable realization of
neutral-atom quantum simulation [FIG.~\ref{fig:placeholder}(c)]
\cite{bernien2017probing}. In these systems, the relevant degrees of freedom are typically encoded in
internal atomic states. For digital operation, one may use long-lived hyperfine
ground states as qubit states, while for analog Rydberg simulation the key
degree of freedom is often the transition between a ground state and a highly
excited Rydberg state. The large electric dipole moments of Rydberg states
generate strong, controllable long-range interactions between atoms, enabling
Hamiltonians that are difficult to realize in conventional optical lattices
based only on tunneling and contact interactions
\cite{saffman2010quantum,jaksch2000fast,isenhower2010demonstration,wilk2010entanglement,urban2009observation}.
Unlike superconducting qubits, the physical degrees of freedom are microscopic
atoms that are intrinsically identical across sites, which helps suppress
device-to-device variability and supports the construction of large-scale arrays
with minimal intrinsic disorder.

A defining feature of neutral-atom platforms is that atoms in their electronic
ground states interact only weakly under typical tweezer-array conditions.
Strong and controllable interactions can be introduced by optically exciting
atoms to Rydberg states, which are highly excited electronic states with large
principal quantum number and spatially extended electronic orbitals
\cite{gallagher1988rydberg,saffman2010quantum,low2012experimental}. These
states possess large polarizabilities and transition dipole moments, giving rise
to strong resonant dipole--dipole or off-resonant van der Waals interactions
between atoms excited to Rydberg levels.

A central mechanism is Rydberg blockade, whereby excitation of one atom to a
Rydberg state shifts the Rydberg excitation energy of nearby atoms and thereby
suppresses additional excitations within a characteristic blockade radius [see
Fig.~\ref{fig:scar}(a)]. This
effect produces effective interactions between qubits that are strong,
controllable, and spatially constrained
\cite{lukin2001dipole,jaksch2000fast,urban2009observation,
gaetan2009observation}. Unlike trapped-ions, where long-range interactions are
mediated by collective phonon modes, interactions in Rydberg platforms arise
from direct interatomic interactions in real space. In gate-based operation,
these Rydberg interactions are activated transiently during entangling gates and
are strongly suppressed after the atoms are returned to their ground-state qubit
levels, enabling the construction of digital neutral-atom quantum circuits
\cite{isenhower2010demonstration,omran2019generation,saffman2016quantum}.

Rydberg interactions naturally realize Ising-type spin Hamiltonians with
tunable coupling strengths and interaction ranges. These features enable the
simulation of ordered states in two-dimensional square and triangular arrays,
as well as frustrated spin systems with competing classical ground-state
manifolds
\cite{samajdar2021quantum,li2022quantum,scholl2021quantum}. Related theoretical
work has further clarified the entanglement structure and phase diagrams of
Rydberg arrays beyond simple nearest-neighbor blockade models
\cite{o2023entanglement}. Moreover, because Rydberg blockade imposes hard local
constraints and the tweezer geometry can be flexibly programmed, Rydberg arrays
are particularly well suited for constrained spin models, quantum dimer
physics, and RVB-type/topological-spin-liquid signatures
\cite{semeghini2021probing,giudici2022dynamical,zeng2025quantum}. 

Importantly, the ability to control large two-dimensional Rydberg simulators
makes them a powerful platform for probing finite-size signatures of critical
and non-equilibrium dynamics [see Figs.~\ref{fig:dtc2},
\ref{fig:non-unitary}, and~\ref{fig:stringdynamics}]
\cite{scholl2021quantum,manovitz2025quantum}. Complementary theoretical work
has proposed Rydberg-array routes to two-dimensional quantum critical dynamics,
deconfined criticality, and non-Hermitian critical phenomena
\cite{schmitt2022quantum,bombieri2025deconfined,shen2023proposal}. These
features make Rydberg atom arrays a leading platform for exploring quantum
criticality, constrained non-equilibrium dynamics, and quantum many-body scars.

In practical terms, trapped-ions excel at simulating models with tunable
long-range couplings and at implementing high-fidelity digital algorithms,
whereas neutral-atom and Rydberg systems are particularly powerful for
lattice-based models, constrained dynamics [see Figs.~\ref{fig:scar},
\ref{fig:dtc2}, \ref{fig:non-unitary}, and~\ref{fig:stringdynamics}],
programmable geometry, and large-scale many-body phenomena in one and two
dimensions.

Despite their rapid progress, several aspects of neutral-atom and Rydberg
platforms remain active areas of development. Finite Rydberg-state lifetimes
and sensitivity to laser phase, intensity, and frequency noise limit coherence
during interacting dynamics
\cite{saffman2010quantum,low2012experimental,browaeys2020many}. Achieving
high-fidelity digital gates across large arrays also remains challenging,
particularly when scaling parallel operations in two-dimensional systems
\cite{wilk2010entanglement,levine2018high,evered2023high,saffman2016quantum}.
Interaction strengths and blockade radii impose constraints on achievable
lattice spacings, interaction graphs, and crosstalk-free parallel operations.
Continued advances in laser stability, Rydberg-state engineering, atom
rearrangement, cooling, and control techniques are therefore central to
extending neutral-atom platforms toward more precise and versatile quantum
simulations of strongly correlated matter.

\subsubsection{Photonic Quantum Simulators }\label{photonic}

In photonic platforms, photons act as carriers of quantum information, with
quantum states encoded in degrees of freedom such as polarization, spatial,
temporal, path, orbital-angular-momentum, or frequency modes
\cite{kwiat1995new,pan1998experimental,marcikic2002time,
de2005long,tanzilli2005photonic,flamini2018photonic,wang2020integrated}.
Unlike matter-based qubits, photons offer weak environmental coupling,
low intrinsic decoherence, high-speed manipulation, and compatibility with
room-temperature operation, although deterministic photon--photon interactions
and scalable loss management remain major challenges
\cite{o2007optical,o2009photonic,slussarenko2019photonic,
kok2007linear,flamini2018photonic}. In these platforms, photons propagate
through engineered optical structures, such as waveguide arrays, interferometric
networks, time-multiplexed fiber loops, frequency-bin circuits, and optical or
microwave cavities
\cite{politi2008silica,perets2008realization,peruzzo2010quantum,
broome2013photonic,spring2013boson,carolan2015universal,chang2014quantum}.
State evolution is governed by the interference of probability amplitudes as
photons traverse the optical network, giving a natural mapping to tight-binding
Hamiltonians, continuous- or discrete-time quantum walks, and noninteracting or
weakly interacting bosonic models
\cite{perets2008realization,peruzzo2010quantum,crespi2013anderson,
broome2013photonic,hamilton2017gaussian,sparrow2018simulating}. Photonic
platforms have therefore become powerful tools for simulating quantum walks,
transport phenomena, topological band structures, non-Hermitian dynamics, and
quantum-enhanced sensing
\cite{rechtsman2013photonic,rechtsman2013strain,hafezi2013imaging,
lu2014topological,Ozawa2019,giovannetti2004quantum}. In particular, photonic
quantum walks have been used to probe topological invariants, parity--time
symmetric edge states, dynamical quantum phase transitions, non-Hermitian skin
effects, and criticality-enhanced sensing
\cite{xiao2017observation,zhan2017detecting,wang2019simulating,
xue2023observation,lin2023manipulating,xiao2026critical}.

Effective interactions can be introduced using optical nonlinearities, measurement-induced effects, or coupling to matter degrees of freedom in atoms, quantum dots, or superconducting circuits~\cite{knill2001scheme,kok2007linear,chang2014quantum,faraon2010fast,peyronel2012quantum}.
For instance, in Ref.~\cite{wang2024realization}, the authors engineered a two-dimensional circuit-QED lattice that supports strongly interacting photons under programmable, Floquet-driven couplings. This achievement bridges topological photonics and correlated quantum matter, enabling the assembly and characterization of a photonic analog of the Laughlin state using a local-potential-assisted adiabatic protocol in a reconfigurable superconducting architecture~\cite{wang2024realization}. Another demonstration of 2D QED is shown in Fig.~\ref{fig:lgt} below. Related progress has also been achieved in alternative strongly interacting photonic platforms, where Laughlin-type states of light and associated many-body correlations have been experimentally demonstrated in a twisted optical cavity setting~\cite{clark2020observation}. Compared to ultracold-atom implementations, circuit-QED and cavity-polariton platforms offer superior local addressability and reconfigurability, as well as natural interfaces to microwave/optical control and measurement; conversely, cold-atom lattices provide exceptionally clean Hamiltonian engineering and microscopic control of interactions.

Despite these developments, we point out that the native strength of photonic simulators lies in the high-fidelity realization of coherent, noninteracting quantum evolution, rather than strongly correlated many-body physics, where interactions are typically weaker and more challenging to scale. Active areas of development include the integration of nonlinear optical elements, hybrid photonic–matter systems, and measurement-based protocols that effectively induce interactions~\cite{knill2001scheme,kok2007linear,chang2014quantum,peyronel2012quantum}. Advances in integrated photonics and on-chip fabrication continue to improve stability and scalability, expanding the scope of photonic simulators toward more complex quantum dynamics~\cite{carolan2015universal,spring2013boson,broome2013photonic,flamini2018photonic}.

\subsubsection{Moir\'{e} Quantum Materials }\label{moir}
Moir\'{e} quantum materials provide a conceptually distinct route to quantum
simulation, based on materials-by-design rather than isolated quantum registers.
When two atomically thin crystals are stacked with a small relative twist angle
or lattice mismatch, a long-wavelength Moir\'{e} superlattice emerges, producing
narrow electronic minibands whose bandwidth, interaction strength, symmetry, and
topology can be tuned by twist angle, carrier density, displacement field,
pressure, dielectric environment, and layer alignment
\cite{bistritzer2011moire,kennes2021moire,balents2020superconductivity}.
Their exceptional tunability and clean separation of energy scales allow them to
directly realize effective lattice Hamiltonians, particularly strongly
correlated models, including Hubbard and extended Hubbard models
\cite{tang2020simulation,regan2020mott,xu2020correlated}. Landmark experiments
include the observation of correlated insulating behavior and unconventional
superconductivity in magic-angle twisted bilayer graphene
\cite{cao2018correlated,cao2018unconventional,yankowitz2019tuning,
lu2019superconductors,cao2021nematicity,oh2021evidence}, correlated
ferromagnetic, orbital-magnetic, Chern-insulating, and quantum anomalous Hall
states in graphene-based Moir\'{e} bands
\cite{sharpe2019emergent,serlin2020intrinsic,choi2021correlation}, and Mott
and generalized Wigner-crystal states in transition-metal-dichalcogenide
Moir\'{e} heterostructures
\cite{tang2020simulation,regan2020mott,xu2020correlated}. These experiments
establish Moir\'{e} materials as tunable solid-state quantum simulators in which
the simulated degrees of freedom are real itinerant electrons in engineered
minibands, rather than atoms, ions, photons, or superconducting qubits.

A central feature of Moir\'{e} systems is the formation of flat or nearly flat
electronic bands, which strongly enhance the role of Coulomb interactions
relative to kinetic energy. This has enabled experimental access to a wide range
of correlation-driven phenomena, including Mott-like insulating states,
symmetry-broken phases, unconventional superconductivity, and
interaction-driven topology
\cite{balents2020superconductivity,yankowitz2019tuning,
cao2018correlated,cao2018unconventional,lu2019superconductors,
sharpe2019emergent,serlin2020intrinsic,choi2021correlation,
regan2020mott,xu2020correlated}. In twisted bilayer graphene and related
Moir\'{e} heterostructures, electrostatic gating allows continuous control of
carrier density, while displacement fields, twist-angle engineering, and
substrate alignment tune band dispersion, symmetry, and topology
\cite{cao2018correlated,yankowitz2019tuning,lu2019superconductors,
cao2021nematicity,oh2021evidence}. These capabilities have led to realizations
of correlated insulators and superconductivity proximate to integer fillings, as
well as fractional Moir\'{e} Chern states and optically controllable Chern
ferromagnets
\cite{zeng2023thermodynamic,cai2026optical}.

From the perspective of quantum simulation, Moir\'{e} quantum materials provide
experimentally accessible realizations of strongly correlated lattice models
with tunable microscopic parameters. The long-period Moir\'{e} superlattice
creates narrow electronic bands whose bandwidth, filling, interaction strength,
topology, and symmetry can be controlled through twist angle, electrostatic
gating, displacement fields, dielectric environment, and heterostrain
\cite{kennes2021moire,balents2020superconductivity,yankowitz2019tuning,
cao2018correlated}. This makes these systems closely aligned with long-standing
model Hamiltonians in condensed-matter physics, including Hubbard and
extended-Hubbard models on triangular or honeycomb superlattices, spin and
valley exchange models, flat-band Chern models, and fractional Chern insulator
models
\cite{tang2020simulation,regan2020mott,xu2020correlated,
serlin2020intrinsic,zeng2023thermodynamic,cai2026optical}. In this sense, Moir\'{e}
materials do not simulate an abstract Hamiltonian through externally programmed
gates, but instead realize strongly interacting electron models directly in a
solid-state setting, with continuously tunable parameters and direct access to
correlated insulating states, unconventional superconductivity, charge-ordered
phases, orbital magnetism, and interaction-driven topological phases
\cite{cao2018correlated,yankowitz2019tuning,zeng2023thermodynamic,cai2026optical}.

Despite these strengths, Moir\'{e} platforms differ fundamentally from programmable quantum simulators. They do not support arbitrary Hamiltonian synthesis, controlled initial state preparation, or coherent real-time unitary evolution over long durations. Moreover, environmental coupling, disorder, and phonons play non-negligible roles, limiting access to clean non-equilibrium dynamics. Nevertheless, Moir\'{e} quantum materials occupy a complementary position in the quantum simulation landscape by bridging model Hamiltonians and experimental condensed-matter systems.

\subsection{Commercially available quantum hardware}

Having discussed various physical platforms for quantum simulation, we now provide an overview of what is currently available for the public. This is of exigent importance as over the past decade, quantum computing has transitioned from a primarily academic endeavor into a rapidly expanding commercial enterprise, driven by sustained industrial investment and significant advances in quantum hardware engineering. Multiple quantum technology companies now design, manufacture, and operate programmable quantum processors that are accessible through cloud platforms. These commercially developed systems play a central role in shaping the practical capabilities of quantum computation.

At present, superconducting qubits constitute one of the most mature and widely
deployed commercial platforms for gate-based quantum computing, with a large
industrial ecosystem spanning cloud access, processor development, control
electronics, software stacks, and application-oriented workflows. Among these
efforts, IBM has established a particularly broad publicly accessible
superconducting-quantum ecosystem, operating a sequence of increasingly large
processors and providing cloud access through the IBM Quantum Platform and the
Qiskit software stack
\cite{devitt2016performing,cross2019validating,abughanem2024ibm,wille2019ibm}.
Google has pursued superconducting quantum processors with an emphasis on
hardware-performance benchmarks, quantum computational advantage, quantum
simulation, topological-state preparation, and quantum error correction
\cite{arute2019quantum,satzinger2021realizing,tazhigulov2022simulating,
google2025quantum}. These systems have played a central role in demonstrating
milestones such as quantum computational advantage and in advancing calibration,
error suppression, and scalable chip-design strategies.

Rigetti Computing represents another commercial provider of superconducting
hardware, with a vertically integrated development model spanning qubit
fabrication, cryogenic and control infrastructure, cloud access, and hybrid
quantum--classical workflows~\cite{karalekas2020quantum}. IQM Quantum Computers
has also emerged as a major European full-stack superconducting-quantum provider,
emphasizing both on-premise systems for research centers and high-performance-
computing environments, as well as cloud access through its Resonance platform
\cite{iqmresonance2024}.

Across these companies, the current industrial status of superconducting
quantum hardware is characterized by processors containing tens to hundreds of
physical qubits, operated under continuous calibration and accessed through
cloud-based or on-premise service models. Despite steady improvements in
coherence times, gate fidelities, and calibration automation, algorithmic
performance remains constrained by noise accumulation, crosstalk, limited
connectivity, and device-specific error correlations. As of May 2026,
representative systems include IBM's 156-qubit Heron family and 120-qubit
Nighthawk processor, Google's 105-qubit Willow processor, Rigetti's 84-qubit
Ankaa-3 system, and IQM's Radiance line of superconducting processors, including
20- and 54-qubit systems with a 150-qubit variant targeted for larger
deployments
\cite{ibm_quantum_2026,google2025quantum,rigetti2024ankaa3,iqmresonance2024}.
Reported single-qubit gate errors are commonly in the $10^{-4}$--$10^{-3}$
range, whereas two-qubit gate errors generally remain in the
$10^{-3}$--$10^{-2}$ range, with substantial variation across devices, gate
types, calibration cycles, and benchmarking protocols.

Trapped-ion quantum computers represent an alternative commercial approach,
distinguished by long coherence times, high-fidelity operations, and flexible
connectivity. IonQ platforms are accessed through cloud services and are often
evaluated using application-level performance metrics in addition to raw qubit
counts; for example, IonQ Forte has been benchmarked as a 30-qubit trapped-ion
processor with all-to-all operations using both component-level randomized
benchmarking and Algorithmic Qubit benchmarks
\cite{chen2024benchmarking,teegarden2026three}. Quantinuum, formed through the
merger of Honeywell Quantum Solutions and Cambridge Quantum, operates
trapped-ion hardware based on QCCD-style architectures, with a strong emphasis
on high-fidelity control, system-level benchmarking, error characterization,
mid-circuit measurement, and conditional logic
\cite{pino2021demonstration,moses2023race,iqbal2024topological,
hothem2025measuring}. Its commercial systems support features such as qubit
reset, feed-forward operations, high-fidelity measurement, and qubit reuse,
enabling sophisticated protocols that combine unitary gates with
measurement-conditioned dynamics. From an industrial perspective, the current
status of trapped-ion quantum hardware is defined by moderate physical-qubit
counts combined with long coherence times, high connectivity, high-fidelity
measurement, and strong circuit-level control. Although gate operations are
typically slower than those in superconducting processors, the reduced routing
overhead associated with high connectivity, together with high fidelities and
long coherence times, can enable deep effective circuits for selected algorithms
and quantum-simulation protocols.

Neutral-atom quantum processors based on optical tweezer arrays and Rydberg
interactions have also entered the commercial domain. QuEra has focused on
large-scale neutral-atom processors that emphasize high qubit counts,
reconfigurable two-dimensional atom arrays, and programmable analog dynamics.
Its Aquila system is designed as a field-programmable neutral-atom processor
that allows users to implement time-dependent Rydberg Hamiltonians relevant to
quantum simulation and combinatorial optimization
\cite{wurtz2023aquila,ebadi2021quantum,ebadi2022quantum}. Pasqal has pursued a
complementary full-stack strategy based on neutral-atom hardware and
pulse-level software abstractions for programming analog and digital--analog
neutral-atom protocols~\cite{silverio2022pulser}. In addition to analog
simulation capabilities, such software interfaces allow users to design,
simulate, and optimize pulse sequences that map optimization and simulation
problems onto Rydberg-mediated interactions. This industrial approach
highlights the potential of neutral-atom hardware to bridge analog and digital
paradigms. From a commercial perspective, the current status of neutral-atom
quantum hardware is defined by rapid scaling in qubit number, geometric
reconfigurability, and strong native interactions, while deep universal
gate-model operation across large arrays remains an active area of development.
As a result, these platforms occupy a distinct niche within the quantum
computing industry.

Photonic quantum computing represents another class of commercially developed
quantum hardware. Xanadu has focused on photonic processors based primarily on
continuous-variable encodings and Gaussian optical states, with hardware
demonstrations optimized for sampling tasks such as Gaussian boson sampling
\cite{madsen2022quantum}. Its photonic hardware has been closely integrated with
software tools such as Strawberry Fields and PennyLane, which support the design,
simulation, optimization, and hybrid quantum--classical programming of photonic
and more general quantum workflows
\cite{killoran2019strawberry,bergholm2018pennylane}. PsiQuantum 
is pursuing a long-term architecture centered on fault-tolerant photonic quantum
computing using single-photon qubits, fusion-based computation, and large-scale
integrated photonic circuits
\cite{knill2001scheme,bartolucci2023fusion,psiquantum2025manufacturable}.
Rather than emphasizing near-term cloud access to small processors, this strategy
focuses on manufacturable photonic modules and the engineering challenges needed
for scale, including heralded photon generation, ultra-low-loss photonic
circuits, high-performance detection, chip-to-chip interconnects, switching, and
scalable error correction.

\subsection{Basic building blocks of digital quantum simulation}
The experimental platforms discussed above support a spectrum of quantum-simulation paradigms, ranging from analog Hamiltonian emulation to fully programmable digital computation.  In this review, however, we place particular emphasis on  {digital} quantum simulation, since most algorithmic tools for condensed-matter applications are naturally formulated in the language of gate-based quantum simulation.  In the digital setting, the target time evolution or state-preparation protocol is compiled into a sequence of elementary operations acting on well-defined two-level (qubit) systems, allowing one to implement model Hamiltonians in a controlled manner ~\cite{feynman2018simulating,suzuki1992general,berry2007efficient}.

Thus, to establish a common language for this review, we briefly introduce the basic primitives of digital quantum simulation. In brief, qubits serve as the computational degrees of freedom used to encode spins, fermionic modes, and auxiliary registers. Quantum gates are elementary operations, typically unitary transformations, together with measurement and reset primitives, that implement local dynamics and generate entangling interactions. By composing these gates in order, one obtains quantum circuits: ordered sequences that approximate continuous-time evolution, prepare many-body ground states, or implement hybrid protocols with measurements and feedforward control. In this subsection, we introduce these concepts in more detail.

Although this language is most naturally associated with gate-based digital
quantum simulation, it is also useful more broadly. Some analog and
digital-analog platforms can also be described in terms of effective qubits,
elementary control operations, measurement primitives, and circuit-like
sequences of pulses or quenches.  A simple example is provided by Rydberg atom arrays: although their native
dynamics are often described by an analog Hamiltonian with laser driving and
blockade interactions, the blockade mechanism can also realize effective
controlled operations \cite{jaksch2000fast,isenhower2010demonstration,wilk2010entanglement,
saffman2010quantum,saffman2016quantum,levine2018high,evered2023high}. This illustrates how gate-level language can emerge from analog interaction
physics. Therefore, the circuit language introduced here should not be viewed as
restricted to universal gate-based processors. Rather, it provides a common
operational framework for comparing digital, analog, and hybrid approaches to
quantum simulation.

\subsubsection{Qubits}
In quantum hardware, the basic unit of information is a qubit, which can occupy a superposition of two basis states. A single-qubit can be represented as the state vector
\begin{equation}
	|\psi\rangle=a|0\rangle+b|1\rangle.
\end{equation}
where $1$ or $0$ label two well-defined physical states of the device—for example, two internal atomic levels in neutral-atom (Rydberg) platforms, or two quantized energy levels of an artificial atom (a superconducting qubit)~\cite{nielsen2010quantum,preskill2018quantum}. {The complex amplitudes $a$ and $b$ encode probability amplitudes. When the qubit is measured in the $1$ or $0$ basis (i.e. computational basis), the outcomes occur with probabilities $|a|^{2}$ or $|b|^{2}$. Beyond these probabilities, the relative phase between $a$ and $b$ is also a uniquely quantum feature. For a single isolated qubit measured only in the computational basis, this phase may not be directly visible. However, it becomes crucial when qubits interfere or interact with others. This relative phase governs quantum interference, allowing superpositions to add constructively or destructively in multi-qubit circuits. This interference determines how amplitudes combine in multi-qubit circuits, and, in the context of condensed-matter simulation, it controls how coherent superpositions faithfully encode many-body dynamics and correlation patterns of the target model.

A qubit is mathematically equivalent to a spin-$1/2$ degree of freedom. Its
computational basis states, $\ket{0}$ and $\ket{1}$, are conventionally chosen
as eigenstates of the Pauli-$Z$ operator. Depending on the simulation
problem, these two basis states may represent different physical degrees of
freedom: the up and down states of a localized spin, the empty and occupied
states of a hard-core bosonic mode, or the occupation of a fermionic mode after
a fermion-to-qubit mapping. For example, in one-dimensional fermionic models
such as the Kitaev chain, the Jordan--Wigner transformation maps fermionic
creation and annihilation operators to strings of Pauli operators acting on
qubits; more general encodings include parity and Bravyi--Kitaev mappings~\cite{jordan1928paulische,somma2003quantum,bravyi2002fermionic,tranter2015b}.
Thus, qubits provide a universal discrete representation of many-body degrees
of freedom once the appropriate encoding is specified. The state of a single-qubit can be visualized on the Bloch sphere. The basis states
$\ket{0}$ and $\ket{1}$ occupy the north and south poles, while superposition
states lie elsewhere on the sphere.

The effect of measurements fundamentally distinguishes quantum from classical information. Measuring a qubit in the computational basis yields a probabilistic outcome: $|0\rangle$ and $|1\rangle$ with probability $|a|^2$ and $|b|^2$ respectively, and the post‑measurement state collapses to the observed outcome. This departure from classical behavior becomes even more striking when multiple qubits are considered. In classical systems, the state of an $n$-bit register is fully specified by listing each bit’s value. In contrast, the state of an $n$-qubit system lives in a $2^n$-dimensional complex vector space, where qubits can exhibit correlations that are not just probabilistic, but fundamentally nonlocal. These non-factorizable states are known as entangled states. They can encode joint properties that cannot be reduced to individual qubits, and underlie the exponential expressive power of quantum systems~\cite{nielsen2010quantum,horodecki2009quantum,jozsa2003role}.

Concretely, a general two-qubit pure state can be written as
\begin{equation}
    \sum_{i,j\in\{0,1\}}\alpha_{ij}|i\rangle\otimes|j\rangle=\alpha_{00}|00\rangle+\alpha_{01}|01\rangle+\alpha_{10}|10\rangle+\alpha_{11}|11\rangle
\end{equation}
where the complex amplitudes, $\alpha_{ij}$, encode both probabilities and relative phases across joint basis states. This tensor‑product structure enables entanglement, a defining feature of multi‑qubit quantum systems. An entangled state cannot be expressed as a product of single‑qubit states, reflecting correlations that have no classical analog. A paradigmatic example is the Bell state~\cite{bohm2012quantum,bennett1993teleporting}
\begin{equation}
    \frac{1}{\sqrt{2}}(|00\rangle+|11\rangle)\label{bell}
\end{equation}
which cannot be factorized into a product of two single-qubit states, i.e. $(c_1|0\rangle+c_2|1\rangle)\otimes(c_3|0\rangle+c_4|1\rangle)$. The measurement outcomes for the Bell state on the two qubits are perfectly correlated, even though neither qubit possesses a definite state on its own before measurement. Importantly, these correlations are not simply statistical, but arise from coherent superpositions across the composite Hilbert space.

Beyond a few qubits, quantum entanglement is a crucial concept in characterizing many-body condensed matter~\cite{amico2008entanglement,eisert2010colloquium,chen2010local,kitaev2006topological,peschel2009reduced,herviou2019entanglement,lee2022exceptional}. Ground states and low‑energy excitations of interacting many‑body Hamiltonians are generically entangled, and their physical properties, such as correlation and response functions, are encoded in this entanglement structure. This makes quantum entanglement a central resource in quantum simulation, which hinges on the ability to prepare, manipulate, and measure entangled multi‑qubit states. A detailed description of how quantum entanglement is defined and probed on quantum hardware is given in Sect.~\ref{sec:EE}.

\subsubsection{Quantum gates}

To simulate quantum systems algorithmically, we need precise control over how quantum states evolve. This control is achieved through quantum gates, which are physical operations that implement unitary transformations on one or more qubits. These gates form the building blocks of quantum circuits, just as logic gates assemble to form classical digital circuits~\cite{deutsch1989quantum,barenco1995elementary,lloyd1996universal}.

We start by describing single-qubit gates, which modify the state of a single
qubit and are represented by $2\times 2$ unitary matrices. The simplest example
is the identity gate,
\begin{equation}
I =
\begin{pmatrix}
1 & 0 \\
0 & 1
\end{pmatrix},
\end{equation}
which leaves every quantum state unchanged. The three Pauli gates are
\begin{equation}
X =
\begin{pmatrix}
0 & 1 \\
1 & 0
\end{pmatrix},
\qquad
Y =
\begin{pmatrix}
0 & -i \\
i & 0
\end{pmatrix},
\qquad
Z =
\begin{pmatrix}
1 & 0 \\
0 & -1
\end{pmatrix}.
\end{equation}
Up to an overall global phase, these gates correspond to $180^\circ$ rotations
about the $x$, $y$, and $z$ axes of the Bloch sphere, respectively. Their action
on the computational basis is
\begin{align}
X\ket{0} &= \ket{1}, &
X\ket{1} &= \ket{0}, \\
Y\ket{0} &= i\ket{1}, &
Y\ket{1} &= -i\ket{0}, \\
Z\ket{0} &= \ket{0}, &
Z\ket{1} &= -\ket{1}.
\end{align}
Thus, the $X$ gate acts as a bit-flip gate, interchanging $\ket{0}$ and
$\ket{1}$. The $Z$ gate acts as a phase-flip gate, leaving $\ket{0}$ unchanged
while multiplying $\ket{1}$ by a relative phase $-1$. The $Y$ gate combines a
bit flip with a phase factor, mapping $\ket{0}$ to $i\ket{1}$ and $\ket{1}$ to
$-i\ket{0}$.

Another important single-qubit gate is the Hadamard gate,
\begin{equation}
H =
\frac{1}{\sqrt{2}}
\begin{pmatrix}
1 & 1 \\
1 & -1
\end{pmatrix}.
\end{equation}
It creates equal superpositions of computational-basis states:
\begin{equation}
H\ket{0}
=
\frac{\ket{0}+\ket{1}}{\sqrt{2}},
\qquad
H\ket{1}
=
\frac{\ket{0}-\ket{1}}{\sqrt{2}}.
\end{equation}
Equivalently, the Hadamard gate changes between the $Z$ and $X$ bases, since
\begin{equation}
H Z H = X,
\qquad
H X H = Z.
\end{equation}
More generally, arbitrary rotations about the Bloch-sphere axes are defined as
\begin{align}
R_x(\theta)
&=
e^{-i\theta X/2}
=
\cos\frac{\theta}{2} I
-
i\sin\frac{\theta}{2} X,
\\
R_y(\theta)
&=
e^{-i\theta Y/2}
=
\cos\frac{\theta}{2} I
-
i\sin\frac{\theta}{2} Y,
\\
R_z(\theta)
&=
e^{-i\theta Z/2}
=
\cos\frac{\theta}{2} I
-
i\sin\frac{\theta}{2} Z.
\end{align}
For $\theta=\pi$, these rotations reproduce the Pauli gates up to global
phases:
\begin{equation}
R_x(\pi)=-iX,
\qquad
R_y(\pi)=-iY,
\qquad
R_z(\pi)=-iZ.
\end{equation}

In physical quantum processors, arbitrary single-qubit rotations are usually
not implemented as independent primitive operations. Instead, they are compiled
into the hardware-native gate set. In superconducting qubits, rotations around
axes in the equatorial plane of the Bloch sphere are typically generated by
resonant microwave pulses with controlled amplitude, duration, and phase, while
$R_z$ rotations are commonly implemented virtually by updating the phase
reference frame of subsequent microwave pulses. In trapped-ion systems,
single-qubit rotations are similarly produced by resonant laser or microwave
drives between the qubit states. At the circuit level, an arbitrary single-qubit
unitary can be decomposed into Euler rotations, for example $e^{i\alpha}
R_z(\phi) R_y(\theta) R_z(\lambda),$
and then synthesized from the calibrated native operations of the specific
hardware platform.

Beyond single-qubit gates are two-qubit gates which allow qubits to interact. Frequently, they are given by controlled gates where the state of one qubit determines whether an operation is applied to another qubit. The most common type is the controlled-NOT gate, also known as CNOT or CX gate, which only performs the X-gate on the second qubit if the first qubit is $\ket{1}$. In the basis of $(\ket{00},\ket{01},\ket{10},\ket{11})$, this gate is represented as 
\begin{equation}
	\mathrm{CX}=\left[\begin{array}{llll}
		1 & 0 & 0 & 0 \\
		0 & 1 & 0 & 0 \\
		0 & 0 & 0 & 1 \\
		0 & 0 & 1 & 0
	\end{array}\right].
\end{equation}
Multi-qubit gates are essential to create entanglement, which single-qubit gates cannot do on their own. Let's consider a product state of two qubits
\begin{equation}
    \frac{1}{\sqrt{2}}(|0\rangle+|1\rangle)\otimes|0\rangle
\end{equation}
There is no entanglement yet. Applying the cNOT gate yields the Bell state (Eqn. \ref{bell}), which is entangled. Any sequence of single-qubit gates only acts independently on each qubit. Such product unitaries cannot create non-separable states from product inputs. Hence, two-qubit gates are the minimum requirement for encoding physical interactions.

Another important two-qubit gate is the SWAP gate, which exchanges the quantum
states of two qubits. In the computational basis
$\{\ket{00},\ket{01},\ket{10},\ket{11}\}$, its action is
\begin{equation}
\mathrm{SWAP}\ket{a,b}
=
\ket{b,a},
\qquad
a,b\in\{0,1\},
\end{equation}
and its matrix representation is
\begin{equation}
\mathrm{SWAP}
=
\begin{pmatrix}
1 & 0 & 0 & 0 \\
0 & 0 & 1 & 0 \\
0 & 1 & 0 & 0 \\
0 & 0 & 0 & 1
\end{pmatrix}.
\end{equation}
Although the SWAP gate is conceptually simple, it plays an important role in
hardware compilation because it allows quantum information to be routed between
qubits that are not directly connected by the native device topology.

Other multi-qubit gates can generally be decomposed into basic single- and
two-qubit gates. A widely used example is the {Toffoli gate}, or
controlled-controlled-NOT (CCX), which flips the state of a target qubit if and
only if two control qubits are both in the state $\ket{1}$. More generally, a
$k$-controlled unitary gate, denoted $\mathrm{C}^k U$, applies the operation
$U$ to the target subsystem conditioned on all $k$ control qubits being in the
state $\ket{1}$. It can be written as
\begin{equation}
\mathrm{C}^k U
=
\ket{1}\bra{1}^{\otimes k}\otimes U
+
\left(
I^{\otimes k}
-
\ket{1}\bra{1}^{\otimes k}
\right)
\otimes I_{\rm tar},
\end{equation}
where $I_{\rm tar}$ is the identity operator acting on the target subsystem.
This expression makes clear that $U$ is applied only in the control subspace
$\ket{11\cdots 1}$, while the target subsystem is left unchanged for all other
control configurations.

The quantum gates described above are by no means exhaustive, and different linear combinations of them are physically implemented as native entangling gates on different hardware platforms:
\begin{itemize}
    \item \textbf{ECR (Echoed Cross-Resonance) Gate}: This is a hardware-native
CNOT-like entangling gate used in fixed-frequency superconducting-transmon
architectures. In the idealized description, the desired cross-resonance
interaction generates a conditional rotation of the target qubit,
\begin{equation}
    U_{\mathrm{CR}}(\theta)
    =
    \exp\!\left(-i\frac{\theta}{2} Z\otimes X\right).
\end{equation}
In practice, the driven cross-resonance Hamiltonian contains additional
single-qubit and two-qubit terms. Echoed cross-resonance sequences use
interleaved $\pi$ pulses and calibrated pulse phases to reinforce the desired
$ZX$ interaction while canceling or suppressing unwanted coherent terms,
thereby producing a maximally entangling native gate that is locally equivalent
to a CNOT gate
\cite{chow2011simple,sheldon2016procedure,sundaresan2020reducing}.

    \item \textbf{iSWAP Gate}: This typs of gates can be engineered by a system with $XY$-type coupling
    \begin{equation}
        H_{XY} = g(X\otimes X + Y\otimes Y),
    \end{equation}
    where $g$ is the coupling strength. Evolving for a time $t=\pi/(4g)$ leads to
    \begin{equation}
        \mathrm{iSWAP} =
        \begin{bmatrix}
        1 & 0 & 0 & 0 \\
        0 & 0 & i & 0 \\
        0 & i & 0 & 0 \\
        0 & 0 & 0 & 1
        \end{bmatrix},
    \end{equation}
    which swaps the $|01\rangle$ and $|10\rangle$ states while adding a phase $i$.
    \item \textbf{Controlled-Z (CZ) Gate:} This type of gate is commonly implemented in Rydberg atom arrays. The CZ gate in the computational basis $(|00\rangle, |01\rangle, |10\rangle, |11\rangle)$ is
    \begin{equation}
        \mathrm{CZ} =
        \begin{bmatrix}
        1 & 0 & 0 & 0 \\
        0 & 1 & 0 & 0 \\
        0 & 0 & 1 & 0 \\
        0 & 0 & 0 & -1
        \end{bmatrix}.
    \end{equation}
\end{itemize}

\subsubsection{Quantum circuits}\label{circuit}
While individual quantum gates provide the building blocks of quantum control, it is their orchestrated combination into circuits that enables practical computation. Much like how classical programs are built from logic gates assembled into circuits, quantum algorithms are implemented as sequences of gates acting on an initialized state, followed by measurement. In standard circuit diagrams, each qubit is depicted as a horizontal line, and quantum gates are placed along these lines in temporal order from left to right.

In the context of quantum simulation, these circuits are designed to approximate processes such as Hamiltonian time evolution, ground-state preparation, or measurement of observables. Designing efficient circuits is a central challenge in practical quantum simulation given hardware limitations such as gate fidelity and qubit connectivity. Very often, one has to perform a tradeoff between system size or particle number with the cost of incurring more noise and computational time (see Sect.~\ref{error} for a detailed discussion).

To use a quantum computer as a simulator for condensed matter systems, we must first represent the physical degrees of freedom (spins, fermions, bosons) in terms of qubit states~\cite{bravyi2002fermionic,jordan1928paulische,tranter2015b,somma2003quantum}. This mapping is not unique and depends on the structure of the model and the simulation goals. The goal is to construct a qubit Hamiltonian whose dynamics and eigenstates correspond to those of the physical system.

Below, we first illustrate a minimal implementation of models built from spin-$1/2$ degrees of freedom, such as the transverse field Ising (TFI) model or Heisenberg chains, where the mapping to qubits is direct. Each spin is naturally encoded as a single-qubit. Hamiltonian terms correspond directly to single- or two-qubit Pauli operators, and can be implemented using standard gate sets. Many-body spin models thus form the most natural testbed for near-term quantum simulation~\cite{blatt2012quantum,aspuru2012photonic,smith2019simulating}. As a representative example, the TFI model on a 1D chain reads
\begin{equation}
\hat H_{\mathrm{TFIM}}
=
- J\sum_{i=1}^{L-1}Z_iZ_{i+1}
- h\sum_{i=1}^{L}X_i ,
\label{eq:TFIM}
\end{equation}
where the interaction term is a two-qubit operator, and the transverse field is a single-qubit rotation generator.  A
first-order product-formula step (Trotter decomposition) for time evolution
\begin{equation}
e^{-i\hat H_{\mathrm{TFIM}} t}
\approx
\left[
\prod_{i=1}^{L-1} e^{\,i J \Delta t\,Z_i Z_{i+1}}
\prod_{i=1}^{L} e^{\,i h \Delta t\,X_i}
\right]^{n},
\Delta t=t/n,
\label{eq:TFIM_trotter}
\end{equation}
reduces to a sequence of single-qubit $X$-rotations and two-qubit $ZZ$ entangling gates (see Section.~\ref{trotter} for more details).

Next we discuss the realization of fermionic systems~\cite{koh2022stabilizing}, where extra care has to be taken to account for the fermionic anti-commutation relations. To simulate such systems with qubits, one must use fermion-to-qubit mappings that preserve the algebra of the creation and annihilation operators
\begin{equation}
\{c_i,c_j\}=0,\qquad
\{c_i^\dagger,c_j^\dagger\}=0,\qquad
\{c_i,c_j^\dagger\}=\delta_{ij},
\label{eq:fermion_anticomm}
\end{equation}
while rewriting the Hamiltonian in terms of Pauli strings. The most widely used transformations are the
Jordan--Wigner (JW) mapping~\cite{jordan1928paulische,bravyi2002fermionic} and the Bravyi--Kitaev (BK) mapping~\cite{bravyi2002fermionic,tranter2015b}.
In the JW mapping, a fermionic mode $j$ is encoded into a qubit $j$ via
\begin{equation}
c_j
=
\left(\prod_{\ell<j} Z_\ell\right)\frac{X_j+iY_j}{2},
\qquad
c_j^\dagger
=
\left(\prod_{\ell<j} Z_\ell\right)\frac{X_j-iY_j}{2},
\label{eq:JW_map}
\end{equation}
and the occupation operator becomes
\begin{equation}
n_j=c_j^\dagger c_j=\frac{\mathbb{I}-Z_j}{2}.
\label{eq:number_op}
\end{equation} The nonlocal parity string $\prod_{\ell<j}Z_\ell$ ensures that fermionic sign changes are correctly reproduced,
but leads to Pauli strings whose length scales linearly with system size. This non-locality leads to greater circuit depth, but is intrinsically necessary for capturing the antisymmetric fermionic statistics.

In general, by combining multiple quantum gates into a quantum circuit, a
gate-based hardware platform can approximate a desired unitary transformation.
A platform is said to support universal quantum computation if its available
operations form a  universal gate set, meaning that arbitrary unitaries on
many qubits can be decomposed, exactly or approximately, into a sequence of
gates from that set~\cite{barenco1995elementary,nielsen2010quantum}.   This is a
central distinction between a programmable gate-based quantum computer and a
more restricted quantum simulator.

A standard universal construction consists of arbitrary single-qubit rotations
together with one entangling two-qubit gate, such as CNOT or CZ, for example $\{R_x(\theta),R_y(\theta),R_z(\theta),\mathrm{CNOT}\}$. Equivalently, one may use CZ instead of CNOT. In fault-tolerant quantum computing, it is often useful to use a finite
approximately universal gate set~\cite{boykin1999universal}. A widely used example is the Clifford+$T$
gate set, $\{H,S,T,\mathrm{CNOT}\},$
where
\begin{equation}
S=
\begin{pmatrix}
1&0\\
0&i
\end{pmatrix},
\qquad
T=
\begin{pmatrix}
1&0\\
0&e^{i\pi/4}
\end{pmatrix}.
\end{equation}

\section{Phenomena for quantum simulation}\label{sec1}

Having previously reviewed the key physical platforms and concepts for quantum simulation, we now describe the physical phenomena and states that have been simulated digitally on quantum hardware platforms. Broadly, they can be divided into state-preparation [Sect.~\ref{sec1:groundstate}] or dynamical evolution problems, which we further classify based on their key mechanism: topological [Sect.~\ref{sec1:topological}] or non-equilibrium [Sect.~\ref{sec1:noneq}], which can include time-modulated, non-Hermitian or measured protocols. This section will be mainly focused on the physical results from quantum simulation; the next section [Sect.~\ref{sec:methodology}] will elaborate on the more involved aspects of the methodology.

\subsection{Ground-State Problems}
\label{sec1:groundstate}
The ground-state problem involves identifying the lowest-energy state of a
quantum many-body Hamiltonian and is central to understanding low-temperature
phases, correlation effects, and emergent order in interacting quantum systems
\cite{sachdev1999quantum,white1992density,schollwock2011density,
verstraete2009quantum,vidal2004efficient,cirac2021matrix,
hastings2007area,orus2014practical}. At the same time, it is one of the
canonical hard problems in quantum many-body physics: even deciding properties
of the ground state of a local Hamiltonian can be computationally intractable in
general
\cite{kempe2006complexity,osborne2012hamiltonian,gharibian2015quantum,
cubitt2015undecidability}. It encompasses key topics of current interest,
including quantum spin liquids
\cite{anderson1973resonating,balents2010spin,zhou2017quantum,
savary2016quantum,knolle2019field,broholm2020quantum,kitaev2006anyons},
topologically ordered states
\cite{wen1990topological,wen2004quantum,levin2006detecting,
kitaev2006topological,levin2005string,kitaev2003fault}, including
non-Abelian topological orders~\cite{moore1991nonabelions,nayak2008non}, and
symmetry-protected topological phases
\cite{affleck1987rigorous,gu2009tensor,pollmann2010entanglement,
pollmann2012symmetry,schuch2011classifying,chen2013symmetry,
senthil2015symmetry}.

From the standpoint of quantum simulation, ground-state preparation problems can
be broadly separated into structured states with efficient exact circuit
descriptions, or efficient idealized preparation procedures, and generic
strongly correlated states that require approximate preparation. The first class
includes stabilizer and graph states, such as cluster states
\cite{raussendorf2003measurement,walther2005experimental}, experimentally
prepared symmetry-protected or Floquet symmetry-protected states
\cite{choo2018measurement,zhang2022digital}, ground states of
commuting-projector Hamiltonians such as the toric code and related topological
codes
\cite{kitaev2003fault,dennis2002topological,nussinov2009symmetry,
iqbal2024topological}, and certain one-dimensional matrix-product states that
can be generated sequentially or represented by finite-bond-dimension tensor
networks
\cite{fannes1992finitely,schon2005sequential,perezgarcia2007matrix,
cirac2021matrix}. Recent quantum-processor experiments have further extended
this direction to the preparation and manipulation of Abelian and non-Abelian
topological states, including measurement-and-feed-forward toric-code
preparation, $D_4$ non-Abelian topological order on trapped-ions, Fibonacci
anyon braiding on a superconducting processor, and Floquet non-equilibrium
topological order
\cite{iqbal2024topological,iqbal2024non,xu2024non,will2025probing}.

The second class consists of more complicated ground states that are not known
to admit simple exact stabilizers, commuting-projector, or low-bond-dimension
tensor-network constructions. Such states arise in strongly correlated systems
with frustration, fermionic sign structures, long-range correlations, competing
orders, or noncommuting and chiral forms of intrinsic topological order. In
practice, their realization on quantum hardware typically relies on approximate
state-preparation strategies, including variational quantum eigensolvers and
adaptive ansatz-construction methods
\cite{peruzzo2014variational,mcclean2016theory,kandala2017hardware,
cerezo2021variational,tilly2022variational,grimsley2019adapt}, as well as
quantum imaginary-time-evolution approaches and their variational, adiabatic,
and double-bracket extension
\cite{motta2020determining,mcardle2019variational,hejazi2024adiabatic,
gluza2026double}.

\begin{figure*}
    \centering
    \includegraphics[width=0.99\linewidth]{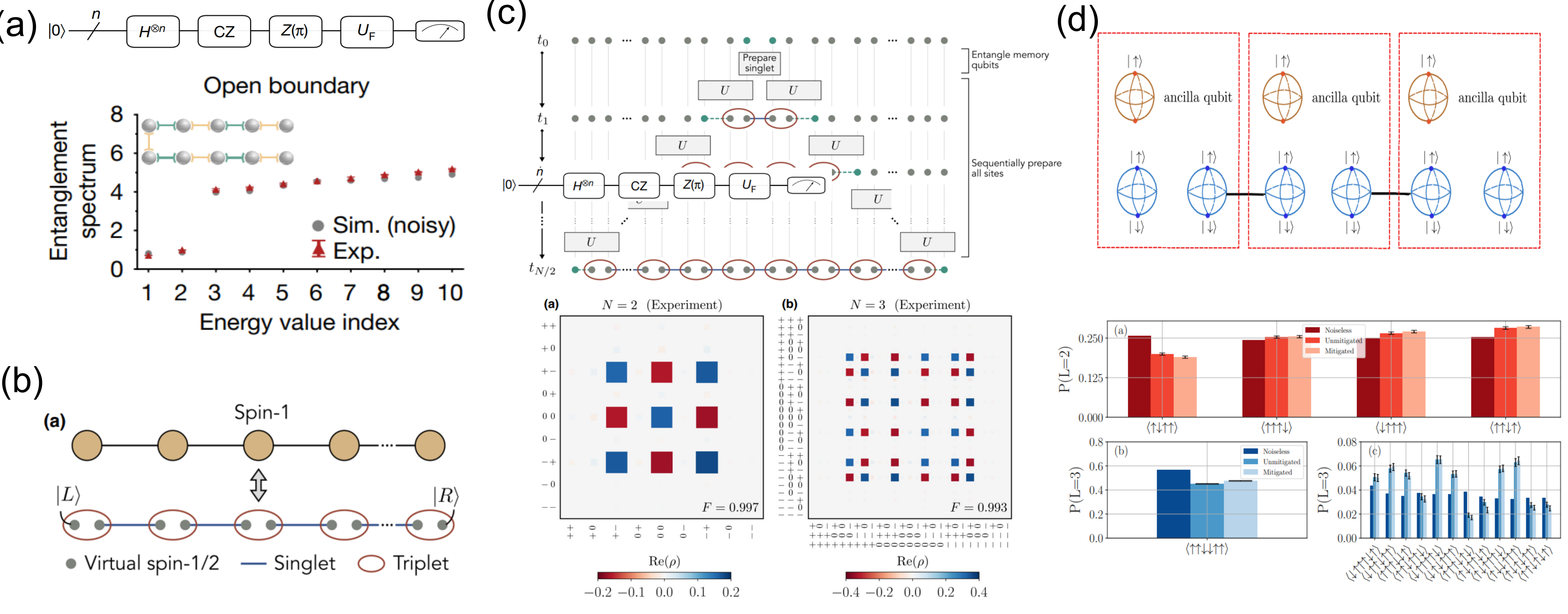}
    \caption{ Experimental demonstrations of exactly solved SPT and AKLT-type quantum states on noisy quantum hardware. (a) Preparation and characterization of a one-dimensional symmetry-protected topological (SPT) state using a circuit composed of global Hadamard gates and controlled-Z gates. The lower plot shows the corresponding entanglement spectrum extracted from experimental measurements (red triangles) and noisy simulations (gray circles), revealing the expected low-lying degeneracies associated with the open-boundary SPT phase.~\cite{zhang2022digital}. (b) Conceptual illustration of the Affleck–Kennedy–Lieb–Tasaki (AKLT) construction, where each physical spin-1 is represented by two virtual spin-1/2 degrees of freedom forming singlet bonds. (c) Sequential protocol for preparing and probing larger AKLT chains using entangled memory qubits (top) together with site-by-site initialization and controlled-unitary operations (bottom). The experimentally reconstructed density matrices illustrate the robustness of this method~\cite{smith2023deterministic} (d)  Ancilla-assisted projector-based preparation of AKLT chains. Top: circuit-level schematic of the local building block used to implement the non-unitary projection.  Bottom: experimental benchmarks comparing the ideal noiseless reference with hardware executions~\cite{chen2023high}. Panel (a) and (b) are adopted from Ref.~\cite{zhang2022digital}. Panel (c) is adopted from Ref.~\cite{smith2023deterministic}.  Panel (d) is adopted from~\cite{chen2023high}.}
    \label{fig:spt}
\end{figure*}

\subsubsection{Ground-state preparation for solvable models}\label{solve}

For certain models, the ground-state wavefunction is known exactly, or has an
exact tensor-network, stabilizer, or commuting-projector description, making it
possible to design efficient state-preparation protocols on quantum hardware.
Among such ``exactly solvable'' paradigms, the cluster state, the toric code, and the
Affleck--Kennedy--Lieb--Tasaki (AKLT) chain stand out as canonical examples of illustrating measurement-based quantum computation, intrinsic topological
order, and symmetry-protected phases
\cite{raussendorf2003measurement,briegel2009measurement,kitaev2003fault,
dennis2002topological,affleck1987rigorous,haldane1983nonlinear,
pollmann2010entanglement}. Their algebraic structure leads to compact but physically distinct circuit
realizations. Cluster states are stabilizer states that can be prepared by
low-depth local Clifford circuits. Toric-code ground states are also stabilizer
states, but their intrinsic topological order generally prevents preparation by
constant-depth local unitary circuits from a product state; instead, they are
prepared using stabilizer-measurement protocols, circuits with system-size-
dependent depth, or measurement-and-feed-forward constructions (see Sect.~\ref{anyon} below). The AKLT ground
state admits an exact finite-bond-dimension matrix-product-state representation
and can be prepared using local isometries or sequential preparation circuits;
more recent measurement-assisted fusion protocols further enable deterministic
constant-depth preparation on quantum processors
\cite{gross2007measurement,smith2023deterministic,chen2023high}.

\noindent{\textbf{Symmetry-protected topological order.}} Cluster states form a canonical family of stabilizer states obtained by
entangling a lattice, or more generally a graph, of qubits with controlled-$Z$
gates along its edges. Starting from a product of $\ket{+}$ states, a cluster
state $\ket{C_N}$ can be prepared as
\begin{equation}
\ket{C_N}
=
\prod_{\langle i,j\rangle\in E} CZ_{i,j}\;
\bigotimes_{k=1}^{N}\ket{+}_k ,
\label{eq:cluster_state}
\end{equation}
where $E$ denotes the set of edges of the underlying graph. Cluster states have
been realized across multiple hardware platforms. Early photonic experiments
demonstrated one-way quantum computation using four-photon cluster states and
subsequently implemented gate operations with four-photon six-qubit cluster
states
\cite{walther2005experimental,gao2010experimental}.

On superconducting and cloud-accessible quantum processors, cluster and graph
states have been used as benchmarks of multipartite entanglement,
measurement-based quantum computation, and symmetry-protected topological
order. Notable examples include the measurement of the entanglement spectrum of
a one-dimensional SPT state on an IBM quantum computer
\cite{choo2018measurement}, protocols for identifying SPT order in noisy
cluster-state circuits on IBM devices~\cite{azses2020identification},
graph-state and whole-device entanglement preparation on 20- and 65-qubit IBM
devices~\cite{mooney2019entanglement,mooney2021whole}, superconducting-processor
realizations of 12-qubit linear cluster states~\cite{gong2019genuine}, digital
simulation of Floquet SPT phases~\cite{zhang2022digital}, and large-scale
generation of one- and two-dimensional superconducting cluster states
\cite{cao2023generation}.

In particular, in Ref.~\cite{zhang2022digital}, Zhang \emph{et al.}
experimentally studied Floquet SPT physics on a programmable flip-chip
superconducting quantum processor developed at Zhejiang University. As part of
their characterization, one-dimensional SPT states were initialized using a
shallow stabilizer circuit consisting of global Hadamard gates followed by
nearest-neighbor controlled-$Z$ gates, which prepares the fixed-point
cluster-state wavefunction, as shown in Fig.~\ref{fig:spt}(a). Random local
$Z$ operators were then applied to selected sites to create stabilizer
excitations, thereby generating highly excited eigenstates of the cluster
stabilizer Hamiltonian. These random SPT eigenstates were subsequently evolved
under the Floquet drive.

The protected structure was revealed through entanglement spectroscopy [see Sect.~\ref{sec:EE} for method details]. After
one driving period, the reduced density matrix of half of the system,
$\rho_{\rm half}$, was reconstructed by quantum-state tomography, and the
entanglement spectrum was obtained from the eigenvalues of
$-\ln \rho_{\rm half}$. The experimental data, shown as red triangles with
error bars, agree with noisy numerical simulations that include experimental
imperfections, shown as gray circles. For open boundary conditions, the
low-lying entanglement levels exhibit an approximate two-fold degeneracy, while
for periodic boundary conditions they exhibit an approximate four-fold
degeneracy. These degeneracies reflect effective virtual edge modes associated
with the entanglement cut and provide evidence that the evolved states retain
the characteristic topological structure of the SPT phase. A complementary
route to preparing related SPT ground states uses quantum imaginary-time
evolution on IBM Quantum processors, as demonstrated in
Ref.~\cite{shen2025robust} and discussed below.

\noindent{\textbf{AKLT states.~}} The AKLT state is a paradigmatic exactly solvable many-body wavefunction and a
canonical representative of a one-dimensional symmetry-protected topological phase
\cite{affleck1987rigorous,pollmann2010entanglement}. It arises as the unique
gapped ground state of a spin-1 antiferromagnetic chain with Haldane-type
structure: the bulk is short-range entangled, while open boundaries host
emergent fractionalized spin-$1/2$ edge degrees of freedom protected by
symmetry
\cite{haldane1983nonlinear,affleck1987rigorous,pollmann2010entanglement}.
From the quantum-simulation perspective, the AKLT state provides a particularly
transparent set of diagnostics, most notably a nonlocal string-order parameter
and boundary-mode signatures that distinguish the SPT phase from trivial product
states and remain accessible through correlation measurements on quantum
hardware.

A convenient way to understand the AKLT state is the {valence-bond}  picture shown in Fig.~\ref{fig:spt}(b)~\cite{smith2023deterministic}. Each physical spin-1 at site $i$ is represented by two \emph{virtual} spin-$1/2$ degrees of freedom. Neighboring virtual spins form singlets (valence bonds), and then the two virtual spins on each site are projected onto the symmetric (triplet) subspace, which defines the physical spin-1 Hilbert space. This construction explains the edge physics: for open boundary conditions, one virtual spin-$1/2$ remains unpaired at each end, producing effective spin-$1/2$ edge modes.

On digital quantum processors, AKLT physics can be accessed through \emph{state preparation and verification} protocols
that directly mirror the valence-bond construction. The key ingredients are precisely those suggested by the
valence-bond picture: (i) engineer singlet-like correlations between neighboring virtual spin-$1/2$ degrees of freedom
(e.g., by preparing Bell singlets on inter-site virtual-qubit pairs), and (ii) implement an on-site encoding or
projection that maps two virtual qubits into the symmetric (triplet) subspace, thereby realizing the physical spin-1
Hilbert space.

Fig.~\ref{fig:spt}(b) highlights a particularly hardware-friendly,
measurement-assisted preparation strategy for the AKLT state. Smith
\emph{et al.} proposed and experimentally demonstrated a deterministic
constant-depth protocol on an IBM Quantum superconducting processor
\cite{smith2023deterministic}. Instead of preparing the full AKLT chain
through a long coherent circuit, the protocol first prepares short AKLT
fragments locally and then connects them using two-qubit fusion
measurements. These fusion measurements act as entangling connectors,
effectively teleporting the valence-bond correlations from one fragment
to the next while keeping the required coherent circuit depth small. The
lower panels benchmark the prepared states through local reduced density
matrices reconstructed on small subsystems. The observed block structure
and off-diagonal coherences are consistent with the valence-bond
correlations of the ideal AKLT state, providing a local verification of
successful state preparation under realistic device noise.

A complementary gate-based route to AKLT-state preparation on IBM
Quantum hardware was demonstrated by Chen \emph{et al.}
\cite{chen2023high}. This approach follows the valence-bond
construction more directly. One first prepares spin singlets between
neighboring virtual spin-$1/2$ degrees of freedom, and then enforces the
AKLT constraint by projecting each pair of on-site virtual qubits into
the symmetric spin-$1$ manifold. Since these projections are intrinsically
non-unitary, the experiment realizes them through an ancilla-assisted
unitary dilation, as discussed in Sect.~\ref{ancilla}. In this
implementation, each local projector is embedded into a larger unitary
acting on the two system qubits and an ancilla, and the desired
projection is obtained by post-selecting the target ancilla-measurement
outcome. The results summarized in Fig.~\ref{fig:spt}(d) show the
experimentally measured computational-basis probability weights of the
prepared state, with the agreement between experiment and theory
quantified by the fidelity between the measured and ideal probability
distributions.

\begin{figure*}
    \centering
    \includegraphics[width=0.9\linewidth]{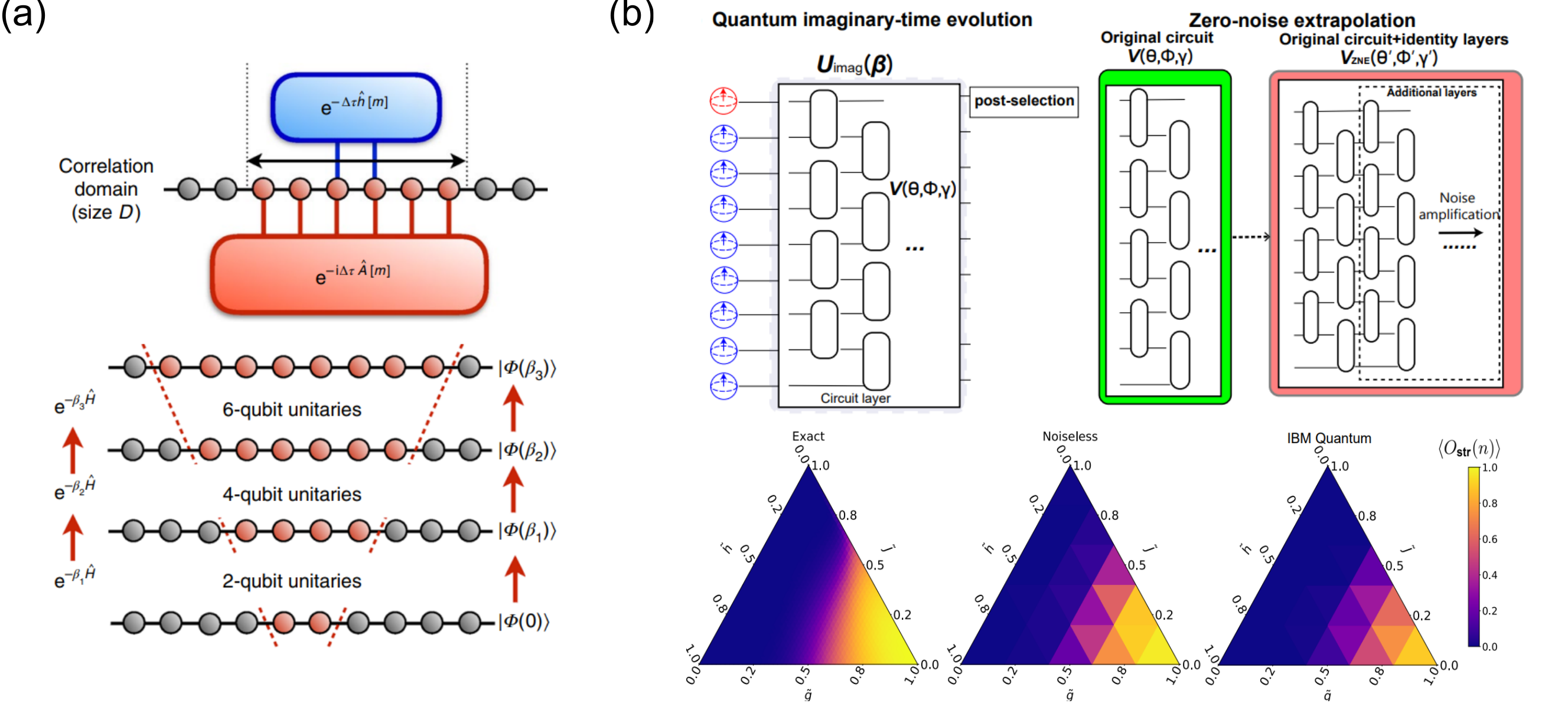}
    \caption{Quantum imaginary-time evolution and error-mitigated simulation of quantum phases. (a) Schematic illustration of the quantum imaginary-time evolution (QITE) algorithm. The non-unitary imaginary-time operator is approximated by a sequence of local unitary transformations~\cite{motta2020determining}. (b) Experimental workflow combining QITE with zero-noise extrapolation (ZNE).
Left: the parametric unitary ansatz. Right: ZNE evaluates observables using the original circuit (green) and its noise-amplified versions with extra identity layers (red) inserted. Bottom: comparison of the string-order parameter, predicted by exact theory, a noiseless simulator, and data collected on IBM Quantum hardware.  The triangular phase diagrams illustrate the reconstruction of the phase structure across coupling parameters, demonstrating the combined effectiveness of QITE and ZNE for probing ground-state order in noisy devices~\cite{shen2025robust}. Panel (a) is adopted from ~\cite{motta2020determining}. Panel (b) is adopted from ~\cite{shen2025robust}.
}
    \label{fig:qite}
\end{figure*}

\subsubsection{Obtaining ground states through quantum imaginary-time evolution (QITE)}\label{qite}

Previously, we described how ground states with known ansatz structures can be
prepared on quantum hardware. In general, however, the form of the ground state
is either unknown or too complicated to encode directly. In such cases, a
broadly applicable route, used both in classical many-body numerics and in
quantum algorithms, is to evolve toward the ground state through
\emph{imaginary-time evolution}
\cite{ceperley1980ground,sandvik2010computational,motta2020determining,
mcardle2019variational}. Starting from an initial state $\ket{\psi_0}$ with
nonzero overlap with the ground-state subspace, one considers the normalized
imaginary-time flow
\begin{equation}
\ket{\psi(\tau)} =
\frac{e^{-\tau H}\ket{\psi_0}}
{||e^{-\tau H}\ket{\psi_0}||},
\label{psibeta}
\end{equation}
which suppresses excited-state components relative to the ground state as
$e^{-\tau(E_n-E_0)}$. In the limit $\tau\to\infty$, and in the absence of
symmetry or degeneracy obstructions, the state is projected onto the
ground-state component contained in $\ket{\psi_0}$.

The central difficulty on gate-based quantum computers is that
$e^{-\tau H}$ is nonunitary, whereas quantum circuits natively implement
unitary operations. The most established near-term QITE strategy does not try to
implement $e^{-\Delta\tau H}$ as an operator identity on all possible input
states. Instead, following the state-dependent construction of
Ref.~\cite{motta2020determining}, each small imaginary-time update is replaced
by a unitary operation that approximately maps the current state to the
normalized imaginary-time-updated state. This makes QITE useful for ground-state
and thermal-state preparation on small and intermediate system sizes, and it
has motivated several algorithmic variants and applications in chemistry, spin
systems, open-system simulation, and finite-temperature calculations
\cite{motta2020determining,sun2021quantum,kamakari2022digital,
yeter2020practical}.

To construct the local QITE update, we can decomposes the Hamiltonian into local
terms,
\begin{equation}
H=\sum_{\ell}h_{\ell},
\end{equation}
and applies a first-order imaginary-time product formula,
\begin{equation}
e^{-\Delta\tau H}
=
e^{-\Delta\tau\sum_{\ell}h_{\ell}}
\approx
\prod_{\ell} e^{-\Delta\tau h_{\ell}}
+
\mathcal{O}(\Delta\tau^2),
\end{equation}
where the product-ordering dependence and the leading error are controlled by
commutators among the local terms. For each local term $h_{\ell}$, QITE seeks a
state-dependent Hermitian generator $A_{\ell}$, supported on a chosen domain
around $h_{\ell}$, such that
\begin{equation}
\frac{e^{-\Delta\tau h_{\ell}}\ket{\psi}}
{\left\|e^{-\Delta\tau h_{\ell}}\ket{\psi}\right\|}
\approx
e^{-i\Delta\tau A_{\ell}}\ket{\psi}.
\label{eq:qite_local_unitary}
\end{equation}
The operator $A_{\ell}$ is expanded in a basis of Hermitian Pauli strings on
that domain,
\begin{equation}
A_{\ell}
=
\sum_{\mu} a_{\ell,\mu} P_{\ell,\mu},
\end{equation}
where $P_{\ell,\mu}$ denotes a Pauli string and the coefficients
$a_{\ell,\mu}$ are real. These coefficients are determined approximately from
measurement data by solving a linear system obtained by matching the
imaginary-time update to the unitary update to first order in $\Delta\tau$
\cite{motta2020determining}. Once $A_{\ell}$ has been determined, the unitary
$e^{-i\Delta\tau A_{\ell}}$ is decomposed into the native gate set of the
hardware.

Motta \textit{et al.} introduced quantum imaginary-time evolution
(QITE) as an ansatz-independent route to preparing low-energy eigenstates and
thermal states on near-term quantum devices~\cite{motta2020determining}. In this
approach, each small imaginary-time step is replaced by a state-dependent local
unitary whose generator is obtained from a linear system, chosen so that the
unitary approximately reproduces the normalized action of the corresponding
nonunitary imaginary-time propagator on the current state within a finite
correlation domain [Fig.~\ref{fig:qite}(a)]. This formulation avoids both the
deep circuits and ancilla requirements of phase estimation and the large
nonconvex parameter searches typical of direct variational energy-minimization
workflows. Motta \textit{et al.} benchmarked the method through exact classical
emulations for several model Hamiltonians, including one- and two-dimensional
transverse-field Ising models, and implemented proof-of-principle circuits on
Rigetti's quantum virtual machine and Aspen-1 superconducting quantum processing
unit.

Related variational imaginary-time methods were developed to project
imaginary-time dynamics onto a parameterized ansatz, thereby updating circuit
parameters according to variational equations of motion rather than through
direct energy minimization~\cite{mcardle2019variational}. Subsequent work
improved the practicality of QITE and related imaginary-time algorithms under
near-term hardware constraints, including nonlocal approximations and circuit
compression, step-merging strategies for quantum chemistry, and experimental
demonstrations or assessments on superconducting processors
\cite{nishi2021implementation,gomes2020efficient,yuan2023realizing}. In
particular, step-merged QITE was implemented on Rigetti quantum processing
units, while related variational imaginary-time experiments have been realized
in superconducting-qubit systems.

The QITE framework has also been extended to finite-temperature observables.
Sun \textit{et al.} used QITE to compute finite-temperature energies, static
and dynamical correlation functions, and excitation spectra of few-site spin
systems on IBM Quantum devices, providing a route from finite-temperature
imaginary-time-evolved states at moderate inverse temperature toward
ground-state physics as the inverse temperature increases~\cite{sun2021quantum}.
Further developments include QITE-based quantum Lanczos energy estimation on
IBM-Q hardware, QITE-based digital simulations of open quantum systems governed
by Lindblad-type dynamics on IBM Quantum hardware, fragmented imaginary-time
evolution for early-stage quantum signal processors, adiabatic QITE, and
double-bracket QITE formulations with explicit cooling or fidelity-improvement
guarantees
\cite{yeter2020practical,kamakari2022digital,silva2023fragmented,
hejazi2024adiabatic,gluza2026double}.

Moreover, QITE has recently been used to go beyond ``energy benchmarks'' and
to access phase-diagnostic observables associated with topological matter on
noisy superconducting hardware. In particular, Shen \textit{et al.}
reported a fully digital experiment demonstrating robust simulation of
many-body symmetry-protected topological (SPT) phase transitions by combining
Quantum Imaginary-Time Evolution (QITE) with enhanced zero-noise extrapolation
(ZNE) on IBM Quantum hardware, as depicted in Fig.~\ref{fig:qite}(b)
\cite{shen2025robust}.

Technically, the QITE update is implemented through an enhanced ancilla-based
construction, in which the desired nonunitary imaginary-time step is embedded
into a larger circuit acting on the system and ancilla degrees of freedom.
This circuit-level embedding makes imaginary-time projection compatible with
gate-based hardware, while enhanced ZNE and related error-mitigation procedures
reduce the impact of coherent and stochastic gate errors by extrapolating
measured observables toward an effective zero-noise limit. Using this
QITE--ZNE hybrid protocol, the authors reconstructed the ground-state phase
diagram of the Ising--cluster Hamiltonian and identified the transition between
trivial and cluster SPT phases by measuring the nonlocal string-order parameter
and related edge/entanglement diagnostics. These results illustrate how QITE,
when paired with practical error mitigation, can make digitally prepared
ground states sufficiently accurate to probe SPT diagnostics and critical
behavior on present-day superconducting quantum processors.

Another important development is the introduction of a variational-based quantum imaginary-time evolution (VQITE) protocol, experimentally realized on a superconducting-qubit platform~\cite{yuan2023realizing}. The authors highlighted that variationally parameterized QITE can achieve faster convergence and higher numerical stability than standard energy-minimization schemes. In contrast to conventional VQE approaches that rely solely on energy minimization, VQITE directly approximates imaginary-time propagation through a parameterized unitary ansatz whose parameters are updated using a local McLachlan variational principle. The experiment demonstrated that this strategy offers markedly faster convergence. By efficiently steering the quantum state toward the ground state, the VQITE method provides a powerful alternative to standard variational algorithms to be described below.

\begin{figure*}
    \centering
    \includegraphics[width=0.99\linewidth]{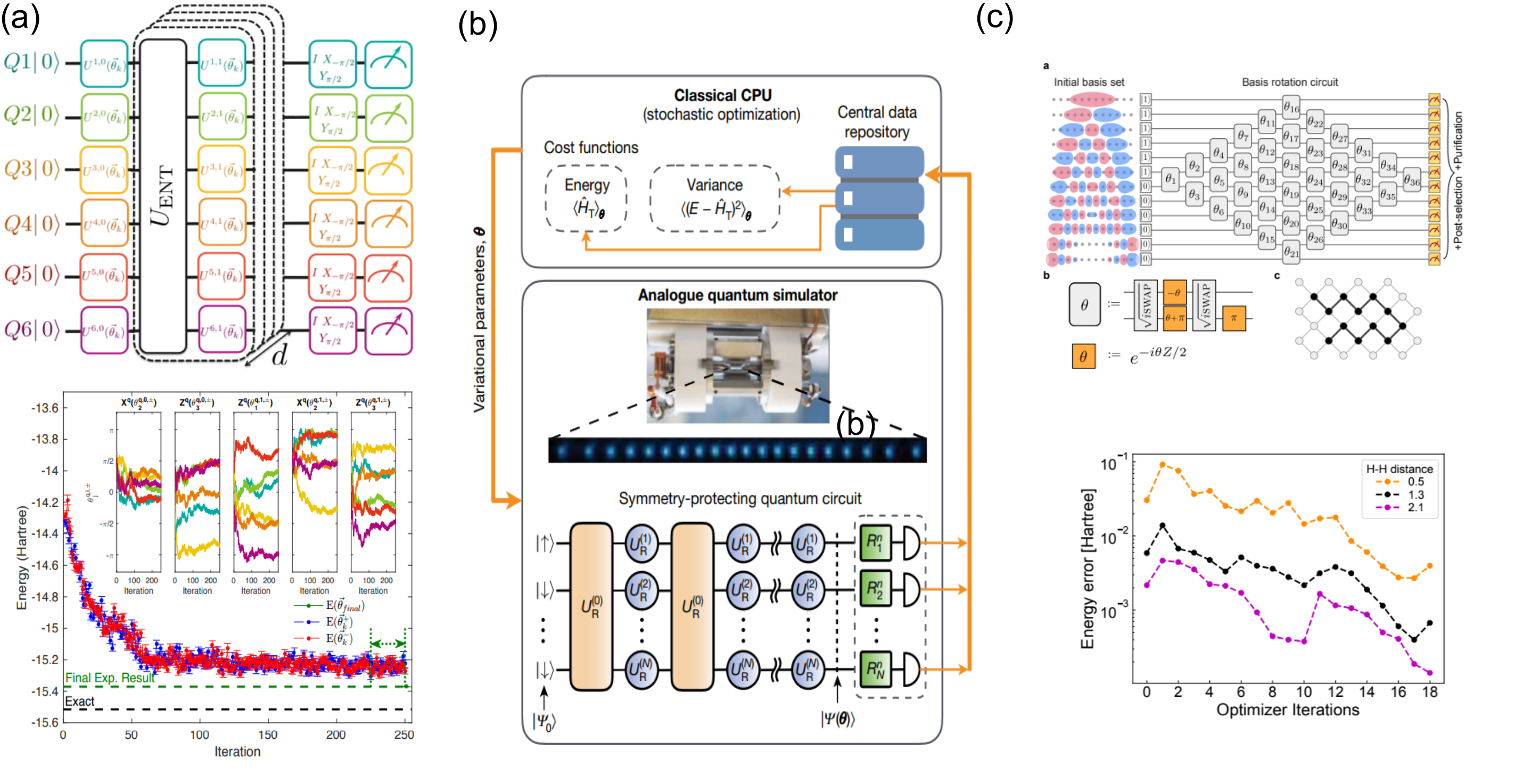}
    \caption{Representative quantum hardware demonstrations of variational quantum solvers. (a) First proof-of-principle hardware implementation of the variational quantum eigensolver (VQE) using a programmable superconducting-qubit processor. An entangling unitary followed by local single-qubit rotations prepares parameterized trial states. The lower panel shows measured energy trajectories compared with exact theoretical values.~\cite{kandala2017hardware}. (b) Demonstration of a hybrid analog–digital VQE approach. A classical optimizer updates the variational parameters, while an analog trapped-ion quantum simulator evaluates cost functions, such as the energy. A symmetry-protecting quantum circuit enhances the stability of the variational landscape. ~\cite{kokail2019self} (c) Realization of a unitary coupled-cluster VQE for molecular hydrogen using a trapped-ion quantum computer. The circuit implements a basis-rotation ansatz with post-selection–based purification, yielding chemically accurate energies across different interatomic distances. The lower right panel displays the energy-error convergence under iterative optimization~\cite{google2020hartree}. Panel (a) is adopted from ~\cite{kandala2017hardware}. Panel (b) is adopted from ~\cite{kokail2019self}. Panel (c) is adopted from ~\cite{google2020hartree}.}
    \label{fig:vqe}
\end{figure*}

\subsubsection{Obtaining ground-states through variational approaches}

For generic condensed-matter models, the ground state is rarely known in closed
form, and exact preparation by a finite-depth analytically known circuit is
typically impossible. The QITE and VQITE approaches discussed above address
this problem by mimicking imaginary-time projection to the ground state. In particular, VQITE
uses a parameterized ansatz but updates its parameters according to a
variational form of imaginary-time evolution. 
A related but distinct strategy is the variational quantum eigensolver (VQE) approach, which is one of the most
prominent and powerful strategies for approximating ground states of quantum many-body Hamiltonians on near-term
quantum devices~\cite{cerezo2021variational}. This
approach does not exploit imaginary-time dynamics at all; rather, it prepares
\begin{equation}
\ket{\psi(\boldsymbol{\theta})}
=
U(\boldsymbol{\theta})\ket{\psi_{\rm ref}},
\end{equation}
through a trial unitary $U(\boldsymbol{\theta})$ and the measured energy
\begin{equation}
E(\boldsymbol{\theta})
=
\bra{\psi(\boldsymbol{\theta})}H\ket{\psi(\boldsymbol{\theta})}
\end{equation}
is estimated on hardware and minimized (via parameters $\boldsymbol{\theta}$) in a classical optimization loop. 
Compared to QITE or VQITE, this workflow replaces explicit imaginary-time
propagation by direct variational energy minimization. Its accuracy is therefore
controlled by the expressivity of the ansatz, the quality of the classical
optimizer, and the measurement precision of the estimated Hamiltonian
expectation values. This hybrid formulation provides a flexible and
hardware-compatible framework for approximating ground states of interacting
many-body Hamiltonians, with circuit depth and ansatz structure chosen according
to the target model and device constraints. More details of this methodology are
presented in Section~\ref{vqa}.

Ground-state preparation, however, is not specific to condensed-matter systems,
but is a broadly relevant task that also appears in quantum chemistry, materials
simulation, and optimization problems. Many early and influential VQE
demonstrations on quantum processors were developed in the context of molecular
Hamiltonians or chemistry-inspired benchmarks, where the goal is to approximate
the lowest-energy state, molecular spectrum, or optimal configuration within a
variational manifold
\cite{peruzzo2014variational,mcclean2016theory,kandala2017hardware,
colless2018computation,mccaskey2019quantum,google2020hartree}. Nevertheless,
the underlying methodological advances---including ansatz design, symmetry
preservation, measurement strategies, error mitigation, and classical
optimization---are directly transferable to condensed-matter Hamiltonians
\cite{bonet2018low,sagastizabal2019experimental,grimsley2019adapt,
cerezo2021variational,tilly2022variational}.

More recent chemistry- and hardware-oriented developments, including
contextual-subspace VQE and qubit-efficient VQE variants
\cite{weaving2025contextual,ma2025experimental}, as well as sample-based
diagonalization and Krylov-subspace approaches that are not strictly VQE but
remain closely related to variational and subspace quantum simulation
\cite{JavierYunoki2025,piccinelli2025randomizedsqd,
piccinelli2026exploringpathwaysquantumadvantage,
shajan2026molecularquantumcomputationsprotein,
merz2026crossing12000atombarrierheterogeneous}, further illustrate how
algorithmic advances developed for molecular problems can inform broader
many-body simulation. For this reason, we include selected chemistry- and
optimization-oriented VQE works, together with related subspace-diagonalization
and Krylov-type approaches, in this review.

\noindent\textbf{Early hardware demonstrations of variational eigensolvers.}~ Despite hardware constraints, multiple experiments have established that
variational protocols can approximate ground-state energies and, in selected
cases, prepare near-ground-state variational wavefunctions on quantum hardware.
Early proof-of-principle demonstrations include Peruzzo \emph{et al.}, who
introduced the VQE paradigm on a photonic processor by computing the
ground-state molecular energy of He--H$^+$, thereby highlighting the NISQ
strategy of trading coherent circuit depth for a hybrid quantum--classical
optimization loop~\cite{peruzzo2014variational}. Shortly thereafter,
O'Malley \emph{et al.} realized scalable quantum simulation of the H$_2$
molecular energy surface on a superconducting platform using a shallow,
hardware-compatible implementation of variational quantum chemistry
\cite{o2016scalable}. Building on this direction, Colless \emph{et al.}
combined VQE with quantum subspace expansion to extract molecular spectra on a
superconducting processor, underscoring that careful measurement design,
subspace estimation, and noise-aware analysis can be as important as the ansatz
itself in early implementations~\cite{colless2018computation}.

\noindent\textbf{Hardware-efficient ansätze.}~A widely cited milestone in hardware-efficient ground-state preparation is the
experiment by Kandala \emph{et al.}, who implemented VQE on an IBM
superconducting quantum processor to estimate ground-state energies of small
molecular Hamiltonians and quantum-magnetism models
\cite{kandala2017hardware}. More specifically, the experiment used up to six
qubits of a purpose-built seven-qubit IBM superconducting processor, thereby
establishing the hardware-efficient ansatz as a practical, device-native
baseline for NISQ simulation. The corresponding workflow is illustrated in
Fig.~\ref{fig:vqe}(a): a parameterized circuit prepares a trial state
$\ket{\psi(\boldsymbol{\theta})}$, and a classical optimizer updates
$\boldsymbol{\theta}$ to minimize the measured energy expectation value.

Here, in the hardware-efficient construction, $\ket{\psi(\boldsymbol{\theta})}$
is generated by alternating layers of single-qubit Euler rotations and native
entangling operations tailored to the connectivity and available interactions
of the superconducting processor
\cite{kandala2017hardware,mcclean2016theory,cerezo2021variational,
tilly2022variational}. Experimentally, the measured potential-energy curves for
molecular systems, including H$_2$, LiH, and BeH$_2$, reproduced the qualitative
trends expected from exact diagonalization and were well described by noisy
device simulations~\cite{kandala2017hardware}. However, the achieved accuracy
was not yet sufficient for reliable chemical prediction. This limitation
highlighted key practical bottlenecks of early NISQ variational algorithms,
including finite sampling noise, coherent and incoherent gate errors, optimizer
sensitivity, and readout bias
\cite{mcclean2016theory,kandala2017hardware,cerezo2021variational,
tilly2022variational}. It also motivated subsequent advances in noise control
and error-mitigation strategies, discussed in Sect.~\ref{error}, aimed at
pushing variational energy estimation from qualitative demonstrations toward
quantitatively reliable regimes
\cite{temme2016error,kandala2019error,bonet2018low,
sagastizabal2019experimental,endo2021hybrid}.

In parallel, Kokail \emph{et al.} employed a programmable trapped-ion
quantum simulator to variationally prepare eigenstates of the lattice
Schwinger model, a one-dimensional lattice gauge theory, using systems of
up to 20 qubits~\cite{kokail2019self}. This experiment, performed on the
Innsbruck trapped-ion platform rather than on a company-operated quantum
processor, provided a landmark demonstration of variational quantum
simulation beyond small-molecule benchmarks. The protocol determined
ground-state energies, low-lying excitations, and energy gaps, and further
used measurements of the Hamiltonian variance to assign algorithmic error
bars to the variational energies. As illustrated in Fig.~\ref{fig:vqe}(b), a key design principle was to
incorporate the physical constraints and symmetries of the target model
directly into the variational manifold. In the Schwinger-model
formulation used in the experiment, the gauge fields are eliminated using
Gauss's law, yielding an effective spin Hamiltonian with long-range
interactions. The variational state family is then constructed to remain
within the relevant symmetry sector of this Hamiltonian, thereby reducing
the effective search space and avoiding unphysical variational directions
(see Sect.~\ref{gauge} for further discussion of the model). This
symmetry-adapted strategy enabled the preparation of target eigenstates
with comparatively shallow experimental sequences and stable optimization.
More broadly, the experiment illustrates a central advantage of
physics-informed variational ansatz design: by tailoring the trial-state
manifold to the constraints and symmetries of the model, one can reduce
the overhead associated with generic parametrizations and improve the
verifiability of near-term quantum simulations.

\noindent\textbf{Scaling variational methods to many-body-native models.}~More recent experiments have pushed variational ground-state preparation
toward larger and more many-body-native settings. For instance, Stanisic
\textit{et al.} demonstrated a scalable variational strategy for accessing
ground-state properties of medium-scale Fermi-Hubbard instances on Google's
Sycamore superconducting quantum processor~\cite{stanisic2022observing}.
A key element was to exploit the fermionic structure, locality, and symmetries
of the Hubbard Hamiltonian to design low-depth, problem-adapted ans\"atze, so
that the number of variational parameters grows mildly with system size while
still capturing nontrivial correlations. The experiment studied $1\times 8$
and $2\times 4$ Fermi-Hubbard instances encoded on 16 qubits and observed
qualitative signatures such as Friedel oscillations, antiferromagnetic order,
and the onset of a metal-insulator transition.

Complementarily, Ma \emph{et al.} reported an experimental realization of a
qubit-efficient VQE combined with analog error-mitigation primitives on a
superconducting-circuit processor~\cite{ma2025experimental}. By using a
transmon qubit coupled to a high-coherence photonic mode and leveraging an
MPS-compressed representation, the experiment simulated circular transverse-
field Ising models with reduced physical-qubit resources and validated the
method by estimating the ground-state energies of a four-spin Ising model.
The central message is that compact encodings can be paired with
hardware-native mitigation methods to improve the reliability of measured
energies. At the level of two-dimensional hardware connectivity, the Google Quantum AI
team implemented variational Hartree--Fock state preparation on the
superconducting quantum processor~\cite{google2020hartree}
[see Fig.~\ref{fig:vqe}(c)]. Their approach leveraged highly parallel,
hardware-native entangling layers on a planar architecture, together with
post-selection onto the correct particle-number sector, density-matrix
purification, and variational relaxation to improve energy estimation.

\noindent\textbf{Resource-reduced and trainability-aware VQE variants.}~The latest developments further suggest that the frontier is moving beyond
plain hardware-efficient VQE toward resource-reduced, physics-informed, and
subspace-based variants. Weaving \emph{et al.} experimentally demonstrated a
contextual-subspace VQE calculation of the dissociation curve of molecular
nitrogen on IBM superconducting quantum hardware, using contextual subspace
reduction, hardware-aware ansatz construction, dynamical decoupling,
measurement-error mitigation, and zero-noise extrapolation
\cite{weaving2025contextual}. This example is important because it targets a
chemically nontrivial bond-breaking problem while explicitly confronting the
trade-off between the size of the reduced contextual subspace and the circuit
depth that present hardware can tolerate. Moreover, Cao \emph{et al.} showed that initializing
hardware-efficient Floquet variational circuits in a many-body-localized regime
can preserve trainable gradients and mitigate barren plateaus, thereby improving
the scalability of variational simulation
\cite{cao2025exploiting}. Their hardware experiments on the 127-qubit IBM
\texttt{ibm\_brisbane} superconducting processor provided evidence for restored
gradients in kicked Heisenberg-chain circuits, including system sizes up to 31
qubits. A recent development by Wu \emph{et al.} emphasizes input-state design
as a complementary strategy to circuit design: rather than only increasing
ansatz depth, one can modify the initial variational manifold through
linear-combination input states to improve the reachability of target many-body
ground states at fixed gate budget
\cite{wu2026reachability}. These works reinforce the same practical conclusion:
for near-term ground-state preparation, scalability depends not only on the
expressibility of the circuit, but also on whether the ansatz, input state,
measurement strategy, and mitigation pipeline are co-designed with the physics
of the target Hamiltonian and the constraints of the device.

At the same time, recent progress also exposes the limitations of VQE as a
general-purpose route to ground-state physics. As the system size grows,
decoherence, coherent control errors, accumulated state-preparation
imperfections, and finite sampling increasingly distort both the measured cost
function and the optimization landscape
\cite{preskill2018quantum,Wecker:15,mcclean2016theory,tilly2022variational,
cerezo2021variational,bharti2022noisy,mcclean2018barren,cerezo2021cost,
wang2021noise,holmes2022connecting,arrasmith2021effect,larocca2025barren}.
Deeper circuits amplify stochastic noise and coherent over-rotations, while
finite sampling can bias gradient estimates and introduce optimizer-dependent
drift \cite{sweke2020stochastic,gonthier2022measurements}. These effects are
especially severe for two- and three-dimensional systems, where the entangling
depth required to build long-range correlations and the measurement overhead
required to estimate many Hamiltonian terms can both grow rapidly. This has
motivated a broader movement toward hybrid subspace and sampling-based
alternatives \cite{McClean:17,yoshioka2022generalized,JavierYunoki2025,
yu2025skqd,piccinelli2025randomizedsqd}. One example is randomized
sample-based quantum diagonalization, in which the quantum processor supplies
samples from physically motivated states or randomized time-evolution circuits,
and a classical routine diagonalizes the Hamiltonian in the reduced subspace
generated from those samples \cite{piccinelli2025randomizedsqd}. Such
approaches do not replace VQE in all regimes, but they clarify an important
trend: near-term ground-state algorithms are increasingly judged by the
combined cost of state preparation, measurement, mitigation, and classical
post-processing, rather than by circuit depth alone.

\subsubsection {Focus Area I: Digital quantum simulation of fractional quantum Hall physics }

Having reviewed various state-preparation approaches, we now highlight their
application to a prototypical class of strongly interacting states that are
central to condensed-matter physics but challenging to engineer and probe in a
fully programmable manner: fractional quantum Hall (FQH) states. FQH systems
provide one of the cleanest settings for strongly correlated topological
matter. In a partially filled Landau level, the kinetic energy is quenched into
a macroscopically degenerate single-particle manifold, so interactions dominate
the low-energy physics and generate intrinsically nonperturbative order
\cite{tsui1982two,laughlin1983anomalous,aro1984fractional,wen1990topological,
wen1995topological,nayak2008non,goerbig2009quantum}.

Important classes of FQH model states admit exact analytic wavefunctions
specified by clustering or generalized exclusion rules associated with parent
pseudopotential Hamiltonians
\cite{haldane1983fractional,trugman1985exact,moore1991nonabelions,
read1999beyond,bernevig2008model,bernevig2008generalized,
bernevig2009anatomy,lee2015geometric}. Despite their mathematical elegance,
many model states beyond the simplest Laughlin sequence require idealized
few-body parent interactions or finely controlled projected dynamics, making
their direct realization and manipulation challenging in natural materials and
in scalable analog simulators. At the same time, recent analog and near-analog
experiments have begun to realize minimal FQH states with ultracold atoms and
interacting photons
\cite{leonard2023realization,wang2024realization}.  Moreover, gate-based digital quantum simulation provides a complementary
route for accessing FQH phenomena that are difficult to probe directly, including
geometric collective modes, neutral magnetoroton response, charged quasiholes,
and braiding statistics
\cite{GMP85,GMP86,HaldaneViscosity,read2009non,lee2015geometric,kirmani2022probing,
PhysRevB108064303}.

\noindent\textbf{Generation of fractional quantum Hall states} 

FQH states result from long-range interactions between the degenerate states within a fractionally filled Landau level. When the Landau gap between the levels is larger than the interaction energy scale, the specific form of the effective Hamiltonian results from projection of the Coulomb interaction into one Landau level, and can be expressed as a 1D long-ranged interacting Hamiltonian between nondispersive electrons in Landau orbitals~\cite{haldane1983fractional,trugman1985exact,moore1991nonabelions,
read1999beyond,bernevig2008model,bernevig2008generalized,
bernevig2009anatomy,lee2015geometric}. Recent theoretical advancements have shown that with a suitable gauge choice, the truncated interactions between Landau orbitals would lead to ground states that are close to the actual FQH ground state resulting from the long-ranged electronic interactions in the degenerate orbitals, with common topological properties~\cite{Seidel2005,Flavin2011,nakamura_exact}. One example is the interaction potential that leads to the Laughlin wavefunction under torus geometry, given by~\cite{Seidel2005,nakamura_exact}
\begin{equation}\label{intp}
V_{km}\propto (k^2-m^2) e^{-2\pi^2(m^2+k^2)/{L_1^2}},
\end{equation}
where $k$ and $m$ label lowest Landau level orbitals in the Landau gauge on the system torus and $L_1$ is the torus circumference. The above interaction potential exponentially decays with the orbital number spacing, justifying (in the thin torus limit with small $L_1$) truncating the interaction into the following Hamiltonian~\cite{nakamura_exact}:
\begin{align}\label{mainh}
    H_\text{trun} &=\sum_{j=0}^{M-1} V_{10} n_{j+1} n_{j+2} +V_{20} n_j n_{j+2} +V_{30}n_j n_{j+3}\nonumber\\
    &\hspace{1cm} + \sqrt{V_{10}V_{30}}\left( c_j^\dagger c_{j+3}^\dagger c_{j+2} c_{j+1} + h.c.\right).
\end{align}
Here $c_j$ is the annihilation operator for electrons in Landau orbital $j$ and $n_j = c_j^\dagger c_j$.
A FQH many-body ground state wavefunction corresponds to a particular superposition of filled degenerate orbitals in a Landau level that minimizes this interaction energy.

\begin{figure*} 
    \centering
    \includegraphics[width=0.9\linewidth]{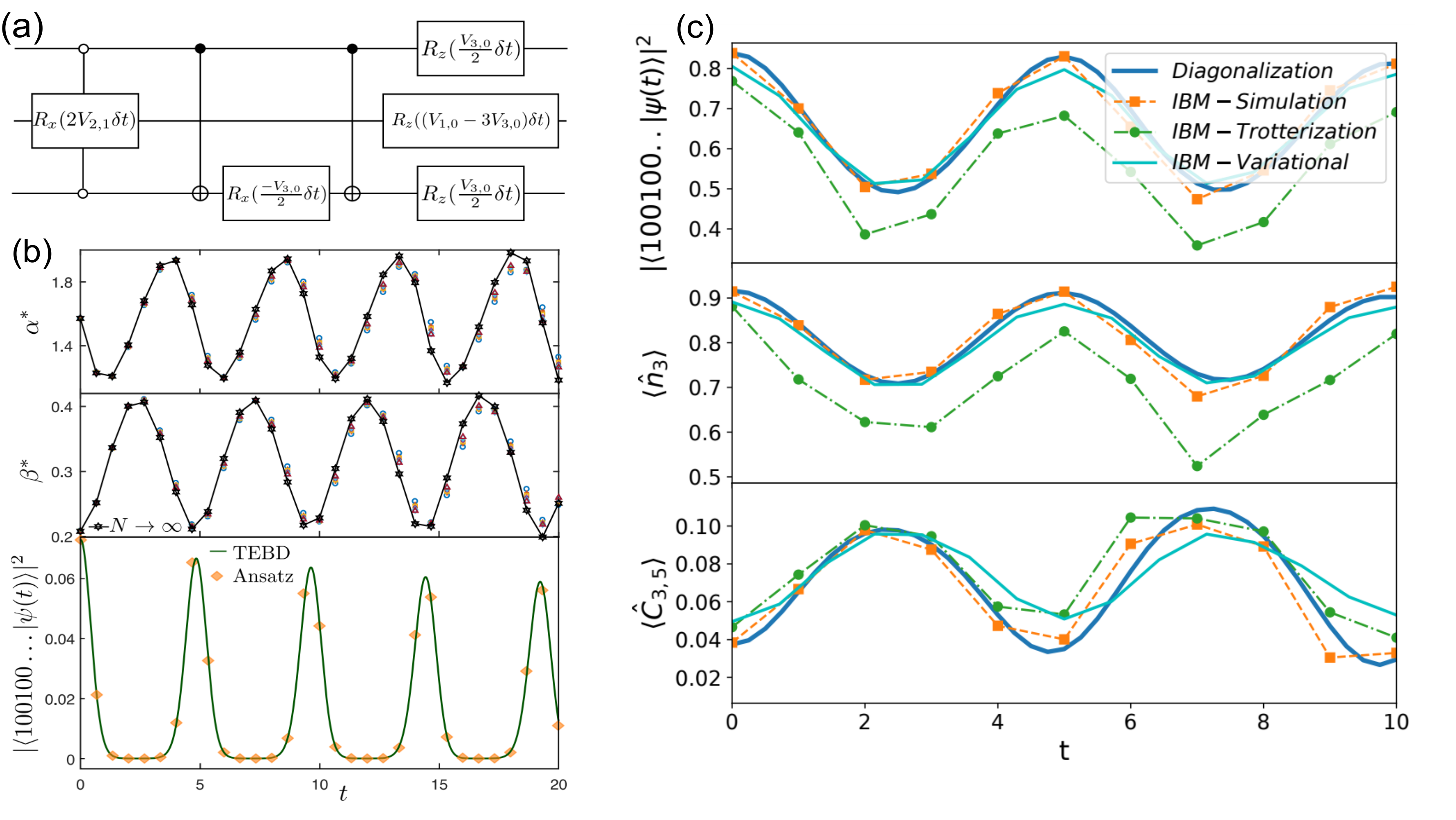}
    \caption{(a) Circuit implementation of the trotterized unitary $U_{\ell}$ representing one local unit of the FQH spin chain in Eq.~\ref{Uprod}.  
    (b) Top two panels: Optimal variational parameters $\alpha^*$ and $\beta^*$ [Eq.~\ref{eq:ansatz}] for the $1/3$-filled FQH ground state for $N=7$ to $13$ particles, and their extrapolation to $N{\to}\infty$ (solid black line). These optimal parameters vary smoothly with time and importantly exhibit negligible variation with system size. Bottom panel: Excellent agreement in the overlap of the time evolved state with the root state between that computed from TEBD simulations for $N{=}60$, and that from the variational ansatz [Eq.~\ref{eq:ansatz}] extrapolated to the same system size. TEBD simulations were performed using a bond dimension 20 with a time step $\Delta t{=}0.01$, resulting in truncation error $10^{-5}$. (c) $N=5$ post-quench evolution results for the fidelity, density and correlation function that capture quantum metric fluctuations. Comparison is made between exact diagonalization results, $k=15$ depth Trotterization circuits on the IBM simulator, gate optimized unitary output from IBM-Perth, and variational ansatz results from IBM-Santiago processors. {All panels of this figure} are adopted from Ref.~\cite{kirmani2022probing}.
    }
    \label{fig:fqhe2}
\end{figure*}

To utilize quantum computers in studying FQH states, the structure of the FQH Hilbert space should be mapped to the available quantum computing hardware. The one-dimensional structure of the effective FQH Hamiltonian and the truncation of the interaction potential in the appropriate gauge are crucial in this regard. Furthermore, fermionic operators can be mapped to spin (and hence qubit) degrees of freedom through the Jordan-Wigner transformation~\cite{jordan1928paulische} [see Sect.~\ref{circuit} above]. 
The exact ground state of the spin Hamiltonian in Eq.~\ref{mainh} at $1/3$ filling is given by~\cite{nakamura_exact}:
\begin{equation}\label{creator}|\psi\rangle={\cal N}\prod_j(1-t
S_j)|100100100\ldots\rangle
\end{equation}
where $t=\sqrt{V_{30}/V_{10}}$, $S_j=X^+_{j+1}X^+_{j+2}X^-_{j+3}X^-_j$ implements correlated spin flips while conserving the center-of-mass and electron filling, and $|100100100\ldots\rangle$ is the root state in the $\sigma_z$ spin basis. Spins $\uparrow$ and $\downarrow$ correspond to the Landau orbital being filled ($|1\rangle$) and empty ($|0\rangle$).

We next describe how to construct an equivalent unitary operator that creates this state. This exploits the idea of encoding $\ket{\psi}$ in Eq.~\ref{creator} by means of reduced qubits $|\mathbb{0}\rangle$. Each block of three consecutive qubits can be represented by a reduced qubit that indicates whether or not a block is squeezed~\cite{PRXQuantum.1.020309}. For any block of three qubits, the squeezing operator acts on it and the first qubit of the next block. i.e. $|1{ 0}0,1\rangle\to|0{ 1}1,0\rangle$.  
The initial state has no squeezed blocks and is represented by a sequence of zeros in the reduced space:  $|100100100100\rangle\to| \mathbb{0000}\rangle$. When the squeezing operator acts on a given block, the first qubit as well as the first qubit of the next three-qubit block will be set to $0$. Since $S_j$ annihilates the state unless the four qubits it acts upon are in the $1001$ configuration, if a block is already squeezed, the application of the squeezing operator on the neighboring left or right block will annihilate the state. Thus, a squeezed block cannot have any neighboring blocks that are also squeezed. In all, we can prepare the ground state defined in Eq.~\ref{creator} using the following sequence of unitary operators
\begin{align}
\label{Uprod}
 \ket{\psi} = U_{N/3-1}(\phi_{N/3-1})\cdots U_{1}(\phi_{1})\, U_{0}(\phi_{0})\,\ket{\mathbb{0}\cdots \mathbb{00}}
\end{align}
where 
\begin{equation}\label{eq:uk}
    U_k=e^{\phi_j(S_{3k}-S^\dagger_{3k})}
\end{equation}
acts on the $k$-th register and the angles $\phi_k$ obey the recursion relation 
\begin{equation}
  \phi_{k-1}=\arctan\left[-t\cos(\phi_k)\right],
\end{equation}
with boundary condition $\phi_{N/3-1}=\arctan(-t)$. As {demonstrated on an IBM quantum processor}~\cite{PRXQuantum.1.020309}, this construction gives a resulting wavefunction that exhibits key properties of the Laughlin-type FQH wave function in terms of the entanglement entropy~\cite{kitaev2006topological,levin2006detecting} as well as correlations of non-
local string operators\cite{Girvin1989}. In the explicit circuit construction of Ref.~\cite{PRXQuantum.1.020309}, the reduced-qubit preparation can be implemented with a quasi-one-dimensional three-leg-ladder layout, and the required controlled rotations act only between neighboring reduced registers and can be decomposed into local single- and two-qubit gates with circuit depth linear in the number of Landau orbitals.

\noindent\textbf{Probing geometric excitations and Hall viscosity} 
Recently, much interest has focused on the universal geometric features of FQH fluids such as Hall viscosity ~\cite{PhysRevLett.75.697, read2009non, HaldaneViscosity} and Girvin-MacDonald-Platzman magnetoroton collective excitations~\cite{GMP85,GMP86}. Of particular interest is the ``FQH graviton"~\cite{yang2012model,Golkar2016}  corresponding to the quadrupole magnetoroton degree of freedom in the long-wavelength limit~\cite{HaldaneGeometry}, so-termed due to its formal similarity with the fluctuating space-time metric in quantum gravity~\cite{bergshoeff2013zwei,bergshoeff2018gravity}. 

Haldane proposed that quantum-metric fluctuations can be induced by breaking rotational symmetry~\cite{HaldaneGeometry,yang2017generalized}. Following up on this idea, theoretical works~\cite{PapicMain, Lapa19} have probed the FQH graviton by quenching the metric of ``space", i.e., by suddenly introducing anisotropy in the FQH state. 
 Such geometric quenches induce coherent dynamics of the FQH graviton~\cite{PapicMain}, even though the graviton mode resides at finite energy densities above the FQH ground state. However, experimental signatures of FQH graviton modes were observed only very recently~\cite{Liang2024}, since the long-wavelength regime of magnetorotons is not easily accessible via inelastic light scattering~\cite{Pinczuk93, Platzman96, Kang01, Kukushkin09} which probes the short wavelength $\sim\ell_B=\sqrt{\hbar/eB}$ limit.

Given the versatility of universal quantum processors in preparing FQH states, the possibility of creating and detecting graviton models on digital quantum devices was investigated in Ref.~\cite{kirmani2022probing}. Geometric distortions are captured by the FQH metric $g_{ab}$ and modify the quasi-1D interaction Hamiltonian Eq.~\ref{intp} to
\begin{eqnarray}\label{Vkm}
V_{k,m} =  (k^2-m^2)e^{-2\pi^2(k^2+m^2-2ikmg_{12})/{L_2^2 g_{11}}}.
\end{eqnarray}
The mass tensor $g_{ab}$, $a,b=1,2$ is symmetric and unimodular ($\mathrm{det} g{=}1$)~\cite{HaldaneGeometry} and can generally be written as $g=\exp(\hat Q)$ where $\hat Q= Q (2\hat{d}_a \hat{d}_b - \delta_{a,b})$ is the Landau-de Gennes order parameter defined by the unit vector $\hat{\mathbf{d}}=(\cos(\phi/2), \sin(\phi/2))$~\cite{maciejko2013field}. Parameters $Q$ and $\phi$ intuitively represent the stretch and rotation of the metric, respectively. The FQH state is invariant under area-preserving deformations of $g$. Numerical exact diagonalization and entanglement entropy calculations verified that $V_{k,m}$ (Eq.~\ref{Vkm}) generates the appropriate ground state for the $1/3$ FQH state with modified geometry~\cite{kirmani2022probing}.

To capture FQH graviton fluctuations in quench dynamics, the ground state $|\psi_0\rangle$ corresponding to the  isotropic metric ($g_{11}{=}g_{22}{=}1$, $g_{12}{=}0$) is first prepared. At time $t = 0$, instantaneously diagonal anisotropy $g_{11}'{=}1/g_{22}'{> }1$ is introduced and the initial ground state evolves under the unitary dynamics generated by the post-quench anisotropic Hamiltonian.  The dynamical fluctuations of the emergent quantum metric 
$\widetilde{g}$, which is related to but not generally identical to the imposed 
post-quench metric $g'$, are associated with the FQH geometric excitation.

In Ref.~\cite{kirmani2022probing}, this quench dynamics was implemented on IBM quantum processors through two different approaches. The first approach is an efficient optimal-control-based~\cite{werschnik_quantum_2007,petersen_quantum_2010,Rahmani13} variational quantum algorithm~\cite{peruzzo2014variational,Wecker:15,wecker,mcclean2016theory}, analogous to the Quantum Approximate Optimization Algorithm (QAOA)~\cite{Farhi, Farhi:3,Yang17,Wang18,Zhou20,green1,ng2024analytical}. It generates the post-quench state using a hybrid classical-quantum approach~\cite{kokail2019self,kandala2017hardware}. The particular advantage of this variational method is favorable scaling with system size, with a linear-depth circuit depth and two variational parameters. Its disadvantage is that it is approximate by nature, and also involves the classical computation of the state, which hence cannot be interpreted as a full quantum evaluation. The second approach directly implements the unitary time evolution in discrete
time steps. Although this direct implementation is less scalable because the
circuit depth grows with simulated time and system size, it corresponds more
closely to the actual real-time quantum dynamics.

Next, we elaborate on how the quench is concretely implemented in the reduced register space of unsqueezed and squeezed blocks $\mathbb 0$ to $ \mathbb 1$ (see previous section). In terms of the reduced registers, squeezing terms in Hamiltonian Eq.~\ref{mainh} act as flips of $\mathbb 0$ to $ \mathbb 1$, so they can be viewed as Pauli $X$ operators. However, there is an important distinction, in that the Hilbert space is not a tensor product of reduced registers, since the squeezing can never generate two neighboring $\ldots\mathbb{1}\mathbb{1}\ldots$ configurations of the reduced registers~\cite{MoudgalyaThinTorus, MoudgalyaKrylov}. This type of constrained Hilbert space arises e.g., in the Fibonacci anyon chain~\cite{Feiguin2007}.  The inverse mapping is constructed as follows: for any $\mathbb 1$ we make a $011$ block. A $\mathbb 0$ that follows a $\mathbb 1$ ($\mathbb 0$) corresponds to a $000$ ($100$) block. With this mapping, the Hamiltonian Eq.~\ref{mainh} maps to a local spin-chain Hamiltonian 
\begin{equation}
\label{eq:spin-Hamiltonian}
\begin{split}
    \hat{H}=&\sum_\ell \left( (V_{1,0}-3V_{3,0}){\cal N}_\ell+V_{3,0}{\cal N}_\ell{\cal N}_{\ell+2} \right.\\
    & \left. +(1-{\cal N}_{\ell-1})[{\rm Re}(V_{2,1})X_\ell-{\rm Im}(V_{2,1}) Y_\ell](1-{\cal N}_{\ell+1}) \right),
    \end{split}
\end{equation}
where the boundary terms are omitted for simplicity. Here $\cal N {\equiv} |\mathbb{1}\rangle \langle \mathbb{1}|$ is the occupation number and $X {\equiv}  |\mathbb{0}\rangle \langle \mathbb{1}| {+}  |\mathbb{1}\rangle \langle \mathbb{0}|$ and $Y {\equiv}  -i |\mathbb{0}\rangle \langle \mathbb{1}| {+} i  |\mathbb{1}\rangle \langle \mathbb{0}|$ are effective Pauli operators. Incorporating the quantum Hall metric, the interaction potential $V_{k,m}$ in Eq.~\ref{Vkm} can be analogously mapped to a qubit chain~\cite{PRXQuantum.1.020309}.

The time evolution implementation $e^{-i \hat Ht}$  follows the standard procedure through the Trotter decomposition, as elaborated in Sect.~\ref{trotter} below. Here $\hat H$ is given by Eq.~\ref{eq:spin-Hamiltonian} with real $V_{2,1}$, and is already split into local terms i.e. $\hat H=\sum_\ell H_\ell$ .  
In the temporal direction, the evolution operator is decomposed into $n$ Trotter steps as $e^{-i\hat Ht} \approx \left[ \prod_l U_\ell(t/n) \right]^n$, where $\delta t=t/n$ and the approximation improves for larger $n$.
Ref.~\cite{kirmani2022probing} implements this dynamics using an error-mitigated Trotter circuit on IBM quantum hardware~\cite{saravanan2022pauli,Qiskit,9259942}. In the bulk decomposition, the next-nearest-neighbor term $N_\ell N_{\ell+2}$ is reindexed as $N_{\ell-1}N_{\ell+1}$, so the local gate $U_\ell(\delta t)$ acts only on the three reduced-register qubits $\ell-1,\ell,\ell+1$, as shown in Fig.~\ref{fig:fqhe2}(a).  Larger systems will be met with more noise in the quantum hardware, since the total circuit depth is expected to scale as $Nt$ (for a 1D local lattice)~\cite{Haah, Childs}. Reasonably accurate results were obtained on IBM quantum processors with 5 qubits after using noise-aware error mitigation methods and optimized compilations~\cite{saravanan2022pauli,Qiskit,9259942}.

To access larger systems, one must transcend the limitations from the significant errors accrued by the large number of entangling gates. One approach is through a hybrid classical-quantum method involving classical optimization, using the following variational ansatz for the final post-quench state
~\cite{kirmani2022probing}:
\begin{eqnarray}\label{eq:ansatz}
 |\psi_{\rm var}(\alpha,\beta)\rangle= \prod_\ell e^{-i\alpha {\cal N}_\ell} e^{-i\beta (1-{\cal N}_{\ell-1})X_\ell}|\mathbb{000}\dots\rangle. \quad 
\end{eqnarray}
Instead of the full exact unitary operation, alternating gates ${\cal N}_\ell$ and $(1-{\cal N}_{\ell-1})X_\ell$ are applied on each reduced register $\ell$ in this ansatz. 
The optimal parameters $\alpha^*, \beta^* \in [0,2\pi)$ are classically optimized at each time step $t$ using a dual annealing algorithm that maximizes the overlap $|\langle \psi_0|U^\dagger(t)|\psi_{\rm var}(\alpha,\beta)\rangle|$ with the exact state. Naively, it appears that the classical optimization needs to be performed for each $t$ and system size. Importantly, however, it was found that the optimal parameters $\alpha^*$, $\beta^*$ exhibit a simple oscillatory behavior with time, but depend only very weakly on the system size as shown in Fig.~\ref{fig:fqhe2}(b)~\cite{kirmani2022probing}. The data for system sizes $N=7,\dots, 13$ almost coincide, suggesting a direct extrapolation to the thermodynamic limit ($N{\to}\infty$), shown as the solid black line. As shown in the bottom panel of this figure, direct time-evolving block decimation (TEBD)~\cite{vidal2004efficient} calculations of $|\psi(t)\rangle$ for larger systems exhibit excellent agreement with the extrapolated parameters. Thus, the weak system-size dependence of the variational parameters eliminates the need for classical optimization at every system size, thereby providing access to system sizes that are classically inaccessible.

The quantum simulated dynamics are presented in Fig.~\ref{fig:fqhe2}(c) in the form of measured root-state fidelity \( |\langle \psi(t)|100100\ldots\rangle|^2 \), local density \( \langle n_j\rangle \) and equal-time density--density correlation function
\begin{equation}
C_{i,j}(t)= \langle n_i(t)\rangle \langle n_j(t)\rangle  -\langle n_i(t)n_j(t)\rangle.
\end{equation}
Although these observables are defined in the original fermionic basis, they can be directly reconstructed from measurements performed in the reduced basis using the mapping rules discussed above Eq.~(\ref{eq:spin-Hamiltonian}). As shown in Fig.~\ref{fig:fqhe2}(c), the variational results are in excellent agreement with the numerical simulations. Likewise, the error-mitigated Trotter approach reproduces the expected graviton oscillations, even though its deeper circuits and longer execution times lead to somewhat larger errors than in the variational implementation. These imperfections mainly produce quantitative deviations, while the essential dynamical features remain clearly captured.

In all, we have seen that quantum-simulation experiments have begun to access FQH-relevant diagnostics and response functions on gate-based platforms~\cite{PRXQuantum.1.020309,kirmani2022probing}. In the small system sizes accessible by the early NISQ demonstrations in Ref.~\cite{kirmani2022probing}, graviton-frequency oscillations can already be resolved from fidelities and density correlators. These results motivate extending digital protocols to larger registers and longer times, where finite-size scaling and more quantitative reconstruction of the emergent metric dynamics become feasible. We note that this approach can also be used to characterize the Hall viscosity~\cite{PhysRevLett.75.697}, a prototypical FQH property elusive in most experimental settings. Also measured through space-time metric modifications, the Hall viscosity has been measured on quantum devices via a similar algorithm~\cite{viscosityqc}. 
The 1D FQH formulation described also enables convenient access to FQH quasiparticle excitations with fractional statistics~\cite{Flavin2011}, as epitomized by the quantum hardware measurements of quasiparticle braiding and fractional anyonic phases in Ref.~\cite{PhysRevB108064303}.

\begin{figure*}
    \centering
    \includegraphics[width=0.9\linewidth]{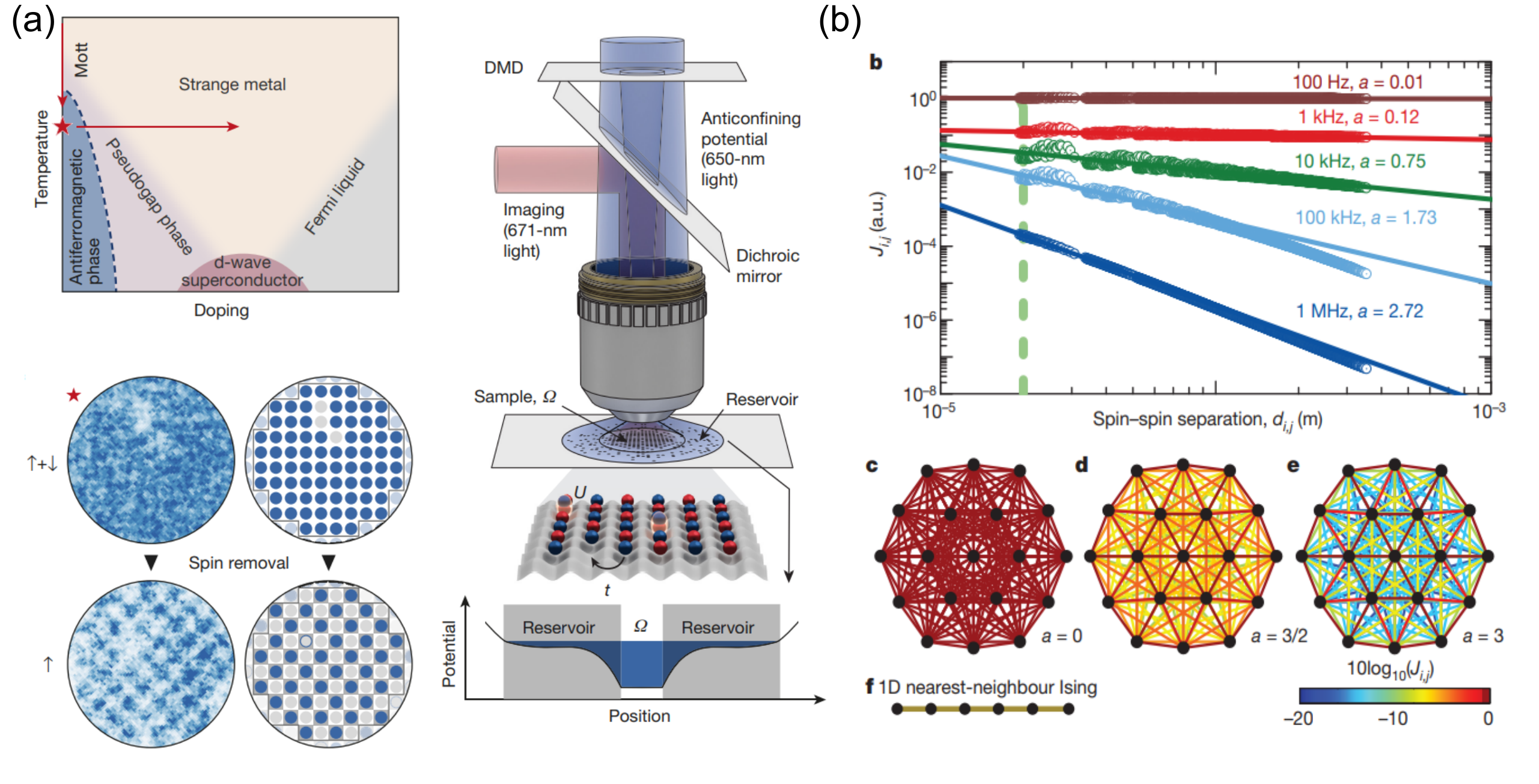}
    \caption{Strongly-correlated phases realized in pioneering analog
    quantum simulators. (a) Strongly correlated electronic phases and their realization in quantum simulators.
Top left of (a): schematic phase diagram of a doped Mott insulator system. Bottom left of (a): spin-resolved fluorescence images of the atomic distribution. Right of (a): experimental platform based on ultracold atoms in an optical lattice, imaged using a high-resolution quantum gas microscope ~\cite{mazurenko2017cold}.  (b) Programmable long-range spin models in trapped-ion simulators~\cite{britton2012engineered}.  Top of (b): Measured spin–spin Ising couplings $J_{i,j}\propto 1/r_{i,j}^{\alpha} $, as a function of ion–ion separation. Bottom of (b): Graph representation of the interaction networks: $\alpha=0$ almost all-to-all couplings with nearly uniform strength;  $\alpha=3/2$ intermediate-range interactions;  $\alpha=3$ interactions with rapidly suppressed long-distance couplings. Panel (a) is adopted from ~\cite{mazurenko2017cold}. Panel (b) is adopted from ~\cite{britton2012engineered}. }
    \label{fig:strong}
\end{figure*}

\subsubsection{Focus Area II: Pioneering quantum simulations for strongly correlated physics}

While gate-based ground-state preparation methods are promising new approaches,
it is important to recognize that analog quantum simulators had already achieved
major successes in realizing strongly correlated, non-equilibrium, and many-body
quantum phenomena before the current generation of programmable digital
processors. Examples include the optical-lattice realization of the
superfluid--Mott-insulator transition, quantum-gas-microscope studies of
Hubbard-regime atoms and antiferromagnetic correlations, non-equilibrium
correlation spreading and many-body localization in cold atoms, programmable
spin dynamics in trapped-ions, and large-scale Rydberg-array simulations of
constrained spin models, ordered phases, and topological spin-liquid
signatures
\cite{greiner2002quantum,bakr2009quantum,cheneau2012light,
schreiber2015observation,choi2016exploring,mazurenko2017cold,
friedenauer2008simulating,kim2010quantum,britton2012engineered,
jurcevic2014quasiparticle,bernien2017probing,ebadi2021quantum,
scholl2021quantum,semeghini2021probing}. These analog experiments provide
direct physical insight because the target Hamiltonian is implemented natively
through controllable tunneling, interactions, confinement potentials, or
long-range couplings. We therefore devote the following subsections to this
historical and physical context. For recent perspectives on analog quantum
simulators, including their architectures, verification, scalability, stability
to noise, and prospects for practical quantum advantage, the reader may refer
to several excellent reviews and perspectives
\cite{georgescu2014quantum,cirac2012goals,hauke2012can,
lewenstein2007ultracold,bloch2008many,bloch2012quantum,gross2017quantum,
Altman2021quantumSimulators,daley2022practical,harley2024going,
trivedi2024quantum,blatt2012quantum,monroe2021programmable,
saffman2010quantum,saffman2016quantum,browaeys2020many,
morgado2021quantum,Henriet2020quantumcomputing,defenu2023long,
goldman2014light,aidelsburger2018artificial,wiese2013ultracold,
dalmonte2016lattice,banuls2020simulating}.

In an analog simulator, the quantum system evolves \emph{natively} according
to the dynamics of a well-controlled physical platform whose effective
Hamiltonian can be tuned through external fields, engineered interactions, or
time-dependent control parameters. Examples include trapped-ion chains
[see Sect.~\ref{ion} above], optical lattices [see Sect.~\ref{optical} above],
and Rydberg-atom tweezer arrays [see Sect.~\ref{rydberg} above]. A common
strategy is \emph{adiabatic state preparation}
\cite{kato1950adiabatic,farhi2000quantum,greiner2002quantum}
(see Sect.~\ref{adiabatic} for more details): one initializes the simulator in
a simple, readily prepared state that is the ground state of an easily
implemented Hamiltonian $H_0$, and then slowly ramps control parameters so that
$H_0$ is continuously deformed into the desired target Hamiltonian $H_1$.
Provided that the ramp is slow compared with the inverse relevant many-body gap
and that diabatic transitions are sufficiently suppressed, the state follows
the instantaneous ground state and approaches the target ground state. In the
following, we introduce how this setup has been exploited to engineer ground
states within strongly correlated phases.

\noindent\textbf{Optical-lattice Hubbard physics and bosonic Mott transitions.} Strongly correlated quantum phases emerge when interparticle interactions compete with or dominate kinetic energy, rendering simple single-particle pictures inadequate. Examples include conventional symmetry-breaking phases, such as superfluids and unconventional magnets~\cite{leggett2001bose,anderson1950antiferromagnetism,sachdev1999quantum}, as well as more subtle forms of quantum
order~\cite{hasan2010colloquium,qi2011topological} such as Mott insulators~\cite{gutzwiller1963effect,imada1998metal,lee2006doping} and quantum spin liquids
\cite{anderson1973resonating,balents2010spin,zhou2017quantum,kitaev2006anyons}. 
Ultracold atoms in optical lattices provide one of the most mature and flexible
analog platforms for realizing and probing such phases [see Sect.~\ref{optical} above], since lattice geometries, tunneling amplitudes, and
interaction strengths can be tuned with high precision via laser configurations, lattice depths, and Feshbach
resonances.

A paradigmatic model for strongly correlated bosons in an optical lattice is the
Bose--Hubbard Hamiltonian
\cite{jaksch1998cold,bloch2008many,lewenstein2007ultracold},
\begin{equation}
H_{\rm BH}
=
-\sum_{i,j} J_{ij}\, a_i^\dagger a_j
+\frac{U}{2}\sum_i n_i(n_i-1)
-\mu\sum_i n_i,
\label{eq:BH}
\end{equation}
where $a_i$ ($a_i^\dagger$) annihilates (creates) a boson at site $i$,
$n_i=a_i^\dagger a_i$ is the number operator, $J_{ij}$ denotes the hopping
matrix element between sites $i$ and $j$ and is typically nonzero only for
nearest neighbors in a deep lattice, $U$ is the on-site interaction energy, and
$\mu$ is the chemical potential. The competition between kinetic delocalization
and interaction-induced number locking is controlled by the ratio $U/J$, where
$J$ denotes the characteristic nearest-neighbor tunneling amplitude. For weak
interactions, $U/J\ll 1$, the ground state is superfluid, exhibiting phase
coherence and off-diagonal long-range order, which in a symmetry-breaking or
mean-field description is captured by a nonzero order parameter
$\langle a_i\rangle\neq 0$. For strong repulsion at commensurate filling,
$U/J\gg 1$, the system enters a Mott-insulating phase
\cite{fisher1989boson,stoferle2004transition,capogrosso2008monte}.

A landmark experiment by Greiner \emph{et al.} provided a direct observation of
the superfluid--Mott-insulator transition, firmly establishing optical-lattice
gases as a controllable setting for quantum many-body ground-state physics
\cite{greiner2002quantum}. In their work, a Bose--Einstein condensate of
$^{87}$Rb atoms was loaded into a three-dimensional optical lattice, and the
transition was identified using time-of-flight interference imaging. In the
superfluid regime, coherent matter-wave interference produced sharp Bragg peaks
in the expanded density distribution. As the lattice depth was increased,
thereby suppressing tunneling and increasing the effective ratio $U/J$, these
interference peaks disappeared, indicating the loss of global phase coherence
and the onset of the Mott-insulating state. Subsequent developments enabled
microscopic access to bosonic Mott physics via single-site-resolved imaging,
including early quantum-gas-microscope demonstrations of fluorescence-based
detection in optical lattices
\cite{bakr2009quantum,sherson2010single}. These capabilities enabled direct
observation of local number statistics, defects, and site-resolved correlations,
and were soon leveraged to reveal nonlocal correlation diagnostics such as
correlated particle--hole pairs and string order in low-dimensional Mott
insulators~\cite{endres2011observation}.

\noindent\textbf{Fermi--Hubbard magnetism and microscopic spin correlations.} Moreover, the fermionic counterpart of Hubbard physics is captured by the
Fermi--Hubbard model,
\begin{equation}
H_{\rm FH}
=
-t \sum_{\langle i, j\rangle, \sigma}
\left(c_{i \sigma}^{\dagger} c_{j \sigma}+\mathrm{h.c.}\right)
+U \sum_i n_{i \uparrow} n_{i \downarrow},
\label{eq:FH}
\end{equation}
where $c_{i\sigma}$ ($c_{i\sigma}^\dagger$) annihilates (creates) a fermion
with spin $\sigma\in\{\uparrow,\downarrow\}$ on site $i$,
$n_{i\sigma}=c_{i\sigma}^\dagger c_{i\sigma}$ is the number operator, $t$ is
the nearest-neighbor tunneling amplitude, and $U$ penalizes double occupation.
Although Eq.~\eqref{eq:FH} is compact, it hosts a rich hierarchy of correlated
phenomena. At half filling and strong repulsion, $U/t\gg 1$, charge
fluctuations are suppressed, placing the system in a Mott regime in which the
low-energy physics is dominated by spin degrees of freedom. This spin sector
emerges from virtual hopping processes: a fermion can hop to a neighboring site
only virtually, and second-order perturbation theory in $t/U$ generates
antiferromagnetic superexchange between neighboring spins. Projecting to the
singly occupied subspace yields the effective exchange Hamiltonian
\begin{equation}
H_{\rm eff}
=
J \sum_{\langle i,j\rangle}
\left(\mathbf{S}_i\cdot\mathbf{S}_j-\frac{1}{4}n_i n_j\right),
\qquad
J=\frac{4t^2}{U},
\label{eq:superexchange}
\end{equation}
where
$\mathbf{S}_i=\frac{1}{2}\sum_{\alpha,\beta}
c_{i\alpha}^\dagger \boldsymbol{\sigma}_{\alpha\beta} c_{i\beta}$
and $n_i=n_{i,\uparrow}+n_{i,\downarrow}$. At half filling,
$n_i=1$ within the low-energy subspace, and the density term contributes only
a constant, leaving the standard antiferromagnetic Heisenberg model with
exchange scale $J=4t^2/U$. This mechanism accounts for the emergence of
short-range antiferromagnetic correlations in the Mott regime
\cite{anderson1950antiferromagnetism,zhang1988effective,
auerbach2012interacting,imada1998metal,esslinger2010fermi}.

Early experimental access to the fermionic Mott-insulating regime in an optical
lattice was achieved by J\"ordens \emph{et al.} using a two-component gas of
$^{40}$K atoms~\cite{jordens2008mott}. By tuning the interaction strength via a
Feshbach resonance and increasing the lattice depth, thereby reducing the
tunneling $t$ relative to the on-site repulsion $U$, they observed a pronounced
suppression of the double occupancy,
$\langle n_{i,\uparrow}n_{i,\downarrow}\rangle$. This directly reflects Mott
physics: at large $U/t$, configurations with two fermions on the same site are
energetically penalized. In the same regime, the thermodynamic response became
increasingly incompressible, as inferred from the reduced response of double
occupancy to atom-number changes, consistent with the formation of an
incompressible Mott region near unit filling. Complementary evidence for
fermionic Mott behavior and finite-temperature thermodynamic response was
obtained by Schneider \emph{et al.} through compressibility and
equation-of-state measurements in optical lattices~\cite{schneider2008metallic}.

Building on this capability, Greif \emph{et al.} then demonstrated short-range magnetic correlations in a thermalized Fermi–Hubbard system~\cite{greif2013short}. The key challenge is that magnetism requires temperatures below the exchange scale, which is typically much smaller than $t$ and $U$. By engineering a dimerized (or otherwise anisotropic) lattice configuration that facilitates entropy redistribution and enhances local spin ordering, the experiment reached a regime where nearest-neighbor spin correlations became measurable. Using singlet–triplet–resolved detection on adjacent sites, they observed an excess of singlet correlations relative to triplets, consistent with the onset of local antiferromagnetic order along the direction with stronger tunneling. A closely related and conceptually important milestone is the dynamical control and direct observation of superexchange processes, which provide the effective low-energy mechanism for magnetic ordering in the Mott regime~\cite{trotzky2008time}.

A particularly direct probe of magnetic ordering was demonstrated by
Hart \emph{et al.}~\cite{hart2015observation}, who measured antiferromagnetic
correlations in a Fermi--Hubbard system using spin-sensitive Bragg scattering.
The use of a compensated optical lattice improved density homogeneity and
reduced heating, enabling the system to access lower-entropy regimes. This
allowed the measured spin correlations to be quantitatively compared with
quantum Monte Carlo predictions, marking one of the earliest demonstrations
that optical-lattice quantum simulators could access strongly correlated
magnetic physics with near-quantitative accuracy.

The subsequent observation of robust antiferromagnetic correlations in the
two-dimensional Fermi--Hubbard model was enabled by quantum gas microscopy. In
a landmark experiment, Mazurenko \emph{et al.} used spin-resolved, single-site
imaging to probe antiferromagnetic order in a 2D optical lattice
\cite{mazurenko2017cold}. The experiment directly measured site-resolved spin
configurations and spin--spin correlation functions [Fig.~\ref{fig:strong}(a)].
The extracted antiferromagnetic correlation length reached the size of the
finite system, accompanied by a peak in the spin structure factor and a
staggered magnetization close to the ground-state value. These observations
provided strong evidence for long-range antiferromagnetic order in a finite
cold-atom Fermi--Hubbard system. Beyond half filling and lower dimensions, quantum gas microscopy has further
enabled direct probes of doped Mott physics. Hilker \emph{et al.} revealed
hidden antiferromagnetic correlations in doped one-dimensional Hubbard chains
through nonlocal string correlators~\cite{hilker2017revealing}, while Chiu
\emph{et al.} identified geometric string patterns in the doped two-dimensional
Hubbard model, providing microscopic evidence for the interplay between hole
motion and spin order~\cite{chiu2019string}. Together, these milestones firmly
established ultracold-atom platforms as powerful quantum simulators capable of
resolving microscopic mechanisms underlying magnetic ordering and doped Mott
physics in the Fermi--Hubbard model.

\noindent\textbf{Frustrated spin models in trapped-ion and Rydberg simulators.} Trapped-ion and Rydberg-atom quantum simulators also provide highly flexible
platforms for engineering quantum spin Hamiltonians with tunable interaction
range and anisotropy, enabling controlled exploration of frustrated magnetism
and quantum phase transitions. A pioneering demonstration was reported by
Kim \emph{et al.}~\cite{kim2010quantum}, who realized the minimal frustrated
Ising triangle using three trapped-ions with programmable Ising couplings.
Through direct measurement of state populations and entanglement-witness
observables, the experiment revealed a transition from separable configurations
to strongly entangled GHZ- and W-type ground states, showing that even the
simplest frustrated motif can generate multipartite entanglement in its
low-energy manifold.

Building on this foundation, Britton \emph{et al.}~\cite{britton2012engineered}
extended trapped-ion spin simulation to a much larger two-dimensional setting:
a triangular crystal of hundreds of $^9{\rm Be}^+$ ions confined in a Penning
trap. By applying a spin-dependent optical dipole force, the experiment
engineered antiferromagnetic Ising interactions with an approximately power-law
form,
\begin{equation}
J_{ij}\propto \frac{1}{d_{ij}^{\alpha}},
\end{equation}
as shown in Fig.~\ref{fig:strong}(b). The tunability of $\alpha$ allowed the
simulator to access different effective interaction ranges, from nearly
infinite-range to more rapidly decaying couplings, thereby establishing a
large-scale platform for studies of long-range quantum magnetism. More recently,
programmable Rydberg-atom arrays have enabled direct access to highly frustrated
and topological spin-liquid-like regimes, including experimental probes of
nonlocal string diagnostics in engineered lattice geometries
\cite{semeghini2021probing}. These results demonstrate that trapped-ion and
Rydberg platforms can access strongly correlated spin physics with quantitative
control, and they lay the groundwork for subsequent studies of frustrated
magnetism, spin liquids, and dynamical phases in programmable atomic arrays.

\noindent\textbf{Digital variational approaches to strongly correlated models.} Digital quantum processors have begun to provide a complementary perspective
on the same class of strongly correlated problems. Early variational-algorithm
work identified strongly correlated electron models, including Hubbard-type
Hamiltonians, as important targets for near-term quantum algorithms
\cite{wecker2015solving}. Subsequent superconducting-processor experiments
established related variational and fermionic-state-preparation workflows in
molecular and spin-model settings
\cite{kandala2017hardware,google2020hartree}. More directly connected to
Hubbard physics, Stanisic \emph{et al.} used a scalable variational strategy to
access ground-state properties of medium-scale Fermi--Hubbard instances on
Google superconducting processor~\cite{stanisic2022observing}, as
discussed above.

\subsection{Quantum simulation of topological Dynamics and Invariants}\label{sec1:topological}

\subsubsection{Digital quantum simulation of topological edge states}
\label{sec:topological_dynamics}

Digital quantum processors provide a flexible route to emulate topological band
structures and many-body topological phases through programmable unitary
circuits. Compared with analog simulators, the digital approach offers
controllable initialization, quenches, and Floquet driving, as well as the
ability to compile high-dimensional single-particle topology into
lower-dimensional qubit layouts
\cite{xiang2023simulating,xiao2023robust,viyuela2018observation,
koh2022stabilizing,koh2022simulation,koh2024realization,roushan2014observation,
schroer2014measuring,zhang2022digital,barends2015digital,mei2020digital,ng2026digital}.
In this sense, gate-based quantum processors are not limited to
direct lattice emulation. They can also realize topological phenomena through
circuit-native resources, including fermion-to-qubit encodings, engineered
interactions, many-body Hilbert-space compression, Floquet unitaries, and
non-unitary simulation primitives.

A broad class of topological models can be mapped to qubit Hamiltonians using
Jordan--Wigner (JW) or Bravyi--Kitaev (BK) transformations for fermions
[see Sect.~\ref{circuit}], or through direct spin encodings for
spin-conserving and particle-conserving models, such as SSH-like chains,
Kitaev wires, and spin-orbit-coupled lattices. Under the JW transform,
quadratic hopping and pairing terms become Pauli strings, which can then be
implemented through gate sequences or compiled into circuit-native evolution
blocks. This paradigm underlies early superconducting-circuit digital
simulations of fermionic models with hundreds of gates~\cite{barends2015digital}
and continues to be used in recent quantum hardware realizations of topological
fermion Hamiltonians
\cite{xiao2021determining,rancic2022exactly,koh2022stabilizing,
stenger2022simulating,sung2023simulating,koh2022simulation,
google2023non,koh2024realization,QianScience2025}.

\begin{figure*}[t]
    \centering
    \includegraphics[width=\textwidth]{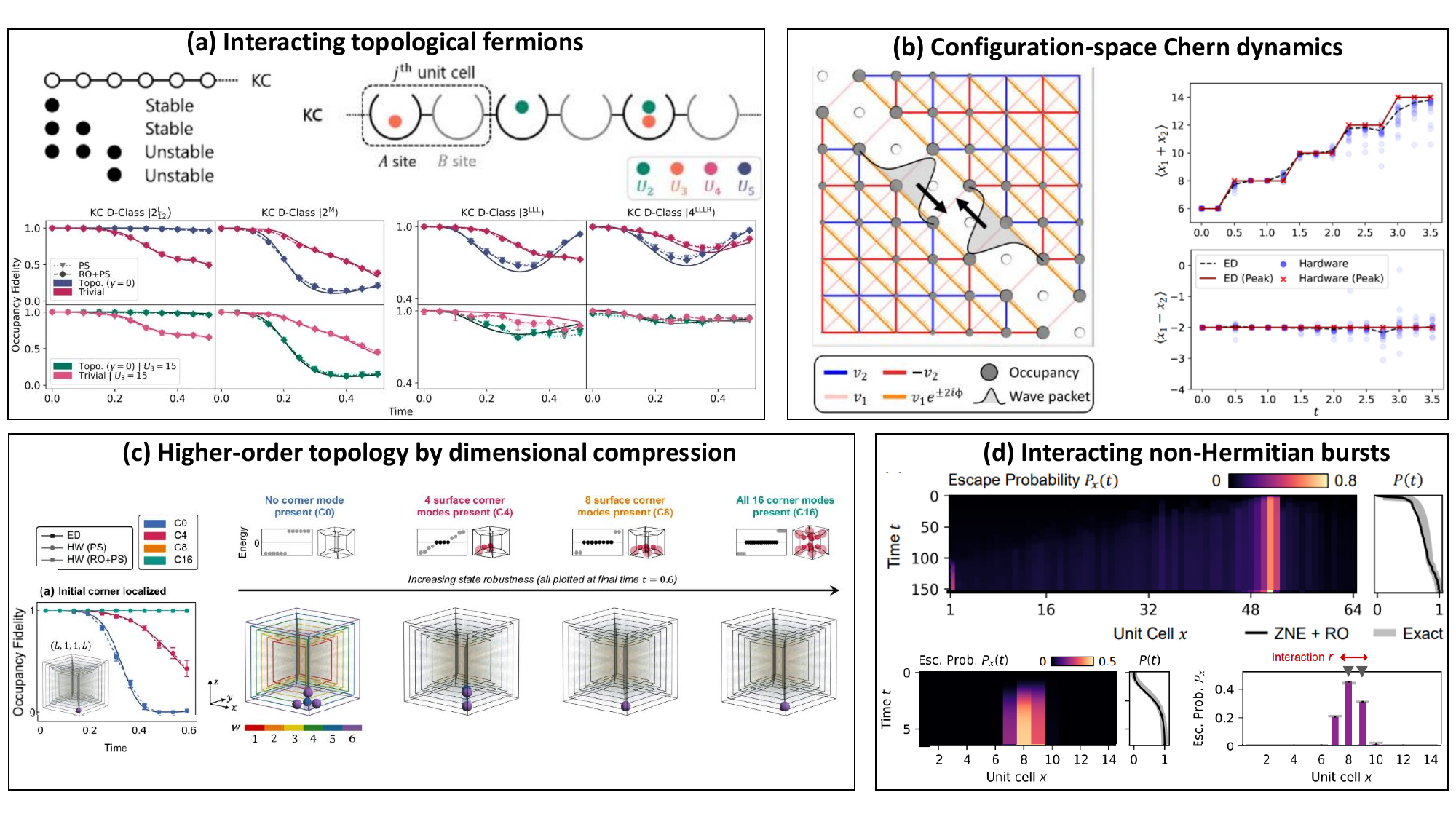}
 \caption{
Circuit-native routes to topological phenomena on digital quantum processors.
\textbf{(a)} Interacting topological fermions are studied on an extended Kitaev-chain model. The schematic shows the finite number of boundary modes and the sublattice-selective density-density interactions used in the digital Hamiltonian. The occupancy-fidelity traces show that engineered interaction can stabilize few-body boundary-localized configurations, including states with more fermions near an edge than expected from single-particle topological-mode counting~\cite{koh2022stabilizing}.
\textbf{(b)} Interaction-induced Chern dynamics in configuration space. A two-dimensional Chern checkerboard lattice is encoded in the two-body configuration space of a one-dimensional interacting chain. The hardcore constraint defines a virtual boundary along $x_1=x_2$, and the measured dynamics show monotonic motion of $\langle x_1+x_2\rangle$ while $\langle x_1-x_2\rangle$ remains approximately fixed, demonstrating chiral propagation along the virtual boundary~\cite{koh2022simulation}.
\textbf{(c)} Dimensional compression of higher-order topological lattices. A $d$-dimensional higher-order topological lattice is mapped to a $d$-species interacting one-dimensional chain, reducing the qubit cost from an explicit $L^d$ site encoding to $dL$. The data illustrate the emergence and robustness of corner modes in compressed higher-dimensional lattices, including the $d=4$ tesseract case~\cite{koh2024realization}.
\textbf{(d)} Interacting non-Hermitian bursts. Digital non-unitary evolution enables non-Hermitian edge-burst dynamics and interaction-induced cluster bursts. The escape-probability profiles show how interactions reshape dissipative topological dynamics beyond the single-particle edge-burst picture, allowing probability to accumulate and escape near an interacting particle cluster
rather than only at a physical boundary~\cite{koh2025interacting}. Panels (a-d) are respectively adopted from Refs.~\cite{koh2022stabilizing}, \cite{koh2022simulation}, \cite{koh2024realization} and \cite{koh2025interacting}.
}
    \label{fig:digital_topological_phenomena}
\end{figure*}

We summarize in
Fig.~\ref{fig:digital_topological_phenomena} four circuit-native strategies that highlight how programmable interactions, configuration-space mappings, dimensional compression, and non-unitary circuit primitives expand the range of topological phenomena accessible to gate-based digital quantum processors.

\noindent\textbf{One-dimensional interacting topological states. } A first example, shown in Fig.~\ref{fig:digital_topological_phenomena}(a), is the simulation of
topological fermion chains where quantum statistics and interactions play an
essential role. These 1D topological chains possess nontrivial windings in momentum space, which leads to nontrivial spectral flow~\cite{qi2011generic, huang2012entanglement,lee2015free} that mandates the existence of topological boundary modes. In Ref.~\cite{koh2022stabilizing}, SSH and extended
Kitaev-chain Hamiltonians in the BDI and D symmetry classes were implemented on
IBM superconducting quantum processors using tensor-network-aided circuit
recompilation, readout-error mitigation, post-selection, and iterative quantum
phase estimation. 
The experiments verified the persistence of
boundary-localized topological fermions and reconstructed the corresponding
few-fermion topological spectra. Significantly, designed few-fermion
interactions were used to stabilize boundary-localized states containing more
fermions than would be allowed by the number of available single-particle
topological boundary modes alone.

At the Hamiltonian level, this example starts from fermionic SSH and extended
Kitaev-chain models, with hopping, pairing, chemical-potential, and interaction
terms written in second quantization. For the extended Kitaev-chain setting,
one may schematically write
\begin{equation}
H_{\rm KC}
=
H_{\mu}+H_{\rm hop}+H_{\rm pair}+H_{\rm int},
\end{equation}
where the hopping and pairing structure determines the BDI or D symmetry class.
The interaction-stabilization physics can be captured by adding
sublattice-selective density-density terms, for example
\begin{equation}
H_{\rm int}^{\rm KC}
=
U_2\sum_j n^A_j n^A_{j+1}
+
U_3\sum_j n^A_j n^A_{j+2},
\qquad
n^A_j=(c^A_j)^\dagger c^A_j .
\label{eq:koh_kc_interaction_review}
\end{equation}
These terms do not increase the number of single-particle topological zero
modes. Rather, after the fermionic Hamiltonian is mapped to Pauli strings by a
JW- or BK-type encoding and compiled into shallow superconducting-qubit
circuits, the interactions reshape the few-body spectrum so that otherwise
unprotected boundary-localized configurations become long-lived. The measured
diagnostic is the occupancy fidelity, which tracks whether the site-occupation
pattern remains close to the initial boundary-localized configuration during
real-time evolution. Thus the quantum processor directly probes how
programmable interactions modify topological boundary robustness in a
many-body fermionic setting. This provides one of the first NISQ-era examples  in which programmable
interactions are not merely a route to implementing a target Hamiltonian, but
become the mechanism by which topological robustness is reshaped in a few-body
quantum setting.

\noindent\textbf{Chern edge states. } In two dimensions or higher, 
nontrivial band topology can give rise to robust edge or corner modes with no lower-dimensional analogs, as is the case of Chern or higher-order topological systems \cite{TKNN1982,haldane1988model,hasan2010colloquium,qi2011topological,koh2024realization,gu2016holographic,QianScience2025}. Chern lattices~\cite{haldane1988model}, protected by nontrivial second-homotopy of the band manifold, in particular possess robust edge modes that circulate around the sample boundary, impervious to disorder.  

To clearly simulate the boundary states of a higher-dimensional topological lattice, at least hundreds or thousands of logical unit cells must be accessible. 
A circuit-native strategy is to use the many-body Hilbert space of a quantum processor as a compressed representation, as illustrated in Fig.~\ref{fig:digital_topological_phenomena}(b). 
In Ref.~\cite{koh2022simulation}, a two-dimensional Chern checkerboard model~\cite{sun2011nearly,regnault2011fractional,lee2013lattice} was mapped exactly onto the two-body
configuration space of a one-dimensional chain. The two spatial coordinates of
the original lattice are encoded as the positions of two distinguishable
hardcore bosons, while diagonal hoppings of the Chern model become engineered
two-body hopping interactions. A key consequence is that the hardcore constraint generates a virtual
boundary along $x_1=x_2$, such that the measured occupancy density $\rho(x_1,x_2)$
exhibits unidirectional chiral propagation along this boundary (despite unbroken translation invariance at the model level), in agreement
with exact diagonalization.

Concretely, the model consists of horizontal, vertical, and diagonal real-space hoppings on a two-dimensional checkerboard lattice, and the compression identifies a basis state
of the original lattice with a two-particle configuration,
\begin{equation}
|x_1,x_2\rangle
\;\longleftrightarrow\;
\mu^\dagger_{x_1}\nu^\dagger_{x_2}|0\rangle ,
\label{eq:chern_basis_mapping_review}
\end{equation}
where $\mu$ and $\nu$ label two distinguishable hardcore-boson species on a
one-dimensional chain. Hoppings that change only $x_1$ or only $x_2$ become
single-particle hopping terms of the corresponding species, while diagonal
hoppings that change both coordinates become programmable two-body hopping
terms. In this representation, the physical edge of the Chern lattice becomes a
virtual boundary in configuration space~\cite{lee2021many}, and the chiral edge mode is detected
through the measured distribution $\rho(x_1,x_2)$ and centroid observables
$\langle x_1+x_2\rangle$ or $\langle x_1-x_2\rangle$. The resulting chiral dynamics therefore arise from designed interactions rather than from a direct implementation of two-dimensional single-particle hoppings.

\noindent\textbf{Higher-order topological states. } Higher-order topological phases, characterized by vanish first-order topological polarization but non-vanishing higher-order polarization, possess boundary states localized at
corners or hinges rather than edges. They pose particular challenges for both
simulation and diagnosis because their defining boundary physics is inherently
higher-dimensional. Quantum processors provide a natural platform for studying
such phases through programmable Hamiltonian engineering, dimensional
encodings, and time-resolved measurements. Early superconducting-qubit
experiments simulated two-dimensional second-order topological phases and
related topological phase transitions by reconstructing bulk pseudo-spin
textures~\cite{USTCARXIV2022}. More recently, USTC experiments demonstrated
both equilibrium and non-equilibrium higher-order topological phases on a
two-dimensional programmable superconducting quantum processor
\cite{QianScience2025}. In that work, programmable circuits on a $6\times 6$
qubit array were used to implement driven higher-order topological dynamics,
and experimentally accessible real-space dynamical invariants were introduced
to identify zero- and $\pi$-quasienergy corner modes. These results establish a
measurement paradigm for higher-order topology that does not rely only on
momentum-space band structure and is naturally suited to finite, driven quantum
systems.

A complementary hardware-efficient strategy, summarized in
Fig.~\ref{fig:digital_topological_phenomena}(c), is to reduce
the qubit overhead needed to represent higher-dimensional higher-order
topological lattices. The experiment in Ref.~\cite{koh2024realization}
introduced an encoding that maps a $d$-dimensional higher-order topological
lattice onto a reduced-dimensional interacting model. In this construction, a
lattice coordinate is represented as a multiparticle configuration on a
one-dimensional chain, reducing the qubit cost from $L^d$ to $dL$ for
single-particle higher-order topology.
This enabled the simulation of square, cubic, and
four-dimensional tesseract higher-order topological lattices on IBM
superconducting processors. The experiments measured the dynamical robustness
of corner- and edge-localized states and probed protected midgap modes using
iterative phase estimation.

The dimensional-compression map generalizes the two-particle construction
above. Starting from a generic $d$-dimensional single-particle lattice
Hamiltonian
\begin{equation}
H
=
\sum_{\mathbf r,\mathbf r'}
\sum_{\gamma,\gamma'}
h^{\gamma\gamma'}_{\mathbf r\mathbf r'}
c^\dagger_{\mathbf r\gamma}c_{\mathbf r'\gamma'},
\label{eq:hot_parent_hamiltonian_review}
\end{equation}
each spatial coordinate is assigned to a distinct particle species on a
one-dimensional chain,
\begin{equation}
c^\dagger_{\mathbf r\gamma}
\;\longrightarrow\;
\prod_{\alpha=1}^{d}
\left(\omega^{\alpha}_{r_\alpha\gamma}\right)^\dagger ,
\qquad
c_{\mathbf r\gamma}
\;\longrightarrow\;
\prod_{\alpha=1}^{d}
\omega^{\alpha}_{r_\alpha\gamma}.
\label{eq:hot_operator_mapping_review}
\end{equation}
The compressed Hamiltonian becomes
\begin{equation}
H_{\rm 1D}
=
\sum_{\mathbf r,\mathbf r'}
\sum_{\gamma,\gamma'}
h^{\gamma\gamma'}_{\mathbf r\mathbf r'}
\prod_{\alpha=1}^{d}
\left(\omega^{\alpha}_{r_\alpha\gamma}\right)^\dagger
\omega^{\alpha}_{r'_\alpha\gamma'} .
\label{eq:hot_1d_hamiltonian_review}
\end{equation}
Thus, a single-particle hopping process in the original $d$-dimensional lattice
is represented as a correlated $d$-body hopping process among distinguishable
particles in one dimension. The resulting interacting chain has only $dL$
sites, while its $d$-particle configuration space reproduces the original
$L^d$ lattice. This is the mathematical reason why square, cubic, and
tesseract higher-order topological lattices can be simulated on available
superconducting-qubit hardware without explicitly laying out the full
hypercubic lattice.

Together, these approaches illustrate two complementary advantages of quantum
processors for higher-order topology: direct implementation of finite driven
lattices, and circuit-level encodings that compress higher-dimensional
topological structure into hardware-accessible qubit layouts.

More broadly, quantum processors motivate a reexamination of topology in
finite, driven, noisy, and open quantum systems. In realistic experimental
settings, topological invariants may not appear as perfectly quantized
integers, but instead as robust finite-size plateaus, real-space markers, or
dynamically stabilized quantities. In Hermitian systems, this motivates
finite-sample formulations such as local Chern markers, Bott indices, and
noncommutative Chern numbers, which remain meaningful in the presence of
boundaries, disorder, and inhomogeneity
\cite{bianco2011mapping,hastings2010almost,prodan2013noncommutative,
xu2020measuring}. In non-Hermitian systems, finite-size effects are even more
pronounced: the bulk--boundary correspondence must often be formulated through
non-Bloch or generalized-Brillouin-zone invariants, while finite chains can
display oscillatory gaps, size-dependent topological modes, critical
skin-effect scaling, and fragmentation of the generalized Brillouin zone
\cite{chen2019finite,yao2018edge,kunst2018biorthogonal,yokomizo2019non,
song2019non,zhang2020correspondence,li2020critical,qin2023universal,
li2025phasespace,li2026gbzfragmentation}. Understanding how these invariants
persist, degrade, or transform under noise and dissipation has therefore
become an active direction
\cite{viyuela2014uhlmann,budich2015topology,bardyn2013topology,
bardyn2018probing,viyuela2018observation,azses2020identification},
with implications for both fundamental physics and quantum information
processing. In particular, the resilience of topological signatures observed
on superconducting, trapped-ion, and neutral-atom platforms has been proposed
as a diagnostic tool for benchmarking quantum hardware beyond conventional
fidelity-based metrics
\cite{choo2018measurement,zhang2022digital,QianScience2025,google2023non,
xu2024non,iqbal2024topological,iqbal2024non,semeghini2021probing}.

\noindent\textbf{Consequences of nontrivial spectral topology. } 
Nontrivial topological winding also manifests in the complex spectra of non-Hermitian systems. One particularly dramatic phenomenon is the so-called non-Hermitian edge burst~\cite{xue2022non}, where a wavepacket undergoes sudden extinction i.e. ``bursts" upon encountering a boundary, due to the lossiness of non-Hermitian boundary modes.  
A recent IBM Quantum experiment, shown in Fig.~\ref{fig:digital_topological_phenomena}(d) and later detailed in Sect.~\ref{nonunitary}, observed the non-Hermitian edge burst and its
interacting generalizations on superconducting quantum hardware
\cite{koh2025interacting}. The protocol combines product-formula time evolution
with a linear-combination-of-unitaries construction for the anti-Hermitian
component, enabling non-unitary dynamics to be embedded into a unitary circuit
with ancillary-qubit reuse. Applied to a lossy quantum ladder, this approach
revealed edge-burst signatures in systems of up to 64 unit cells and showed
that interactions can generate spatially extended edge-burst patterns and
cluster bursts in the bulk. This work extends the digital simulation of
topological dynamics from Hermitian band and Floquet settings to correlated
non-Hermitian systems, where topology, dissipation, and many-body constraints
are intertwined.

The effective model is an interacting lossy ladder whose single-particle part
contains Hermitian hopping together with sublattice-selective loss, and whose
many-body part contains range-dependent density-density interactions. A compact
form is
\begin{equation}
H_{\rm eb}
=
H_{\rm H}+H_{\rm A}
+
\sum_{r\geq 1}U_r\sum_z n_z n_{z+r},
\label{eq:nh_ladder_review}
\end{equation}
where $H_{\rm H}$ is Hermitian and $H_{\rm A}$ is anti-Hermitian. The loss makes
the propagator $V(t)=e^{-iH_{\rm eb}t}$ non-unitary, so the state norm decays
according to the occupation of the lossy sublattice. This gives the
cell-resolved escape probability
\begin{equation}
P_x(t)
=
2\gamma\int_0^t
\langle n_{x,b}(\tau)\rangle\,d\tau,
\qquad
P_x=\lim_{t\rightarrow\infty}P_x(t),
\label{eq:nh_escape_probability_review}
\end{equation}
which records where probability is lost during the evolution. Since
superconducting circuits natively implement unitary gates, the non-unitary step
is embedded algorithmically: the Hermitian part is implemented by ordinary
product-formula evolution, while the anti-Hermitian contribution is realized
through a linear-combination-of-unitaries block using an ancillary qubit,
controlled forward and backward Hermitian evolutions, and mid-circuit reset, methods which are elaborated in more detail in Sect.~\ref{ancilla}. 
The measured densities reconstruct $P_x(t)$, making it possible to distinguish
ordinary edge bursts from interaction-induced spatially extended edge patterns
and bulk cluster bursts.

\noindent\textbf{Topological pumping in synthetic dimensions. } Another route toward the simulation of higher-dimensional band topology employs
\emph{synthetic dimensions}. In this approach, a control parameter or an
internal degree of freedom is promoted to an effective momentum or lattice
coordinate, allowing a lower-dimensional device to emulate aspects of a
higher-dimensional topological band structure
\cite{Ozawa2019,li2019emergence,cooper2019topological,ArguelloLuengo2024SyntheticDimensions}.
This strategy is complementary to recent real-space or Fock-space encodings of
higher-dimensional topological lattices on quantum computers, where the
additional dimensions are represented directly in qubit connectivity or in the
many-body configuration space
\cite{koh2024realization,USTCARXIV2022,QianScience2025}. Synthetic dimensions
instead keep the implemented hardware model closer to a lower-dimensional
single-particle or few-particle lattice.

A representative example is the Aubry--Andr\'e--Harper (AAH) family,
\begin{equation}
\hat H_{\mathrm{AAH}}(\phi)
=
J \sum_{j=1}^{L-1}
\big(\hat c_{j}^\dagger \hat c_{j+1}+{\rm h.c.}\big)
+
\Delta \sum_{j=1}^{L}
\cos(2\pi b j+\phi)\,\hat n_j ,
\label{eq:AAH}
\end{equation}
where $\phi\in[0,2\pi)$ plays the role of a synthetic quasimomentum. By
sweeping $\phi$, one effectively samples a two-dimensional parameter space, so
that the one-dimensional AAH chain can be interpreted as a dimensional
reduction of a two-dimensional Chern insulator. This mapping provides a compact
route to reconstructing Chern-band physics, bulk--edge correspondence, and
topological pumping from a lower-dimensional device. Unlike the
configuration-space mappings discussed above, the synthetic-dimension approach
keeps the hardware model essentially as a one-particle lattice problem, while using external control parameters to supply the missing momentum or spatial
dimension. While it can demonstrate topological pumping~\cite{stegmaier2024realizing,stegmaier2025topological}, the lack of a physical real-space boundary can restrict the prospects of observing topological localization.

Cold-atom Thouless-pump experiments and superconducting AAH simulations use
related dimensional-reduction ideas, but the physical implementations and
observables are different. In cold atom experiments, a one-dimensional Hamiltonian with a cyclic pump parameter defines a
two-dimensional parameter space whose first Chern number determines the
quantized transported charge. Operationally, however, these experiments do not
realize an additional physical synthetic lattice direction. Instead, atoms physically move
in a real one-dimensional optical lattice or superlattice, and the topology is
detected through the quantized center-of-mass displacement over an adiabatic
cycle
\cite{lohse2016thouless,nakajima2016topological}.

Superconducting quantum processors provide a different implementation of the
same dimensional-reduction principle. Instead of measuring adiabatic transport
of atoms in a continuum optical lattice, the processor is programmed to realize
a finite tight-binding lattice directly, and the phase $\phi$ is sampled as a
synthetic quasimomentum or control coordinate. This makes it possible to
reconstruct effective higher-dimensional band structures and diagnose boundary
physics from wavefunction-resolved or site-resolved dynamics. For example,
synthetic-dimension Chern-insulator simulations have reconstructed band
structures through time-domain spectroscopy and diagnosed edge physics from the
dynamical localization of boundary excitations, thereby implementing
bulk--edge correspondence in a programmable superconducting setting
\cite{xiang2023simulating}. Related large-scale experiments on
one-dimensional superconducting arrays have implemented generalized AAH models
and observed Hofstadter-butterfly spectra and topological zero modes
\cite{shi2023quantum}.

The spectral and dynamical probes discussed above provide an alternative to
direct measurements of topological invariants, which are reviewed separately in
the next subsection. Rather than reconstructing a Berry curvature or Chern
number directly, these approaches infer topology from experimentally accessible
features such as boundary-state dynamics and population
transport.

Digital quantum processors offer a natural route to dynamical probes of
topological physics. For static target Hamiltonians, real-time evolution can be
implemented through product formulas, as already demonstrated in early
superconducting simulations of fermionic dynamics~\cite{barends2015digital}.
For periodically driven topological systems, however, the more natural object
is the Floquet unitary $U_F$ defined over one driving period [see Sect.~\ref{trotter} for more details]. This unitary can
be engineered directly at the gate level and then repeated over many cycles,
providing a circuit-native way to access non-equilibrium topological phases. In
particular, such protocols have been used to realize Floquet SPT phases on
programmable superconducting qubits, where subharmonic or edge-encoded
responses remain stable over many driving periods~\cite{zhang2022digital}. A closely related dynamical route is topological pumping. In this case, the
topological invariant is not extracted through full state tomography, but is
converted into a real-space transport response. This makes pumping especially suitable for experiments on current quantum hardware, because the
relevant observable can often be obtained from site-resolved population
dynamics. For instance, Liu \emph{et al.} engineered Thouless-pump cycles with
controlled disorder on a 41-qubit superconducting chain and mapped the
interplay between disorder and topology using only site-resolved populations
\cite{liu2025interplay}. Thus, Floquet evolution and pumping protocols provide
complementary circuit-level approaches to topological physics: the former
emphasizes quasienergy structure and dynamical topological phases, while the
latter converts a topological invariant into a directly measurable transport
response.

Taken together, these advances establish quantum processors as uniquely
versatile platforms for realizing and diagnosing topological phenomena. By
combining precise quantum control, flexible measurement protocols, and
programmable Hamiltonian engineering across diverse hardware architectures,
they enable systematic investigations of topology far beyond equilibrium band
theory and the thermodynamic limit
\cite{Flaschner2016,sun2018uncover,
flurin2017observing,viyuela2018observation,wang2019simulating,
xu2020measuring,mei2020digital,zhang2022digital,dumitrescu2022dynamical,
QianScience2025}. From now and beyond the NISQ era, the study of topological
phenomena on quantum processors not only deepens our understanding of robust
quantum dynamics, including dynamical topology, symmetry-protected phases,
higher-order topology, and non-Hermitian effects, but also informs the design
of future quantum technologies that harness topology and geometry for enhanced
functionality and error resilience
\cite{choo2018measurement,google2023non,xu2024non,iqbal2024non}.
The next subsection focuses on the sharper task of measuring
the topological invariant itself, rather than inferring topology from its
spectral, dynamical, or boundary signatures.

\subsubsection{Measuring topological invariants on a quantum processor}
\label{sec:topological_invariants}

Following the digital simulation platforms discussed above, a central question
is how to certify that the implemented dynamics or prepared states realize the
intended topological phase. Since topological phases are generally not characterized by
local order parameters, such certification requires measurements of global or
geometric quantities, such as Chern numbers, Berry and Zak phases, winding
numbers, $\mathbb{Z}_2$ indices, Wilson loops, or real-space topological markers
\cite{TKNN1982,niu1985quantized,hatsugai1993chern,fukui2005chern,
Berry1984,bloch2012quantum,Atala2013,hasan2010colloquium,qi2011topological,
soluyanov2011computing,yu2011equivalent,wilson1974confinement,
bianco2011mapping,prodan2013noncommutative,li2019geometric}. Quantum
processors are especially useful in this context because they provide direct
access to state overlaps, phase information, nonlocal observables, and
time-resolved dynamics, allowing topological invariants to be reconstructed from
experimentally measurable quantities
\cite{Aidelsburger2015,Flaschner2016,schroer2014measuring,
roushan2014observation,flurin2017observing,viyuela2018observation,
tan2019experimental,zheng2022measuring,xiao2023robust,
xiang2023simulating,QianScience2025}.

\noindent\textbf{Simulations of Berry phases and quantum geometry.} On quantum hardware, a direct way to benchmark topological physics is to
measure the geometric phases and geometric response functions from which
topological invariants are built. A unifying quantity underlying many measurement strategies is the geometric
phase accumulated by a quantum state under cyclic parameter evolution. The
Berry phase and its associated Berry curvature provide a natural bridge between
abstract topological invariants and experimentally accessible quantities
\cite{Berry1984}. In quantum simulation, geometric phases can be accessed in
several complementary ways. In cold-atom optical lattices, adiabatic parameter
cycles and Bloch-band tomography have enabled direct measurements of the Zak
phase, Berry curvature, and Chern number
\cite{Atala2013,Aidelsburger2015,Flaschner2016}. On superconducting circuits,
geometric phases and related topological transitions have been measured using
controlled parameter sweeps and interferometric or quantum-walk protocols
\cite{schroer2014measuring,roushan2014observation,flurin2017observing}.  More generally, on gate-based quantum processors, adiabatic or quasiadiabatic
parameter sweeps can be implemented using time-dependent Hamiltonians or
parametrized circuits, while interferometric techniques such as Ramsey
interferometry, Hadamard-test-type circuits, or ancilla-assisted phase
measurements can extract phase information from state overlaps. Related
photonic quantum-walk and synthetic-gauge-field experiments have also measured
or manipulated topological invariants and synthetic magnetic responses
\cite{zhan2017detecting,lin2023manipulating}.

To measure geometric
phases on current digital quantum platforms,  a representative strategy is to break up the task into shallow state preparation 
followed by low-depth interferometric overlap measurements, from which a topological invariant is assembled from locally defined phases. In the
holonomy-based estimator of Ref.~\cite{xiao2023robust}, the
normalized link variables are
\begin{equation}
U_{\delta\mathbf{k}}(\mathbf{k})
=
\frac{\langle \psi(\mathbf{k}) \mid
\psi(\mathbf{k}+\delta\mathbf{k}) \rangle}
{\big|\langle \psi(\mathbf{k}) \mid
\psi(\mathbf{k}+\delta\mathbf{k}) \rangle\big|},
\label{eq:link}
\end{equation}
which can be obtained on {quantum} hardware through a single-ancilla
Hadamard-test--type primitive [see Sect.~\ref{qae}] that returns the real and
imaginary parts of
$\langle \psi(\mathbf{k}) \mid \psi(\mathbf{k}+\delta\mathbf{k}) \rangle$.
On a discretized Brillouin-zone mesh, defining
$U_x(\mathbf{k})\equiv U_{\delta k_x\hat{\mathbf{x}}}(\mathbf{k})$ and
$U_y(\mathbf{k})\equiv U_{\delta k_y\hat{\mathbf{y}}}(\mathbf{k})$, the
lattice Berry flux is
\begin{equation}
F(\mathbf{k})
=
\ln\!\left[
\frac{
U_x(\mathbf{k})\,U_y(\mathbf{k}+\delta k_x\hat{\mathbf{x}})
}{
U_x(\mathbf{k}+\delta k_y\hat{\mathbf{y}})\,U_y(\mathbf{k})
}
\right],
\qquad
\mathcal{C}
=
\frac{1}{2\pi i}\sum_{\mathbf{k}} F(\mathbf{k}),
\label{eq:chern_link}
\end{equation}
with the principal branch chosen so that
$F(\mathbf{k})\in(-\pi,\pi]$. Since
Eqs.~\eqref{eq:link}--\eqref{eq:chern_link} depend only on overlap phases,
normalization suppresses sensitivity to overall contrast drift, making the
final integer primarily controlled by accumulated phase errors.

Geometric phases also provide a natural route beyond pure-state topology,
motivating mixed-state generalizations of Berry phases. In a superconducting-
qubit experiment, the topological Uhlmann phase was measured by purifying the
system density matrix and extracting the associated holonomy
interferometrically~\cite{viyuela2018observation}. In the minimal
implementation, the phase is obtained from single-qubit observables as
\begin{equation}
\Phi_M
=
\arg\!\big(\langle X\rangle+i\langle Y\rangle\big),
\end{equation}
and it tracks the Uhlmann phase across a transition between topological and
trivial mixed-state regimes.

Beyond Berry phases, recent experiments have shown that the full quantum
geometric tensor (QGT), whose real and imaginary parts correspond respectively
to the quantum metric and Berry curvature, can be directly reconstructed on
gate-based quantum processors
\cite{ProvostVallee1980,Berry1984,tan2019experimental,
yu2020experimental,zheng2022measuring,chen2024direct}. By expressing
geometric quantities in terms of density matrix elements and Pauli expectation
values, topological and geometric information can be extracted without explicit
wavefunction tomography. This strategy has been implemented on IBM
superconducting quantum processors using both variational quantum circuits and
quantum imaginary-time evolution, enabling direct probes of band topology and
quantum geometry in the presence of realistic noise~\cite{chen2024direct}. Such
measurements highlight the suitability of quantum processors for accessing
geometric structures that are often hidden in conventional solid-state
experiments.

\noindent\textbf{Dynamical benchmarks of topological phases.}
Another broad route is to infer topology from dynamical response rather than
from adiabatic transport or full state tomography. After a quench or under
periodic driving, topological information can be encoded in time-dependent spin
textures, dynamical winding numbers, vortices in momentum--time space,
quantized mean displacements, edge-state dynamics, or dynamical topological
order parameters
\cite{Budich2015,Flaschner2016,sun2018uncover,flurin2017observing,
wang2019simulating,xu2020measuring,mei2020digital}. Related driven many-body
experiments have further shown that dynamical topology can also appear as
robust Floquet or emergent dynamical SPT behavior in programmable quantum
simulators~\cite{dumitrescu2022dynamical}. These methods are attractive for
near-term devices because they often require only time-resolved measurements of
local populations, spin textures, edge responses, or correlation functions
rather than full state tomography.

This dynamical philosophy is especially powerful for Floquet and higher-order
topology. Superconducting quantum processors developed at USTC have enabled the
programmable realization and detection of two-dimensional higher-order
non-equilibrium topological phases using Floquet circuits applied for more than
50 cycles on a $6\times 6$ qubit array~\cite{QianScience2025}. In this
experiment, the equilibrium benchmark is the Benalcazar--Bernevig--Hughes
(BBH) second-order topological phase, implemented as a dimerized qubit lattice
with engineered $\pi$ flux per plaquette and chiral symmetry. The key
non-equilibrium advance is the gate-level construction of Floquet higher-order
topological phases supporting both zero- and $\pi$-quasienergy corner modes.
A central diagnostic is a real-space detection protocol based on the dynamics
of chiral density, from which Floquet spectral information and associated
topological winding numbers can be inferred. This provides a hardware-efficient
way to identify zero- and $\pi$-quasienergy corner modes using population
readout, without requiring full state tomography.

\noindent\textbf{Nonlocal and photonic probes of topology.}~Another important class of topological characterization strategies relies on
measuring carefully designed nonlocal or geometric observables. Wilson-loop and
string-operator measurements diagnose topological sectors and anyonic order in
topologically ordered systems, while holonomy-based link variables and
real-space markers encode band-topological information in quantities accessible
to finite-size quantum devices
\cite{kitaev2003fault,levin2006detecting,xiao2023robust,bianco2011mapping}.
These protocols are especially useful for interacting or finite-size systems,
where topology cannot always be inferred from single-particle band structure
alone. On superconducting and trapped-ion platforms, such measurements can be
combined with variational state preparation, time-evolution circuits, or
ancilla-assisted interferometric readout to extract many-body topological
indices, diagnose topological sectors, and verify nonlocal order
\cite{satzinger2021realizing,iqbal2024topological,iqbal2024non}.

Photonic platforms provide a complementary route to measuring topological
structure. In photonic quantum walks and synthetic-dimension architectures,
topological invariants can be inferred from interferometric phases,
wave-packet dynamics, mean displacement, and momentum--time winding patterns.
These methods have enabled measurements of winding numbers, nonunitary
topological invariants, dynamical topological order parameters, and
synthetic-gauge-field responses
\cite{zhan2017detecting,wang2019simulating,xu2020measuring,
lin2023manipulating,yu2025topologicalnetwork}. Continuous-variable
programmable photonic processors, including architectures based on
time-multiplexed squeezed-light modes and photon-number-resolved detection,
further suggest a scalable sampling-based route for probing topological
features of photonic networks
\cite{killoran2019strawberry,madsen2022quantum}.

Overall, the measurement of topological invariants on quantum simulators can be viewed
through three complementary lenses: geometric protocols that measure Berry
phases, Berry curvature, QGTs, and Chern numbers; dynamical protocols that
extract winding or real-space invariants from time evolution; and nonlocal
operator protocols that measure Wilson loops, holonomies, or twisted-boundary
responses.

\subsubsection{Anyon control on quantum hardware}\label{anyon}

\begin{figure*}
    \centering
    \includegraphics[width=0.98\linewidth]{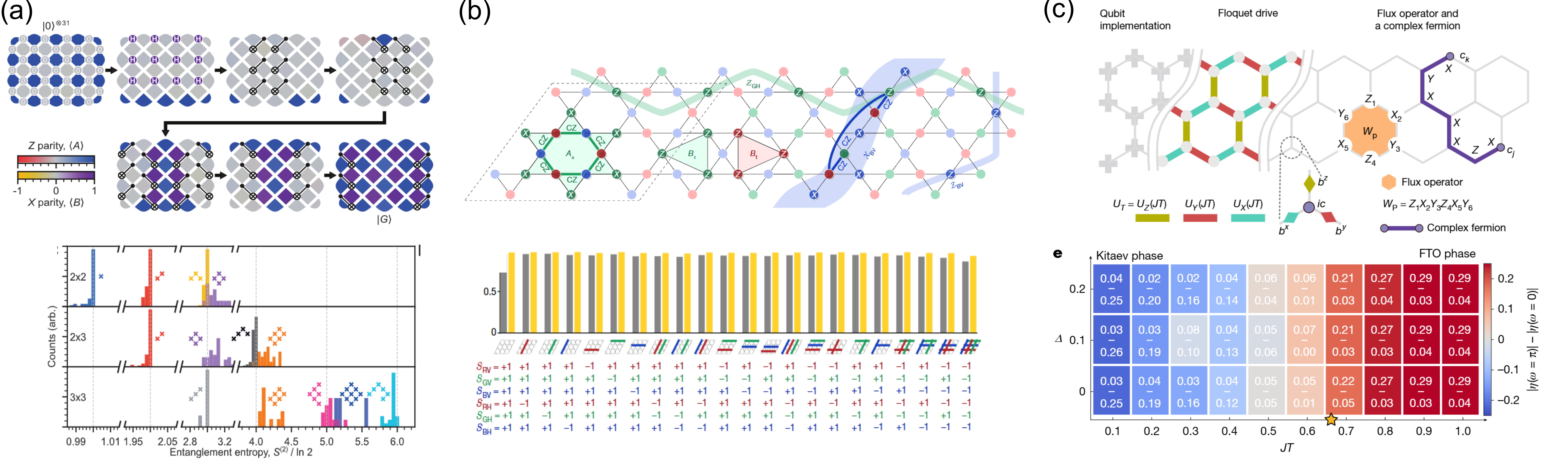}
    \caption{Quantum hardware demonstrations of topological order and anyonic phenomena. (a) Top: Step-by-step preparation of the many-qubit ground state of the toric-code Hamiltonian. Single-qubit superposition layers and entangling CZ operations progressively construct a stabilizer-valid topological state from the trivial product state. Bottom: Measured second-Rényi entanglement entropy for a variety of rectangular subregions, illustrating the characteristic entanglement structure of the prepared topological state~\cite{satzinger2021realizing}. (b) Top: Stabilizers, logical operators, and anyon trajectories in a digitally simulated non-Abelian string-net model.  Bottom: Bottom: Experimental characterization of the logical sectors. Each column corresponds to one target logical sector, marked by the six signs.  The gray bars show the measured energy density, and
The yellow bars show the logical pinning function. These values are close to 1 for the prepared state near the ground-state manifold. 
 (c) Floquet implementation of the Kitaev model and phase diagram. Left: Schematic depiction of the Floquet-engineered Kitaev model. Right:  Experimental phase diagram showing the transition from the equilibrium Kitaev phase (left, blue) to the Floquet topologically ordered (FTO) phase (right, red) as a function of drive strength $JT$~\cite{will2025probing}. Panel (a) is adopted from ~\cite{satzinger2021realizing}. Panel (b) is adopted from ~\cite{iqbal2024non}. Panel (c) is adopted from ~\cite{will2025probing}.}
    \label{fig:anyon}
\end{figure*}

Beyond the topological phenomena discussed above, which primarily hinge on
topological bands at the \emph{single-particle} level, there exist
\emph{topologically ordered} many-body phases whose defining properties cannot
be reduced to band topology
\cite{wen1990topological,wen2004quantum,levin2005string,kitaev2006anyons,
nayak2008non}. A defining feature of intrinsic topological order is that the
low-energy theory supports anyonic quasiparticles---excitations with exchange
statistics beyond those of bosons and fermions
\cite{aro1984fractional,moore1991nonabelions,nayak2008non,qin2026anyon}---together with
ground-state degeneracy that depends on spatial topology and nonlocal operators,
such as Wilson loops, that diagnose the order
\cite{kitaev2003fault,dennis2002topological,nussinov2009symmetry,
levin2006detecting}. These nontrivial orders are founded on mathematical knotted structures that elude the completely classification by any single topological invariant\cite{murasugi1996knot,kontsevich1993vassiliev,kauffman2005mathematics,field2018introduction,lee2020enhanced, bi2017nodal,bode2019constructing,lee2020imaging,carlstrom2019knotted,tai2021anisotropic}. 

Recent quantum-hardware experiments have demonstrated increasingly precise
control over such topological degrees of freedom. Early and recent
superconducting experiments emulated or demonstrated Abelian toric-code anyons
and their braiding statistics
\cite{zhong2016emulating,niu2024demonstrating}, while trapped-ion experiments
have extended toric-code preparation to qudit settings, including a
$\mathbb{Z}_3$ qutrit toric code and parafermion-related defects
\cite{iqbal2025qutrit}. In parallel, superconducting processors have been used
to engineer and braid non-Abelian excitations in synthetic, projective, and
string-net settings
\cite{google2023non,xu2023digital,xu2024non,minev2025realizing}. In the
following, we review these experimental achievements, emphasizing the
preparation of topologically ordered states, the controlled creation and motion
of anyons, and the protocols used to verify their exchange statistics on
quantum processors.

\noindent\textbf{Abelian anyons and toric-code benchmarks.} Early hardware realizations of anyonic statistics focused on the paradigmatic
$\mathbb{Z}_2$ toric code
\cite{kitaev2003fault,dennis2002topological,pachos2009revealing,
lu2009demonstrating,zhong2016emulating}, where Abelian electric ($e$) and
magnetic ($m$) excitations appear as localized defects at the endpoints of
string operators. In this setting, creating and moving anyons amounts to
applying products of Pauli operators along a path: open strings create pairs of
anyons, while closed strings act as Wilson-loop operators that diagnose the
underlying topological sector. At the proof-of-principle level, anyonic
fractional statistics were first emulated in photonic quantum simulators and
later in superconducting circuits, where the associated braiding phase was
extracted through controlled interferometric protocols
\cite{lu2009demonstrating,zhong2016emulating}.

Building on these foundations, a representative Google superconducting-qubit
realization is shown in Fig.~\ref{fig:anyon}(a)~\cite{satzinger2021realizing}.
There, a finite-size toric-code ground state was prepared using an efficient
stabilizer-based circuit and verified by measuring second-Rényi entropies of
subregions, from which a topological entanglement entropy close to the expected
$\ln 2$ contribution was extracted. The experiment further simulated anyon
interferometry to extract the braiding statistics of emergent excitations. The
top panel depicts the circuit construction of the toric-code ground state,
where layers of single-qubit operations and entangling gates build a
stabilizer-valid topological state from a trivial product state. The bottom
panel benchmarks the prepared state through the measured second-Rényi
entanglement entropy for rectangular subregions, verifying the characteristic
entanglement structure expected for a stabilizer topological state. Modular superconducting-circuit experiments further strengthened the braiding
interpretation by explicitly demonstrating path-independent anyonic braiding
phases within a programmable superconducting architecture
\cite{niu2024demonstrating}. More recently, trapped-ion experiments have
realized $\mathbb{Z}_3$ toric-code physics with qutrits and manipulated
parafermion-related defects~\cite{iqbal2025qutrit}. The proposal by
Jovanovi\'{c}, Wille, Timmers, and Simon further outlines a route toward
implementing non-Abelian quantum-double models with richer symmetries, such as
$D_4$, on small-scale quantum devices~\cite{jovanovic2024proposal}.

\noindent\textbf{Synthetic non-Abelian anyons and twist defects.} Subsequent experiments have expanded these capabilities to synthetic
\emph{non-Abelian} anyonic settings, where quasiparticles or defect excitations
carry internal fusion degrees of freedom and braiding can implement
\emph{noncommuting} unitary transformations within a degenerate encoded Hilbert
space. Google Quantum AI and collaborators demonstrated non-Abelian braiding of
graph vertices on a superconducting quantum processor by implementing a
generalized stabilizer code and a unitary protocol to create, braid, and fuse
projective Ising-type anyons~\cite{google2023non}. The experiment verified both
fusion outcomes and braiding-induced transformations, showing that braiding acts
nontrivially on the encoded subspace rather than merely producing an Abelian
phase. Building on this direction, a complementary route is to engineer \emph{twist
defects} in toric-code circuits. Twist defects act as domain-wall endpoints
that exchange electric and magnetic charges when an anyon crosses the defect
line; collections of such defects therefore support nontrivial fusion spaces
and realize Ising-type non-Abelian behavior within an otherwise Abelian
$\mathbb{Z}_2$ code. This mechanism was demonstrated by Xu \emph{et al.} in a
large-scale digital simulation using up to 68 programmable superconducting
qubits, where twist defects were created and manipulated in quantum circuits
and their fusion rules and braiding statistics were benchmarked on hardware
\cite{xu2023digital}.

\noindent\textbf{Intrinsic non-Abelian topological order.} A further qualitative advance is to move beyond defect-based non-Abelian
behavior and prepare \emph{intrinsic} non-Abelian topological order, where the
degenerate ground-state manifold and its nonlocal logical, string, or Wilson
operators are properties of the phase itself rather than being induced by
extrinsic defects~\cite{iqbal2024non}. Fig.~\ref{fig:anyon}(b) illustrates this
progression through an adaptive-circuit implementation of $D_4$ non-Abelian
topological order on Quantinuum's H2 trapped-ion quantum processor. The
\emph{top} panel schematically highlights the local constraints defining the
code space, together with nonlocal string operators that create, transport, and
fuse non-Abelian anyonic excitations. The \emph{bottom} panel reports an
experimental characterization of the logical sectors through nonlocal
measurements, resolving the topological ground-state manifold. In particular,
the experiment used non-Abelian anyon tunnelling around a torus to access the
distinct ground-state sectors, providing an operational benchmark of intrinsic
non-Abelian topological order on programmable trapped-ion hardware.

\noindent\textbf{Floquet topological order and non-equilibrium anyons.} More recent work has extended the study of topological order and anyonic
excitations into the non-equilibrium regime by engineering Floquet topologically
ordered states~\cite{will2025probing}. Will \emph{et al.} realized a
Floquet-induced topological phase on a Google Quantum AI superconducting
quantum processor and probed its emergent anyonic structure, rather than
implementing a direct braiding experiment of individually manipulated anyons.
The key physical ingredient is the stroboscopic unitary evolution over one
drive period, rather than a static Hamiltonian ground state. In this setting,
topological features can be dynamically stabilized by periodic driving and can
exhibit invariants, edge dynamics, and excitations without equilibrium
counterparts.

Fig.~\ref{fig:anyon} (c) summarizes this superconducting-qubit realization of
non-equilibrium topological order~\cite{will2025probing}: the left panel
sketches the Floquet implementation and associated emergent operators, including
flux/Wilson-loop-type diagnostics and anyonic excitations, while the right panel
shows an experimentally reconstructed phase diagram as a function of drive
strength $JT$ and control parameter $\Delta$. The data show a transition from a
weak-drive regime connected to the equilibrium topological phase to a
strong-drive Floquet topologically ordered regime, with the boundary occurring
at intermediate drive strengths as indicated by a rapid change in the measured
topological diagnostic. This demonstrates an important conceptual extension:
programmable superconducting hardware can access topological phenomena not only
as properties of prepared ground states, but also as intrinsically dynamical
phases of driven many-body evolution.

\begin{figure*}
    \centering
    \includegraphics[width=0.99\linewidth]{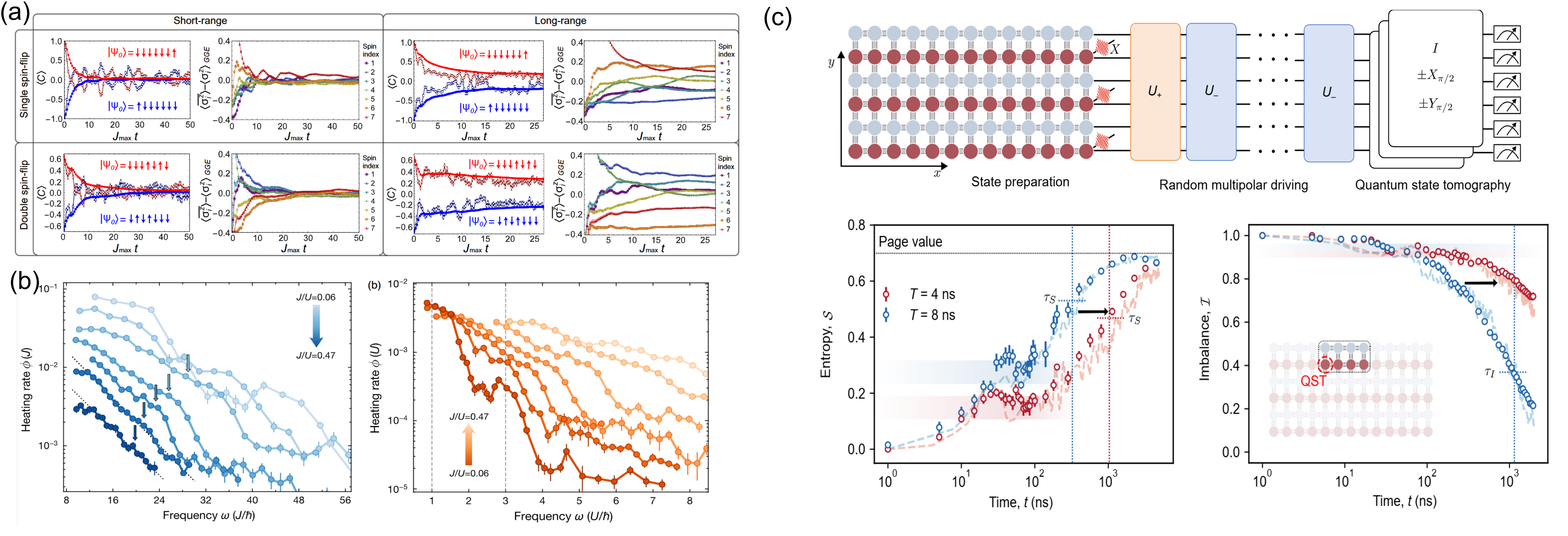}
    \caption{Quantum simulations of prethermalization. (a) Trapped-ion long-ranged Ising dynamics exhibiting a memoryful prethermal plateau. Initial states are quenched under power-law interactions; for shorter-range couplings (larger $\alpha$) the center-of-mass observable relaxes toward its symmetric ensemble value, while for sufficiently long-ranged couplings (smaller $\alpha$) it remains biased~\cite{neyenhuis2017observation}. (b) Floquet prethermalization in a driven 2D Bose--Hubbard system quantified by the heating rate per Floquet cycle. Heating is strongly suppressed as $\omega$ increases across lattice depths~\cite{rubio2020floquet}. (c) Prethermal plateaus under random multipolar driving on a 2D superconducting-qubit array~\cite{liu2025prethermalization}. The subsystem entropy $S(t)$ and imbalance $\mathcal{I}(t)$ display an extended intermediate-time plateau. Panel (a) is adopted from ~\cite{neyenhuis2017observation}. Panel (b) is adopted from ~\cite{rubio2020floquet}. Panel (c) is adopted from ~\cite{liu2025prethermalization}.
    }
    \label{fig:pre}
\end{figure*}

\subsection{Quantum simulation of Non-Equilibrium Physics}
\label{sec1:noneq}

Classical thermodynamics asserts that when a finite object (the ``system'') is weakly coupled to a much larger
reservoir (the ``bath'') at temperature $T$, the system will, given sufficient time, relax to thermal equilibrium at
the bath temperature. In the quantum regime, however, this expectation can fail: isolated or weakly open quantum
many-body systems may exhibit \emph{non-ergodic} dynamics, in which long-time states retain memory of initial
conditions and cannot be described by standard thermal ensembles. The breakdown of ergodicity in quantum
systems~\cite{hardy2014thermodynamics} is central to modern many-body physics~\cite{anderson1972more}, and
underlies a broad hierarchy of phenomena spanning the eigenstate thermalization hypothesis (ETH)~\cite{deutsch1991quantum,srednicki1994chaos,Yunger2023},
many-body localization~\cite{nandkishore2015many,Abanin2019}, quantum many-body scars~\cite{Serbyn2021,Chandran2023},
quantum chaos~\cite{d2016quantum}, and time-crystalline order~\cite{Wilczek2012,Bruno2013,Watanabe2015,Sacha2018,khemani2019brief}.

In this section, we first review representative non-ergodic quantum phenomena that have been realized and characterized
across multiple driven and/or dissipative quantum hardware platforms. Next we highlight how broader classes of non-unitary phenomena -- including non-Hermitian evolution and measurement-based processes -- have been effectively, if not natively, realized in the most recent quantum simulators.

\subsubsection{Prethermalization}

Prethermalization is a hallmark of non-equilibrium quantum many-body dynamics, describing the emergence of long-lived
quasi-stationary states that appear locally equilibrated yet precede eventual thermalization. Following a sudden
quench or under periodic driving, local observables often relax on a short timescale $\tau_{\rm pre}$ to a metastable
plateau, while true thermal equilibration occurs only at much later times $\tau_{\rm th}\gg\tau_{\rm pre}$ \cite{berges2004prethermalization,gring2012relaxation,
rigol2008thermalization,polkovnikov2011colloquium,eisert2015quantum,
d2016quantum}.  Such
separation of timescales typically arises when the dynamics is constrained by approximate conservation laws,
near-integrability, or emergent dynamical symmetries, so that the system first relaxes within a restricted manifold
(e.g., an effective integrable description or a Floquet prethermal regime) before weak integrability-breaking
processes drive it toward a conventional thermal ensemble
\cite{berges2004prethermalization,moeckel2008interaction,kollar2011generalized,lazarides2014equilibrium,abanin2017effective,potter2016classification,else2017prethermal}.

Recent experiments across diverse quantum-simulation platforms have established that prethermalization is not merely
a theoretical construct but a robust and directly observable dynamical phenomenon. A paradigmatic early cold-atom
demonstration was provided by Gring \textit{et al.}, who coherently split a one-dimensional Bose gas and observed a
rapid approach to a long-lived prethermal state~\cite{gring2012relaxation}. The ensuing steady-like regime could not
be captured by a conventional thermal ensemble; instead, its correlations were consistent with an (approximately)
integrable description in terms of a generalized Gibbs ensemble (GGE). This experiment established a clear operational distinction between fast local
equilibration and slow global thermalization, and highlighted how emergent conserved quantities can stabilize
metastable non-equilibrium states over experimentally relevant timescales.

In particular, trapped-ion quantum simulators provide a clean setting in which the interaction range can be tuned. 
For example, in a long-range transverse-field Ising chain of $N$ spins with power-law couplings 
$J_{ij}\propto 1/|i-j|^{\alpha}$, where $i,j=1,\ldots,N$ label the lattice sites and $\alpha$ controls the interaction range, Neyenhuis \textit{et al.} used spatially localized spin-flip product states to probe relaxation and memory retention~\cite{neyenhuis2017observation}. 
The memory of the initial spatial imbalance was quantified by the center-of-mass observable
\begin{equation}
C=\sum_{i}\frac{2i-N-1}{N-1} \frac{Z_i+1}{2}.
\end{equation}
Here, the prefactor $(2i-N-1)/(N-1)$ assigns a normalized position coordinate ranging from the left to the right side of the chain. 
Because both thermal and generalized Gibbs ensemble descriptions are inversion symmetric for the corresponding initial energy and conserved quantities, they predict $\langle C\rangle=0$ ~\cite{neyenhuis2017observation}. As shown in Fig.~\ref{fig:pre}(a), for relatively short-range interactions (larger $\alpha$) the dynamics rapidly lose spatial memory and $\langle C(t)\rangle$ relaxes toward zero. Strikingly, when interactions are sufficiently long-ranged (smaller $\alpha$), the system instead enters a long-lived quasi-stationary regime where $\langle C\rangle$ remains biased toward the side where excitations were initially placed. This memoryful prethermal plateau highlights that long-range couplings can qualitatively reshape the thermalization landscape and stabilize metastable states.

Another view of prethermalization arises in periodically driven (Floquet)
systems, where a sufficiently high-frequency modulation can generate an
effective static Hamiltonian while strongly suppressing energy absorption. In
an optical-lattice Bose--Hubbard realization, Rubio-Abadal
\textit{et al.} directly quantified Floquet heating and identified a broad
prethermal regime in which the energy absorbed per Floquet cycle is strongly
reduced as the drive frequency increases~\cite{rubio2020floquet}. As summarized
in Fig.~\ref{fig:pre}(b), for a range of lattice depths, the measured heating
rate decreases rapidly with increasing drive frequency, consistent with the
generic Floquet expectation that fast modulation suppresses resonant absorption
channels and yields a parametrically long-lived prethermal window.

Prethermalization has also been observed in solid-state spin platforms, providing a complementary setting with excellent coherence and direct access to Floquet heating rates. In particular, an NMR experiment on dipolar spin chains reported a long-lived Floquet-prethermal regime characterized by parametrically slow energy absorption in the high-frequency limit, thereby offering a clean quantitative benchmark for the predicted suppression of heating under rapid driving~\cite{peng2021floquet}. This result is especially valuable from a review perspective because it isolates prethermal physics in a well-controlled many-body setting while enabling precise characterization of the drive-frequency dependence of the heating timescale.

Beyond ions and cold atoms, similar prethermalization phenomenology is now
observable on programmable superconducting processors, where both the drive
protocol and measured diagnostics can be varied with high flexibility
\cite{liu2025prethermalization}. Fig.~\ref{fig:pre}(c) illustrates this in the
setting of random multipolar driving on the 78-qubit superconducting processor
Chuang-tzu 2.0, where a two-dimensional qubit array is initialized in a
spatially patterned product state and then subjected to repeated drive cycles.
Two complementary observables track the approach to thermalization: the
subsystem entanglement entropy $S(t)$, benchmarked against the Page value
expected near an effectively infinite-temperature state, and the decay of a
density-pattern imbalance $\mathcal{I}(t)$, which quantifies how quickly the
system forgets its initial spatial order. Both diagnostics reveal an extended
intermediate-time plateau---with suppressed entanglement growth and a large
imbalance---before a later crossover to strong heating, providing direct
evidence of a prethermal regime in a large-scale programmable quantum
processor. Taken together, these experiments demonstrate that
prethermalization is a platform-agnostic feature of interacting quantum matter,
arising from either near-integrability after quenches or constrained heating in
driven systems, and that it can now be quantitatively diagnosed using a diverse
toolkit ranging from local correlation functions and heating rates to
entanglement-growth and imbalance observables.

\begin{figure*}
    \centering
    \includegraphics[width=0.95\linewidth]{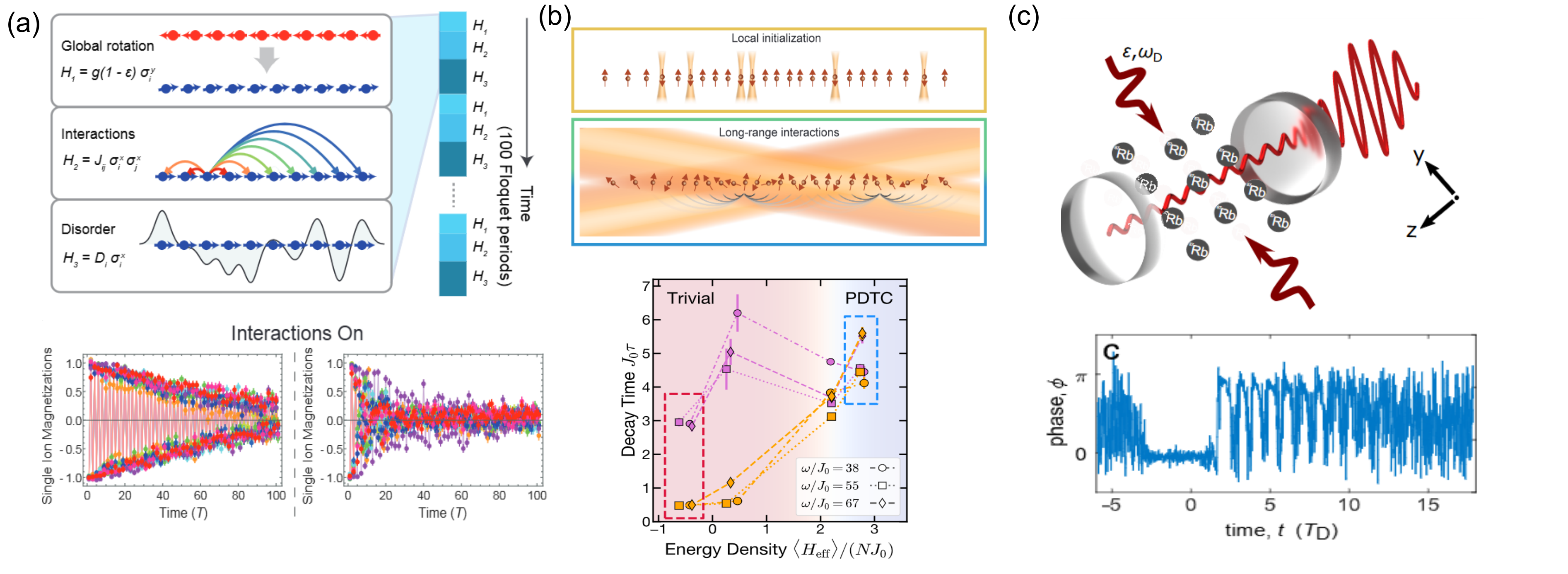}
    \caption{{ {Early} experimental realizations of time-crystalline order in driven many-body systems.} (a) {Disorder-assisted discrete time crystal (DTC)} in a periodically driven trapped-ion Ising chain. The Floquet cycle alternates a near-global spin rotation with interacting evolution and quenched disorder (schematic, top), yielding a rigid subharmonic (period-doubled) response in the stroboscopic magnetization ~\cite{zhang2017observation}. (b) {Prethermal DTC} stabilized without disorder by high-frequency driving in a long-ranged interacting trapped-ion simulator. Local initialization and long-range couplings (schematics, top) produce a long-lived prethermal plateau with robust period doubling. The lower panel summarizes the stability of the prethermal DTC regime versus effective energy density and decay time~\cite{kyprianidis2021observation}. (c) {Dissipative time crystal} in an open atom--cavity platform, where continuous drive and dissipation stabilize persistent oscillations (limit-cycle dynamics) manifested as long-lived phase-coherent time traces (bottom)~\cite{kessler2021observation}. Panel (a) is adopted from ~\cite{zhang2017observation}. Panel (b) is adopted from ~\cite{kyprianidis2021observation}. Panel (c) is adopted from ~\cite{kessler2021observation}.}
    \label{fig:time}
\end{figure*}

\subsubsection{Time-crystalline phases}\label{time}

Time crystals provide a paradigmatic example of non-equilibrium order in driven
quantum many-body systems
\cite{Wilczek2012,Bruno2013,Watanabe2015,Sacha2018,khemani2019brief,
else2016floquet,khemani2016phase,yao2017discrete,
vonkeyserlingk2016absolute,zaletel2023colloquium}. They can appear in both
continuous and discrete forms. In continuous time crystals, persistent
oscillations spontaneously break continuous time-translation symmetry, typically
in open or driven-dissipative settings
\cite{riera2020time,kessler2021observation,wu2024dissipative,
jiao2025observation}. In discrete time crystals, a Floquet system driven with
period $T$ responds with a longer period, such as $2T$ or $nT$, thereby breaking
the discrete time-translation symmetry of the drive
\cite{zhang2017observation,choi2017observation,kyprianidis2021observation,
randall2021many,mi2022time,frey2022realization,chen2023robust,
liu2024higher,xiang2024longlived,switzer2026realization,
shinjo2026unveiling}. A key challenge in isolated Floquet systems is heating
toward an effectively infinite-temperature state
\cite{lazarides2014equilibrium,lazarides2015fate,ponte2015many,
bukov2015universal,kuwahara2016floquet,mori2016rigorous,
abanin2017effective}. Prethermalization can delay this process, and the
prethermal window allows robust subharmonic oscillations to persist before
eventual heating, providing a mechanism for realizing prethermal discrete time
crystals
\cite{else2017prethermal,kyprianidis2021observation,rubio2020floquet,
peng2021floquet,zaletel2023colloquium}.

\noindent\textbf{Floquet formulation and subharmonic response.} We here consider a periodically driven, or Floquet, system whose Hamiltonian satisfies $H(t+T)=H(t)$, where $T$ is the drive period. 
The evolution over one complete period is described by the Floquet unitary
\begin{equation}
U_F
=
\mathcal{T}
\exp\left[
-i\int_0^T H(t)\,dt
\right],
\label{eq:floquet_unitary}
\end{equation}
where $\mathcal{T}$ denotes time ordering. 
Stroboscopic dynamics refers to observing the system only after an integer number of drive cycles. 
For an observable $\hat O$, its stroboscopic Heisenberg evolution is
\begin{equation}
\hat O(mT)
=
(U_F^\dagger)^m \hat O U_F^m .
\label{eq:stroboscopic_observable}
\end{equation}
Discrete time-crystalline (DTC) order is identified by a robust subharmonic response of a suitable many-body order parameter, such as a magnetization or density-wave observable. 
For an initial state $\rho_0$, define the measured stroboscopic signal $\langle O_m\rangle
=
\mathrm{Tr}\!\left[
\rho_0 \hat O(mT)
\right]$.
A period-$nT$ DTC satisfies
\begin{equation}
O_{m+n}
\simeq
O_m,
\qquad
O_{m+q}
\not\simeq
O_m
\quad
\text{for } q=1,\ldots,n-1,
\label{eq:dtc_period_n}
\end{equation}
over a long time window. 
The most common case is period doubling, $n=2$, where the observable alternates sign or amplitude every drive cycle but returns after two periods. Moreover, the response is rigid. 
In the Fourier spectrum of the stroboscopic signal $O_m$, the subharmonic peak remains pinned at $\omega
=
\frac{2\pi}{nT},$
for example $\omega=\pi/T$ for a period-doubled DTC. 
This peak remains locked over a finite range of drive imperfections, such as imperfect $\pi$ pulses, rather than continuously shifting with the microscopic detuning. 
This distinguishes a DTC from a fine-tuned single-spin or mean-field period-doubled trajectory, where subharmonic oscillations may occur but are not protected as a stable many-body dynamical phase.

\noindent\textbf{Localization and prethermalization.} The earliest DTC experiments leveraged disorder, interactions, and slow
heating to stabilize time-crystalline order. A trapped-ion Floquet Ising-chain
experiment by Zhang \emph{et al.}, performed on a trapped-ion
quantum-simulation platform, provided a clean demonstration of rigid period
doubling over a broad range of drive imperfections, consistent with an
interaction-locked dynamical phase [Fig.~\ref{fig:time}(a)]
\cite{zhang2017observation}. A complementary realization in dense dipolar spin
ensembles, based on nitrogen-vacancy centers in diamond, demonstrated that
robust DTC signatures can persist even with long-range interactions,
highlighting that slow thermalization and emergent constraints can stabilize
temporal order beyond idealized short-range many-body-localized settings
\cite{choi2017observation}. Subsequent solid-state spin experiments, based on
individually controllable $^{13}$C nuclear spins in diamond, further
strengthened the phenomenology by mapping the stability and lifetime of the
DTC response across parameter space~\cite{randall2021many}.

A distinct stabilization route relies on {prethermalization} under high-frequency driving, where heating is
parametrically suppressed and the dynamics are governed for long times by an effective quasi-conserved Hamiltonian.
In this disorder-free setting, a DTC can persist throughout a long-lived prethermal plateau, with lifetimes that
increase with the drive frequency [Fig.~\ref{fig:time} (b)]~\cite{kyprianidis2021observation}. Related prethermal time-crystalline phenomena
have also been explored in ensemble platforms, including the observation of a prethermal $U(1)$ DTC
\cite{stasiuk2023observation}.

Time-crystal dynamics are naturally expressible in quantum circuit language through Floquet unitaries composed of imperfect global rotations interleaved with interacting evolution. This makes DTCs attractive as many-body dynamical benchmarks on available quantum hardware: the defining signature, namely a rigid subharmonic response, can be more tolerant to finite circuit depth than precision Hamiltonian simulation. Protocol-level frameworks for realizing and diagnosing DTC physics on quantum processors were developed in Ref.~\cite{ippoliti2021many}. Superconducting-qubit processors subsequently reported time-crystalline dynamics, including eigenstate-ordered time-crystalline behavior and scalable digital-circuit realizations~\cite{mi2022time,frey2022realization}.

\noindent\textbf{Generalized and higher-order temporal order.} More recent work has extended this direction from conventional period doubling
to larger-period and topologically enriched temporal order
~\cite{chen2023robust,liu2024higher,zhang2022digital,xiang2024longlived}.
For example, robust large-period DTC signatures have been proposed and probed
in IBM Quantum digital quantum circuits~\cite{chen2023robust}
[see Fig.~\ref{fig:dtc2}(a)], and a programmable superconducting processor has
been used to observe signatures of long-lived topological time-crystalline
order, where the subharmonic response is encoded in nonlocal logical operators
rather than only local magnetization~\cite{xiang2024longlived}.
Moreover, Fig.~\ref{fig:dtc2}(b) shows higher-order temporal responses in a
Floquet-driven Rydberg-atom platform, where the measured transmission exhibits
oscillations with periods $T$, $2T$, and $3T$~\cite{liu2024higher}. These
results demonstrate that time-crystalline order is not restricted to the
standard period-doubled response, but can be generalized to programmable
larger-period, fractional, and topologically structured forms of temporal
order.

The most recent experiments further broaden the accessible DTC phenomenology.
In Floquet-driven Rydberg atomic gases, higher-order and fractional DTC
responses have been observed on a Rydberg-atom platform, including
integer $n$-DTC responses beyond period doubling and fractional responses
interpolating between adjacent integer time-crystalline regimes
\cite{liu2024higher}. In gate-based digital processors, recent experiments
have extended DTC physics to genuinely two-dimensional lattice geometries and
more general interacting spin models. This marks an important step beyond
earlier demonstrations that were primarily based on effectively Ising-like
Floquet dynamics. Switzer \emph{et al.} realized a two-dimensional DTC with anisotropic
Heisenberg interactions by combining IBM quantum processors with
tensor-network analysis~\cite{switzer2026realization}. Shinjo \emph{et al.}
further demonstrated both a clean two-dimensional DTC and an incommensurately
modulated DTC on the 133-qubit IBM Quantum Heron processor
\texttt{ibm\_torino}, tracking magnetization dynamics for up to 100 Floquet
cycles~\cite{shinjo2026unveiling}. These results show that present digital
quantum processors can access robust subharmonic Floquet responses in
spatially extended two-dimensional many-body systems, with interaction
structures and system sizes closer to those encountered in condensed-matter
spin models.

Recent work has also generalized the notion of temporal order beyond strictly periodic Floquet driving. Experiments on strongly interacting spin ensembles in diamond have realized discrete time quasicrystals under quasiperiodic driving, where robust subharmonic responses occur at multiple incommensurate frequencies~\cite{he2025experimental}. A related development is the observation of a time rondeau crystal, in which long-time stroboscopic order coexists with controlled short-time temporal disorder~\cite{moon2025rondeau}. These extensions indicate that time-crystalline phenomenology is not limited to simple period-$nT$ responses, but belongs to a broader family of non-equilibrium temporal orders stabilized by interactions, drive structure, and slow heating.

\noindent\textbf{Dissipative time crystals and emerging applications.} In open quantum systems, dissipation provides a distinct route to
time-crystalline order by stabilizing a non-equilibrium limit cycle or a
long-lived oscillatory manifold~\cite{riera2020time,zaletel2023colloquium}.
This mechanism differs from isolated Floquet DTCs, where the evolution is
unitary. In a dissipative time crystal, the long-time dynamics are instead
governed by the spectrum of the Lindblad generator or, for periodically driven
open systems, the Floquet-Liouvillian propagator. The relevant slow modes have
nonzero imaginary parts, which set the oscillation frequency, while their real
parts determine the decay rate. Persistent time-crystalline oscillations occur
when these decay rates vanish in the thermodynamic or large-system limit.
Thus, ``dissipation-stabilized'' does not mean that all oscillations decay;
rather, dissipation damps perturbations toward a stable oscillatory attractor. For example, Ke\ss ler \emph{et al.}~\cite{kessler2021observation} reported dissipation-stabilized
time-crystalline dynamics in a driven open atom-cavity system,
arising from the interplay of drive, cavity-mediated interactions, and
controlled loss [Fig.~\ref{fig:time}(c)].
More recently, strongly interacting Rydberg gases have become a key platform
for driven-dissipative time-crystal physics. Experiments have reported
dissipative time-crystalline order in room-temperature Rydberg gases
\cite{wu2024dissipative}, bifurcations between distinct time-crystalline
regimes~\cite{liu2025bifurcation}, and multiple time-crystalline phases in a
single driven-dissipative setting, including continuous, subharmonic, and
high-harmonic time crystals~\cite{jiao2025observation}. Injection locking of
Rydberg dissipative time crystals has also been demonstrated, showing that an
external radio-frequency field can synchronize and stabilize the emergent
temporal oscillation~\cite{arumugam2026injection}.

Time crystals are increasingly being explored not only as exotic
non-equilibrium phases, but also as functional dynamical resources. A recent
experiment demonstrated frequency-selective sensing with a prethermal DTC in
a solid-state spin platform based on strongly driven, dipolar-coupled
$^{13}$C nuclear spins in diamond~\cite{moon2026sensing}. In this setting, an
applied AC magnetic field can extend the time-crystal lifetime and generate a
narrow resonant response. This suggests that the rigidity of the subharmonic
response, originally viewed primarily as a diagnostic for
time-translation-symmetry breaking, may also be exploited for metrology and
coherent dynamical control.

Collectively, these experiments establish time-crystalline order as a practically realizable non-equilibrium phase
across leading quantum-simulation platforms. Importantly, they highlight three broadly useful stabilization mechanisms for
quantum hardware: disorder-assisted suppression of heating, disorder-free prethermal plateaus, and dissipation-stabilized limit cycles.

\begin{figure*}
    \centering
    \includegraphics[width=0.75\linewidth]{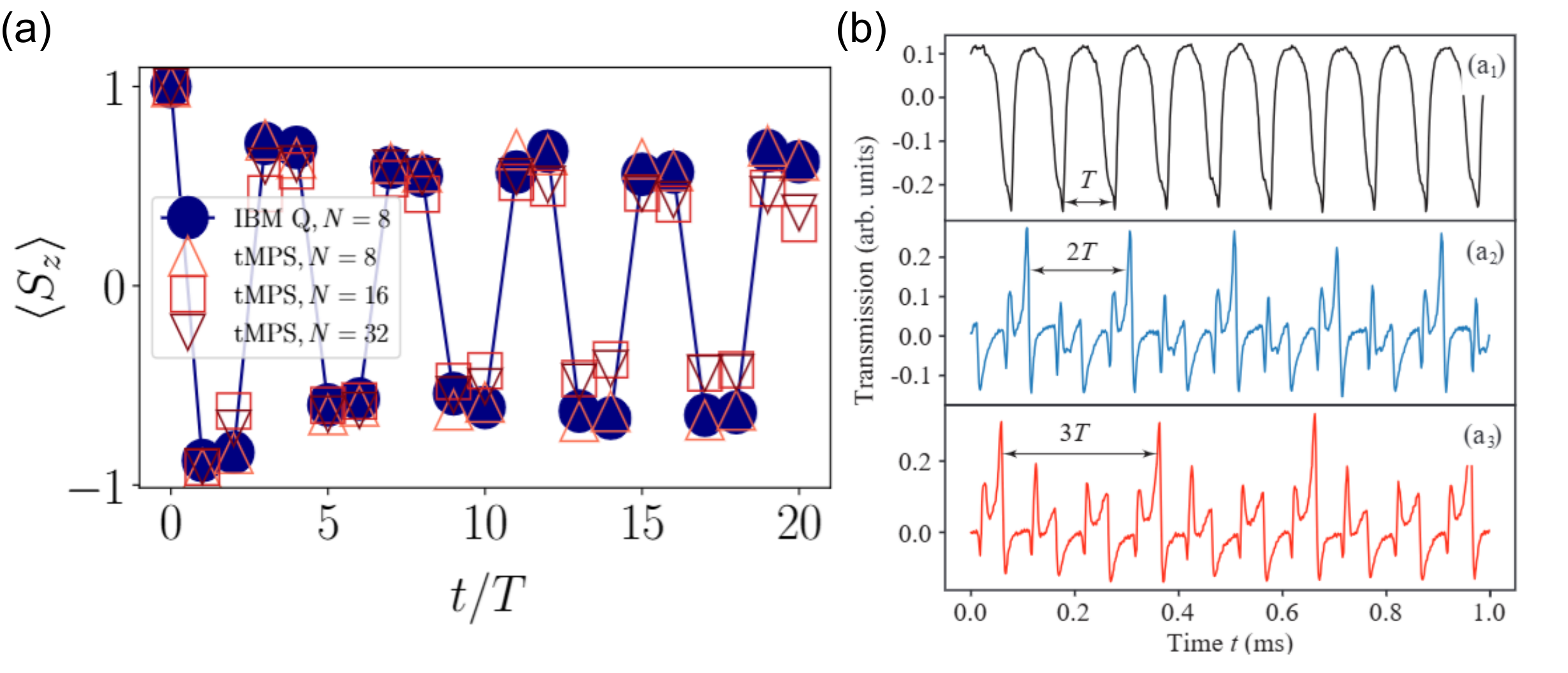}
    \caption{Larger-period temporal order in quantum-simulation platforms.
(a) Robust large-period $4T$-discrete time-crystalline response from a driven Floquet protocol simulated on a
digital quantum processor, with the time-resolved magnetization being
$\langle S_z\rangle$.
(b) Higher-order temporal responses observed in a driven Rydberg-atom system,
showing transmission signals with periods $T$, $2T$, and $3T$. Panel (a) is adopted from Ref.~\cite{chen2023robust}, and Panel (b) is adopted from Ref.~\cite{liu2024higher}}
    \label{fig:dtc2}
\end{figure*}

\subsubsection{Ergodicity breaking and many-body localization}\label{mbl}

Significant research efforts have focused on how non-equilibrium quantum
many-body systems approach thermal equilibrium. Thermalization refers to the
emergence of equilibrium thermodynamics from unitary dynamics in an effectively
closed system, where local observables relax to values described by a thermal
ensemble fixed by energy and other global conserved quantities
\cite{tasaki1998from,goldstein2006canonical,popescu2006entanglement,
reimann2008foundation,linden2009quantum,deutsch1991quantum,
srednicki1994chaos,rigol2008thermalization,polkovnikov2011colloquium,
eisert2015quantum,gogolin2016equilibration,d2016quantum,
mori2018thermalization}. In generic nonintegrable systems, this behavior is
commonly rationalized by the eigenstate thermalization hypothesis (ETH), which
posits that individual many-body eigenstates already encode thermal expectation
values for local observables
\cite{deutsch1991quantum,srednicki1994chaos,rigol2008thermalization,
kim2014testing,beugeling2014finite,deutsch2018eigenstate,d2016quantum}.

A key paradigm that violates this thermalization picture is
\emph{many-body localization} (MBL)
\cite{oganesyan2007localization,pal2010many,bardarson2012unbounded,
nandkishore2015many,altman2015universal,alet2018many,zakrzewski2026many}.
Unlike ergodic systems obeying ETH, MBL systems retain long-time memory of
their initial conditions. Microscopically, this nonergodicity is associated
with an extensive set of emergent quasi-local integrals of motion,
which constrain dynamics and lead to hallmark signatures such as persistent
local autocorrelations, suppressed DC transport, Poissonian level statistics,
and slow logarithmic growth of entanglement entropy following a global quench.

Because MBL concerns interacting disordered or quasiperiodic many-body
systems, quantum platforms provide a natural setting for probing localization
dynamics directly. They enable controlled preparation of initial product
states, tunable disorder and interactions, and time-resolved access to local
densities, correlations, entanglement proxies, and scrambling diagnostics.
Different aspects of MBL physics are more naturally accessed on different
quantum-simulation platforms because each hardware architecture realizes
disorder, interactions, connectivity, measurements, and dissipation in
distinct ways.

For example, cold atoms provide an especially clean analog setting for
Hubbard- and Aubry--Andr\'e--Harper-type models, where quasiperiodic or random
potentials, interaction strengths, and lattice geometries can be tuned with
high precision. Optical-lattice experiments have measured imbalance decay and
spatially resolved relaxation in one- and two-dimensional localization
settings
\cite{schreiber2015observation,choi2016exploring,luschen2017signatures}.
Together with quantum gas microscopy, ultracold atomic platforms have also
enabled probes of entanglement growth, correlation spreading, and critical
behavior near the MBL crossover
\cite{lukin2019probing,rispoli2019quantum}. More generally, scrambling
diagnostics such as out-of-time-ordered correlators provide a complementary
language for characterizing information spreading and its suppression in
localized or slowly thermalizing systems~\cite{XuSwingle2024Scrambling}.

Superconducting quantum processors offer a complementary gate-based route, in
which disordered spin dynamics and Floquet driving can be programmed directly
at the circuit level. Early experiments emulated MBL-like dynamics and
energy-resolved/Stark localization on superconducting devices
\cite{xu2018emulating,guo2021observation}, while Google superconducting
experiments developed circuit-level scrambling diagnostics
\cite{Mi2021ScramblingGoogle}. More recent digital simulations have used IBM
superconducting processors to probe MBL crossovers in disordered or
quasiperiodic Floquet circuits through autocorrelation functions,
out-of-time-ordered correlators, and related diagnostics
\cite{hayata2025digital,nagao2026probing}. These developments make
superconducting hardware particularly attractive for testing circuit-based
diagnostics such as correlations and OTOCs.

\noindent\textbf{Disorder-induced MBL.}
Disorder-induced MBL arises when randomness in local fields or couplings
competes with interaction-driven transport, thereby suppressing thermalization
and localizing the dynamics in many-body Fock space
\cite{oganesyan2007localization,pal2010many,bardarson2012unbounded,
nandkishore2015many,alet2018many}.

On superconducting and trapped-ion quantum simulators, MBL and MBL-like
dynamics are commonly studied by encoding disordered spin-chain Hamiltonians
either into gate-based circuits or into analog interaction graphs
\cite{smith2016many,xu2018emulating,guo2021observation}. A useful starting
point is an interacting disordered spin-exchange model,
\begin{equation}
H^{\rm 1D}_{\rm MBL}
=
\sum_{i\neq j} J_{ij}
\left(
\sigma_i^{+}\sigma_j^{-}
+
\sigma_i^{-}\sigma_j^{+}
\right)
+
\sum_{i\neq j} V_{ij} n_i n_j
+
\sum_i h_i n_i ,
\label{eq:H_MBL_generic}
\end{equation}
where $\sigma_i^{\pm}=(X_i\pm iY_i)/2$ are spin raising and lowering
operators and $n_i=(\mathbb{I}-Z_i)/2$ denotes the local spin-down
occupation. Here, $J_{ij}$ describes tunable spin-exchange processes, including
the long-range profiles naturally available in trapped-ion systems, $V_{ij}$
denotes interactions between spin excitations, and $h_i$ are site-dependent
fields that act as disorder. In the nearest-neighbor limit, this class of
models is equivalent, up to constant energy shifts and coefficient conventions,
to the random-field XXZ chain,
\begin{equation}
H_{\rm XXZ}
=
\sum_{i=1}^{L-1}
\left[
J_{\perp}
\left(
X_iX_{i+1}\!+\!Y_iY_{i+1}
\right)
\!+\!
J_z Z_iZ_{i+1}
\right]
\!+\!\!
\sum_{i=1}^{L} h_i Z_i ,
\label{eq:H_XXZ_random}
\end{equation}
where $J_{\perp}$ drives delocalizing spin exchange, $J_z$ supplies the
interaction responsible for genuinely many-body dynamics, and the random fields
$h_i$ favor localization by energetically pinning local spin configurations.

Early programmable quantum-simulation experiments established key aspects of
this MBL phenomenology in both trapped-ion and superconducting platforms. In
trapped-ions, Smith \emph{et al.} implemented an interacting long-range Ising
model with programmable random disorder in a chain of ten ions
\cite{smith2016many}. The experiment observed several essential signatures of
MBL, including memory retention of the initial state, Poissonian level
statistics, and slow entanglement growth. In superconducting circuits,
Xu \emph{et al.} emulated MBL dynamics on a 10-qubit programmable
superconducting processor by implementing a spin-$1/2$ XY model with
programmable disorder and long-range spin--spin interactions
\cite{xu2018emulating} [Fig.~\ref{fig:mbl}(a)]. The experiment observed
persistent memory of initial states, violation of ETH, and direct evidence for
long-time logarithmic growth of entanglement entropy.

\begin{figure*}
    \centering
    \includegraphics[width=0.9\linewidth]{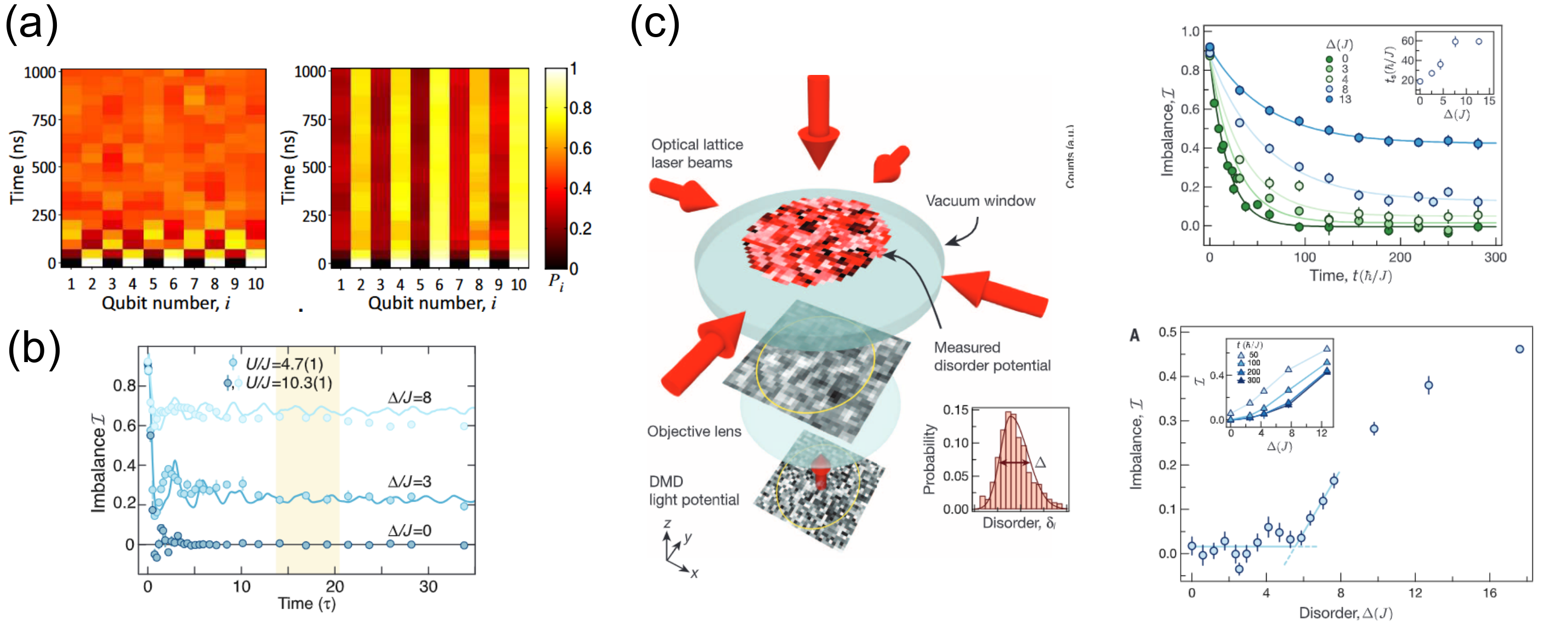}
    \caption{Pioneering experimental probes of many-body localization (MBL) in quantum simulators. (a) Spatio-temporal evolution of local excitations in a chain of ten qubits. Left: In the ergodic regime, excitations rapidly spread across all sites, leading to featureless dynamics and fast decay of spatial structure. Right: In the localized regime, excitations remain confined to their initial sites, producing persistent site-resolved patterns~\cite{xu2018emulating}.  (b) Time evolution of the  {CDW imbalance [Eq.~\ref{eq:imbalance}]}, showing the transition from thermalizing dynamics to a stable non-zero late-time imbalance~\cite{schreiber2015observation}.  (c)  Ultracold-atom realization of disordered Hubbard systems exhibiting MBL crossover. 
Left: Schematic of the experiment, where a digital micromirror device (DMD) is used to project a programmable, site-resolved disorder potential onto a fermionic optical lattice~\cite{choi2016exploring}.   Right (top): Imbalance decay for increasing disorder strengths, demonstrating slower relaxation and emergent localization as $\Delta/J$ grows.  Right (bottom): Steady-state imbalance as a function of disorder strength. The rise of imbalance identifies the crossover from thermal diffusion to many-body localized behaviour. Panel (a) is adopted from ~\cite{xu2018emulating}. Panel (b) is adopted from ~\cite{schreiber2015observation}. Panel (c) is adopted from ~\cite{choi2016exploring}.
}
    \label{fig:mbl}
\end{figure*}

Subsequent experiments broadened the scope of localization physics accessible
on quantum hardware. Superconducting-qubit experiments using a 19-qubit
programmable processor accessed energy-dependent localization behavior,
thereby probing mobility-edge-related phenomenology in a finite-size setting
\cite{guo2021observation}. Phase-sensitive measurements on a Google
superconducting-qubit processor further enabled a direct characterization of
nonlocal effective interactions in the localized phase, probing the microscopic
structure expected from quasi-local integrals of motion
\cite{chiaro2022direct}. In parallel, Stark-MBL experiments in an academic
trapped-ion quantum simulator showed that localization-like nonergodic dynamics
can arise from a strong linear potential gradient even without quenched random
disorder~\cite{morong2021observation}. These results expanded the experimental
view of localization beyond random-field MBL to include inhomogeneous,
gradient-induced, and disorder-free forms of nonergodic dynamics.

More recently, gate-based superconducting processors have enabled a
complementary strategy: rather than requiring high-precision simulation of a
static Hamiltonian over long physical times, one can implement surrogate
disordered or quasiperiodic Floquet circuits. In these circuits,
autocorrelations, imbalance, out-of-time-ordered correlators, and quantum
Fisher information can distinguish thermalizing and nonergodic regimes.
Hayata \emph{et al.} proposed and demonstrated a digital simulation of the MBL
crossover in a disordered kicked-Ising model on the 156-qubit IBM Heron r2
superconducting processor \texttt{ibm\_fez}, using 60 qubits of the heavy-hex
device~\cite{hayata2025digital}. Nagao \emph{et al.} subsequently investigated
the ergodic--MBL crossover in quasiperiodic Floquet Ising systems using up to
144 qubits on an IBM Quantum processor, with Floquet circuits reaching up to
5000 cycles~\cite{nagao2026probing}. Their measurements of autocorrelation
functions and quantum Fisher information revealed persistent correlations and
slow growth of an entanglement-sensitive diagnostic in the
strong-quasiperiodic-potential regime, including localization-like behavior in
both one- and two-dimensional geometries.

The hallmark observables used to diagnose MBL include the \emph{imbalance}, which quantifies the memory of an
initial charge-density-wave (CDW) pattern, and the growth law of entanglement entropy. For a typical CDW initial
state with occupations on even sites (or ``N\'eel'' order in a spin mapping), the imbalance is defined as
\begin{equation}
\mathcal{I}(t)
=
\frac{N_{\rm e}(t)-N_{\rm o}(t)}{N_{\rm e}(t)+N_{\rm o}(t)},
~
N_{\rm e/o}(t)=\sum_{i\in \mathrm{even/odd}}\langle n_i(t)\rangle,
\label{eq:imbalance}
\end{equation}
so that $\mathcal{I}(0)\simeq 1$ for a perfect CDW and $\mathcal{I}(t)\to 0$ signals loss of memory and approach to
thermal equilibrium [Fig.~\ref{fig:mbl} (b)]. In the ergodic regime (weak disorder), the system rapidly thermalizes: local observables relax
toward equilibrium values and entanglement typically grows approximately linearly in time after a global quench
~\cite{xu2018emulating}. By contrast, in the localized regime (strong disorder), the imbalance
remains finite at long times, reflecting persistent memory of the initial state, while entanglement growth is
strongly slowed and is often consistent with a logarithmic law
\begin{equation}
S_A(t)\sim \log t,
\label{eq:log_ent}
\end{equation}
as expected from dephasing induced by quasi-local integrals of motion.

In circuit-QED settings, the disordered Bose--Hubbard model is often realized
as a Kerr--Hubbard model, because the onsite interaction originates from the
Kerr nonlinearity, or anharmonicity, of each bosonic mode
\cite{gong2021experimental}:
\begin{equation}
H_{\rm BH}
=
-\sum_{\langle i,j\rangle}
J_{ij}
\left(
a_i^\dagger a_j+a_j^\dagger a_i
\right)
+
\frac{U}{2}
\sum_i n_i(n_i-1)
+
\sum_i \epsilon_i n_i ,
\label{eq:BH_disordered}
\end{equation}
where $a_i^\dagger$ and $a_i$ are bosonic creation and annihilation operators,
$n_i=a_i^\dagger a_i$ is the local occupation, $J_{ij}$ is the hopping
amplitude, $U$ is the onsite interaction or Kerr nonlinearity, and $\epsilon_i$
is a site-dependent onsite potential. In the hard-core limit, where occupations
are restricted to $n_i=0,1$, Eq.~\eqref{eq:BH_disordered} maps onto a
disordered spin-$1/2$ $XY$ model with longitudinal fields, up to constant
energy shifts and coefficient conventions. A full XXZ form such as
Eq.~\eqref{eq:H_XXZ_random} is obtained only when additional intersite
density--density interactions, or equivalently $Z_iZ_j$ couplings, are present.
Localization then manifests as suppressed transport and long-lived memory of
initially prepared occupation or spin configurations under the combined action
of disorder, interactions, and finite connectivity.

In optical lattices, many-body localization has been realized using ultracold
atoms subjected to quasiperiodic potentials or controlled random disorder
\cite{schreiber2015observation,choi2016exploring,luschen2017signatures,
bordia2016coupling,kondov2015disorder}. A paradigmatic lattice model for
quasiperiodic localization is the interacting Aubry--Andr\'e--Harper-type
Hamiltonian
\cite{harper1955single,schreiber2015observation,luschen2017signatures},
\begin{equation}
\begin{aligned}
H_{\rm AAH}
&=
-J
\sum_i
\left(
c_{i+1}^{\dagger}c_i
+
{\rm H.c.}
\right)
+
\Delta
\sum_i
\cos(2\pi\beta i+\phi)\, n_i
\\
&\quad
+
U
\sum_i n_i n_{i+1},
\end{aligned}
\label{eq:AAH_interacting}
\end{equation}
where $c_i$ and $c_i^\dagger$ annihilate and create a particle on site $i$,
$n_i=c_i^\dagger c_i$, $J$ is the nearest-neighbor tunneling amplitude,
$\Delta$ is the quasiperiodic potential strength, $\beta$ is an incommensurate
modulation wave number, $\phi$ is a tunable phase, and $U$ parameterizes
interactions. In the noninteracting limit, $U=0$, Eq.~\eqref{eq:AAH_interacting}
exhibits the Aubry--Andr\'e localization transition at $\Delta/J=2$ for this
choice of convention, while finite interactions produce an interacting
localization crossover or MBL regime. In spinful-fermion optical-lattice
experiments, the interaction is often instead an onsite Hubbard interaction
between two spin components, but the same quasiperiodic-potential mechanism
provides the single-particle localization backbone.

A landmark cold-atom experiment by Schreiber \emph{et al.} observed MBL of
interacting fermions in a one-dimensional quasirandom optical lattice by
preparing an initial charge-density-wave state and monitoring the imbalance
over time~\cite{schreiber2015observation}. In the ergodic regime, the imbalance
rapidly decayed, whereas in the localized regime it remained finite, establishing
cold atoms as a benchmark platform for non-equilibrium localization physics.
Subsequent experiments accessed complementary diagnostics beyond imbalance
decay, including probes of entanglement growth through number fluctuations and
correlations~\cite{lukin2019probing}, quantum-gas-microscope measurements of
spatially resolved relaxation and critical behavior near the MBL crossover
\cite{rispoli2019quantum}, and controlled coupling to an environment to test how
weak dissipation destabilizes localization and drives eventual relaxation
\cite{luschen2017signatures}.

Particularly for quantum devices with higher connectivity and larger qubit
counts, it becomes feasible to explore localization physics beyond one
dimension, including finite-size two-dimensional localization dynamics and
MBL-like crossovers
\cite{deroeck2017stability,potirniche2019exploration,hur2025stability,
li2025many}. A generic two-dimensional random-field XXZ model can be written as
\begin{equation}
H^{\rm 2D}_{\rm MBL}
=
\sum_{\langle i,j\rangle}
J_{ij}
\left(
X_iX_j+Y_iY_j+\Delta Z_iZ_j
\right)
+
\sum_i h_iZ_i ,
\label{eq:H_MBL_2D}
\end{equation}
where the sum extends over nearest-neighbor bonds on a planar lattice or
programmable coupling graph, $J_{ij}$ sets the bond-dependent exchange scale,
$\Delta$ controls the Ising anisotropy or interaction strength, and $h_i$ are
site-dependent random longitudinal fields. This model captures the competition
between delocalizing spin exchange, interaction-induced dephasing, and
disorder-induced pinning. On finite two-dimensional devices, such Hamiltonians
provide a controlled setting for studying disorder-averaged relaxation,
autocorrelation decay, transport suppression, and the crossover between
localized and delocalized many-body dynamics.

The status of true asymptotic MBL in two and higher dimensions remains subtle.
Thermal avalanches seeded by locally ergodic regions are expected to destabilize
random-disorder MBL in sufficiently large systems
\cite{deroeck2017stability,potirniche2019exploration}. Nevertheless, current
quantum simulators operate in finite-size and finite-time regimes, where
long-lived nonergodic dynamics can still be observed. Early optical-lattice work
by Choi \emph{et al.} explored localization physics in a two-dimensional
bosonic system using an academic ultracold-atom platform with single-site
resolution, observing slow relaxation and incomplete thermalization consistent
with MBL-like behavior over experimentally accessible timescales
\cite{choi2016exploring} [Fig.~\ref{fig:mbl}(c)].

More recent cold-atom experiments have sharpened this question by comparing
random and quasiperiodic disorder in two dimensions. In optical-lattice systems
with variable sizes up to $24\times24$ sites, Hur \emph{et al.} found that the
apparent crossover for random disorder shifts to stronger disorder with
increasing system size, consistent with avalanche-driven instability, whereas
the quasiperiodic case shows no clear size dependence over the accessible range
\cite{hur2025stability}. These results suggest that finite-size localization
phenomenology in higher dimensions depends sensitively on the structure of the
disorder potential, with random disorder more susceptible to rare-region
avalanches than quasiperiodic disorder.

Moreover, superconducting processors provide a complementary route to dimensional-scaling tests of localization. Recent experiments using a 70-qubit two-dimensional superconducting quantum simulator observed that imbalance decay becomes more pronounced as the system size is increased from 21 to 42 and 70 qubits, providing evidence for many-body delocalization in two-dimensional disordered systems and supporting the finite-size avalanche picture~\cite{li2025many}. This result is important because it uses the scalability of superconducting hardware not only to observe long-lived localization signatures, but also to test how those signatures evolve systematically with system size.

\noindent\textbf{Disorder-free MBL.~} In contrast to disorder-induced many-body localization, disorder-free localization can arise in clean systems when a strong spatial potential gradient suppresses transport. A paradigmatic example is \emph{Stark many-body localization} (Stark MBL), where a linear tilt, generated for example by an electric-field gradient, gravitational tilt, or programmable site-dependent detuning, produces Wannier--Stark localization of single-particle orbitals even in the absence of quenched randomness~\cite{schulz2019stark,taylor2020experimental,guo2021stark,morong2021observation}. Interactions then couple the localized Wannier--Stark orbitals and generate a many-body nonergodic regime with persistent memory of initial density patterns, slow relaxation, and constrained information spreading. However, Stark MBL should not be viewed as identical to conventional random-field MBL: because the potential is spatially structured rather than random, the dynamics can show strong initial-state dependence and is closely related to Hilbert-space fragmentation and emergent dipole-conserving constraints~\cite{doggen2021stark,kohlert2023exploring}.

A minimal interacting Stark-localized model for spinless fermions is
\begin{equation}
H_{\rm Stark}
=
-J
\sum_i
\left(
c_{i+1}^{\dagger}c_i
+
{\rm H.c.}
\right)
+
F
\sum_i i\, n_i
+
U
\sum_i n_i n_{i+1},
\label{eq:H_Stark}
\end{equation}
where $c_i^\dagger$ and $c_i$ create and annihilate a fermion on site $i$, $n_i=c_i^\dagger c_i$ is the local density, $J$ is the nearest-neighbor hopping amplitude, $F$ is the linear potential gradient, and $U$ is the nearest-neighbor interaction. The second term imposes an energy offset that grows linearly with site index, as illustrated in Fig.~\ref{fig:mbl2}(a). In the noninteracting limit, Eq.~\eqref{eq:H_Stark} supports Wannier--Stark ladder states whose wave functions are localized even without disorder. For finite $U$, interactions generate dephasing and resonant couplings between localized orbitals, leading to a many-body regime that shares several phenomenological signatures with MBL, including persistent imbalance and slow entanglement growth, while retaining qualitative differences from disorder-induced localization~\cite{taylor2020experimental,doggen2021stark}.

\begin{figure*}
    \centering
    \includegraphics[width=0.9\linewidth]{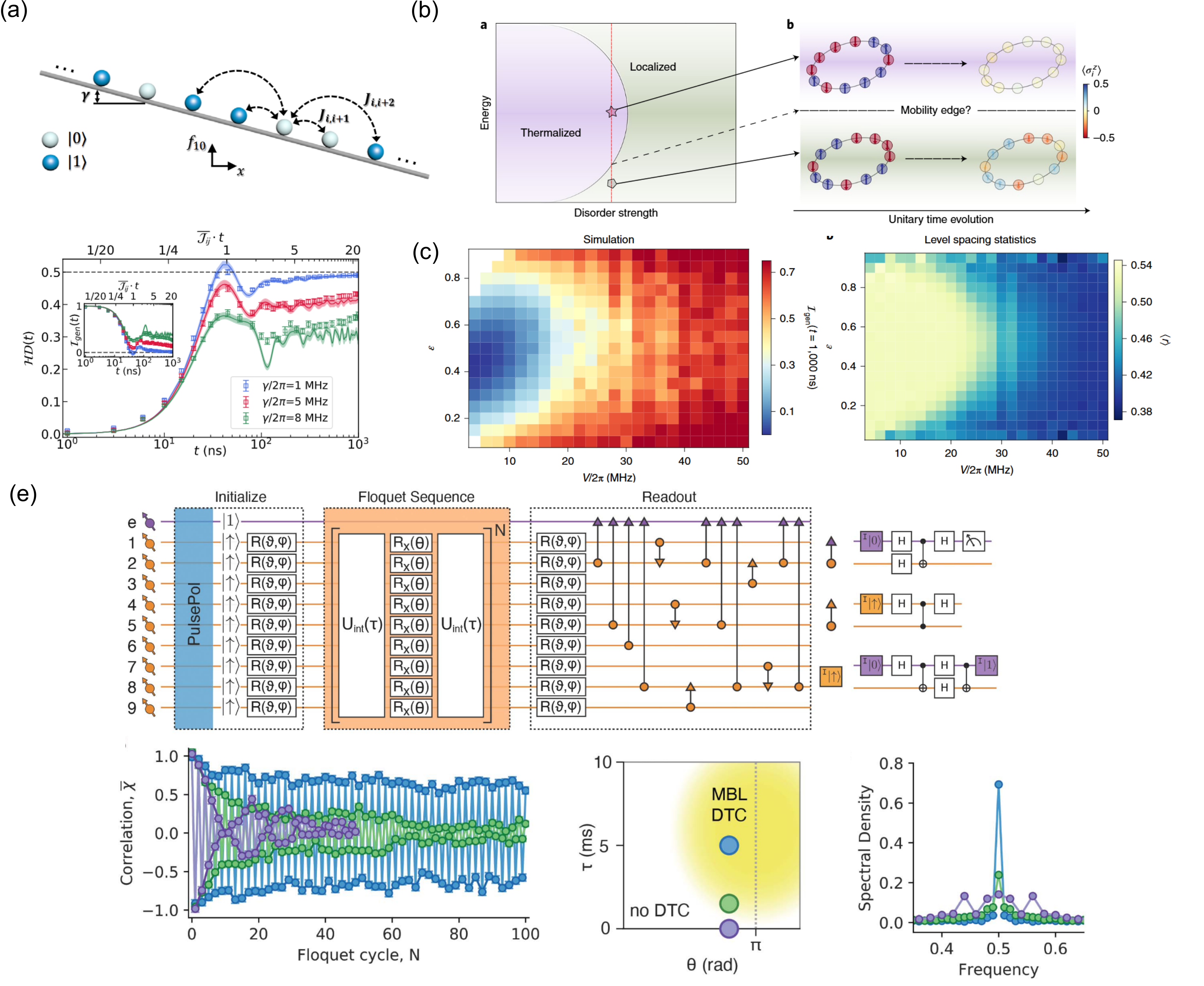}
    \caption{ Further experimental demonstrations of many-body localizations (MBL). (a) Stark (tilt-induced) MBL in a disorder-free superconducting qubit array.
A strong linear potential gradient is applied on a one-dimensional chain of qubits. The bottom panel shows the  Hamming distance measured over time dynamics: in the strong-tilt regime, dynamics can retain the pattern of initial states, so that the measured Hamming distance shows a lower value (lower panel of (a))~\cite{guo2021stark}. (b) Energy-resolved MBL and the emergence of mobility edges. Schematic phase diagram showing how interacting disordered systems can exhibit coexistence of thermal and localized eigenstates depending on energy density~\cite{guo2021observation}. (c) Experimental signatures of energy-resolved MBL~\cite{guo2021observation}. Left heatmap: Simulated long-time imbalance as a function of drive frequency and disorder amplitude. Right heatmap: Level-spacing statistics as a function of interaction strength, showing a transition from ergodic to localized behaviour. (d) MBL-stabilized discrete time crystal~\cite{randall2021many}. Top: Floquet sequence used to implement the periodic unitary. Bottom left: Correlation versus Floquet cycle, showing persistent oscillations in the MBL regime, in contrast to rapid decay in the thermal regime (see the bottom-middle phase diagram). Bottom right: Fourier spectrum of the time-trace exhibiting a pronounced peak at half the drive frequency, a hallmark of discrete time-crystalline order.   Panel (a) is adopted from ~\cite{guo2021stark}. Panel (b) and (c) are adopted from ~\cite{guo2021observation}. Panel (d) is adopted from ~\cite{randall2021many}.
}
\label{fig:mbl2}
\end{figure*}

Experimentally, Stark MBL has been demonstrated in both trapped-ion and
superconducting quantum simulators. In trapped-ions, Morong \emph{et al.}
observed disorder-free Stark MBL on an academic trapped-ion quantum simulator
by engineering a strong field gradient in an interacting long-range spin
system~\cite{morong2021observation}. The experiment showed that increasing the
tilt suppresses thermalization, slows correlation spreading, and preserves
memory of the initial spin configuration over experimentally accessible times.

In superconducting circuits, Guo \emph{et al.} implemented Stark-localized
dynamics on a programmable superconducting quantum processor with 29 functional
qubits~\cite{guo2021stark}. By applying a linear frequency gradient across the
qubit array and initializing the system in a charge-density-wave configuration,
they monitored the density imbalance as a function of evolution time and tilt
strength. At weak tilt, interactions induce relaxation and the imbalance
decays. At strong tilt, the imbalance remains finite for long times, indicating
robust memory of the initial CDW pattern and suppression of thermalization. The
measured imbalance dynamics in Fig.~\ref{fig:mbl2}(a) therefore provide a
direct dynamical signature of disorder-free localization.

A useful complementary viewpoint is obtained by expanding
Eq.~\eqref{eq:H_Stark} in the strong-tilt regime. When the tilt scale $F$ is
large compared with the bare tunneling $J$, ordinary hopping processes become
off-resonant because they change the dipole moment of the configuration. The
low-energy dynamics is then governed by higher-order resonant processes that
approximately conserve both particle number and dipole moment. This emergent
constraint decomposes the many-body Hilbert space into dynamically disconnected
or only weakly connected sectors, producing Hilbert-space fragmentation. As a
result, the late-time state can depend strongly on the initial configuration,
and relaxation can be anomalously slow even without spatial randomness
\cite{doggen2021stark,kohlert2023exploring}. This fragmentation perspective
explains why Stark-localized systems can exhibit MBL-like phenomenology while
differing microscopically from conventional random-field MBL. Cold-atom experiments provide an additional route to this disorder-free
mechanism. Strongly tilted Fermi--Hubbard chains have been used to explore the
fragmentation regime directly, showing constrained relaxation dynamics
controlled by the interplay of tunneling, onsite interactions, and the linear
potential gradient~\cite{kohlert2023exploring}. In particular, near resonances
between the tilt energy and the Hubbard interaction, the tilted Fermi--Hubbard
model realizes distinct effective Hamiltonians whose fragmented structure
captures the observed transient dynamics in experimentally accessible regimes.

Recent theoretical and numerical work further sharpens this picture. Periodically driven Stark systems have been studied as Floquet extensions of disorder-free localization, where the competition between tilt-induced localization and drive-induced resonances can either preserve or destabilize the localized regime~\cite{duffin2024stark}. Open-system studies have also suggested that Stark localization can protect coherence against environmental relaxation in certain quantum-dot simulator settings~\cite{sarkar2024protecting}. More recently, rigorous progress on interacting-particle Stark localization has established strong localization properties for finite numbers of interacting particles in a linear potential, clarifying the mathematical basis of Stark localization while still leaving the thermodynamic many-body limit distinct from conventional random MBL~\cite{deroeck2026stark}. Quantum-information diagnostics such as stabilizer R\'enyi entropy and nonstabilizerness have been proposed as probes of disorder-free ergodicity breaking, showing that Stark-localized systems can suppress transport while still generating nontrivial computational resources over slow dynamical timescales~\cite{li2026slow}.

Another intriguing development in the study of localization phenomena is the emergence of energy-resolved many-body localization (MBL), which has been explored experimentally in superconducting-qubit processors~\cite{guo2021observation}. 
In the conventional dynamical diagnosis of MBL, one usually fixes an initial state and tunes a global parameter such as the disorder strength. 
The energy-resolved viewpoint instead asks whether localization occurs uniformly across the many-body spectrum at fixed Hamiltonian parameters.  As illustrated schematically in Fig.~\ref{fig:mbl2}(b), this distinction can be accessed by preparing different initial product states whose energy densities with respect to the same disordered Hamiltonian are different. 
States with strong overlap on localized parts of the spectrum can retain spatial memory under unitary time evolution, whereas states overlapping predominantly with thermal eigenstates relax more rapidly and lose memory of their initial configuration.

Such an energy-selective response was probed in superconducting-qubit
experiments by Guo \emph{et al.}, who measured local observables and
imbalance-like memory diagnostics for initial states with different effective
energy densities on a programmable superconducting processor
\cite{guo2021observation}. The experiments show that states near the middle of
the many-body spectrum relax more rapidly and approach thermal behavior,
whereas states near the spectral edges retain longer-lived memory of their
initial spin configuration. The corresponding energy-resolved localization
phenomenology is reflected both in dynamical observables, such as the imbalance
shown in Fig.~\ref{fig:mbl2}(c), left, and in spectral diagnostics, such as the
level-spacing statistics shown in Fig.~\ref{fig:mbl2}(c), right. These
simulations show that localization in interacting quantum systems can be
strongly energy selective, enriching the standard disorder-driven MBL picture.

MBL also plays a crucial role in stabilizing discrete time crystals (DTCs) in
periodically driven, strongly interacting systems; see Sect.~\ref{time}. In an
isolated driven many-body system, the drive would normally heat the system
toward an infinite-temperature state, washing out any long-lived temporal
order. Strong disorder can prevent this heating by inducing MBL, thereby
providing a nonergodic background in which Floquet eigenstates retain local
memory. The resulting MBL-DTC phase is a genuine non-equilibrium phase of matter
in which localization protects subharmonic oscillations against perturbations,
pulse imperfections, and slow drifts. A prototypical model exhibiting this behavior is a one-dimensional disordered
spin chain subjected to a periodic sequence of global rotations and interacting
Ising terms, as schematically shown in the Floquet circuit of
Fig.~\ref{fig:mbl2}(d)~\cite{randall2021many}. Experimentally, this physics was
realized on a programmable solid-state spin simulator based on individually
controllable $^{13}$C nuclear spins in diamond. The hallmark signature is the
persistence of subharmonic spin autocorrelations over many Floquet cycles, as
shown in the lower-left panel of Fig.~\ref{fig:mbl2}(d), together with a sharp
subharmonic peak in the Fourier spectrum of the time trace, as shown in the
lower-right panel. These features provide direct evidence that periodic driving,
interactions, and localization can combine to produce a robust dynamical phase
beyond equilibrium statistical mechanics.

\begin{figure*}
    \centering
    \includegraphics[width=0.9\linewidth]{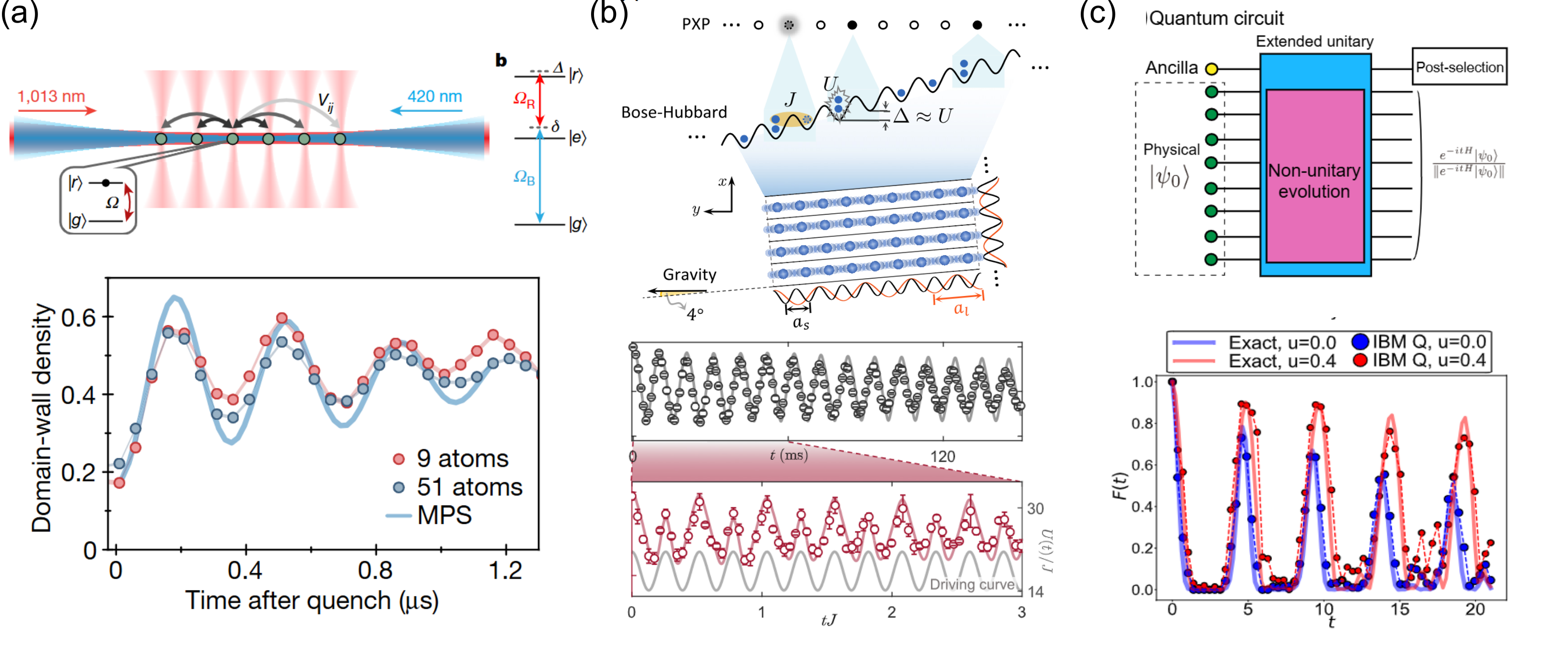}
    \caption{Experimental realizations of quantum many-body scars (QMBS) across different quantum platforms. (a)Rydberg-atom scars. Top: Schematic of a one-dimensional Rydberg array, where strong Rydberg blockade enforces the PXP constraint.Bottom: After a global quench, the system exhibits robust, long-lived oscillations of the domain-wall density (bottom), a hallmark of quantum many-body scarring~\cite{bernien2017probing}. (b) Quantum many-body scars in Bosonic Hubbard systems. Top: Illustration of a tilted Bose–Hubbard chain with onsite interactions and tilt. Middle and bottom: Experimentally measured oscillations of the imbalance and quench dynamics reveal strong revivals at specific tilt values, consistent with energy-resolved scar states embedded in a thermal spectrum~\cite{su2023observation}. (c)  Non-unitary circuit simulations of the enhanced scars. Top: Quantum-circuit architecture in which an extended unitary evolution on system plus ancilla implements an effective non-unitary operator. Bottom: Measured fidelity revivals for different non-unitary parameters, giving rise to stronger periodic revivals~\cite{shen2024enhanced}.  Panel (a) is adopted from ~\cite{bernien2017probing}. Panel (b) is adopted from ~\cite{su2023observation}. Panel (c) is adopted from ~\cite{shen2024enhanced}.
}
    \label{fig:scar}
\end{figure*}

\subsubsection{Quantum many-body scars}

Unlike many-body localization, which yields a strong and disorder-stabilized
violation of ergodicity, quantum many-body scars (QMBS) represent a \emph{weak}
form of ergodicity breaking in otherwise clean, nonintegrable systems
\cite{bernien2017probing,turner2018weak,turner2018quantum,
Serbyn2021,Chandran2023,moudgalya2022scars}. In many
canonical examples, a small set, or tower, of atypical nonthermal eigenstates is
embedded within an otherwise thermal spectrum. Although these scarred states do
not prevent thermalization for generic initial conditions, they can dominate the
dynamics of specially prepared states, leading to long-lived revivals,
anomalously slow entanglement growth, and persistent memory of the initial
configuration
\cite{turner2018quantum,ho2019periodic,khemani2019signatures,
lin2019exact,choi2019emergent,bull2020quantum,
chattopadhyay2020quantum}. In this sense, QMBS provide a distinct route to
nonergodic dynamics that does not rely on quenched disorder
\cite{shiraishi2017systematic,moudgalya2018exact,
moudgalya2018entanglement,schecter2019weak,mark2020unified,
MoudgalyaKrylov,Serbyn2021,moudgalya2022scars}. Recent experiments and digital
simulations have further extended this idea beyond the original Rydberg-chain
setting, including Bose--Hubbard simulators, superconducting processors,
trapped-ion platforms, and quantum-circuit state-preparation protocols
\cite{su2023observation,zhang2023manybodyscarring,
chen2022errormitigatedscars,gustafson2023preparing,
desaules2024robust,zhao2025observation}.

Although DTCs and QMBS can both produce robust or long-lived periodic
many-body dynamics, their physical origins are distinct. A DTC is a Floquet
phase of matter characterized by subharmonic response and rigidity against
perturbations, and it is typically stabilized by many-body localization,
prethermalization, or dissipation
\cite{khemani2016phase,else2016floquet,yao2017discrete,else2017prethermal,
randall2021many,wu2024dissipative}. By contrast, scar revivals arise from a
small set of atypical nonthermal eigenstates embedded in an otherwise thermal
spectrum, and are therefore strongly dependent on the choice of initial state
\cite{bernien2017probing,turner2018weak,turner2018quantum,Serbyn2021}. This
distinction is especially important in finite-size quantum simulations, where scar
revivals and DTC oscillations may appear similar over limited observation times,
even though they reflect different mechanisms of ergodicity breaking.

\noindent\textbf{Rydberg-blockade realization and the PXP scar mechanism.} The experimental discovery of QMBS that motivated this subfield came from a
one-dimensional chain of 51 Rydberg atoms assembled with optical tweezers
\cite{bernien2017probing} [Fig.~\ref{fig:scar}(a)]. The microscopic dynamics is
governed by the Rydberg Hamiltonian
\begin{equation}
H_{\rm Ryd}
=
\sum_i
\left(
\frac{\Omega}{2} X_i
-\Delta n_i
\right)
+
\sum_{i<j} V_{ij} n_i n_j ,
\label{eq:H_Ryd}
\end{equation}
where $\Omega$ is the Rabi frequency of the laser drive, $\Delta$ is the
detuning, $n_i=\ket{r_i}\!\bra{r_i}$ projects onto the Rydberg-excited state at
site $i$, and $V_{ij}$ denotes the interaction between Rydberg excitations. The
term proportional to $\Omega$ coherently drives transitions between the atomic
ground and Rydberg states, while the detuning term controls the energy cost of
Rydberg excitation. In the strong-blockade regime, nearest-neighbor double
excitations are energetically suppressed. Projecting the Rabi drive into this
blockade-constrained Hilbert space gives the constrained PXP model,
\begin{equation}
H_{\rm PXP}
=
\frac{\Omega}{2}
\sum_i
P_{i-1} X_i P_{i+1}
-\Delta \sum_i n_i ,
\qquad
P_i = 1-n_i ,
\label{eq:H_PXP}
\end{equation}
which acts only within the subspace with no adjacent Rydberg excitations.

When the system is initialized in the $\mathbb{Z}_2$ charge-density-wave,
or N\'eel-like, state and quenched near resonance, it exhibits unexpectedly
persistent, large-amplitude oscillations of the density-wave order parameter.
Rather than rapidly relaxing to thermal equilibrium, the system repeatedly
returns close to its initial configuration. Turner \emph{et al.} subsequently
interpreted these coherent revivals in terms of a tower of special many-body
eigenstates embedded in the otherwise thermal PXP
spectrum
\cite{turner2018weak,turner2018quantum,ho2019periodic,choi2019emergent,
lin2019exact}. A central theoretical insight is that the revival dynamics can
be understood using a forward-scattering approximation (FSA) built from the
$\mathbb{Z}_2$ initial state. In this picture, the constrained Hamiltonian
organizes a small Krylov-like subspace of basis states at increasing Hamming
distance from the initial product state, and the dynamics remains anomalously
confined to this subspace for long times. The approximate regular spacing of
the scarred eigenstates, together with their large overlap with the FSA
subspace, gives rise to the emergent revival periodicity.

\noindent\textbf{Extensions beyond the original PXP setting.} Theoretical developments have further broadened the landscape of quantum scar
physics. Exact scar eigenstates have been identified beyond the original
Rydberg/PXP setting, including in Bose--Hubbard models with additional
constraints such as a three-body constraint~\cite{kaneko2024quantum}. In
open-system settings, scar towers can be embedded into decoherence-free
subspaces of Lindblad dynamics, suggesting that controlled dissipation can
stabilize, rather than necessarily destroy, scarred oscillations
\cite{wang2024embedding}. Related work on Lindblad many-body scars further
extends this perspective to chaotic many-body systems coupled to Markovian
baths~\cite{garciagarcia2026lindblad}. In parallel, scarred dynamics has been
argued to be much more widespread in many-body spin systems, where unstable
periodic orbits can generate genuine quantum scars even within otherwise
thermal spectra~\cite{pizzi2025genuine}. These developments indicate that QMBS
should be viewed as a broader organizing principle for weak ergodicity breaking
in constrained, open, and generic many-body settings.

Recent experiments have demonstrated that scarred dynamics is not confined to
Rydberg-blockaded spin chains. A particularly notable achievement is the
observation of many-body scarring in a Bose--Hubbard quantum simulator
\cite{su2023observation} [Fig.~\ref{fig:scar}(b)]. In that platform,
ultracold bosonic atoms are loaded into a one-dimensional optical lattice and
quenched into a regime with strong on-site repulsion and a large linear tilt,
so that ordinary tunneling becomes off-resonant and the effective dynamics is
strongly constrained. In this tilted Bose--Hubbard regime, the experiment
emulates PXP-like constrained dynamics and observes several characteristic
signatures of scarring: local densities exhibit long-lived oscillatory behavior,
the system retains pronounced memory of the initial state, and relaxation into
the thermalizing many-body Hilbert space is strongly delayed.  A central advance of Ref.~\cite{su2023observation} is that scarred dynamics was
observed not only from the conventional $\mathbb{Z}_2$-type initial state, but
also from previously unexpected initial conditions such as the unit-filling
state at finite detuning. In addition, the experiment used a quantum
interference protocol to measure entanglement entropy, showing that the scarred
dynamics remains confined to a low-entropy subspace rather than rapidly
spreading over the full many-body Hilbert space.

\noindent\textbf{Scar dynamics on programmable quantum processors.} Shen \emph{et al.} proposed and analyzed a mechanism for enhancing quantum many-body scar dynamics in open or non-Hermitian many-body systems by exploiting non-Hermitian asymmetric transition probabilities [see Sect.~\ref{nonunitary} below] within the Fock space~\cite{shen2022non,shen2024enhanced}. The starting point is a non-Hermitian generalization of the PXP model\cite{shen2024enhanced}. By reorganizing the many-body Hilbert space into the FSA basis, grouped by Hamming distance from the N\'eel initial state, the dynamics maps onto an effective non-Hermitian 1D tight-binding chain in Fock space, albeit with each ``site" being a macroscopically large collection of basis states. Asymmetric hopping between FSA layers produces a unidirectional flow that biases the evolution back toward the initial-state manifold, suppressing leakage into the thermal bulk of the spectrum, dubbed as the \emph{Fock-space non-Hermitian skin effect}. Digital simulations on IBM quantum processors showed that the non-Hermitian extension can stabilize and enhance coherent scar oscillations compared with the Hermitian PXP dynamics under realistic hardware noise; see Fig.~\ref{fig:scar} (c). This provides a concrete route to engineering more robust scarred dynamics on noisy quantum devices through non-Hermitian or dissipative control. The following subsection further reviews related non-Hermitian demonstrations on quantum hardware.

QMBS have also been explored directly on other
gate-based quantum processors. Zhang \emph{et al.} observed many-body
Hilbert-space scarring on a superconducting processor by engineering scarred
dynamics in a programmable qubit array and resolving coherent revivals,
anomalous confinement in Hilbert space, and nonthermal entanglement dynamics
\cite{zhang2023manybodyscarring}. Chen \emph{et al.} demonstrated an
error-mitigated simulation of scarred dynamics using pulse-level control,
showing that hardware-aware compilation, noise tailoring, zero-noise
extrapolation, dynamical decoupling, and post-selection can extend the
observable lifetime of mixed-field-Ising scar dynamics on noisy quantum
computers~\cite{chen2022errormitigatedscars}. Finite-temperature scar dynamics
has also been probed on IBM hardware, where PXP scar-induced revivals were
shown to remain visible after mixing with a thermal background
\cite{desaules2024robust}. Most recently, trapped-ion hardware has provided
access to dynamical signatures of conventional and asymptotic QMBS in a
two-local Floquet model, exploiting the high connectivity of the Quantinuum
processor to prepare scarred states and probe their relaxation dynamics
\cite{logaric2026dynamical}.

\subsubsection{Non-unitary dynamics in non-Hermitian open Quantum Systems}\label{nonunitary}

\begin{figure*}
    \centering
    \includegraphics[width=0.9\linewidth]{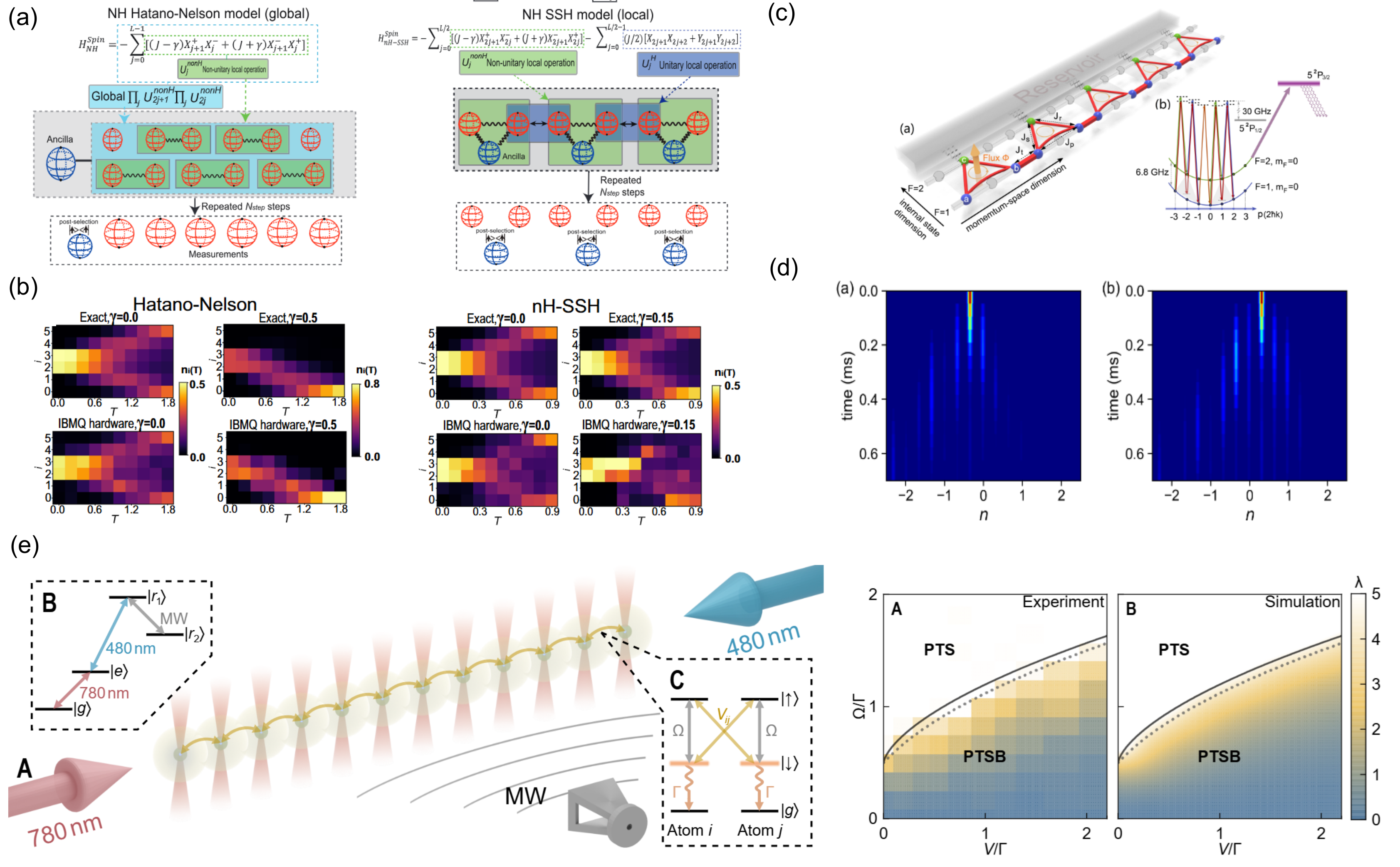}
    \caption{Experimental realizations of non-Hermitian quantum physics on digital and analog quantum platforms. (a) Digital non-unitary simulation on superconducting qubits. Schematic of two ancilla-based dilation architectures used to implement non-Hermitian time evolution.  The left panel shows the global-ancilla protocol for the non-Hermitian Hatano–Nelson model. The right panel shows the local-decomposition protocol for the non-Hermitian SSH model~\cite{shen2025observation}. (b) The corresponding measured time-resolved density evolution for the Hatano–Nelson and non-Hermitian SSH chains, implemented as spin models on the IBM Quantum platform. At nonzero non-Hermitian hopping asymmetry $\gamma$, salient directional pumping of population can be observed, characteristic of the NHSE~\cite{shen2025observation}. The steady-state Fermi-Dirac-like spatial ``Fermi skin" profile arises from many-body exclusion, and cannot be realized in classical implementations of the NHSE. (c) Analog realization of non-Hermitian dynamics in ultracold atoms. Illustration of the momentum-space lattice. Asymmetric hopping is engineered by applying controlled dissipation to one internal state, producing a non-Hermitian Aharonov–Bohm chain with tunable flux and loss~\cite{liang2022dynamic}.   (d) Experimental signature of asymmetric transport in cold atoms ~\cite{liang2022dynamic}. (e) Experimental realization of non-Hermitian XY model in Rydberg-atom arrays, and the measured phase diagram between: PT symmetry and PT-symmetry breaking regions~\cite{zhang2025nonhermitianrydberg}.  Panel (a) and (b) are adopted from ~\cite{shen2025observation}.  Panel (c) and (d) are adopted from ~\cite{liang2022dynamic}. Panel (e) is adopted from  ~\cite{zhang2025nonhermitianrydberg}.}
    \label{fig:non-unitary}
\end{figure*}

As showcased above, non-Hermitian transition amplitudes have the potential to drastically alter state dynamics. 
In general, non-Hermiticity has been recognized as a powerful framework for describing quantum systems with asymmetric couplings, effective gain and loss, measurement backaction, or postselected dynamics \cite{bender1998real,hatano1996localization,elganainy2018non,
ashida2020non,bergholtz2021exceptional,lin2023topological}. 
In open quantum systems, an effective
non-Hermitian Hamiltonian naturally appears in the no-jump evolution of quantum
trajectories, while post-selection and measurement backaction provide operational
routes for realizing nonunitary dynamics on quantum simulators and quantum
processors
\cite{carmichael1993open,dalibard2011colloquium,daley2014quantum,
li2019measurement,google2023measurement}. 
 Much of its most striking phenomena can be traced to the spectral and eigenstate structure of the non-Hermitian effective Hamiltonian, which are respectively not constrained to be real or orthogonal unlike in Hermitian systems. 
In lattice systems, non-Hermitian
couplings can produce phenomena without Hermitian analogs, including
exceptional points, complex spectra, non-Bloch band topology, and the
non-Hermitian skin effect
\cite{yao2018edge,qin2025nonlinear,miri2019exceptional,lee2019anatomy,kawabata2019symmetry,
okuma2020topological,li2020critical,zhang2022review,lin2023topological}.

Two key classes of non-Hermitian phenomenology are the (i) non-Hermitian skin effect (NHSE) associated with nontrivial spectral winding~\cite{kawabata2019symmetry,okuma2020topological,li2021quantized}, in which generic states are asymmetrically amplified and accumulate against system boundaries or impurities~\cite{yao2018edge,song2019non,okuma2020topological,lee2019anatomy,li2020critical,lee2020ultrafast,zhang2021tidal,jiang2022filling,yang2022designing,longhi2022non,tai2023zoology,wan2023observation,qin2023non, wang2024amoeba}, and (ii) exceptional points, where eigenstates coalesce at complex spectral branch points, leading to enhanced sensitivity, multi-valued adiabatic evolution~\cite{heiss2012physics,dembowski2001experimental,qin2026critical} as well as unconventional entanglement behavior~\cite{lee2022exceptional,meng2024exceptional,chang2020entanglement,tu2022renyi,chen2024quantum,liu2025non,xue2026topologically,yang2026beyond}.

While non-Hermiticity is often associated with environmental coupling in an
open-quantum-system setting, it can arise more broadly whenever the effective
dynamics does not conserve probability, norm, or energy. In classical wave
systems, non-Hermitian terms enter directly at the level of the coupled-mode
equations or dynamical matrix describing macroscopic degrees of freedom. This
has enabled direct experimental probes of non-Hermitian physics in photonic
systems with engineered gain, loss, or radiative leakage
\cite{ruter2010observation,regensburger2012parity,peng2014parity,
hodaei2014parity,zeuner2015observation,weidemann2020topological,
xiao2020observation}, classical stochastic~\cite{shi2025general,hao2025interacting} or active mechanical metamaterials with nonreciprocal
couplings \cite{brandenbourger2019non,ghatak2020observation}, and electrical
or topolectrical circuits engineered to realize non-Hermitian band topology and
skin accumulation \cite{helbig2020generalized,liu2021non}. In quantum settings, which we
focus on here, non-Hermiticity can arise from environmental coupling described
by the Lindblad master equation 
\cite{gorini1976completely,lindblad1976generators,carmichael1993open,
dalibard2011colloquium,daley2014quantum, allington2025distinct}, or from explicitly non-unitary
operations generated by measurement, post-selection, or no-jump quantum
trajectories \cite{dalibard2011colloquium,li2019measurement,
google2023measurement,wen2019experimental}.

\noindent\textbf{Non-Hermitian Hamiltonian
simulation.~}{Experimentally, many early and highly intuitive demonstrations of
non-Hermitian physics were realized in analog or effectively analog settings,
where loss, gain, dissipation, or asymmetric hopping can be engineered directly
at the level of the physical platform. 
Recent advances in digital quantum
hardware have made it possible to simulate genuinely quantum many-body
non-Hermitian dynamics, including postselected non-unitary evolution,
non-reciprocal propagation, Fermi-skin accumulation, exceptional-point dynamics,
and  supersonic modes
\cite{shen2025observation,zhang2025supersonic,zhang2025nonhermitianrydberg}.}

\begin{figure*}
    \centering
    \includegraphics[width=0.99\linewidth]{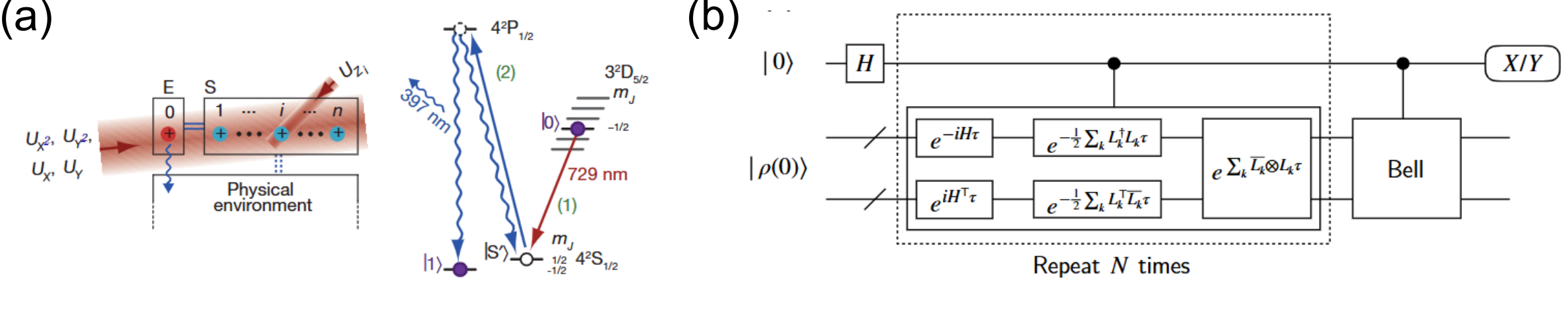}
    \caption{ Digital and analog approaches to simulating open quantum systems. (a) Analog realization of engineered dissipation in trapped-ions~\cite{barreiro2011open}. Schematic of a trapped-ion open-system simulator where internal electronic states of the ions encode the system qubits, and optical pumping channels implement controlled coupling to an engineered environment. (b) Digital simulation of Lindblad evolution using Trotterized non-unitary maps.  Each Trotter step consists of coherent unitary evolution, dissipative channels implemented via ancilla-assisted operations, followed by the effective evolution controlled by jump operators~\cite{kamakari2022digital}.  (a) is adopted from ~\cite{barreiro2011open}.   (b) is adopted from  ~\cite{kamakari2022digital}. }
    \label{fig:open}
\end{figure*}

A recent milestone achievement in the digital quantum simulation of non-Hermitian physics is the observation of the NHSE and its many-fermion analog, the \emph{Fermi skin}, on the IBM digital quantum processor~\cite{shen2025observation}. Shen \emph{et al.} showed that present-day quantum hardware can emulate non-unitary dynamics in paradigmatic non-reciprocal lattice models [Fig.~\ref{fig:non-unitary} (a)] by embedding the desired non-Hermitian evolution into a larger unitary circuit with ancilla qubits, and subsequently doing post-selection on them [see Sect.~\ref{dilation} for details]. This construction enabled the realization of effective asymmetric hoppings, which are necessary for the boundary accumulation characteristic of the NHSE [Fig.~\ref{fig:non-unitary} (b)]. Importantly, this demonstration went beyond the single-particle level: by encoding fermionic many-body states as correlated spins, it revealed how the interplay between the NHSE and Pauli exclusion results in a Fermi-Dirac-like density profile~\cite{mu2020emergent} i.e. ``Fermi skin" in \emph{real} space.

Moreover, Ref.~\cite{koh2025interacting} further expands this framework by
measuring the so-called NHSE-induced ``edge burst'', where NHSE-driven
probability accumulation leads to anomalously enhanced loss at the boundary of
a dissipative non-Hermitian system. Taking advantage of the ability of the IBM
superconducting quantum processors to encode multiple interacting spins, edge
burst physics was generalized to the concept of ``cluster bursts'', where
interparticle repulsion creates effective boundaries in many-body configuration
space [see  Sect.~\ref{sec:topological_dynamics} for simulation details and Sect.~\ref{lcu} for method details]. A complementary trapped-ion
experiment~\cite{zhang2025supersonic} on the Quantinuum H1 processor used variational compilation to
simulate non-Hermitian interacting dynamics without post-selection, observing a
supersonic mode after a non-Hermitian fermionic quench.

Parallel to these demonstrations of non-Hermitian dynamics on gate-based quantum processors are analogous recent experiments in analog quantum simulators, most notably in ultracold atomic gases. There, the non-Hermiticity has to be implemented physically as a form of loss, since measurement operators are not so straightforward to implement in a gate-based-like manner. 
A striking example is the experimental observation of the NHSE in ultracold atoms, reported by Liang \emph{et al.}~\cite{liang2022dynamic}. In this work, the authors engineer a momentum-space lattice with asymmetric hopping amplitudes by selectively introducing controlled dissipation into one internal atomic state, as demonstrated in Fig.~\ref{fig:non-unitary} (c). This realizes an effective non-Hermitian Aharonov–Bohm chain whose dynamics exhibit the hallmark features of non-reciprocal hopping (see Fig.~\ref{fig:non-unitary} (d)). Following a quench, the atomic population undergoes a directional drift and boundary accumulation.

Beyond momentum-space implementations, cold-atom experiments have now demonstrated NHSE physics in higher-dimensional settings. In particular, Zhao \emph{et al.} realized a two-dimensional non-Hermitian topological band in an ultracold Fermi gas with spin--orbit-coupled optical lattices and tunable dissipation~\cite{zhao2025two}. This experiment directly accessed non-Hermitian topology and boundary accumulation in a genuinely two-dimensional geometry, including spectral winding, real-space center-of-mass drift, and exceptional-point physics.
While present cold-atom implementations remain closer to weakly interacting or few-body regimes than to fully generic strongly correlated non-Hermitian matter, they substantially broaden the experimental reach of non-Hermitian topological phenomena and sharpen the pathway toward many-body non-Hermitian phases in tunable atomic settings, particularly in native higher-dimensional settings where non-Hermiticity leads to various counter-intuitive nonlocal effects~\cite{lee2019hybrid,li2020topological,zhang2022universal,wang2023experimental,xu2023two,jiang2023dimensional,qin2024geometry,yang2025beyond,yang2025non,cheng2025stochasticity,yang2025tailoring,yang2026reversing,yang2026algebraic}.

\begin{figure*}
    \centering
    \includegraphics[width=0.99\linewidth]{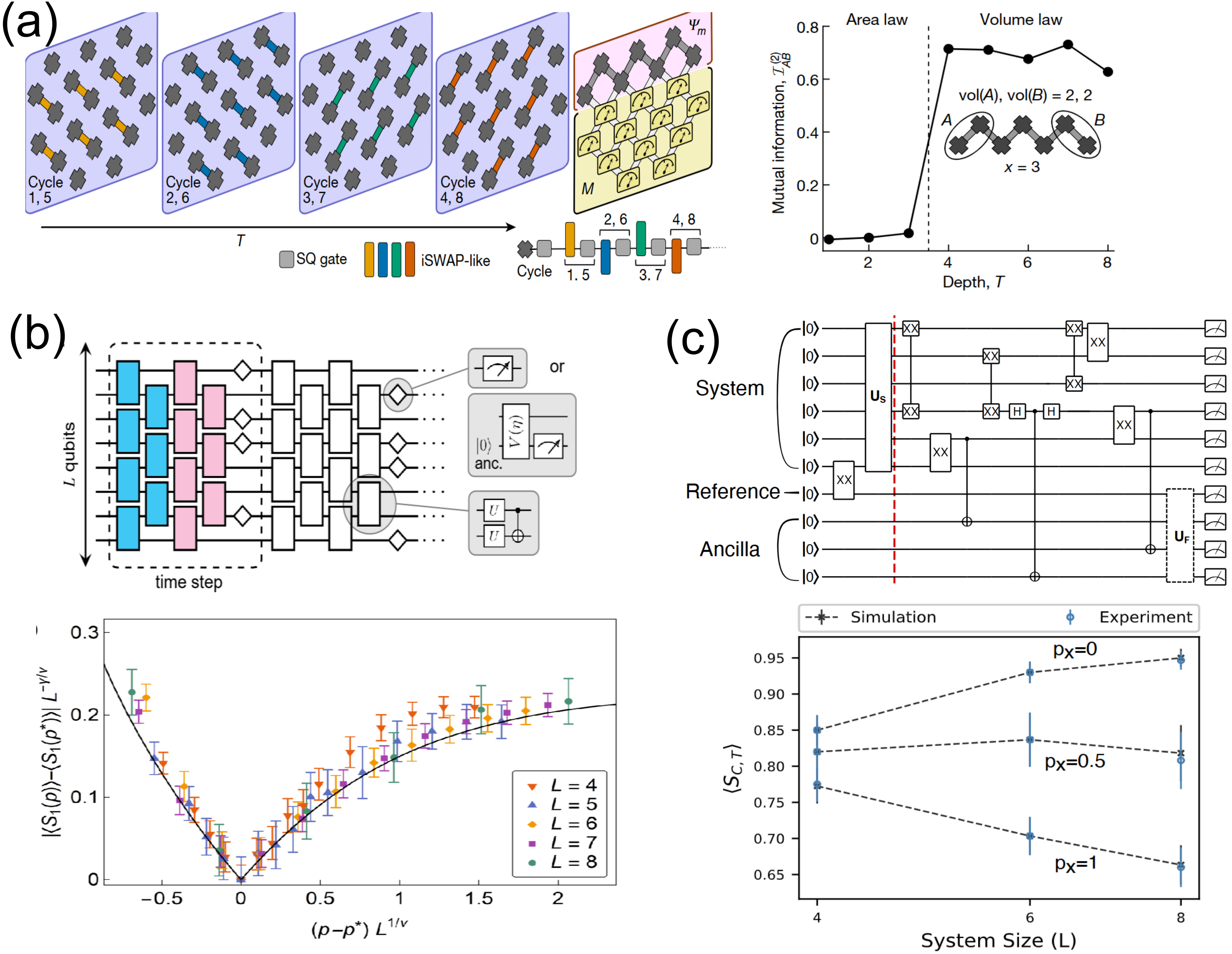}
    \caption{{Experimental demonstrations of measurement-induced phase transitions in monitored quantum circuits.} (a) Digital simulation of large-scale monitored dynamics using repeated scrambling cycles. The transition in the information structure is diagnosed via the mutual information between two separated subregions $A$ and $B$, which exhibits a crossover from an area-law regime (weak long-range correlations) to a volume-law regime (strong nonlocal correlations) as the circuit depth $T$ is increased~\cite{google2023measurement}.  (b) Direct observation of the measurement-induced entanglement transition on a superconducting processor using hybrid random circuits: each discrete time step consists of a brickwork layer of random two-qubit unitaries (blue/pink blocks) followed by stochastic single-qubit measurements applied with probability $p$ (diamonds). Finite-size scaling across system sizes $L$ provides evidence for a critical measurement rate~\cite{koh2023measurement}. (c) Trapped-ion implementation of a purification (coding) transition. The system undergoes alternating scrambling unitaries and measurement operations implemented with ancilla assistance. The measured quantity (lower panel) shows distinct scaling with system size $L$ as the measurement rate $p_x$ increases, showing a transition between a coding phase that retains quantum information ($p_x\!\approx\!0$) and a purifying phase~\cite{noel2022measurement}. Panel (a) is adopted from ~\cite{google2023measurement}.   Panel (b) is adopted from  ~\cite{koh2023measurement}. Panel (c) is adopted from  ~\cite{noel2022measurement}.}
    \label{fig:mes}
\end{figure*}

Rydberg-atom arrays provide a complementary route to non-Hermitian quantum simulation that is distinct from the weakly interacting cold-atom NHSE experiments discussed above. The key point is that Rydberg platforms naturally combine engineered loss with strong blockade-induced many-body constraints. In Ref.~\cite{zhang2025nonhermitianrydberg}, $^{87}$Rb atoms in an optical tweezer array are coupled to a fast-decaying level to realize controlled dissipation, while Rydberg blockade generates strong nearest-neighbour interactions. 
By tuning the coherent drive relative to the loss rate, the experiment observes a dynamical transition between PT-symmetric and PT-broken regimes, as shown in Fig.~\ref{fig:non-unitary}~(e). 
Thus, Rydberg arrays extend non-Hermitian quantum simulation from mostly band-structure and boundary-accumulation physics toward collective, interaction-dominated non-Hermitian dynamics.

\noindent\textbf{
 Simulation of open quantum systems.~}As discussed above, open quantum many-body simulation differs fundamentally from
conventional Hamiltonian simulation. In closed systems, the target dynamics are
generated by a Hermitian Hamiltonian and are described by (coherent) unitary time
evolution on a state vector. In open systems, by contrast, the evolution is generally described
by quantum channels, Lindblad generators, measurement backaction, reset
operations or, in engineered systems--reservoir couplings. The evolution acts at the level of the density matrix, which can also encode incoherent processes among mixed states. Coherent dynamics
therefore compete with dissipation, dephasing, and measurement-induced
nonunitarity, giving rise to phenomena with no direct analog in closed
unitary evolution. Examples include dissipation-driven phase transitions,
quantum Zeno dynamics, dissipative state preparation, and
reservoir-stabilized entangled phases
\cite{sieberer2025universality,facchi2008quantum,biella2021many,
verstraete2009quantum,diehl2008quantum,weimer2021simulation}.

A foundational milestone was achieved by Barreiro \emph{et al.}
\cite{barreiro2011open}, who demonstrated that dissipation can be engineered as
a useful resource rather than merely treated as an obstacle. Using trapped
ions, they constructed controlled open-system quantum maps by combining
coherent multi-qubit gates with optical pumping [Fig.~\ref{fig:open}(a)]. This
toolbox was used to autonomously steer the system toward entangled steady
states, simulate coherent many-body spin interactions, and perform
quantum-nondemolition measurements of multi-qubit observables. The experiment
therefore established a central principle of engineered open-system quantum
simulation: carefully tailored system--environment couplings can be used to
prepare, stabilize, and manipulate many-body quantum states.

A major application of nonunitary dynamical evolution is quantum
imaginary-time evolution (QITE), previously introduced in the context of
ground-state and thermal-state preparation in Sec.~\ref{qite}
\cite{motta2020determining,mcardle2019variational,nishi2021implementation,
sun2021quantum,gomes2020efficient,yeter2020practical,hejazi2024adiabatic,
gluza2026double}. In the open-system setting, QITE ideas can be adapted to
digitally simulate Lindblad dynamics by first vectorizing the density matrix
into an enlarged system--ancilla Hilbert space. The nonunitary evolution
generated by the Liouvillian is then approximated, step by step, by locally
determined unitary updates in this vectorized representation
[Fig.~\ref{fig:open}(b)]~\cite{kamakari2022digital}. This construction avoids
the need to directly implement the nonunitary density-matrix evolution as a
physical channel on the quantum processor, and enables small- and
intermediate-scale simulations of dissipative dynamics with circuit depths
compatible with present noisy hardware.

Kamakari \emph{et al.} demonstrated this approach on IBM Quantum
superconducting hardware for spontaneous emission in a two-level system and for
dissipative transverse-field Ising dynamics, showing accurate relaxation
behavior within the accessible circuit depth~\cite{kamakari2022digital}.
Related QITE-based and variational nonunitary simulation methods have also been
developed for finite-temperature observables, quantum Lanczos energy
estimation, fragmented imaginary-time evolution, adiabatic QITE, and
double-bracket QITE
\cite{sun2021quantum,yeter2020practical,silva2023fragmented,
hejazi2024adiabatic,gluza2026double}. More general approaches to open-system
simulation on universal quantum computers have also been proposed, including
methods based on mixed-unitary sampling of an adjoint density matrix
\cite{liu2025simulationopen}.

The broader algorithmic landscape in nonunitary quantum simulation is
continuing to develop. Recent approaches include adjoint-density-matrix and
mixed-unitary sampling methods, which reduce certain open-system simulations to
stochastic sampling over unitary channels and can avoid auxiliary qubits in
favorable settings~\cite{liu2025simulationopen}; quantum-trajectory-inspired
Lindbladian simulation algorithms, which improve the scaling with the number of
jump operators~\cite{peng2025trajectory}; and quantum-trajectory-based
algorithms that achieve improved query complexity for broad classes of Lindblad
dynamics~\cite{borras2026trajectory}. These developments complement
QITE-based methods by emphasizing different resource tradeoffs, including
ancilla overhead, sampling cost, dependence on the number of Lindblad jump
operators, and access to long-time steady-state behavior. Together, engineered
dissipation in analog platforms and digital algorithms for Lindblad dynamics
establish open quantum systems as a central frontier for quantum simulation.

\subsubsection{Measurement-based quantum simulation and monitored Quantum Circuits}\label{sec5}

\begin{figure*}
    \centering
    \includegraphics[width=0.99\linewidth]{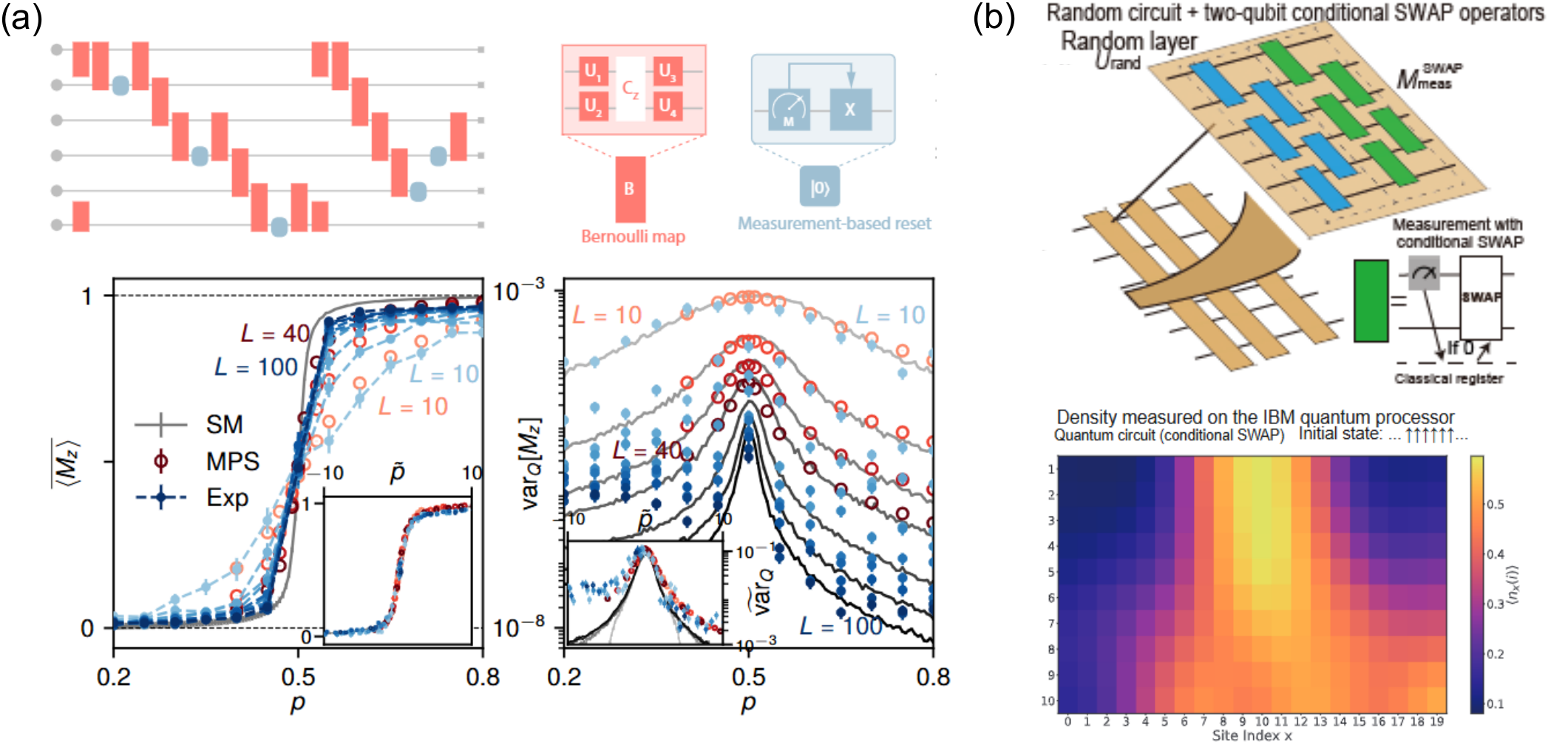}
    \caption{ {Simulations of adaptive quantum circuits on the IBM quantum processor} (a)  Top: schematic of the adaptive Bernoulli circuit implemented ~\cite{pokharel2025order}. Red blocks denote two-qubit entangling “scrambler’’ gates, while blue circles indicate single-qubit measurements followed by measurement-based reset.  Bottom left: Steady-state magnetization versus measurement probability. Experimental data (blue) show a sharp crossover near $p=0.5$, between chaotic and controlled phases. Bottom right: Measured magnetization variance, exhibiting a pronounced peak at the transition point. (b) Simulation of feedback-based adaptive circuit in random scrambling layers with two-qubit conditional-SWAP modules~\cite{shen2026observation}. The SWAP operation is selected in real time according to an intermediate
measurement outcome. The lower panel shows the resulting
density dynamics measured on an IBM quantum processor, illustrating that
measurement-conditioned gates can steer the propagation of local occupation. Panel (a) is adopted from ~\cite{pokharel2025order}.   Panel (b) is adopted from~\cite{shen2026observation}}
    \label{fig:adaptive}
\end{figure*}

Measurement-based quantum simulation provides a distinct paradigm for
engineering and probing nonunitary quantum dynamics, particularly on digital
quantum processors equipped with mid-circuit measurement, reset, and
feed-forward capabilities
\cite{raussendorf2001one,gross2007measurement,briegel2009measurement,
daley2014quantum,muller2012engineered,barreiro2011open,
riste2013deterministic,iqbal2024topological,noel2022measurement,
google2023measurement,koh2023measurement}. In this framework, measurements are
not merely tools for final readout, but dynamical operations that can project,
condition, stabilize, or steer quantum states during the evolution. Repeated
measurements of selected degrees of freedom can generate effective
measurement-induced constraints, quantum-Zeno dynamics, dissipative evolution,
or engineered steady states for the remaining system
\cite{misra1977zeno,itano1990quantum,facchi2008quantum,
diehl2008quantum,kraus2008preparation,verstraete2009quantum,
rossini2020measurement,biella2021many}. More generally, monitored and hybrid
unitary-projective circuits have revealed measurement-induced entanglement
transitions, purification transitions, and dynamical phases that have no direct
counterpart in purely unitary evolution
\cite{li2019measurement,skinner2019measurement,chan2019unitary,
choi2020quantum,potter2022entanglement,fisher2023random}.

\noindent\textbf{Measurement-induced phases on quantum processors.} A central example is the measurement-induced entanglement transition, or more
broadly measurement-induced quantum information phases, in hybrid random
circuits. In these systems, unitary gates generate entanglement while local
projective measurements remove information from the evolving state. Varying the
measurement rate can therefore drive a transition between a weak-measurement
volume-law entangled phase and a strong-measurement area-law or disentangled
phase in the post-measurement ensemble. Such physics has recently been explored
on superconducting and trapped-ion processors: Koh \emph{et al.} directly
observed a measurement-induced entanglement transition on superconducting
qubits with mid-circuit readout, Google Quantum AI and collaborators probed
measurement-induced entanglement and teleportation on a noisy superconducting
processor using a space--time-duality approach, and Noel \emph{et al.} observed
measurement-induced purification and coding phases in a trapped-ion quantum
computer
\cite{koh2023measurement,google2023measurement,noel2022measurement}.

Fig.~\ref{fig:mes} summarizes representative experimental realizations
across platforms. Fig.~\ref{fig:mes}(a) highlights a superconducting-qubit
experiment by Google Quantum AI and collaborators
\cite{google2023measurement}, who leveraged space--time duality to diagnose
measurement-induced quantum information phases on up to 70 superconducting
qubits while avoiding the most demanding mid-circuit-measurement and
post-selection requirements. Another superconducting realization was achieved by Koh, Sun, Motta, and
Minnich on IBM Quantum superconducting devices~\cite{koh2023measurement}, who
implemented hybrid random circuits with explicit mid-circuit readout on up to
14 superconducting qubits and directly observed the measurement-induced
entanglement transition via finite-size scaling [Fig.~\ref{fig:mes}(b)]. By
varying the measurement rate, they observed extensive and sub-extensive scaling
of the measured entanglement entropy in the volume-law and area-law regimes,
respectively, and demonstrated phenomenological critical behavior through data
collapse.

Moreover, the trapped-ion experiment by Noel \emph{et al.}
\cite{noel2022measurement} focused on the closely related purification, or
coding, transition. Instead of diagnosing spatial entanglement within the
system, the protocol tracks how quantum information about an initial state
survives under monitored dynamics. Operationally, this was implemented by
initially entangling the system with a reference qubit and then measuring how
much information about that reference remains protected during the subsequent
hybrid evolution. In the strong-measurement regime, trajectories rapidly purify
conditioned on the measurement outcomes. In the weak-measurement regime, a
finite fraction of the initial information persists and can be interpreted as
being encoded into an emergent quantum-error-correcting codespace generated by
the interplay of scrambling and measurements. As shown in Fig.~\ref{fig:mes}(c), the experimental protocol begins by
entangling a subset of system qubits with a reference qubit. The system then
undergoes a sequence of discrete time steps consisting of a scrambling unitary
layer $U_s$ acting on the system and a measurement layer applied stochastically
in space and time. In the trapped-ion architecture, mid-circuit measurement and
feedback primitives allow the monitored dynamics to be implemented while
maintaining coherent control over the remaining qubits.

The lower panel in Fig.~\ref{fig:mes} (c) shows the experimentally extracted reference diagnostic as a function of system size $L$ for different measurement rates. In the none-measurement regime ($p_x=0$), the system retains substantial correlations with the reference, and the corresponding entropy-like diagnostic remains large and tends to increase with $L$, consistent with a \emph{coding} phase where information is protected by the interplay of scrambling and incomplete monitoring. In contrast, in the strong-measurement regime ($p_x\!\approx\!1$), the reference correlations are strongly suppressed and the diagnostic decreases with system size, indicating a \emph{pure} phase where measurement back-action rapidly projects the system toward a trajectory-conditioned state.

A major conceptual shift in quantum information processing and quantum
simulation has been driven by the emergence of adaptive and dynamic quantum
circuits
\cite{pokharel2025order,li2019measurement,choi2020quantum,
allocca2024statistical,iadecola2025concomitant,decross2023qubit}. Such
architectures support mid-circuit measurements, real-time classical
feed-forward, qubit reset, and measurement-conditioned gate operations. By
processing measurement outcomes during circuit execution, they allow subsequent
operations to be selected conditionally, thereby extending programmable quantum
evolution beyond purely unitary dynamics toward the controlled implementation
of measurement-conditioned quantum channels.

\noindent\textbf{Adaptive and dynamic quantum circuits.} On superconducting processors, adaptive monitored circuits and dynamic-circuit
primitives are now experimentally accessible. For example, recent IBM
experiments have used local mid-circuit measurements, reset, and conditional
feedback to realize adaptive monitored quantum circuits on systems up to
100 qubits~\cite{pokharel2025order}. At the same time, mid-circuit-measurement
errors and measurement-induced crosstalk have become important hardware-level
quantities that can be benchmarked directly on both trapped-ion and
superconducting processors~\cite{hothem2025measuring}. Trapped-ion platforms
likewise provide high-fidelity mid-circuit measurement and feed-forward, and
have demonstrated deterministic nonunitary protocols on programmable hardware;
a representative example is the measurement-and-feed-forward preparation of
topological order on Quantinuum's trapped-ion quantum computer
\cite{iqbal2024topological}.

The recent experiment provides a striking demonstration of the power of adaptive
and dynamic circuits for realizing measurement-induced dynamics on quantum
hardware~\cite{pokharel2025order}. As shown in Fig.~\ref{fig:adaptive}(a),
Pokharel \emph{et al.} implemented an adaptive monitored circuit on an IBM
superconducting quantum processor, composed of alternating scrambling layers of
two-qubit entangling gates, stochastic mid-circuit measurements, resets, and
conditional feedback operations. The circuit realizes a quantum version of the
Bernoulli map, in which scrambling dynamics compete with measurement-and-reset
feedback that steers the system toward a fixed-point-like configuration. The
measured steady-state behavior reveals a transition between two distinct
dynamical regimes: a low-measurement \emph{active} phase, where scrambling
dominates, and a high-measurement \emph{frozen} or controlled phase, where
repeated measurements, resets, and feedback stabilize an ordered configuration.
This experiment opens a route toward exploring adaptive non-equilibrium quantum
phases and measurement-controlled critical dynamics in large-scale hybrid
quantum simulators.

Moreover, Fig.~\ref{fig:adaptive}(b) illustrates a complementary
feedback-directed protocol on IBM superconducting quantum processors
\cite{shen2026observation}, in which intermediate measurement outcomes are used
as real-time control signals for subsequent operations, including conditional
local gates such as SWAP modules. This measurement-conditioned gate selection
generates the feedback-directed many-body evolution shown at the bottom of
Fig.~\ref{fig:adaptive} (b), including robust signatures of feedback-induced
asymmetry and directional information flow. This protocol highlights a
different use of measurements from passive monitoring: the measurement record is
processed during the circuit and becomes an active dynamical input. More
broadly, these examples show that adaptive circuits are not merely a technical
extension of standard gate-based computation, but a qualitatively new framework
in which measurement, reset, and feedback can be used as programmable resources
for engineering many-body quantum dynamics.

\subsection{Phenomena on spin lattices beyond condensed matter }\label{gauge}

While this review primarily focuses on quantum simulations of condensed-matter
systems, it is natural to also include lattice gauge theory (LGT) simulations because they are
deeply connected to condensed-matter physics at both the conceptual and
implementation levels
\cite{kogut1975hamiltonian,wilson1974confinement,zohar2015quantum,
banuls2020simulating}. Gauge structures do not arise only in high-energy
physics: they also appear as emergent descriptions of quantum spin liquids,
frustrated magnets, dimer models, topological order, fractionalized
quasiparticles, and confinement--deconfinement phenomena in strongly correlated
materials
\cite{wen1990topological,wen2004quantum,levin2005string,balents2010spin,
zhou2017quantum,kitaev2003fault,nayak2008non}. For example, the toric code is
a $\mathbb{Z}_2$ lattice gauge theory, and quantum dimer models can be viewed as
constrained gauge theories in which local Gauss-law constraints encode the
allowed dimer configurations
\cite{kitaev2003fault,dennis2002topological,nussinov2009symmetry,
semeghini2021probing,zeng2025quantum,bombieri2026u1}.

From the hardware perspective, LGT simulations also use many of the same tools
as condensed-matter simulations: spin-chain mappings, qubit encodings of matter
and gauge fields, constrained Hilbert spaces, Trotterized real-time evolution,
variational ground-state preparation, and measurements of nonlocal string or
Wilson-loop observables
\cite{zohar2015quantum,banuls2020simulating,martinez2016real,kokail2019self,
hayata2024floquet,meth2025simulating}. Thus, although LGTs are often motivated
by high-energy physics, their mathematical structure, physical phenomena, and
quantum-hardware implementations strongly overlap with condensed-matter quantum
simulation. For this reason, we include recent developments in LGT quantum
simulation as a closely related direction that both draws on and enriches the
broader condensed-matter simulation program
\cite{cochran2025visualizing,cobos2025realtime,xu2026glueball,
joshi2026observation}.

\noindent\textbf{Gauge constraints and the Schwinger-model benchmark.}  LGTs were introduced by K. G. Wilson as a nonperturbative formulation of
quantum field theories, providing discretized spacetime or Hamiltonian lattice
frameworks that capture strongly interacting phenomena such as confinement and
the phase structure of quantum chromodynamics (QCD)
\cite{wilson1974confinement,creutz1983quarks,rothe2012lattice}. Despite their
success, classical simulations of LGTs face severe computational barriers,
including the fermion sign problem in Monte Carlo sampling at finite density or
real time, the high cost of approaching continuum and large-volume limits, and,
in Hamiltonian formulations, the rapid growth of the many-body Hilbert space
with system size
\cite{chandrasekharan1999meron,de2010simulating}. Quantum simulation offers a
promising route to address some of these constraints by encoding gauge and
matter degrees of freedom directly into controllable quantum platforms and by
enforcing gauge invariance at the level of the dynamics, either exactly through
symmetry-preserving encodings or effectively through energetic penalties and
constraint-preserving circuits
\cite{zohar2015quantum,rico2014tensor,banuls2020simulating,kokail2019self}.

A LGT couples matter fields on lattice sites to gauge fields on the links
connecting them. A standard example is the $(1\!+\!1)$-dimensional
$\mathrm{U}(1)$ Schwinger model, which in the Kogut--Susskind Hamiltonian
formulation with staggered fermions reads
\cite{schwinger1962gauge,kogut1975hamiltonian,susskind1977lattice}
\begin{equation}
\begin{aligned}
\hat{H}_{\rm Sch}
&=
-iw
\sum_{n=1}^{N-1}
\left[
\hat{\Phi}_n^\dagger
e^{i\hat{\theta}_n}
\hat{\Phi}_{n+1}
-
\mathrm{h.c.}
\right]
+
J\sum_{n=1}^{N-1}\hat{L}_n^2
\\
&\quad
+
m\sum_{n=1}^{N}
(-1)^n
\hat{\Phi}_n^\dagger\hat{\Phi}_n .
\end{aligned}
\label{eq:schwinger_H}
\end{equation}
Here, $\hat{\Phi}_n$ and $\hat{\Phi}_n^\dagger$ annihilate and create a
staggered fermion on site $n$, with number operator
$\hat n_n=\hat{\Phi}_n^\dagger\hat{\Phi}_n$. The operator
$e^{i\hat{\theta}_n}$ is the compact $\mathrm{U}(1)$ gauge-link operator on
the bond $(n,n+1)$, while $\hat L_n$ is the corresponding electric-field
operator. The coefficient $w$ sets the matter--gauge hopping strength, $J$ is
the electric-field energy scale, and $m$ is the staggered fermion mass.
Equivalently, in standard lattice-gauge-theory notation one may identify
$U_{n,n+1}=e^{i\hat{\theta}_n}$ and $E_{n,n+1}=\hat L_n$. The link variables obey
\begin{equation}
[\hat L_n,e^{i\hat{\theta}_m}]
=
\delta_{nm}e^{i\hat{\theta}_m},
\end{equation}
so that $\hat L_n$ is the generator conjugate to the compact link phase, while
$e^{i\hat{\theta}_n}$ acts as the parallel transporter that preserves local
gauge covariance of the matter hopping term. Gauge invariance imposes Gauss's
law on physical states. For an open chain, the local gauge generator can be
written as
\begin{equation}
\hat G_n
=
\hat L_n-\hat L_{n-1}-\hat\rho_n,
\qquad
\hat\rho_n
=
\hat{\Phi}_n^\dagger\hat{\Phi}_n
-
\frac{1-(-1)^n}{2},
\end{equation}
with the boundary electric field fixed by the chosen background sector.
Physical states satisfy
\begin{equation}
\hat G_n\ket{\psi}_{\rm phys}=0
\end{equation}
for all sites $n$, up to possible background-charge sectors.

A milestone in the quantum simulation of gauge theories was achieved with the
trapped-ion realization of this $\mathrm{U}(1)$ Schwinger model
\cite{martinez2016real}. This experiment demonstrated that nontrivial
real-time phenomena in lattice gauge models can be accessed on a programmable
quantum device at the few-qubit scale available at the time. A key step was to
use Gauss's law to eliminate the dynamical gauge links in
Eq.~\eqref{eq:schwinger_H}, thereby mapping the gauge theory to a spin
Hamiltonian with structured long-range interactions,
\begin{equation}
\begin{aligned}
\hat{H}_{S}
&=
\frac{m}{2}
\sum_{n=1}^{N}
(-1)^n Z_n
+
w
\sum_{n=1}^{N-1}
\left(
\sigma_n^{+}\sigma_{n+1}^{-}
+
\mathrm{h.c.}
\right)
\\
&\quad
+
J
\sum_{n=1}^{N-1}
\left[
\epsilon_0
+
\frac{1}{2}
\sum_{m=1}^{n}
\left(
Z_m+(-1)^m
\right)
\right]^2 .
\end{aligned}
\label{eq:H_spin}
\end{equation}
Here, $Z_n$ and $\sigma_n^\pm=(X_n\pm iY_n)/2$ act on the spin degree of
freedom obtained after the staggered-fermion-to-spin mapping, and $\epsilon_0$
specifies the background electric-field sector. Because Gauss's law has been
used to integrate out the gauge links analytically, this spin-only
representation acts directly within the gauge-invariant physical sector. It is
therefore efficiently implementable through digital gate sequences on the
trapped-ion architecture while preserving the gauge-invariant dynamics of the
original Schwinger model.

For gate-based simulation, a key technical ingredient is an efficient
gate-decomposition protocol for realizing the structured long-range spin
interactions generated by the electric-field term in Eq.~\eqref{eq:H_spin}.
Expanding the squared Gauss-law term produces nonlocal spin--spin couplings
whose strengths depend on the positions of the two spins. These couplings are
long-ranged and spatially inhomogeneous, and their efficient implementation is
essential for reproducing the gauge-invariant Schwinger dynamics on an ion-chain
quantum processor. In the trapped-ion architecture of
Ref.~\cite{martinez2016real}, collective entangling operations make it possible
to implement this structured interaction efficiently, avoiding a naive
decomposition into all individual pairwise couplings.
The central result, summarized in Fig.~\ref{fig:lgt}(a), is the observation of
Schwinger pair-production dynamics. Starting from the bare vacuum, the measured
particle-number density exhibits the characteristic rise and oscillations
associated with coherent particle--antiparticle creation, while increasing the
fermion mass suppresses pair production by raising its energetic cost
\cite{martinez2016real}. Closely related trapped-ion and gate-based experiments
have since extended quantum-simulation studies of Schwinger-model and
lattice-gauge-theory dynamics, including variational gauge-invariant
eigensolvers, gauge-protection protocols, and larger circuit-based
implementations on digital quantum devices
\cite{kokail2019self,halimeh2024spin}.

\noindent\textbf{Digital and analog routes to confinement dynamics.} A recent milestone in the experimental study of LGTs is the digital quantum
simulation of confinement dynamics on a superconducting-qubit processor
\cite{mildenberger2025confinement}. Using up to 21 qubits, the authors realized
a nonperturbative, fully gauge-invariant decomposition of a
$(1\!+\!1)$-dimensional $\mathbb{Z}_{2}$ LGT, enabling real-time access to
confinement physics within a native gate-model framework. The simulated
Hamiltonian can be written as
\begin{equation}
\begin{aligned}
H_{\mathbb{Z}_2}
&=
-J
\sum_i
\left(
\sigma_i^{+}\tau_{i,i+1}^{z}\sigma_{i+1}^{-}
+
\mathrm{h.c.}
\right)
-
\\&f
\sum_i
\tau_{i,i+1}^{x}
+
\frac{\mu}{2}
\sum_i
(-1)^i Z_i ,
\end{aligned}
\label{eq:HZ2}
\end{equation}
where $\sigma$ acts on matter degrees of freedom on sites and $\tau$ acts on
gauge fields on links. The first term describes gauge-invariant matter hopping
dressed by the link field $\tau^z$, the second term provides gauge-field
dynamics through a transverse field on the links, and the last term is a
staggered mass for the matter sector. Gauge invariance is enforced by the local
$\mathbb{Z}_2$ Gauss-law generator
\begin{equation}
G_i^{\mathbb{Z}_2}
=
Z_i
\tau_{i-1,i}^{x}
\tau_{i,i+1}^{x},
\qquad
G_i^{\mathbb{Z}_2}\ket{\mathrm{phys}}
=
\ket{\mathrm{phys}},
\label{eq:gauss_Z2}
\end{equation}
up to possible convention-dependent background signs. This generator commutes
with the Hamiltonian, $[H_{\mathbb{Z}_2},G_i^{\mathbb{Z}_2}]=0$, and constrains
the dynamics to the physical gauge sector. This fully gauge-invariant
implementation is conceptually distinct from gauge-eliminated
$\mathrm{U}(1)$ Schwinger-model mappings, where gauge links are removed using
Gauss's law and the resulting spin model contains structured long-range
interactions. From an implementation standpoint, the $\mathbb{Z}_2$ theory is
well matched to existing quantum hardware because the gauge field is intrinsically two-level
and the Hamiltonian decomposes into few-qubit interactions.

The corresponding experiment combines gauge-invariant circuit design, deep Trotterization, and post-selected error mitigation to probe confinement physics. Efficient gate decompositions enable deeply layered Trotter circuits, reaching more than 200 entangling gates and 25 digital time steps. Starting from a configuration with a localized matter excitation in a polarized background gauge field, the measured dynamics reveal a clear transition between two regimes. As shown in Fig.~\ref{fig:lgt}(b), for weak electric fields the charge delocalizes across the chain, accompanied by correlated flips of adjacent gauge fields. For strong fields, by contrast, the excitation remains confined: its motion is suppressed, and the electric flux string remains rigid~\cite{mildenberger2025confinement}.

Cold-atom quantum simulators also have made
substantial progress toward native gauge constraints and Floquet-engineered
gauge dynamics. Floquet protocols have been used to implement minimal
$\mathbb{Z}_2$ lattice-gauge-theory building blocks and to probe
gauge-invariant dynamics in driven optical-lattice settings
\cite{schweizer2019floquet}. Complementary experiments have realized key
microscopic ingredients of matter--gauge coupling, such as density-dependent
Peierls phases and density-assisted tunneling, thereby providing a route toward
dynamical gauge fields rather than purely gauge-eliminated mappings
\cite{gorg2019realization}. A further milestone is the scalable realization of
local $U(1)$ gauge invariance in ultracold-atom mixtures, where gauge
constraints are engineered at the level of the effective Hamiltonian and
verified experimentally~\cite{mil2020scalable}. Together, these results
establish a systematic experimental pathway from static constraints and
effective gauge invariance toward dynamical gauge fields and more complex
lattice-gauge-theory architectures. Cold-atom experiments also incorporate additional
nonperturbative field-theory ingredients, such as a tunable topological
$\theta$ angle, within controlled quantum-simulation architectures. Zhang
\emph{et al.} reported microscopic confinement dynamics controlled by a
topological $\theta$ angle in a Bose--Hubbard gauge-theory quantum simulator
\cite{zhang2025observation}. In this experiment, a tilted superlattice
potential induces an effective background electric field, enabling the
realization of a tunable $\theta$ angle and the observation of
confinement--deconfinement dynamics in a $(1\!+\!1)$D quantum-electrodynamics
setting.

\noindent\textbf{Higher-dimensional gauge dynamics across platforms.} Progress in quantum hardware has also enabled studies of gauge theories in
higher spatial dimensions, where new nonperturbative phenomena emerge.
Gonz\'alez-Cuadra \emph{et al.} demonstrated the observation of string breaking
in a $(2\!+\!1)$-dimensional lattice gauge model implemented on a programmable
Rydberg-atom quantum simulator~\cite{gonzalez2025observation}. In this
platform, both Gauss's law and a linearly confining potential arise natively
from the interplay between Rydberg blockade constraints and long-range van der
Waals interactions. Gauge-invariant configurations are mapped onto allowed
blockade patterns on a kagome geometry, enabling the preparation of confined
flux-tube states using a quasi-adiabatic sweep of the global detuning in a
system of 59 atoms. The authors characterize equilibrium string states, as
shown in Fig.~\ref{fig:lgt}(c), and probe real-time string breaking by locally
shifting the detuning on selected sites to resonantly enhance transitions
between unbroken and broken strings.

Moreover, superconducting hardware has enabled direct imaging of
charge--string dynamics in a $(2\!+\!1)$D $\mathbb{Z}_2$ gauge theory using a
two-dimensional Google superconducting-qubit array
\cite{cochran2025visualizing}. In this experiment, low-energy states of the
gauge theory are prepared variationally, local charge excitations are created
by targeted gates, and their dynamics are tracked under Trotterized
gauge-invariant evolution. The experiment resolves two regimes inside the
confining phase: for weak confinement, the string fluctuates strongly in the
transverse direction, whereas for strong confinement, transverse fluctuations
are effectively frozen. It also identifies a resonance condition at which
dynamical string breaking is enhanced. This result is important because it
moves superconducting-qubit LGT simulation from one-dimensional charge
confinement to directly imaged two-dimensional gauge-field dynamics.

Another significant advance is the demonstration of two-dimensional lattice QED on a trapped-ion qudit quantum processor, reported by Meth \emph{et al.}~\cite{meth2025simulating}. This work directly addresses two major challenges in digital LGT simulation: implementing gauge theories in higher dimensions, where gauge bosons possess nontrivial dynamics, and efficiently encoding gauge fields whose Hilbert spaces naturally require more than two levels. By using trapped-ion qudits, each ion encodes a $d$-level truncated gauge field, enabling a compact and local representation of electric-field degrees of freedom while substantially reducing the circuit depth relative to qubit-only encodings. This hardware-efficient mapping supports the digital simulation of a full two-dimensional plaquette, the minimal building block of magnetic-flux dynamics absent in one-dimensional models. The ground state of the plaquette Hamiltonian is prepared using a variational quantum eigensolver, and the authors systematically improve the gauge-field truncation by increasing the qudit dimension from $d=3$ to $d=5$. Here, a qutrit encoding with $d=3$ retained electric-field states to a ququint encoding with $d=5$ retained electric-field states. The corresponding energy levels and qubit--qutrit variational circuit are shown in Fig.~\ref{fig:lgt} (d). This work establishes qudits as an invaluable resource for simulating higher-dimensional gauge theories on universal quantum processors.

\noindent\textbf{Real-time scattering and thermalization.} Beyond confinement and string breaking, recent experiments and algorithms have
begun to access broader real-time phenomena in lattice gauge theories. Farrell
\emph{et al.} simulated hadron wave-packet dynamics in the Schwinger model
using 112 qubits on IBM's 133-qubit Heron processor \texttt{ibm\_torino},
preparing localized hadronic excitations and evolving them with Trotterized
real-time circuits~\cite{farrell2024quantum}. Related work by Davoudi
\emph{et al.} developed and experimentally benchmarked efficient preparation
circuits for scattering wave packets of hadrons in gauge theories, including a
trapped-ion implementation on Quantinuum H1-1, thereby addressing a key
state-preparation bottleneck for future quantum simulations of scattering
processes~\cite{davoudi2024scattering}.

Other recent directions concern thermalization, localization, and constrained
dynamics in gauge theories. Mueller \emph{et al.} demonstrated quantum
computation of universal thermalization dynamics in a $(2\!+\!1)$D
$\mathbb{Z}_2$ lattice gauge theory using a fully connected trapped-ion digital
quantum computer, thereby probing relaxation beyond simple few-body gauge
models~\cite{mueller2025thermalization}. Complementarily, Datla \emph{et al.}
observed statistical localization in a Rydberg simulator of a $U(1)$ lattice
gauge theory, where strong Hilbert-space fragmentation prevents typical charge
configurations from locally thermalizing even though the conserved quantities
are encoded in nonlocal string-like operators~\cite{datla2026statistical}.
These results connect LGT simulation to the broader non-equilibrium themes of
thermalization, fragmentation, and constrained many-body dynamics.

\noindent\textbf{Large-scale and non-Abelian lattice gauge simulation.} The latest large-scale experiments further push LGT simulation along two
complementary frontiers: higher-dimensional real-time gauge dynamics and
non-Abelian gauge theories. Cobos \emph{et al.} reported real-time dynamics in
a $(2\!+\!1)$D gauge theory on IBM Heron heavy-hex superconducting processors
using up to 144 qubits, directly resolving longitudinal and transverse string
motion, string fragmentation, and recombination in a $\mathbb{Z}_2$-Higgs model
\cite{cobos2025realtime} [see Fig.~\ref{fig:stringdynamics}(a)]. Xu
\emph{et al.} observed glueball-like closed-loop excitations and string
breaking in a $(2\!+\!1)$D $\mathbb{Z}_2$ LGT on the Quantinuum System Model H2
trapped-ion quantum computer, using a tunable plaquette term to access genuine
two-dimensional gauge dynamics~\cite{xu2026glueball}. Joshi \emph{et al.}
further implemented a $U(1)$ quantum-link model with a tunable plaquette term
on the Quantinuum System Model H2 trapped-ion quantum computer, reaching
large-scale simulations of genuine $(2\!+\!1)$D string dynamics and observing
string breaking accompanied by matter-pair production across the lattice plane
\cite{joshi2026observation}. 

On the analog side, triangular Rydberg arrays have also been proposed as a
route to a $(2\!+\!1)$D $U(1)$ LGT in which plaquette terms arise at first
order, enabling studies of string roughening, Lüscher corrections, and
string-breaking dynamics near a deconfined critical point
\cite{bombieri2026u1} [see Fig.~\ref{fig:stringdynamics}(b)]. Complementing
these higher-dimensional Abelian gauge-theory experiments and proposals,
Il{\v c}i{\'c} \emph{et al.} reported a large-scale simulation of coherent
non-Abelian hadron dynamics in a $(1\!+\!1)$D $\mathrm{SU}(2)$ LGT on a
156-qubit IBM Heron superconducting processor, using a loop-string-hadron
encoding to preserve gauge invariance efficiently~\cite{ilcic2026observation}.
These works show that the frontier of LGT quantum simulation is rapidly
expanding from Abelian one-dimensional benchmarks toward many-body scattering,
thermalization, glueball dynamics, higher-dimensional string dynamics, and
non-Abelian gauge theories.

\noindent\textbf{False-vacuum decay.} We also highlight false-vacuum decay as a closely related real-time field-theory problem, in which a metastable state decays into a lower-energy vacuum through the nucleation, growth, and interaction of true-vacuum bubbles~\cite{coleman1977fate,callan1977fate}. 
In the semiclassical treatment, the decay rate is commonly obtained by Wick rotating the real-time path integral, $t\rightarrow -i\tau$, and identifying a Euclidean bounce solution that controls the leading tunneling exponent~\cite{coleman1977fate,callan1977fate}. 
This use of imaginary time is conceptually related to quantum imaginary-time evolution~\cite{mcardle2019variational,motta2020determining}. 
However, the two settings should not be conflated: Wick rotation is an analytic tool. 
The physical false-vacuum decay process itself is a real-time non-equilibrium process involving bubble nucleation, expansion, and interaction, which quantum simulators can probe directly.

Recent experiments have begun to probe this physics in cold-atom and superconducting quantum platforms. 
Zhu \emph{et al.} investigated false-vacuum decay in a cold-atom gauge-theory quantum simulator of a $(1\!+\!1)$D $U(1)$ quantum link model with a tunable background electric field~\cite{zhu2024probing}. 
By controlling the background field, the experiment accesses a large pair-production-rate regime and directly probes decay from an infinite-mass false-vacuum through Schwinger-type pair creation. 
In a different but closely related analog setting, Vodeb \emph{et al.} simulated false-vacuum decay on a 5564-qubit superconducting quantum annealer by realizing a ferromagnetic Ising chain in transverse and longitudinal fields~\cite{vodeb2025stirring}.  
These experiments show that quantum simulators can access not only static confinement and string-breaking phenomena, but also the real-time decay of metastable vacua, thereby connecting lattice-gauge quantum simulation to broader questions in non-equilibrium quantum field theory and cosmology-inspired dynamics.

\begin{figure*}
    \centering
    \includegraphics[width=0.95\linewidth]{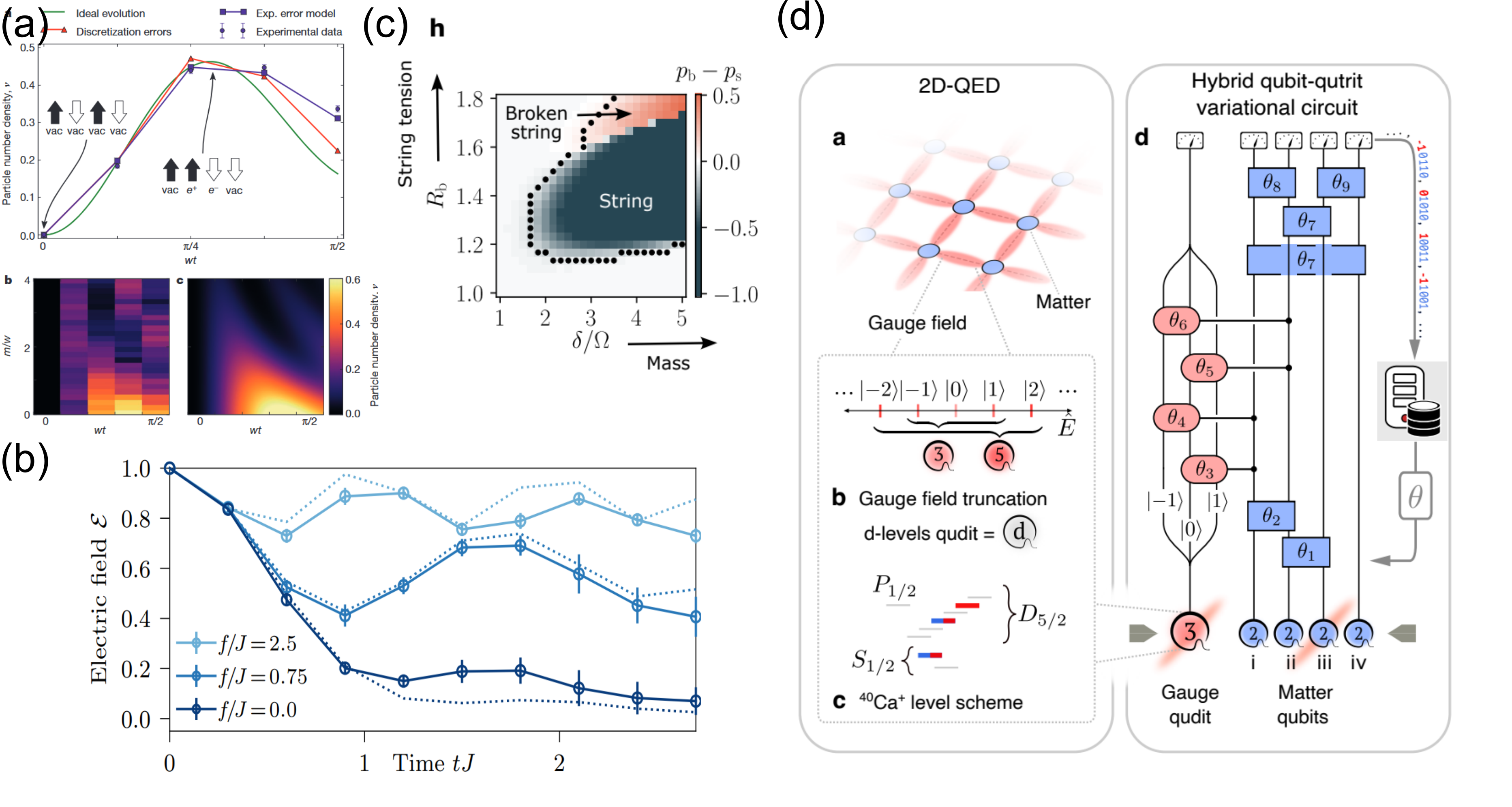}
    \caption{Experimental milestones in digital and analog quantum simulation of lattice gauge theories. (a) Real-time particle production in the Schwinger model. Top: Particle-number density dynamics measured after a global quench. The oscillatory evolution reflects coherent vacuum decay and Schwinger pair production. Bottom: Density plots of particle-number evolution versus fermion mass, highlighting that increasing mass suppresses real-time particle creation~\cite{martinez2016real}.  (b) Confinement dynamics in a superconducting-qubit simulator. Time evolution of the electric-field order parameter for increasing background field. Strong fields stabilize a linear flux string connecting charges, preventing charge delocalization~\cite{mildenberger2025confinement}.  (c) Observation of string breaking in a (2+1)D gauge theory with Rydberg atoms.  Phase diagram in the plane of string tension and bare mass. The system exhibits two regimes: a stable “string” phase and a “broken string” region where pair creation leads to screening~\cite{gonzalez2025observation}.  (d) Two-dimensional lattice QED on a trapped-ion qudit processor. Left: Plaquette geometry for 2D-QED. Middle: Energy-level structure of qudits. Right: Hybrid qubit–qutrit variational circuit used for VQE preparation of the 2D plaquette ground state~\cite{meth2025simulating}.  Panel (a-d) are adopted from ~\cite{martinez2016real},~\cite{mildenberger2025confinement},~\cite{gonzalez2025observation} and \cite{meth2025simulating} respectively.
}
    \label{fig:lgt}
\end{figure*}

\begin{figure*}
    \centering
    \includegraphics[width=0.85\linewidth]{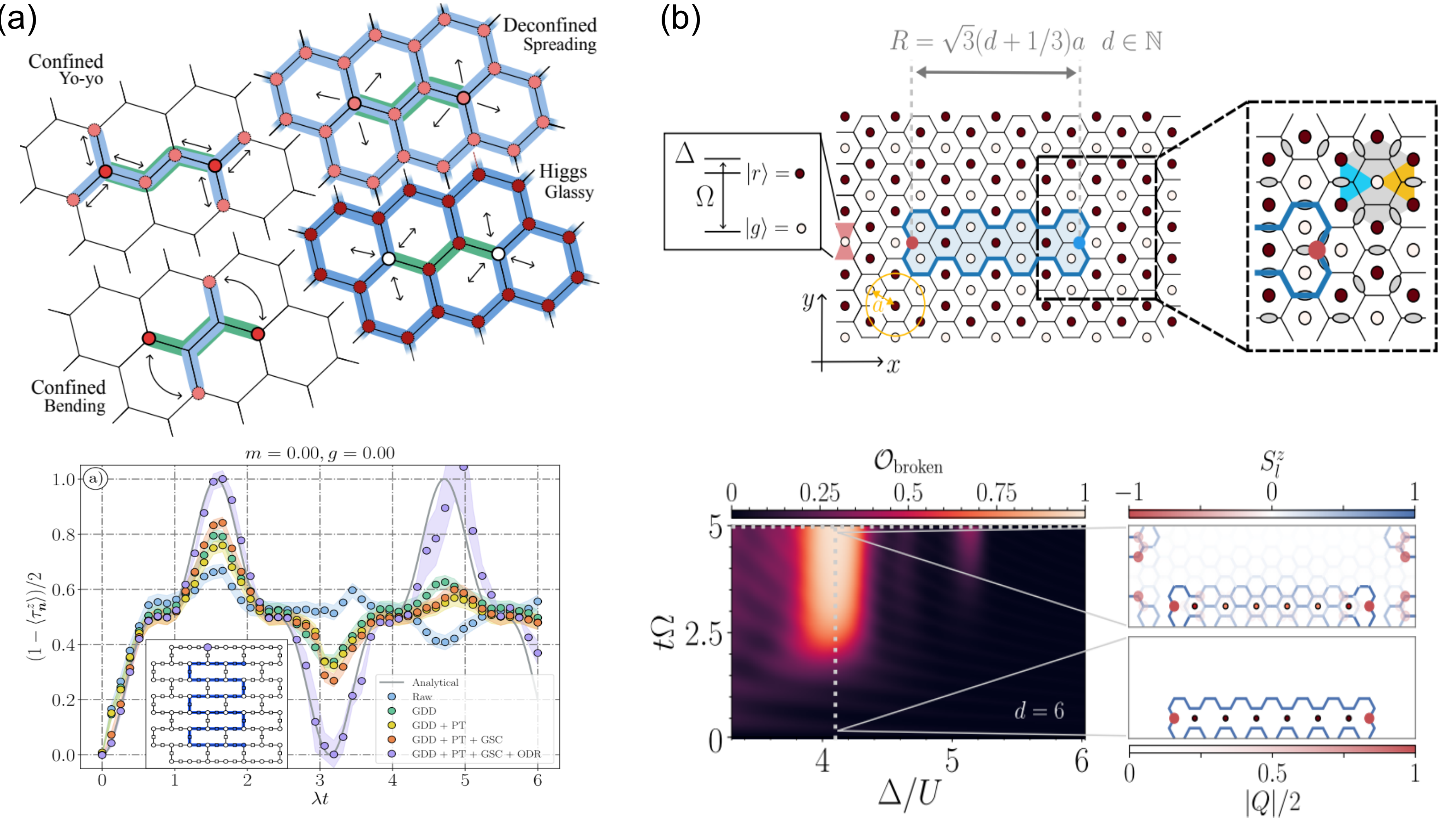}
    \caption{Real-time string dynamics in higher-dimensional lattice gauge theories. (a) Digital quantum simulation of real-time string dynamics in a $(2+1)$D $\mathbb{Z}_2$-Higgs lattice gauge theory on a superconducting quantum processor \cite{cobos2025realtime}. Matter and gauge degrees of freedom are mapped directly onto vertex and link qubits of a heavy-hex superconducting architecture. The measured gauge-field dynamics, quantified through the link excitation density $(1-\langle \tau^x\rangle)/2$, reveal longitudinal oscillations and transverse string fluctuations, providing a direct probe of the string-like nature of confinement in real time. (b) Analog route to a $(2+1)$D $U(1)$ lattice gauge theory using a triangular Rydberg array~\cite{bombieri2026u1}. 
The blockade-constrained Rydberg Hilbert space is mapped to gauge-invariant configurations of a compact $U(1)$ quantum link model. 
The lower color plot displays the broken-string order parameter $\mathcal{O}_{\rm broken}$ after a quench as a function of detuning $\Delta/U$ and evolution time $t\Omega$. Panel (a) is adopted from~\cite{cobos2025realtime}. Panel (b) is adopted from~\cite{bombieri2026u1}. }
    \label{fig:stringdynamics}
\end{figure*}

\section{Methodology } 
\label{sec:methodology}
{The preceding section surveyed the extensive body of work applying quantum simulation techniques to a plethora of condensed matter systems. We now turn to a pedagogical overview of the standard techniques used in quantum simulation. Having already reviewed specific physical realizations or applications, the goal of this section is to introduce the core algorithmic and conceptual tools that form the foundation of quantum simulation.}

\subsection{Real-time dynamical evolution}
\subsubsection{Trotter–Suzuki product formula for Hamiltonian simulation}\label{trotter}

A central task in quantum simulation is to implement the real-time evolution
generated by a target Hamiltonian. Product-formula methods, also known as
Trotter--Suzuki product formulas, provide one of the most direct and historically
foundational approaches to this problem~\cite{suzuki1992general,lloyd1996universal}.
In gate-based quantum simulation, these methods approximate the continuous
time-evolution operator by decomposing it into a sequence of elementary quantum
gates generated by local or few-body Hamiltonian terms.

For a target Hamiltonian $H$, the
exact time evolution operator is
\begin{equation}
U(t)=e^{-iHt}.
\end{equation}
Even if $H$ consists of only local couplings or interactions, its exponentiation $U(t)$ will in principle correspond to a large number of non-local couplings or gates. The key idea behind the Totter--Suzuki formula is to approximate $U(t)$ in terms of a product of many manageable time-step operators. 

Concretely, we first divide the total evolution time into $n$ intervals of duration
$\delta t=t/n$, as exactly given by
\begin{equation}
U(t)=\left(e^{-iH\delta t}\right)^n .
\end{equation}
The challenge is not the time discretization itself, but the
implementation of the short-time propagator $e^{-iH\delta t}$ on a quantum
device.

For a local many-body Hamiltonian, one typically writes
\begin{equation}
H=\sum_j H_j,
\end{equation}
where each term $H_j$ acts nontrivially only on a small subset of degrees of
freedom, such as neighboring spins, local fermionic modes, or plaquette
operators. If all terms commute, the short-time propagator would factorize
exactly as
\begin{equation}
e^{-iH\delta t}
=
\prod_j e^{-iH_j\delta t}.
\end{equation}
In interacting many-body systems, however, the local terms generally do not
commute, $[H_j,H_k]\neq0$, so this factorization becomes approximate. The
first-order product formula replaces each short-time propagator by
\begin{equation}
e^{-iH\delta t}
=
e^{-i\sum_j H_j\delta t}
\approx
\prod_j e^{-iH_j\delta t},
\end{equation}
leading to the approximate unitary
\begin{equation}
U(t)
\approx
\left(
\prod_j e^{-iH_j\delta t}
\right)^n .
\end{equation}
The approximation error arises from the noncommutativity of the terms in
$H$. For two terms, $H=H_A+H_B$, the Baker--Campbell--Hausdorff expansion gives
\begin{equation}
\begin{aligned}
 &e^{-iH_A\delta t}e^{-iH_B\delta t}
\\
&=\exp\!\left[
-i(H_A+H_B)\delta t
-\frac{\delta t^2}{2}[H_A,H_B]
+
\mathcal{O}(\delta t^3)
\right].   
\end{aligned}
\end{equation}
Thus, the local error of a first-order Trotter step scales as
$\mathcal{O}(\delta t^2)$, and the accumulated global error over
$n=t/\delta t$ steps scales as $\mathcal{O}(t\delta t)$, up to constants set by
commutator norms. This first-order construction is simple and broadly
applicable, but its accuracy improves only linearly with the number of time steps.

A standard way to reduce the leading commutator error is to use a symmetric
second-order, or Strang, product formula. For $H=H_A+H_B$, one such Trotter step is
\begin{equation}
U_2(\delta t)
=
e^{-iH_A\delta t/2}
e^{-iH_B\delta t}
e^{-iH_A\delta t/2}.
\end{equation}
Because this sequence is time-reversal symmetric, the leading
$\mathcal{O}(\delta t^2)$ error cancels, and the effective evolution takes the
form
\begin{equation}
U_2(\delta t)
=
\exp\!\left[
-i(H_A+H_B)\delta t
+
\mathcal{O}(\delta t^3)
\right].
\end{equation}
Consequently, the local error scales as $\mathcal{O}(\delta t^3)$ and the
global error scales as $\mathcal{O}(t\delta t^2)$. Suzuki further showed that
higher even-order product formulas can be constructed recursively by composing
symmetric lower-order formulas with carefully chosen time steps ~\cite{childs2012hamiltonian}. These
higher-order formulas systematically suppress nested-commutator errors, but
they require longer sequences of local exponentials.

Product-formula Hamiltonian simulation therefore reduces many-body dynamics to
a sequence of experimentally implementable local gates or short-time
interaction blocks. Its practical performance is governed by a tradeoff:
smaller time steps and higher-order formulas reduce Trotter error, but they
also increase circuit depth, exposure to gate noise, and sampling cost. This
tradeoff is central to the use of product-formula simulation on near-term
quantum hardware.

\subsubsection{Floquet dynamics}

In programmable quantum simulators, implementing inequivalent Trotter steps at each time step is no more difficult than implementing constant Trotter steps. This makes them particularly amenable for simulating time-dependent Hamiltonian evolution. A key arena is that of time-periodic Hamiltonians, where the effective ``Floquet" stroboscopic evolution operator across one entire period can possess features i.e. extensive non-locality not easily obtained from static evolution operators ~\cite{goldman2014periodically,bukov2015universal,eckardt2017colloquium,li2018realistic,rudner2013anomalous}.

This strategy, known as Floquet engineering, provides a versatile route
to non-equilibrium quantum matter beyond the original setting of periodically
driven optical lattices. Representative applications include synthetic gauge
fields in cold atoms [see Sect.~\ref{optical}],
Floquet prethermalization [see Fig.~\ref{fig:pre}], discrete time-crystalline
order [see Sect.~\ref{time}], and Floquet-engineered lattice-gauge dynamics
[see Sect.~\ref{gauge}]
~\cite{kuwahara2016floquet,mori2016rigorous,abanin2017effective,khemani2016phase,else2017prethermal,potter2016classification,zhang2022digital,zaletel2023colloquium}.

Suppose the Hamiltonian of interest is periodic in time, $H(t+T)=H(t)$. The evolution over one driving
period is described by the Floquet operator
\begin{equation}
U_F
=
\mathcal{T}
\exp\!\left[
-i\int_0^T H(t)\,dt
\right],
\label{eq:floquet_operator}
\end{equation}
where $\mathcal{T}$ denotes time ordering. Stroboscopically, the same one-period
evolution can be written in terms of an effective Floquet Hamiltonian $H_F$ as
\begin{equation}
U_F=e^{-iT H_F}.
\label{eq:floquet_hamiltonian}
\end{equation}
Thus, although the time-dependent
Hamiltonian $H(t)$ may vary strongly within each
period, the state observed after an integer number of periods only depends explicitly on $H_F$, i.e. varies stroboscopically as
\begin{equation}
\ket{\psi(nT)}
=
U_F^n\ket{\psi(0)}
=
e^{-i nT H_F}\ket{\psi(0)} .
\end{equation}
In the high-frequency regime, the effective Hamiltonian $H_F$ can furthermore be obtained
perturbatively via the Magnus expansion~\cite{magnus1954exponential}. The
first terms are
\begin{equation}
\begin{aligned}
& H_F
=
\frac{1}{T}
\int_0^T H(t_1)\,dt_1\\
&+
\frac{1}{2iT}
\int_0^T dt_1
\int_0^{t_1} dt_2\,
[H(t_1),H(t_2)]
+\cdots .
\end{aligned}
\label{eq:magnus_expansion}
\end{equation}
The leading part is just the time-averaged Hamiltonian. But the subleading part contains the commutator of the Hamiltonian at different times, which can contain terms that do not physically exist in the original time-dependent Hamiltonian. As a minimal illustration, for $H(t)$ containing only $\sigma_x,\sigma_y$ spin operators, $H_F$ would also contain $\sigma_z$ since $[\sigma_x,\sigma_y]\propto \sigma_z$.  
This mechanism
underlies many Floquet-engineering protocols:
periodic modulation can be used not only to approximate a target Hamiltonian, but
also to generate new effective interactions, Peierls phases, synthetic gauge
fields, spin-orbit couplings, and topological band structures that are difficult
to realize directly in static hardware implementations
~\cite{goldman2014periodically,bukov2015universal,eckardt2017colloquium}.

Beyond possibly replicating otherwise obscure static phenomena, the Floquet unitary $U_F$ defines Floquet quasienergy bands that are periodic in frequency space modulo $2\pi/T$, as seen from its eigenvalue equation
\begin{equation}
U_F\ket{\phi_\alpha}
=
e^{-i\varepsilon_\alpha T}\ket{\phi_\alpha}.
\end{equation}
These quasienergies define the evolution of a generic initial state at  stroboscopic intervals viz.
\begin{equation}
\ket{\psi(nT)}
=
\sum_\alpha
c_\alpha
e^{-in\varepsilon_\alpha T}
\ket{\phi_\alpha}.
\label{eq:floquet_eigen_expansion}
\end{equation}
Because of this spectral periodicity, ``Floquet anomalous" band gaps occur between periodic copies of the same band, in addition to the existing band gaps in $H(t)$. These Floquet anomalous quasienergy gaps can harbor new classes of protected in-gap topological modes~\cite{rudner2013anomalous,QianScience2025}, which are protected by new symmetry-enriched Floquet topological invariants~\cite{potter2016classification,zhang2022digital}.
Beyond robust Floquet topology, discrete time crystals provide another paradigmatic example: the Floquet
operator leads to emergent discrete time-translation symmetry of a different periodicity
~\cite{else2016floquet,khemani2016phase,else2017prethermal,zaletel2023colloquium}
[see Sect.~\ref{time}].

Importantly, trotterized or Floquet-style circuits with modular gate layers can be executed at scales where classical verification is difficult. A compelling demonstration of how digital quantum processors can support large-scale quantum simulation was provided in the superconducting-qubit experiment \cite{kim2023evidence}. In this work, a 127-qubit heavy-hex superconducting device was used to implement deep, strongly entangling circuits (see Fig.~\ref{fig:floquet1} (a)). This work showed that such layer-structured circuits, when optimized for hardware connectivity and calibrated through extensive noise characterization, remain stable even at large system sizes, as evidenced by the close agreement between hardware results and tensor-network benchmarks  [Fig.~\ref{fig:floquet1} (b)]. Generally, benchmarking simulations at this scale requires highly specialized classical algorithms capable of exploiting the structure of the quantum circuit. As illustrated in Figs.~\ref{fig:floquet1}(c) and (d), tensor-network ansatze must be carefully engineered to respect the hardware structure and entanglement growth under Floquet layers~\cite{tindall2024efficient}. Even with advanced methods such as isoTNS or large-bond-dimension MPS, classical resources rapidly become the limiting factor. As digital processors grow in scale, Floquet engineering thus emerges as a robust strategy for realizing non-equilibrium phases and driven quantum matter at classically inaccessible scales.

\begin{figure*}
    \centering
    \includegraphics[width=0.9\linewidth]{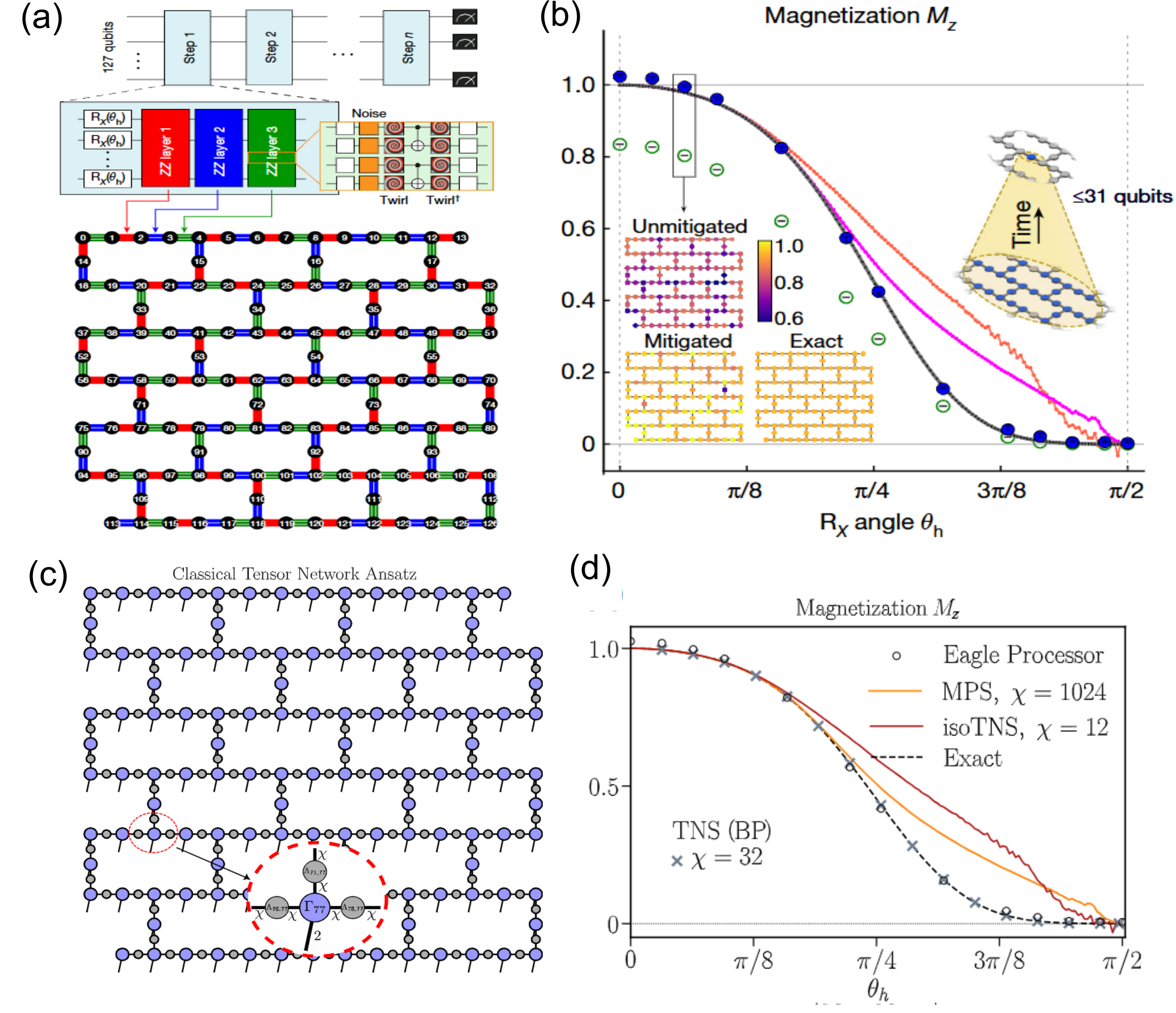}
    \caption{Large-scale Floquet quantum simulation on a superconducting quantum processor. (a) Schematic of the Floquet circuit architecture implemented on the 127-qubit Eagle processor. Each Floquet cycle consists of parameterized single-qubit rotations, followed by two alternating layers of entangling $R_{zz}$ gates. Noise tailoring through randomized twirling is employed to mitigate coherent errors~\cite{kim2023evidence}. (b) Measurement of magnetization obtained from the circuit in (a). The experimental data (blue) track the exact theoretical prediction (black), and agreement improves significantly when error-mitigation techniques are applied (yellow). (c) Classical tensor-network ansatz used to validate and benchmark the quantum experiment. (d) Comparison of magnetization between processor outcomes with tensor-network simulations. As shown in (c)–(d), accurate large-scale benchmarking requires sophisticated classical algorithms with carefully optimized tensor-network architectures~\cite{tindall2024efficient}, whereas the high-fidelity quantum processor, combined with advanced error mitigation, delivers reliable results. Panel (a) and (b) are adopted from ~\cite{kim2023evidence}.  Panel (c) and (d) are adopted from ~\cite{tindall2024efficient}.} 
    \label{fig:floquet1}
\end{figure*}

\subsection{Ground-state preparation}

While dynamical simulation is essential for probing non-equilibrium phenomena
and time-dependent observables, many questions of central interest in
condensed-matter physics require access to ground or low-lying eigenstates as
starting points. These include the characterization of phase structure,
symmetry-breaking or topological order parameters, correlation functions, and
low-energy response
\cite{sachdev1999quantum,white1992density,schollwock2011density,
verstraete2009quantum}. Ground-state preparation is also a computationally
nontrivial task in its own right, since properties of local-Hamiltonian ground
states can be difficult to determine in general
\cite{kempe2006complexity,osborne2012hamiltonian,gharibian2015quantum} (see Sect.~\ref{vqa} above).

\subsubsection{Variational Quantum Eigensolver (VQE)}\label{vqa}

For ground-state preparation, the variational quantum eigensolver (VQE) is a widely used hybrid quantum--classical approach. In VQE, a parametrized quantum
circuit prepares a trial state on a quantum processor, while a classical
optimizer updates the circuit parameters to minimize the measured energy
expectation value~\cite{peruzzo2014variational,mcclean2016theory,
kandala2017hardware} (see Sect.~\ref{vqa}). Specifically, an ansatz circuit
$U(\boldsymbol{\theta})$ generates the variational state
\begin{equation}
\ket{\psi(\boldsymbol{\theta})}
=
U(\boldsymbol{\theta})\ket{\psi_0},
\end{equation}
where $\ket{\psi_0}$ is an initial reference state and
$\boldsymbol{\theta}$ denotes the set of tunable circuit parameters, typically
rotation angles in single- or multi-qubit gates. The variational energy is then
defined as
\begin{equation}
E(\boldsymbol{\theta})
=
\bra{\psi(\boldsymbol{\theta})}
H
\ket{\psi(\boldsymbol{\theta})},
\label{eq:vqe_energy}
\end{equation}
where $H$ is the target Hamiltonian.

In practice, $H$ is decomposed into measurable operator terms,
$H=\sum_\ell h_\ell P_\ell$, where $P_\ell$ are Pauli strings or other
hardware-measurable observables. At each optimization step, the quantum
processor prepares $\ket{\psi(\boldsymbol{\theta})}$ and estimates the required
expectation values $\langle P_\ell\rangle_{\boldsymbol{\theta}}$. These
measurement outcomes are combined to evaluate $E(\boldsymbol{\theta})$, which
is then passed to a classical optimizer. The optimizer proposes updated
parameters $\boldsymbol{\theta}$, and the quantum circuit is executed again.
This quantum--classical loop is repeated until the energy converges to an
approximation of the ground-state energy within the chosen ansatz family.

An essential component of the VQE framework is the classical optimization loop,
which updates the variational parameters to minimize the measured expectation
value of the Hamiltonian. The choice of optimizer strongly affects convergence
speed, shot efficiency, and robustness to hardware noise. Early VQE
implementations commonly used gradient-free optimizers, such as COBYLA and
Nelder--Mead-type methods, because they do not require explicit gradient
evaluation and can be implemented directly from noisy energy estimates
\cite{peruzzo2014variational,o2016scalable,kandala2017hardware,
claudino2020benchmarking}. This feature is advantageous on near-term hardware,
where each cost-function evaluation is affected by finite sampling, readout
errors, and gate noise. However, gradient-free methods can become inefficient
as the number of variational parameters grows, because the number of function
evaluations required to navigate a high-dimensional landscape may increase
substantially. Moreover, gradient-based and quasi-Newton strategies have also
been explored in VQE. Methods such as L-BFGS-B can converge faster on smooth
and well-conditioned landscapes, especially when reliable analytic,
parameter-shift, or finite-difference gradients are available
\cite{cerezo2021variational,tilly2022variational}. In practice, the optimal
choice of optimizer is hardware- and problem-dependent: stochastic noise,
barren plateaus, parameter correlations, and ansatz expressibility can all
change which optimization strategy is most effective.

The main advantage of VQE lies in its hardware compatibility with current
quantum devices, as it requires only shallow circuits compared with more
resource-intensive algorithms. Nonetheless, several challenges hinder its
scalability and accuracy. A central issue is the emergence of barren plateaus,
where the gradients of the cost function vanish exponentially with system size,
making variational training increasingly difficult
\cite{cerezo2021variational}. This trainability problem can be
further exacerbated by hardware noise, which itself can induce barren plateaus,
and by the intrinsic computational hardness of optimizing generic variational
landscapes~\cite{bittel2021training}. Furthermore, the power of VQE is strongly tied to the choice of ansatz. While hardware-efficient circuits are much easier to implement, they may fail to capture the entanglement and symmetry structures inherent in condensed matter systems with different structures from hardware layouts.

Moreover, we remark that several recent extensions of the VQE framework have been developed to improve its trainability, expressibility, and practical performance on near-term quantum hardware. A prominent example is the Adaptive Derivative-Assembled Problem-Tailored Variational Quantum Eigensolver (ADAPT-VQE), in which the ansatz is constructed iteratively by selecting operators from a predefined pool according to their energy gradients~\cite{grimsley2019adapt}. This problem-adaptive construction avoids committing to a fixed circuit architecture in advance and can achieve accurate ground-state preparation with fewer parameters and shallower circuits than generic hardware-efficient ans\"atze. A related adaptive strategy is Fermionic Adaptive Sampling Theory VQE (FAST-VQE), a method for selecting operators based on importance metrics derived solely from the populations of Slater determinants in the wave function~\cite{majland2023fastvqe}. By avoiding repeated gradient evaluations over a large operator pool, FAST-VQE reduces the operator-selection overhead of ADAPT-VQE while retaining a chemistry
motivated adaptive ansatz construction. Another complementary direction is Filtering VQE (F-VQE), where the cost function is modified by applying spectral filtering transformations that suppress excited-state contributions and enhance the low-energy component of the trial state~\cite{amaro2021filtering}. These adaptive and filtering-based methods provide promising routes toward more scalable preparation of ground and low-lying eigenstates in strongly correlated quantum many-body systems, where conventional VQE may suffer from barren plateaus, poor ansatz expressibility, or costly optimization.

\subsubsection{Adiabatic State Preparation}\label{adiabatic}
Adiabatic State Preparation (ASP) harnesses the adiabatic theorem of quantum mechanics to prepare ground states of complex Hamiltonians by slowly evolving from an initial Hamiltonian $H_0$ whose ground state is known and easy to prepare~\cite{kato1950adiabatic,farhi2000quantum,roland2002quantum}. The adiabatic theorem states that a quantum system initialized in the ground state of a Hamiltonian $H(0)$ will remain in the instantaneous ground state of $H(t)$ if the Hamiltonian changes sufficiently slowly ~\cite{kato1950adiabatic,farhi2000quantum}. The process can be expressed in its simplest form:
\begin{equation}
  H(s)=(1-s)H_{0}+sH_{p}
\end{equation}
with $s\in[0,1]$. where $H_{0}$ is a simple Hamiltonian with a known ground state $\ket{\psi(0)}$ or its ground state can be easily prepared. $H_{p}$ is the problem Hamiltonian. Then, the desired ground state can be dynamically generated as
\begin{equation}
e^{-i\int_{0}^{1} H(s) ds}\ket{\psi(0)},
\end{equation}
This approach is ansatz-free, in contrast to VQE, and does not require explicit parameter optimization. {However, one needs to evolve the state slowly compared to the inverse square of the minimum spectral gap, which may close polynomially or exponentially with system size. For gate-based implementations, the corresponding
long real-time evolution must be decomposed into many circuit layers, leading
to deep circuits and substantial accumulated gate error and decoherence. This makes straightforward digitized adiabatic evolution difficult on
present noisy gate-based quantum hardware.

This method has been widely studied in the context of preparing many-body ground states on the analog platform, such as cold atoms or Rydberg arrays, where time-dependent Hamiltonians can be engineered
directly through external control fields~\cite{jaksch1998cold,greiner2002quantum,bloch2008many}. In optical lattices, ultracold bosonic or fermionic atoms are trapped in periodic potentials formed by interfering laser beams, and the effective Hamiltonian is typically described by the Hubbard-type model. ASP is realized experimentally by starting with a weak lattice, where the ground state approximates a Bose–Einstein condensate~\cite{greiner2002quantum,bloch2008many}. The lattice depth is then slowly ramped up, and we switch on the optical lattice potential adiabatically. Once atoms are loaded in optical lattices, the system Hamiltonian is tuned toward the target regime of a certain phase. The same principle can also be implemented on digital quantum hardware, for example, using
Trotterized time evolution. Compared with analog implementations, the digital version
does not require the target Hamiltonian to coincide with the native physical
interactions of the device. Instead, the desired terms can be encoded into
qubits. In this sense, adiabatic state preparation is applicable as a common conceptual
framework for both analog and digital platforms.

\subsection{Extracting spectral and observable information}

Having discussed methods for approximating unitary time evolution via
Trotterization, as well as algorithms designed to prepare ground states, we now
shift gears to examine extracting information from an already prepared state. Specifically, we focus on two approaches for extracting  spectral information directly from
coherent quantum dynamics: quantum phase estimation (QPE) and quantum amplitude estimation (QAE).

\subsubsection{Quantum Phase Estimation (QPE)}

Quantum phase estimation (QPE) encodes energy eigenvalues into
quantum phases through controlled time evolution~\cite{cleve1998quantum,abrams1999quantum,svore2013faster}. It is a canonical algorithm showing how Hamiltonian simulation can be used to resolve energy
spectra and, when the input state has nonzero overlap with an eigenstate,
project the system onto that eigenstate. In quantum simulation, this capability is important because many physical
questions are intrinsically spectral. For example, QPE and related spectral
algorithms provide a natural interface between Hamiltonian simulation and
many-body spectroscopy: after implementing controlled real-time evolution
$e^{-iHt}$, the accumulated phase gives access to eigenenergies, while repeated
or refined measurements can resolve low-lying spectra and dynamical response
features~\cite{somma2003quantum,childs2012hamiltonian,low2019hamiltonian}.

This method provides a route to spectral information. Let $H$ be the target Hamiltonian acting on an $n$-qubit system,
and let $\ket{\psi_\alpha}$ be an eigenstate with energy $E_\alpha$. For a
chosen evolution time $\tau$, the unitary $U(\tau)=e^{-iH\tau}$
has eigenvalue
\begin{equation}
U(\tau)\ket{\psi_\alpha}
=
e^{-iE_\alpha\tau}\ket{\psi_\alpha}
=
e^{2\pi i\phi_\alpha}\ket{\psi_\alpha},
\end{equation}
with $\phi_\alpha
=
-\frac{E_\alpha\tau}{2\pi}
(\mathrm{mod}\;1)$. Estimating the phase $\phi_\alpha$ therefore gives the corresponding energy,
up to the usual $2\pi/\tau$ phase ambiguity. The choice of $\tau$ must be made
so that the relevant spectral window is mapped unambiguously into the unit
circle, or else additional phase-unwrapping information is required.

In standard quantum phase estimation, the phase is extracted using an ancilla
register, controlled powers of $U(\tau)$, and an inverse quantum Fourier
transform. A more economical variant is iterative quantum phase estimation
(IQPE) \cite{kitaev1995quantum,griffiths1996semiclassical,dobsicek2007arbitrary}, which recycles a single ancilla qubit and determines the binary digits
of the phase sequentially. In each iteration, the ancilla is prepared in
$\ket{+}$, coupled to the system through a controlled power $U(\tau)^{2^k}$,
and then measured after a classically chosen feedback rotation. If the system
is in the eigenstate $\ket{\psi_\alpha}$, the controlled operation maps
\begin{equation}
\ket{+}\ket{\psi_\alpha}
\longrightarrow
\frac{1}{\sqrt{2}}
\left(
\ket{0}
+
e^{2\pi i 2^k\phi_\alpha}\ket{1}
\right)
\ket{\psi_\alpha}.
\end{equation}

A single-qubit phase shift on the ancilla, $R_z(2\pi\beta)
=
\ket{0}\!\bra{0}
+
e^{2\pi i\beta}\ket{1}\!\bra{1}$, 
shifts the relative phase in the ancilla interference signal to
$2^k\phi_\alpha+\beta$. Measuring the ancilla in the $X$ basis then gives
\begin{equation}
P(0)
=
\cos^2\!\left[\pi(2^k\phi_\alpha+\beta)\right],
P(1)
=
\sin^2\!\left[\pi(2^k\phi_\alpha+\beta)\right].
\end{equation}
By choosing the feedback phase $\beta$ using the bits already obtained in
earlier iterations, IQPE isolates the next unknown binary digit of
$\phi_\alpha$. After $p$ iterations, the first $p$
binary digits $b_1,\ldots,b_p$ provide the estimate
\begin{equation}
\phi_\alpha
\simeq
\sum_{j=1}^{p}
\frac{b_j}{2^j}.
\end{equation}
and the corresponding energy estimate is $E_\alpha'
=
-\frac{2\pi}{\tau}\phi_\alpha'
\quad
(\mathrm{mod}\;2\pi/\tau)$.

QPE can be seen as an application of dynamical quantum simulation, in which the coherent real-time evolution yields spectroscopic
information. One example is shown in  Fig.~\ref{fig:digital_topological_phenomena} earlier. If the input state has appreciable overlap with several
eigenstates, phase estimation samples the corresponding eigenphases with
probabilities set by those overlaps. In this way, QPE can be used to estimate spectral features. Its main cost is the need for controlled long-time evolution
$U(\tau)^{2^k}$ and sufficiently coherent circuits, which makes full QPE increasingly feasible as the fault-tolerant quantum simulation nears.

\subsubsection{Quantum Amplitude Estimation}\label{qae}
While IQPE provides a powerful route to extracting eigenvalues from coherent
time evolution, many quantum-simulation tasks require only expectation values,
transition probabilities, or state overlaps rather than full spectral
information. These quantities can often be encoded as the probability of a
specified measurement outcome, motivating quantum amplitude estimation (QAE)
\cite{brassard2000quantum,nakaji2020faster,grinko2021iterative,
suzuki2020amplitude}. In its original form, QAE uses amplitude amplification
together with phase-estimation features to estimate this probability with
quadratically improved precision scaling compared with direct sampling in the
ideal coherent setting. More recent variants reduce some of the circuit
requirements, making QAE and related amplitude-estimation methods useful
building blocks for estimating observables and overlaps in quantum simulation.

A standard primitive for estimating expectation values on a quantum processor is
the Hadamard-test circuit. Suppose first that the observable $O$ is a unitary
Hermitian operator, such as a Pauli string, with $O^2=\mathbb{I}$. The system is
prepared in the state $\ket{\psi}$, while an ancilla qubit is initialized in
$\ket{+}
=
\frac{\ket{0}+\ket{1}}{\sqrt{2}}$. 
The joint state is therefore
\begin{equation}
\ket{+}\ket{\psi}
=
\frac{1}{\sqrt{2}}
\left(
\ket{0}\ket{\psi}
+
\ket{1}\ket{\psi}
\right).
\end{equation}
Applying a controlled-$O$ gate, with the ancilla as the control, gives
\begin{equation}
\frac{1}{\sqrt{2}}
\left(
\ket{0}\ket{\psi}
+
\ket{1}O\ket{\psi}
\right).
\end{equation}
A final Hadamard gate on the ancilla maps this state to
\begin{equation}
\frac{1}{2}
\left[
\ket{0}
\left(
\mathbb{I}+O
\right)
\ket{\psi}
+
\ket{1}
\left(
\mathbb{I}-O
\right)
\ket{\psi}
\right].
\end{equation}
The probability of measuring the ancilla in $\ket{0}$ is then
\begin{equation}
P(0)
=
\frac{1}{2}
\left(
1+
\mathrm{Re}\,\langle\psi|O|\psi\rangle
\right),
\end{equation}
since $O^2=1$. For Hermitian observables such as Pauli strings, the expectation value is real,
so that
\begin{equation}
\langle O\rangle
=
2P(0)-1 .
\end{equation}

For a general Hamiltonian or observable decomposed as
$O=\sum_\ell o_\ell P_\ell$, where $P_\ell$ are Pauli strings, the same
primitive can be applied term by term to estimate $\langle P_\ell\rangle$ and
then reconstruct
\begin{equation}
\langle O\rangle
=
\sum_\ell o_\ell \langle P_\ell\rangle .
\end{equation}

\subsection{Ancilla-assisted algorithmic primitives}\label{ancilla}
The algorithms discussed so far are presented as distinct procedures tailored
to specific tasks. However, from the perspective of circuit implementation,
they share a common structural challenge: several of the operations they require
are not natively deterministic unitary evolutions on the system alone, for example,
imaginary-time evolution and projective operations
\cite{motta2020determining,brassard2000quantum,childs2012hamiltonian,
berry2015simulating}. A unifying resolution is provided by ancilla-assisted constructions, in which
the desired nonunitary or conditional operation on the system is realized as
part of a larger unitary acting on an extended Hilbert space that includes
auxiliary qubits. Depending on the protocol, the desired branch is selected by
post-selection, coherently amplified, or combined with other branches to realize
a target linear map. This viewpoint underlies several central primitives in
modern quantum simulation, including QITE-style nonunitary evolution, linear combinations of unitaries, and block
encodings
\cite{motta2020determining,brassard2000quantum,childs2012hamiltonian,
berry2015simulating,gilyen2019quantum,low2019hamiltonian}. In this section, we
introduce two foundational examples of this broader idea: ancilla-based
dilation with post-selection and linear combinations of unitaries.

\subsubsection{Ancilla-based dilation and post-selection techniques}\label{dilation}
Ancilla-based dilation provides a standard way to realize nonunitary operations
within an otherwise unitary circuit model. Suppose that the desired operation on
the system is a linear map $M$.  If $M$ is a contraction, meaning $\|M\ket{\psi}\|^2\leq \|\ket{\psi}\|^2$ for every $\ket{\psi}$, then $M$ can be
embedded into a larger unitary acting on the system together with an ancilla
register~\cite{nielsen2010quantum,childs2012hamiltonian,
gilyen2019quantum}.

Preparing the ancilla in a reference state
$\ket{0_a}$, we can construct a unitary $U$ such that
\begin{equation}
U\left(\ket{0_a}\otimes\ket{\psi}\right)
=
\ket{0_a}\otimes M\ket{\psi}
+
\ket{1_a}\otimes \ket{\psi_{\rm fail}},
\label{eq:dilation_basic}
\end{equation}
where $\ket{\psi_{\rm fail}}$ denotes the component associated with the
orthogonal ancilla outcome. Measuring the ancilla and postselecting on the
outcome $\ket{0_a}$ prepares the system in the normalized state
\begin{equation}
\ket{\psi_M}
=
\frac{M\ket{\psi}}{\sqrt{p_{\rm succ}}},
\qquad
p_{\rm succ}
=
\|M\ket{\psi}\|^2
=
\bra{\psi}M^\dagger M\ket{\psi}.
\label{eq:dilation_success}
\end{equation}
Thus, from the perspective of the system alone, the conditional dynamics is
nonunitary even though the joint system--ancilla evolution is unitary. 
This construction is useful because many operations appearing in quantum
simulation are naturally nonunitary or conditional. Imaginary-time evolution,
for example, suppresses high-energy components of a trial state through the
nonunitary map $e^{-\tau H}$ [see Sect.~\ref{qite} above].

The measurement step introduces post-selection, whereby only experimental runs yielding the desired ancilla outcome are retained. Post-selection is thus probabilistic, with a success rate dependent on both the norm of the evolved state and the specific embedding chosen~\cite{nguyen2022block,sze2025hamiltonian}. While this inevitably adds to the resource costs, it enables the simulation of otherwise inaccessible dynamics, including imaginary-time propagation and non-unitary dynamics. The primary advantage of ancilla-based dilation is its universality: any non-unitary operation can, in principle, be implemented through a sufficiently large unitary extension with multiple ancilla qubits. 
In practice, one encounters a trade-off between various resource costs: while breaking up the non-unitary problem into many small blocks, each served by one ancilla, can significantly reduce the classical overhead in computing the appropriate unitary extension, having more ancilla also incurs significantly more quantum simulation runs due to exponentially lower success probability.

\subsubsection{Linear Combinations of Unitaries}\label{lcu}
Many operators of interest in quantum simulation are not by themselves unitary, but
can be expressed as weighted sums of unitary components. Examples
include imaginary-time-evolution operators and response
operators appearing in correlation functions ~\cite{motta2020determining,
mcardle2019variational}. While a quantum circuit can apply
a single unitary transformation deterministically, it cannot directly apply an
arbitrary weighted sum of different unitaries, because such a linear combination
is generally neither unitary nor physically realizable as a single gate. This
issue is closely related to the nonunitary imaginary-time and post-selected
constructions discussed in Sects.~\ref{qite} and~\ref{dilation}.

To encode relative weights and select unitary
components, the linear-combination-of-unitaries (LCU) framework introduces an
ancilla register that labels which operation is applied. The ancilla provides the
additional Hilbert-space dimension needed to represent the desired linear map as
a block of a larger unitary operator~\cite{childs2012hamiltonian,
berry2015simulating,low2019hamiltonian,gilyen2019quantum}. Suppose that the
target operator admits a decomposition
\begin{equation}
A=\sum_k \alpha_k U_k ,
\end{equation}
where each $U_k$ is unitary. For simplicity, we first take
$\alpha_k\geq 0$ and define
\begin{equation}
\alpha=\sum_k \alpha_k .
\end{equation}
If some coefficients are negative or complex, their phases can be absorbed into
the corresponding unitaries $U_k$, and $\alpha$ should be replaced by
$\sum_k |\alpha_k|$.

The basic LCU construction uses two ingredients. The first is a preparation
unitary ${\rm Pre}$ acting on the ancilla register,
\begin{equation}
{\rm Pre}\ket{0}
=
\sum_k
\sqrt{\frac{\alpha_k}{\alpha}}
\ket{k}.
\end{equation}
The second is the controlled select operation
\begin{equation}
{\rm Sel}(U)
=
\sum_k
\ket{k}\bra{k}\otimes U_k ,
\end{equation}
which applies $U_k$ to the system when the ancilla is in state $\ket{k}$.
Combining these operations gives the block encoding
\begin{equation}
W
=
({\rm Pre}^{\dagger}\otimes I)
{\rm Sel}(U)
({\rm Pre}\otimes I).
\end{equation}
Projecting the ancilla back onto $\ket{0}$ yields
\begin{equation}
(\bra{0}\otimes I)W(\ket{0}\otimes\ket{\psi})
=
\frac{1}{\alpha}A\ket{\psi},
\end{equation}
which recovers the action of the desired non-unitary $A$ on $\ket{\psi}$. In other words, $W$ is a unitary embedding whose upper-left block in the ancilla basis is $A/\alpha$. The
desired transformation is obtained probabilistically by postselecting the
ancilla outcome $\ket{0}$, as in the dilation picture of
Sect.~\ref{dilation}. This LCU construction gives a compact language for both unitary and nonunitary
quantum simulation. For
nonunitary dynamics, the LCU framework connects
standard Hamiltonian simulation with the nonunitary simulation protocols
discussed in Sect.~\ref{sec:topological_dynamics}.

\begin{figure*}
    \centering
    \includegraphics[width=0.9\linewidth]{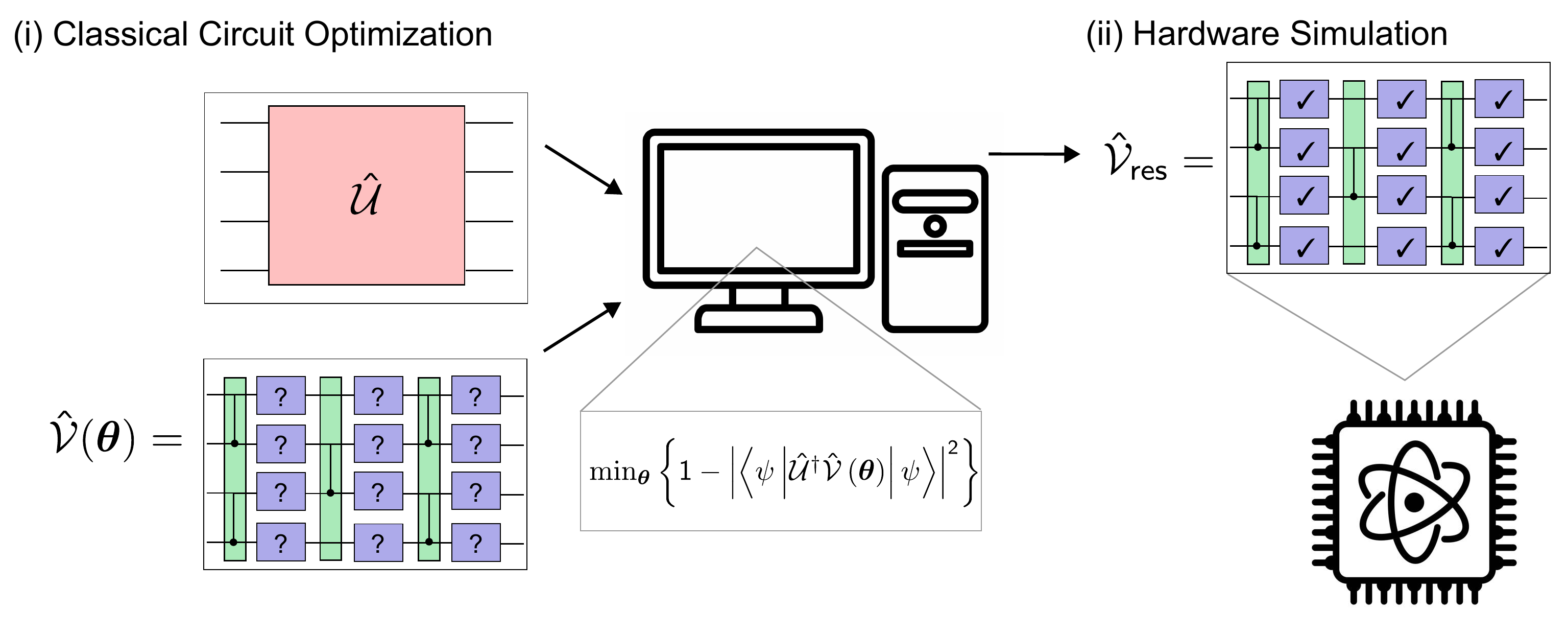}
    \caption{Illustration of the quantum circuit recompilation. (i) Classical optimization stage. A target circuit $\hat{\mathcal{U}}$ is
approximated by a hardware-compatible parametrized circuit
$\hat{\mathcal{V}}(\boldsymbol{\theta})$. Parameters in the single-qubit rotation blocks are optimized on a
classical computer by minimizing a state-dependent recompilation cost, such as
Eq.~\eqref{eq:state_recompilation_cost}.  (ii) Hardware execution stage. After optimization, the resulting recompiled
circuit $\hat{\mathcal{V}}_{\rm res}$ is implemented on the quantum processor.  This procedure reduces the depth or number of 
entangling gates while preserving the target action of
$\hat{\mathcal{U}}$ on the input state. }
    \label{fig:circuit recompilation illustration}
\end{figure*}

\subsection{Quantum Circuit Recompilation}\label{recom} 

In this section, we briefly introduce the basic idea of quantum circuit
recompilation. Circuit compilation broadly refers to the transformation
of an abstract quantum circuit into an implementation compatible with the native
gate set, qubit connectivity, and calibration constraints of a target hardware
platform. Circuit recompilation is a more specialized task: given a target
unitary circuit $\hat{\mathcal{U}}$, one seeks an alternative circuit
$\hat{\mathcal{V}}(\boldsymbol{\theta})$, built from hardware-native gates and
parameterized by variational parameters $\boldsymbol{\theta}$, such that
$\hat{\mathcal{V}}(\boldsymbol{\theta})$ approximates the action of
$\hat{\mathcal{U}}$ on the relevant input states while using fewer costly
operations or a shallower circuit depth~\cite{khatri2019quantum}. This strategy is directly relevant to the recompiled topological-dynamics
examples shown in Fig.~\ref{fig:digital_topological_phenomena}, where circuit
depth reduction is essential for implementing long-time dynamics on present
gate-based quantum hardware~\cite{koh2022simulation,koh2022stabilizing,
koh2024realization}. More broadly, related variational recompilation,
automated circuit optimization, and noise-aware circuit-learning strategies
have also been developed ~\cite{heya2018variational,khatri2019quantum}.

The motivation is straightforward. On present gate-based superconducting
quantum processors, the dominant cost of a circuit is often set by its
entangling-gate content rather than by its total gate count. Single-qubit
rotations are typically much faster and more accurate, whereas two-qubit gates
are more susceptible to calibration errors, crosstalk, and decoherence during
the entangling operation. Representative IBM Heron-class devices report median
two-qubit gate errors of order $10^{-3}$, while single-qubit gate
errors are typically at $10^{-4}$
\cite{abughanem2024ibm,ibm_quantum_2026}. As a result, entangling gates often
dominate the accumulated error in deep circuits.

In addition, hardware
connectivity constraints may introduce the need for additional SWAP operations to route logical qubits
onto adjacent physical qubits, further increasing the number of entangling
gates. Although logical gates such as CNOTs can be expressed in the native gate
basis of the processor, each additional entangling operation contributes to
decoherence, crosstalk, leakage, and readout-correlated errors. Consequently,
circuits containing many entangling gates can lose fidelity rapidly on
near-term devices. Recompilation addresses this bottleneck by replacing a
target circuit with a shorter, hardware-adapted circuit that preserves the
desired transformation to sufficient accuracy for the observables of interest
[see Fig.~\ref{fig:digital_topological_phenomena} for a representative
simulation using recompiled circuits]. This strategy has become useful not only
for quantum-state preparation~\cite{chen2023high,jaderberg2025variational}, but
also for real-time quantum simulation, where long Trotterized circuits can
sometimes be compressed into shallower variational or tensor-network-optimized
circuits~\cite{shen2025observation,frey2024direct,zhang2024scalable,
tepaske2024optimal,gibbs2025deep,guo2025efficient,jaderberg2025variational,
d2025circuit}. Related dynamic parameterized-circuit ansätze provide another
route to expressive trainable circuits with favorable optimization properties
~\cite{deshpande2024dynamic}. Importantly, recompilation need not always
reproduce the full unitary operator with high global fidelity. For many
near-term simulation tasks, it is sufficient to reproduce selected local
observables, reduced density matrices, correlation functions, or dynamics
within a physically relevant subspace.

A general state-dependent formulation trains the recompiled circuit by
minimizing the mismatch between the target and variational circuits on a chosen
input state $\ket{\psi}$ [see Fig.~\ref{fig:circuit recompilation illustration}],
\begin{equation}
\mathcal{C}_{\psi}(\boldsymbol{\theta})
=
1-
\left|
\left\langle
\psi
\left|
\hat{\mathcal{U}}^{\dagger}
\hat{\mathcal{V}}(\boldsymbol{\theta})
\right|
\psi
\right\rangle
\right|^2 .
\label{eq:state_recompilation_cost}
\end{equation}
This cost function compares the two circuits only on the state of interest. It
can also be generalized to an ensemble of input states or to costs defined
directly from observables. Such state-dependent recompilation naturally fits
within the broader variational-quantum-algorithm framework, where a
parameterized circuit is optimized using a classical routine and then executed
on quantum hardware~\cite{mcclean2016theory}. A stricter alternative is operator-level recompilation, where the goal is to
approximate the full unitary action of $\hat{\mathcal{U}}$. A standard global
overlap cost is
\begin{equation}
\mathcal{C}_{\rm op}(\boldsymbol{\theta})
=
1-
\frac{1}{d^2}
\left|
\operatorname{Tr}
\left[
\hat{\mathcal{U}}^{\dagger}
\hat{\mathcal{V}}(\boldsymbol{\theta})
\right]
\right|^2,
\qquad
d=2^M ,
\label{eq:operator_recompilation_cost}
\end{equation}
where $M$ is the number of qubits. This quantity is related to the
Hilbert--Schmidt overlap between the target and recompiled unitaries and is
insensitive to an overall global phase. For moderate system sizes or circuits
with favorable entanglement structure, such overlaps can be evaluated using
tensor-network contractions, including matrix-product-state or
matrix-product-operator representations~\cite{gray2018quimb}. Tensor-network
methods can therefore provide a classical training backend for recompilation,
although their cost still grows rapidly when the operator entanglement becomes
large.

Recompilation protocols differ in the choice of ansatz, cost
function, and optimizer. Typical ans\"atze are constructed from
hardware-native entangling layers interleaved with parameterized single-qubit
rotations. The parameters may be optimized using gradient-free methods,
quasi-Newton algorithms such as L-BFGS-B, or global-search heuristics such as
basin hopping
\cite{Malouf2002,Cao2007,koh2022stabilizing,li1987monte,Wales1997,
Scheraga1999,wales_2004}. The practical objective is to find a circuit that is
shallow enough to reduce two-qubit-gate errors while accurate enough to
reproduce the target dynamics or observables. This creates a central tradeoff:
increasing ansatz depth improves expressibility, but also increases exposure to
hardware noise.

Recent developments have broadened recompilation from a generic circuit
compression tool into a physics-aware method for quantum simulation. Local
compilation strategies exploit the locality of Hamiltonian dynamics and avoid
optimizing a large global unitary when only local observables are required
\cite{Fujii2022localcompilation}. Tensor-network-assisted methods based on
matrix product operators, Pauli propagation, or related classical
representations have demonstrated substantial compression of time-evolution
circuits for lattice Hamiltonians, including transverse-field Ising and
Heisenberg models, while remaining compatible with realistic hardware
geometries~\cite{gibbs2025deep,d2025circuit}. Other approaches formulate
dynamical recompilation as a learning problem, where a circuit is trained on a
selected set of input states and then tested on more complex many-body states
\cite{zhang2024scalable}.

Experimentally, the recompilation-based circuit compression have extended the accessible time
scales of non-equilibrium quantum simulation on noisy processors. For example, such
circuit-compression methods have been used to study dynamical
quasi-condensation and relaxation in hard-core boson systems, with improved
noise resilience characterized through quantum process tomography
\cite{frey2024direct,dinca2025quantumprocesstomographycompressed}. More
broadly, circuit recompilation provides a bridge between classical many-body
structure and hardware-aware quantum control: classical optimization or
tensor-network contraction is used to identify a shorter circuit, while the
quantum processor executes the compressed dynamics. In this way, recompilation
helps reduce the depth bottleneck that limits near-term quantum simulation of
many-body dynamics.

\subsection{Entanglement Entropy Estimation}\label{sec:EE}

Entanglement entropy is an important diagnostic in quantum many-body systems,
playing a key role in identifying quantum phase transitions, topological order,
and critical phenomena\cite{amico2008entanglement,horodecki2009quantum,eisert2010colloquium,
calabrese2004entanglement,kitaev2006topological,
levin2006detecting,chen2010local,jiang2012identifying}. It is also closely connected to the entanglement-growth discussed in
Sects.~\ref{mbl}. Yet the measurement of
entanglement entropy is nontrivial, because it is a nonlinear function of the
reduced density matrix rather than the expectation value of a single ordinary
observable. As a result, experimental protocols typically require tomographic reconstruction, direct estimation of nonlinear functionals using
multiple copies or controlled-SWAP-type measurements or randomized-measurement
protocols~\cite{ekert2002direct,daley2012measuring,islam2015measuring,
elben2018renyi,brydges2019probing,google2023measurement}.

To define the entanglement entropy, we consider a bipartition of the system into
regions $A$ and $B$. The reduced density matrix of region $A$ is obtained by
tracing out the degrees of freedom in region $B$,
\begin{equation}
\rho_A=\Tr_B(\rho).
\end{equation}
The von Neumann entanglement entropy is then defined as
\begin{equation}
S_A=-\Tr(\rho_A\log\rho_A).
\end{equation}
$S_A$ can also be obtained as the $\alpha\rightarrow 1$ limit of the R\'{e}nyi entropy
\begin{equation}
S_A^{(\alpha)}
=
\frac{1}{1-\alpha}
\log \Tr(\rho_A^\alpha),
\end{equation}
which contains a power of the density matrix, and may sometimes more straightforward to compute. 

The reduced density matrix can be explicitly obtained via quantum state tomography. For a subsystem of $n$ qubits, the reduced density matrix can be expanded in the Pauli-operator basis as
\begin{equation}
\rho_A
=
\frac{1}{2^n}
\sum_{P\in\mathcal{P}_n}
\langle P\rangle P,
\end{equation}
where
$\mathcal{P}_n=\{I,X,Y,Z\}^{\otimes n}$ is the set of $n$-qubit Pauli strings,
and
\begin{equation}
\langle P\rangle=\Tr(P\rho_A)
\end{equation}
is the corresponding Pauli expectation value. Reconstructing $\rho_A$ therefore
requires estimating, in general, an exponentially large number of Pauli
coefficients, scaling as $4^n-1$ number of nontrivial observables.  
At limited system sizes (say $n \lesssim 8$), quantum state tomography (QST) can be used to reconstruct the density matrix of a prepared quantum state. While resource intensive, the benefit offered by tomography over more specialized measurements (e.g.~small sets of local operators) is \emph{complete} diagnostic access to the description of the quantum state, allowing, for example, computation of even nonlinear quantities in the density matrix.
That said, full
state tomography is feasible only for small subsystems, although compressed
sensing and classical-shadow protocols can reduce the
measurement cost for restricted classes of states or observables
\cite{nielsen2010quantum,james2001measurement,paris2004quantum,
gross2010quantum,cramer2010efficient,huang2020predicting}.

Randomized measurement protocols provide an alternative route to entanglement entropy
estimation that avoids full state tomography and does not require preparing two
coherent copies of the state
\cite{elben2018renyi,brydges2019probing}. The key idea is to apply random local
unitaries before measurement and use statistical correlations between the
resulting bitstring probabilities to reconstruct nonlinear functions of the
reduced density matrix. The protocol is organized as follows:
\begin{enumerate}
    \item Apply a local random unitary on subsystem $A$,
    \begin{equation}
    U_A=\bigotimes_{i\in A} u_i ,
    \end{equation}
    where each $u_i$ is drawn independently from a unitary 2-design, such as
    the single-qubit Clifford group, or from the Haar measure.

    \item Measure all qubits in $A$ in the computational basis to obtain a
    bitstring $\mathbf{z}\in\{0,1\}^{|A|}$.

    \item Repeat the experiment for an ensemble of random unitaries, with many measurement
    shots per unitary, to estimate the probability distributions $P_U(\mathbf{z})$.
\end{enumerate}
From the obtained $P_U$ distribution, one can write down certain functions of the reduced density matrix. For instance, the purity of subsystem $A$
$S_A^{(2)}=-\ln\Tr(\rho_A^2)$ i.e. second R\'{e}nyi entropy can be estimated from the 
randomized measurements via~\cite{elben2018renyi,brydges2019probing}
\begin{equation}
\Tr(\rho_A^2)
=
2^{|A|}
\sum_{\mathbf{z},\mathbf{z}'}
\mathbb{E}_{U}
\left[
(-2)^{-D(\mathbf{z},\mathbf{z}')}
P_U(\mathbf{z})P_U(\mathbf{z}')
\right],
\end{equation}
where $D(\mathbf{z},\mathbf{z}')$ is the Hamming distance between the two
bitstrings, and $\mathbb{E}_U$ denotes the average over the random local
unitaries.
Because the protocol only requires single-copy randomized measurements, it can
be applied to larger systems, although the number of random unitaries and measurement
shots still grows with subsystem size and target precision. A similar approach, with some modifications, may be adapted to measure the Stabilizer R\'{e}nyi entropy~\cite{leone2022stabilizer,huang2025fast}.

huang2025fast

In the
Google Quantum AI experiment on measurement-induced entanglement [see Fig.~\ref{fig:mes}],
the quantum processor generated monitored many-body dynamics using layers of
entangling gates and intermediate measurements. To estimate the entanglement with 
a chosen subsystem $A$, the experiment then applied random local basis rotations
to the qubits in $A$, measured the resulting bitstrings, and repeated this
procedure over many random choices and measurement shots. The correlations
between the measured bitstring probabilities were used to estimate
$\Tr(\rho_A^2)$ and hence the second R\'{e}nyi entropy,
$-\log\Tr(\rho_A^2)$~\cite{google2023measurement}.

There is also a more direct way to measure the second R\'{e}nyi entropy. To compute $\Tr(\rho_A^2)$, one prepares two
identical copies of the target state and measures the expectation value of the
subsystem-swap operator,
\begin{equation}
\operatorname{SWAP}_A:
\ket{i}_A\otimes\ket{j}_A
\mapsto
\ket{j}_A\otimes\ket{i}_A .
\end{equation}
For two identical reduced density matrices, this gives
\begin{equation}
\Tr(\rho_A^2)
=
\Tr\!\left[
(\rho_A\otimes\rho_A)\operatorname{SWAP}_A
\right].
\end{equation}
Equivalently, if the two-copy state is written as $\rho\otimes\rho$ on the full
bipartite Hilbert space, $\operatorname{SWAP}_A$ is understood to act only on
the two copies of subsystem $A$, being the identity on the complementary subsystem.
Experimentally, the swap expectation value can be accessed either by applying
controlled-SWAP gates between the two copies and measuring an ancilla qubit, or
by interfering two identical many-body copies and measuring parity-resolved
observables~\cite{ekert2002direct,daley2012measuring,islam2015measuring}.
A gate-based example using entanglement-spectrum measurements to diagnose
SPT phases is shown in
Fig.~\ref{fig:qite} (b).

\subsection{Error mitigation and suppression methods}\label{error}
Now and in the foreseeable future, quantum devices remain limited by decoherence, gate infidelities, leakage, crosstalk, and measurement noise, all of which can strongly affect the accuracy of quantum simulations~\cite{preskill2018quantum,bharti2022noisy,temme2016error,shen2025circuit,shen2025benchmarking}. 
Here, we review error-mitigation and error-suppression methods developed to improve simulation fidelity on noisy hardware. 
Although these methods are often discussed as NISQ-era tools, their necessity and applicability often extends beyond near-term devices.

On the path to fully fault-tolerant quantum
computation, a probably intermediate regime  will consist of higher-quality processors, early logical qubits (limited in scales and circuit depth),
and limited-depth fault-tolerant circuits, where residual logical errors,
state-preparation errors, and measurement imperfections remain non-negligible
\cite{bluvstein2024logical,google2025quantum,bluvstein2025fault}. In this
regime, error mitigation will remain a useful complementary tool. Quantum error
correction suppresses errors at the hardware and logical-circuit level, whereas error mitigation, calibration, symmetry verification, and validation techniques help
improve measured observables and quantify residual bias
\cite{temme2016error,endo2021hybrid,cai2023quantum,bonet2018low}. Therefore,
the methods reviewed in this section should not be viewed as tools only for
imperfect NISQ devices; they also provide practical methodology for
benchmarking, validating, and improving quantum simulations on the route toward
fault-tolerant hardware.

\subsubsection{Zero-Noise Extrapolation and noise amplification}

A widely adopted strategy to reduce such bias \emph{without} full error
correction is \emph{zero-noise extrapolation} (ZNE)
~\cite{temme2016error,li2017efficient,kandala2019error,endo2021hybrid,
takagi2022fundamental,koh2025interacting}.  The idea is to run circuits that
implement the same ideal computation, but with  different effective
noise levels. The measured expectation values are then treated as points on a
noise-dependent curve and extrapolated back to the idealized zero-noise limit. ZNE is best understood as a method for reducing noise-induced bias in
observables.  In practice, increasing the noise level also increases
the uncertainty of the measured data, and the final extrapolated value can be
sensitive to the fitting model. The method works best when the noise is stable,
the noise-scaling procedure is well controlled, and the chosen observable
changes smoothly with the effective noise strength.

Concretely, we can consider a circuit intended to prepare
$\rho = U\rho_0 U^\dagger$ and measure an observable $O$. On quantum hardware, the
implemented evolution is noisy and can be modeled phenomenologically by a noise
channel $\mathcal{E}_\lambda$ with an effective strength parameter $\lambda$.
The measured expectation value is then 
\begin{equation}
O(\lambda)=\mathrm{Tr}\!\left[O\,\mathcal{E}_\lambda(\rho)\right].
\end{equation}
ZNE evaluates $O(\lambda)$ at several values
$\lambda_1,\lambda_2,\ldots$ and extrapolates to $\lambda\to 0$ using a
low-order fit, such as linear, quadratic, exponential, or Richardson
extrapolation. For example, a quadratic fit follows,
\begin{equation}
O(\lambda)=O_0+a\lambda+b\lambda^2 .
\end{equation}
For observables dominated by decay-like noise, one may instead use an
exponential model
\begin{equation}
O(\lambda)=O_0+A e^{-b\lambda}.
\end{equation}
Another common choice is Richardson extrapolation, which forms a weighted
linear combination of the measured values,
\begin{equation}
O_0 \approx \sum_j c_j O(\lambda_j),
\qquad
\sum_j c_j \lambda_j^m = \delta_{m0},
\end{equation}

The central experimental requirement is therefore the ability to \emph{amplify}
noise in a controlled manner. That is, the modified circuits should
have the same target expectation value in the absence of noise, and the resulting
noise-scaled expectation values can be properly fitted and extrapolated to the
zero-noise limit. A common digital approach is \emph{unitary folding} that a gate
$G$ is replaced by a longer sequence such as
\begin{equation}
G \;\longrightarrow\; G\,G^\dagger\,G
\qquad \text{or} \qquad
G \;\longrightarrow\; \bigl(G\,G^\dagger\bigr)^k\,G.
\end{equation}
In the noiseless limit, these folded sequences implement the same operation as
the original gate, $G G^\dagger=I$~\cite{giurgica2020digital}. On real quantum hardware, the folded circuit has the same ideal
logical action but a larger accumulated error.
A related method is \emph{global folding}, where an entire circuit $U$ is
replaced by
\begin{equation}
U \;\longrightarrow\; U(U^\dagger U)^k ,
\end{equation}
again preserving the ideal transformation while increasing the effective noise
level~\cite{giurgica2020digital}. In platforms where decoherence is a leading
contribution, an alternative is to stretch gate durations, keeping the same
target unitary while increasing exposure to relaxation and dephasing
~\cite{kandala2019error}. More recently, practical digital-ZNE workflows often
amplify noise by inserting identity layers or compiling ``do-nothing'' blocks
that increase circuit depth while preserving the logical circuit, which can be
simpler to deploy within a transpilation pipeline~\cite{giurgica2020digital,
majumdar2023best,shen2025robust}. A representative application of this method was previously illustrated in Fig.~\ref{fig:qite} (b).

\subsubsection{Readout error mitigation}

Another dominant error source on quantum processors is readout error: the detection of the measurement outcome can misidentify $\ket{0}$ as $\ket{1}$ and vice versa. This can be due to a variety of physical mechanisms: on superconducting devices, imperfect separation of $\ket{0}$ and $\ket{1}$ readout data on the phase-quadrature (IQ) plane can lead to a fraction of misclassified shots even with nonlinear binary classifiers, and on neutral atom and trapped-ion devices, overlapping photon count distributions at finite integration time and imaging noise leads to the same effect. Readout mitigation aims to remove this confusion by calibrating the confusion probabilities and then correcting the observed outcome statistics during post-processing~\cite{BravyiGambetta2021,maciejewski2020mitigation,nation2021scalable,nachman2020unfolding,smith2021qubit}.

At the level of measured bitstring distributions, readout-error mitigation is
usually formulated as a classical response-matrix unfolding problem
\cite{BravyiGambetta2021,nation2021scalable}.  
For $n$ measured qubits, let
$\mathbf{p}_{\rm ideal}$ denote the ideal distribution over computational-basis
outcomes and $\mathbf{p}_{\rm obs}$ the experimentally observed distribution. Readout errors are modeled by a calibration, or confusion, matrix $M$, defined as: 
\begin{equation}
    \mathbf{p}_{\rm obs} \simeq M \mathbf{p}_{\rm ideal},
\end{equation}
where $M_{yx}$ is the probability of recording outcome $y$ under
the ideal outcome $x$.  The mitigated distribution,
$\mathbf{p}_{\rm mit}$ can be is then obtained by approximately inverting $M$
\begin{equation}
    \mathbf{p}_{\rm mit} \simeq M^{+}\mathbf{p}_{\rm obs}.
    \label{eq:readout_error_mitigation_linear_inversion}
\end{equation}
Note that $M^{+}$ here denotes an inverse only when the calibration matrix is
well-conditioned. When calibration matrix is ill-conditioned, $M^{+}$ can still be obtained using the regularized inverse. 

The main practical difficulty of this method is scaling. A full calibration matrix has size
$2^n\times 2^n$ and requires preparing and measuring all computational-basis
states, which quickly becomes prohibitive for large setups. Large-scale experiments typically replace the full response matrix by
structured approximations, such as tensor-product models.  A method is based on the tensor-product readout model, in which the measured register
is divided into smaller partitions of sizes $n_1,n_2,\ldots,n_k$. Under this approximation, the full response matrix is
written as
\begin{equation}
    M \approx \bigotimes_{i=1}^k M^{(i)},
\end{equation}
where $M^{(i)}$ is the calibration matrix for the $n_i$ partition, each to be individually estimated. The
corresponding mitigation step can be applied using the blockwise inverse,
\begin{equation}
    M^{+} \approx \bigotimes_{i=1}^k \left(M^{(i)}\right)^{+}.
\end{equation}
Thus, instead of preparing
all $2^n$ computational-basis states of the full register, one only needs to
calibrate the smaller partitions~\cite{BravyiGambetta2021,nation2021scalable}.

Moreover, another limitation of standard readout-error mitigation is that it is usually applied
only after the circuit has finished. This is sufficient when measurements are
used only to estimate final output probabilities. However, this
post-processing picture is no longer sufficient for dynamic circuits with
mid-circuit measurements and feedforward, such as circuits using conditional
resets or measurement-conditioned gates. Recent work has begun to address this problem by developing mitigation methods
that account for readout errors inside dynamic circuits, rather than only at
the final measurement layer. Examples include an order-by-order error
cancellation scheme~\cite{hashim2025quasiprobabilistic} and an unbiased
quasi-probabilistic sampling method~\cite{koh2026readout}, which is designed to
handle circuits with multiple layers of mid-circuit measurement and feedforward
while adding negligible gate overhead.

\subsubsection{Dynamical decoupling }
Dynamical decoupling (DD) is another hardware-level mitigation technique
that suppresses decoherence by applying short sequences of control pulses during
idle intervals of a quantum circuit. In many existing experiments on quantum hardware, qubits spend significant
time waiting while other operations are executed; during these intervals,
environmental noise causes dephasing and relaxation that degrade simulation
accuracy. DD reduces this effect by inserting pulse sequences whose net logical
action is the identity, but which refocus slow noise accumulated during the idle
window~\cite{souza2011robust,das2021adapt,coote2025resource}.

Operationally, suppose a qubit has an idle interval of duration $\tau$ between
two logical circuit operations. Without DD, this interval is implemented as an
identity operation, $I(\tau)$.
With DD, the idle identity is replaced by a sequence of physical pulses and
shorter idle segments,
\begin{equation}
I(\tau)
\;\longrightarrow\;
I(\tau_{m+1})P_m I(\tau_m)\cdots P_2 I(\tau_2)P_1 I(\tau_1),
\end{equation}
where $P_j$ are control pulses and
\begin{equation}
\sum_{j=1}^{m+1}\tau_j+\sum_{j=1}^{m}t_{P_j}=\tau .
\end{equation}
The pulses are chosen so that their ideal product is the identity $P_mP_{m-1}\cdots P_1=I$,
and therefore the logical circuit is unchanged. For example, a simple
two-pulse echo-type insertion may use
\begin{equation}
I(\tau)
\;\longrightarrow\;
I(\tau/4)\,X\,I(\tau/2)\,X\,I(\tau/4).
\end{equation}
Thus, DD acts as a compiler-level replacement
of idle identity windows by physically active identity sequences. On near-term
devices, this is useful when the reduction in idle error outweighs the
additional pulse errors, crosstalk, and timing constraints introduced by the DD
sequence.

\subsubsection{Randomized compiling and twirling}
\emph{Randomized compiling} and \emph{twirling} are techniques that reduce the
impact of coherent, device-specific errors by converting them into an effectively
stochastic noise model that is easier to characterize, simulate, and mitigate
~\cite{wallman2016noise,hashim2020randomized,ware2021experimental,
erhard2019characterizing,cai2020mitigating}. Coherent errors, such as small
systematic over-rotations or unwanted phase shifts, can add constructively over
many circuit layers and produce large, circuit-dependent biases. By contrast,
stochastic Pauli errors behave more like random bit-flip, phase-flip, or
combined Pauli faults, whose effects are typically easier to model and combine
statistically.

The basic idea of Pauli twirling is to surround a noisy operation by randomly
chosen Pauli gates, while choosing compensating Pauli gates so that the ideal
logical operation is unchanged. For a noisy gate described by a channel
$\mathcal{E}$, a Pauli-twirled channel has the form \begin{equation}
\mathcal{E}_{\rm twirl}(\rho)
=
\frac{1}{|\overline{\mathcal{P}}_n|}
\sum_{P\in\overline{\mathcal{P}}_n}
P^\dagger\,
\mathcal{E}\!\left(P\rho P^\dagger\right)
P ,
\label{eq:pauli_twirl}
\end{equation}
where $\overline{\mathcal{P}}_n=\{I,X,Y,Z\}^{\otimes n}$ denotes the
phase-free $n$-qubit Pauli set. This averaging removes off-diagonal components of the noise in the Pauli
basis and yields an effective Pauli channel,
\begin{equation}
\mathcal{E}_{\rm twirl}(\rho)
=
\sum_{P'\in\overline{\mathcal{P}}_n}
p_{P'}\,P'\rho P'^\dagger ,
\label{eq:pauli_channel}
\end{equation}
where $P'$ labels the Pauli error that occurs after twirling, and
$p_{P'}\geq 0$.
Thus, the ideal circuit is preserved, but coherent features of the noise are
converted into a stochastic distribution over Pauli errors.

Randomized compiling applies this idea at the circuit level. The circuit is
divided into cycles, and random single-qubit ``twirling'' gates are inserted
around each cycle. Averaging
the measurement outcomes over many randomized circuit instances then produces an
effective stochastic Pauli-noise description, reducing the effect of coherent
miscalibrations, drift, and crosstalk-induced coherent bias
~\cite{wallman2016noise,hashim2020randomized,ware2021experimental,cai2020mitigating}.

\subsubsection{Probabilistic error cancellation (PEC)}
\emph{Probabilistic error cancellation} (PEC) is an error-mitigation strategy
that aims to recover ideal, noise-free expectation values by effectively
inverting the noise, without requiring full quantum error correction
~\cite{temme2016error,pashayan2015estimating,song2019quantum,
takagi2022fundamental,cai2023quantum}. The core idea is to represent the ideal
operation, or equivalently the inverse-noise-corrected operation, as a
quasiprobability mixture of experimentally implementable noisy operations. One
then samples from this mixture to generate an ensemble of modified circuits and
combines the measurement outcomes with signed weights, such that the weighted
average gives an unbiased estimator of the target zero-noise observable.

Here, we provide a basic picture of probabilistic error cancellation (PEC). Let
$\mathcal{U}$ denote the ideal operation to be implemented, for example
$\mathcal{U}(\rho)=U\rho U^\dagger$, and let
$\widetilde{\mathcal{U}}$ denote the corresponding noisy operation realized on
hardware. The implemented operation is
modeled as
\begin{equation}
\mathcal{U}
=
\mathcal{N}^{-1}\circ \widetilde{\mathcal{U}} .
\end{equation}
where $\mathcal{N}$ is an effective noise channel calibrated for this gate or
short circuit block. Formally, the ideal operation can then be recovered by
applying the inverse noise map.

In practice, the inverse is not implemented directly,
because it is generally not a physical quantum channel. Instead, one can find a
quasiprobability decomposition of the corrected operation in terms of noisy
operations that can actually be executed on the device,
\begin{equation}
\mathcal{N}^{-1}\circ\widetilde{\mathcal{U}}
=
\sum_k q_k\,\widetilde{\mathcal{O}}_k ,
\qquad
\sum_k q_k=1 .
\end{equation}
Here, $\widetilde{\mathcal{O}}_k$ denotes calibrated operations that can
actually be run on the device. Operationally, the experiment samples operation $\widetilde{\mathcal{O}}_k$
with probability $p_k=\frac{|q_k|}{\gamma},
\gamma=\sum_k |q_k|,$
and rescales the measured outcome by the factor
$\gamma\,{\rm sgn}(q_k)$. Averaging these weighted outcomes over many shots
reproduces the action of $\mathcal{N}^{-1}$ statistically.

PEC can remove bias from a broad class of noise processes and does not rely on
extrapolation, but it is most practical for relatively shallow circuits with
accurate noise characterization and modest quasiprobability overhead
~\cite{temme2016error,pashayan2015estimating,takagi2022fundamental}.
Experimental demonstrations and scalable variants have shown its practical
utility on near-term processors, including implementations based on gate-level
noise characterization and Pauli-noise learning
~\cite{song2019quantum,berg2023probabilistic,kim2023scalable}.

\subsubsection{Symmetry-informed post-selection}\label{symmetry}

Symmetry-informed post-selection mitigates errors by exploiting the fact that many target models and quantum algorithms are confined to a known symmetry sector~\cite{bonet2018low,sagastizabal2019experimental,cai2021quantum,o2023purification,kakkar2022characterizing,gonzales2023quantum}. In condensed-matter simulations, these symmetries may include conserved particle number, fixed fermion parity, total spin (or spin parity), translational or reflection symmetries (when implemented), and gauge constraints. Ideally, the quantum state prepared by the circuit remains in the corresponding symmetry subspace; in practice, hardware noise can populate unphysical sectors, producing biased estimates of observables. The central idea is to  {discard} outcomes that violate the certain symmetry, or equivalently to {project} measured observables onto the desired symmetry sector.

Symmetry-informed post-selection works best when the required symmetry checks
are simple to perform and when the main errors tend to push the state out of
the allowed physical sector~\cite{bonet2018low,sagastizabal2019experimental,
cai2021quantum,cai2023quantum}. In that case, outcomes that violate the known constraint are unlikely to
come from the ideal evolution and can be discarded. This strategy is widely used in digital simulations of fermionic systems, where
particle number and fermion parity provide natural checks\cite{sagastizabal2019experimental,google2020hartree}, in
stabilizer-based preparations of topological states, where stabilizer outcomes
identify whether the state remains in the target code space\cite{satzinger2021realizing,google2023non,iqbal2024non}. Because some measurement shots are rejected, fewer samples
remain for estimating observables, which can increase the uncertainty of the
final result.

\subsubsection{Constraining and purification of tomographic states}

A physically valid density matrix must have unit trace and be positive semidefinite,
but these conditions may not be exactly satisfied in actual measurements. In practice, simple linear-inversion tomography applied to finite-shot noisy data can yield estimates with nonphysical eigenvalues, motivating maximum-likelihood or projection-based reconstruction methods that enforce positivity and trace normalization~\cite{hadril1997quantum,smolin2012efficient}.

In particular, when the tomographed quantum state is expected to be pure (with density matrix eigenvalues $0$ or $1$ only) or otherwise low-rank, McWeeny purification can be used. It imposes a prior that the ideal reconstructed object is idempotent, $\rho^2=\rho$, and therefore close to a projector~\cite{mcweeny1960some}. Given a Hermitian, trace-normalized tomographic estimate, McWeeny updating  $\rho_{k+1}=3\rho_k^2-2\rho_k^3$, possibly followed by trace normalization or a physically constrained variant~\cite{truflandier2016communication,cai2023quantum}, gives rise to a sequence of density matrices $\rho_1,\rho_2,...$ whose eigenvalues converge quadratically to either $0$ or $1$ i.e. become progressively closer to a pure state. The form of this polynomial is chosen such that the update retains the same form upon exchanging $\rho_k$ with $1-\rho_k$. Because the update is a polynomial in $\rho_k$, iterating the process preserves the eigenvectors of $\rho_k$ and acts only on its eigenvalues; the scalar map drives well-conditioned eigenvalues toward the stable fixed points $0$ and $1$, with an unstable threshold at $\lambda=1/2$~\cite{truflandier2016communication}. This makes the method a spectral sharpening procedure: small eigenvalues associated with incoherent admixture are suppressed, while the dominant eigenspace is pushed toward a purified, low-rank estimate. The purified density matrix can then be used to compute desired quantities from the experiment~\cite{truflandier2016communication,mccaskey2019quantum,sun2024quantum}.

\section{Conclusion \& Outlook} 
\label{sec:outlook}

In this review, we have surveyed recent progress in the quantum simulation of
condensed-matter and many-body systems across physical platforms, target
phenomena, and algorithmic methodologies. We have reviewed representative physical targets of quantum simulation,
including correlated and topological matter, non-equilibrium dynamics,
measurement-induced phenomena, nonunitary evolution, and gauge-theory-inspired
systems. These examples highlight how quantum simulation has expanded from
static Hamiltonian emulation to a broader framework for probing dynamical,
topological, open-system, and strongly interacting quantum phenomena. The broader scope also sharpens the hardware challenge. Problems such as
long-time dynamics and critical behavior in large-scale systems cannot be reduced to
short-depth or small-scale circuits. They require larger circuit volume, longer
coherent evolution, lower two-qubit and readout errors, and more stable
calibration.

Near-term quantum utility remains limited by the noise
levels of present devices. For a quantum circuit with many noisy locations and a
representative physical error rate $p$, the probability of an error-free
execution decreases roughly as $\exp(-pN_{\rm loc})$, where $N_{\rm loc}$ is
the number of noisy operations. Thus, without error correction or highly
effective mitigation, error rates near $p\sim10^{-3}$ restrict the useful
circuit volume to the order of $1/p$, up to architecture- and
observable-dependent factors~\cite{preskill2018quantum,preskill2025beyond,
eisert2025mind}.

This limitation is directly relevant to current superconducting processors.
As of early 2026, IBM Quantum devices have reached the 100-qubit scale and beyond: Eagle-class
processors contain 127 qubits, while Heron-class processors contain 156
physical qubits~\cite{kim2023evidence,ibm_quantum_2026}. At the same time,
their physical gate errors are still non-negligible at the circuit volumes
needed for many-body simulation. Recent IBM Heron data report a median
two-qubit gate error rate of $1.17\times 10^{-3}$. These numbers are excellent
for present superconducting hardware, but they imply that circuits containing
thousands of entangling operations can accumulate substantial error. Thus, current processors are already large enough to
explore nontrivial many-body dynamics, but the results remain sensitive to accumulated
hardware errors.  Achieving scalable and robust quantum simulation will require continued
reductions in two-qubit and readout errors, improved 
mitigation strategies, and ultimately fault-tolerant logical qubits.

Moreover, while error mitigation can extend the useful regime of NISQ devices, it
is not a scalable substitute for fault tolerance. Generic mitigation protocols
often require rapidly growing, and in many cases exponential, sampling overheads
with circuit size, depth, or target precision
~\cite{cai2023quantum,takagi2022fundamental,takagi2023universal,
quek2024exponentially}. Complexity-theoretic results further suggest that, if
noisy quantum circuits are classically simulable under suitable noise
assumptions, then an efficient classical mitigation procedure for recovering the
ideal outputs would undermine the source of their presumed classical hardness
~\cite{schuster2025polynomial}. These results do not diminish the value of
near-term experiments, but they indicate that error mitigation alone is unlikely
to provide a scalable route to generic quantum advantage.

A natural route beyond these NISQ limitations is the emergence of early
fault-tolerant quantum processors, in which computations are performed on
partially protected logical qubits rather than on bare physical qubits. Recent
experiments on superconducting, trapped-ion, and neutral-atom platforms have
demonstrated increasingly mature components of this transition, including
logical memories, encoded logical operations, real-time syndrome extraction, and
architectures for scalable error correction
~\cite{google2025quantum,lacroix2025scaling,bluvstein2025fault,perlin2026fault,wang2026multi}.
These developments do not yet constitute fully fault-tolerant quantum
computation in the asymptotic sense, but they indicate a realistic intermediate
regime in which NISQ-style compilation, error mitigation, and hardware-aware
control are combined with partial error correction. This early fault-tolerant regime is likely to require close co-design between
hardware, codes, compilers, and target applications. Rather than minimizing only
asymptotic scaling, near-term logical simulations will also be constrained by
constant-factor overheads, syndrome-extraction latency, logical gate sets,
connectivity, and decoding speed. Several recent proposals address
this regime directly, including efficient partially fault-tolerant logical
rotations~\cite{akahoshi2024partially}, code constructions with low-cost logical
entangling operations~\cite{koh2026entangling,yang2026spacetime}, and
stack-level optimizations of syndrome extraction, decoding, compilation, and
control~\cite{tan26syndrome,guo2026toward,roberts2026cored}.

For quantum simulation, even modest logical protection would already be
transformative. Lower logical error rates would extend accessible real-time
evolution, improve measurements of nonlocal observables and entanglement
diagnostics, and reduce the dependence on costly post-processing. This would
particularly benefit problems that remain difficult to treat quantitatively on
current NISQ devices, including long-time transport and thermalization in
strongly correlated systems~\cite{campbell2021early,kan2025resource},
real-time lattice-gauge dynamics beyond short-time demonstrations
~\cite{banuls2020simulating,gonzalez2025observation,
Spagnoli2026faulttolerant}, robust preparation and manipulation of
topological order~\cite{kitaev2003fault,dennis2002topological,
iqbal2024topological}, deep Floquet dynamics over many driving periods~\cite{else2016floquet,khemani2016phase,else2017prethermal,
potter2016classification,zhang2022digital,hu2026boundary,hayata2024floquet}, and dynamics with repeated measurement, dissipation, and feedback
~\cite{barreiro2011open,google2023measurement,liu2025simulationopen,
iqbal2024topological}. Thus, early fault tolerance is not only a step toward
universal quantum computation, but also a practical route to quantum simulations
requiring larger circuit volume, longer coherent evolution, and higher-precision
observables.

\section{Acknowledgements} 
R.S. and C.H.L. acknowledge support from the Singapore Ministry of Education (MOE) Academic Research Fund Tier-I preparatory grant (WBS: A-8002656-00-00) and  MOE's Tier-II grant MOE-T2EP50224-0007 (WBS: A-8003505-00-00 and A-8003505-01-00). T.C. acknowledges support by the National Research Foundation, Singapore through the National Quantum Office, hosted in A*STAR, under the Advanced Quantum Algorithms and Solutions (AQAS) Funding Initiative (S25Q9DA001). T.T.~and J.M.K.~acknowledge support from the A*STAR Graduate Academy. PG Acknowledges support from U.S. National Science Foundation grant number DMR-2315063.

\bibliography{references_combined_v4}

\onecolumngrid
\flushbottom
\newpage
\appendix
\setcounter{equation}{0}
\setcounter{figure}{0}
\setcounter{table}{0}
\setcounter{section}{0}
\renewcommand{\theequation}{S\arabic{equation}}
\renewcommand{\thefigure}{S\arabic{figure}}
\renewcommand{\thesection}{S\arabic{section}}
\renewcommand{\thepage}{S\arabic{page}}
\renewcommand{\thetable}{S\arabic{table}}
%

\flushbottom

\end{document}